\ifx\pdfoutput\undefined
\newcommand\texorpdfstring[2]{#1}
\else
\fi
\documentclass[prd,twocolumn,showpacs,showkeys,preprintnumbers,floatfix,nofootinbib,superscriptaddress,10pt,aps]{revtex4-1}
\usepackage{amsfonts} 
\usepackage{amssymb} 
\usepackage{amsmath} 
\usepackage{graphicx} 
\usepackage{subfigure} 
\usepackage{array} 
\usepackage{dcolumn} 
\usepackage{bm} 
\let\bold=\bm 
\usepackage{latexsym} 
\usepackage{longtable} 
\usepackage{hyperref} 
\usepackage{bm}
\usepackage{slashed}
\usepackage{bbold}
\usepackage{color}
\usepackage{multirow}
\usepackage{rotating}
\usepackage{comment}

\graphicspath{{./figsX/}}

\usepackage[margin=1in]{geometry}
\usepackage [english]{babel}
\usepackage [autostyle, english = american]{csquotes}
\MakeOuterQuote{"}
\usepackage{graphicx}
\usepackage{subfigure}
\usepackage{xcolor}
\usepackage[toc,page]{appendix}
\usepackage{setspace}
\usepackage{amsmath, amssymb}
\usepackage{slashed}
\newcommand{\be}{\begin{eqnarray}}
\newcommand{\ee}{\end{eqnarray}}

\newcommand{\gf}{\gamma^5}
\newcommand{\ra}{\rangle}
\newcommand{\la}{\langle}
\newcommand{\ripmom}{${\rm RI}^\prime{\rm -MOM}$ }
\newcommand{\MSbar}{\overline{\rm MS} }

\def\ol{\overline}

\def\nn{\nonumber}
\def\dag{\dagger}

\makeatletter
\newcommand{\overleftrightsmallarrow}{\mathpalette{\overarrowsmall@\leftrightarrowfill@}}
\newcommand{\overrightsmallarrow}{\mathpalette{\overarrowsmall@\rightarrowfill@}}
\newcommand{\overleftsmallarrow}{\mathpalette{\overarrowsmall@\leftarrowfill@}}
\newcommand{\overarrowsmall@}[3]{%
  \vbox{%
    \ialign{%
      ##\crcr
      #1{\smaller@style{#2}}\crcr
      \noalign{\nointerlineskip}%
      $\m@th\hfil#2#3\hfil$\crcr
    }%
  }%
}
\def\smaller@style#1{%
  \ifx#1\displaystyle\scriptstyle\else
    \ifx#1\textstyle\scriptstyle\else
      \scriptscriptstyle
    \fi
  \fi
}
\makeatother
\newcommand{\olra}[1]{\overleftrightsmallarrow{#1}}

\providecommand{\abs}[1]{\lvert#1\rvert}
\providecommand{\matrixe}[3]{\langle#1\lvert#2\rvert#3\rangle}

\definecolor{green}{rgb}{0.1, 0.8, 0.1}

\newcolumntype{.}[1]{D{.}{.}{#1}}

\allowdisplaybreaks

\begin{document}


\title{Nucleon Momentum Fraction, Helicity and Transversity from \texorpdfstring{$2+1$}{2+1}-flavor Lattice QCD}
%
%


\author{Santanu Mondal}
\email{santanu@lanl.gov}
\affiliation{Los Alamos National Laboratory, Theoretical Division T-2, Los Alamos, NM 87545}

\author{Rajan Gupta}
\email{rajan@lanl.gov}
\affiliation{Los Alamos National Laboratory, Theoretical Division T-2, Los Alamos, NM 87545}

\author{Sungwoo Park}
\email{sungwoo@lanl.gov}
\affiliation{Los Alamos National Laboratory, Theoretical Division T-2, Los Alamos, NM 87545}

\author{Boram Yoon}
\email{boram@lanl.gov}
\affiliation{Los Alamos National Laboratory, Computer Computational and Statistical Sciences, CCS-7, Los Alamos, NM 87545}

\author{Tanmoy Bhattacharya}
\email{tanmoy@lanl.gov}
\affiliation{Los Alamos National Laboratory, Theoretical Division T-2, Los Alamos, NM 87545}

\author{B\'alint~Jo\'o}
\email{joob@ornl.gov}
\affiliation{Oak Ridge Leadership Computing Facility, Oak Ridge National Laboratory, Oak Ridge, TN 37831, USA}



\author{Frank~Winter}
\email{fwinter@jlab.org}
\affiliation{Jefferson Lab, 12000 Jefferson Avenue, Newport News, Virginia 23606, USA}

\collaboration{Nucleon Matrix Elements (NME) Collaboration}
\noaffiliation
\preprint{LA-UR-20-28586}
%
\pacs{11.15.Ha, 
      12.38.Gc  
}
\keywords{nucleon structure, momentum distribution, helicity and transversity moments, lattice QCD}
\date{\today}
\begin{abstract}
High statistics results for the isovector momentum fraction, $\la x
\ra_{u-d}$, helicity moment, $\la x \ra_{\Delta u-\Delta d}$, and the
transversity moment, $\la x\ra_{\delta u-\delta d}$, of the nucleon
are presented using seven ensembles of gauge configurations generated
by the JLab/W\&M/LANL/MIT collaborations using $2+1$-flavors of
dynamical Wilson-clover quarks. Attention is given to understanding
and controlling the contributions of excited states.  The final result
is obtained using a simultaneous fit in the lattice spacing $a$, pion
mass $M_\pi$ and the finite volume parameter $M_\pi L$ keeping leading
order corrections. The data show no significant dependence on the
lattice spacing and some evidence for finite-volume corrections. The main 
variation is with $M_\pi$, whose magnitude depends on the mass
gap of the first excited state used in the analysis. Our final
results, in the $\MSbar$ scheme at 2~GeV, are $\la x \ra_{u-d} =
0.160(16)(20)$, $\la x \ra_{\Delta u-\Delta d} = 0.192(13)(20)$ and
$\la x \ra_{\delta u-\delta d} = 0.215(17)(20)$, where the first error
is the overall analysis uncertainty assuming excited-state
contributions have been removed, and the second is an additional
systematic uncertainty due to possible residual excited-state
contributions. These results are consistent with other recent lattice
calculations and phenomenological global fit values.

\end{abstract}
\maketitle
%
%
%
%
\section{Introduction}
\label{sec:into}

Steady progress in both experiment and theory is providing an
increasingly detailed description of the hadron structure in terms of
quarks and gluons.  The distributions of quarks and gluons within
nucleons are being probed in experiments at the Relativistic Heavy
Ion Collider (RHIC) at BNL~\cite{Djawotho:2013pga,Adare:2014hsq},
Jefferson Lab~\cite{Dudek:2012vr} and the Large Hadron Collider at CERN. Experiments
at the planned electron-ion collider~\cite{Accardi:2012qut} will
significantly extend the range of Bjorken $x$ and $Q^2$ and further
improve our understanding.  From these data, and using higher order
calculations of electroweak and strong corrections, the
phenomenological analyses of experimental data (global fits) are 
providing parton distribution functions
(PDFs)~\cite{Brock:1993sz,Ji:2020ect}, transverse momentum dependent
PDFs (TMDs)~\cite{Yoon:2017qzo}, and generalized parton distributions
(GPDs)~\cite{Diehl:2003ny}.  These distributions are not measured
directly in experiments~\cite{Cichy:2018mum,Karthik:2019},
necessitating phenomenological analyses that have involved different
theoretical inputs.

Lattice QCD calculations are beginning to
provide such input, and a review of the cross-fertilization between
the two efforts has been presented in Refs.~\cite{Lin:2017snn,Lin:2020rut}. With
increasing computing power and advances in algorithms, the precision
of lattice QCD calculations has increased significantly and there now
exist many quantities for which there is good agreement with
experimental results, and for some, the lattice results are the most
precise as reviewed in the recent Flavor Averaging Group (FLAG) 2019
report~\cite{Aoki:2019cca}.  

In this work, we present high-statistics
lattice data for the isovector momentum fraction, and the helicity and
transversity moments, whose calculation is now reaching precision
comparable to that for nucleon charges, which are the zeroth moments
of the distributions and obtained from the matrix elements of local
quark bilinear operators~\cite{Gupta:2018qil,Aoki:2019cca}.

Calculations for the three first moments, $\la x \ra_{u-d}$, $\la
x \ra_{\Delta u-\Delta d}$ and $\la x \ra_{\delta u-\delta d}$, have
been done on seven ensembles generated using 2+1-flavors of
Wilson-clover quarks by the JLab/W\&M/LANL/MIT
collaborations~\cite{JLAB:2016}.  The data at three values of lattice
spacings $a$, two values of the pion mass, $M_\pi \approx 170$ and
$\approx 270$~MeV, and on a range of large volumes, characterized by
$M_\pi L$, allow us to carry out a simultaneous fit in these three
variables to address the associated systematic uncertainties.  In the
analysis of the two- and three-point correlation functions from each
ensemble, we carry out a detailed investigation of the dependence of
the results on the spectra of possible, $a\ priori$ unresolved,
excited states included in the fits to remove excited-state
contamination (ESC). Concretely, the full analysis is carried out
using three strategies to estimate the mass gap of the first excited
state from the two- and three-point correlation functions, and we use
the spread in the results to assign a second systematic uncertainty to
account for possible remaining contributions from excited-states.

Our final results, given in Eq.~\eqref{eq:finalresults}, are $\la
x \ra_{u-d} = 0.160(16)(20)$, $\la x \ra_{\Delta u-\Delta d} =
0.192(13)(20)$ and $\la x \ra_{\delta u-\delta d} = 0.215(17)(20)$ in
the $\MSbar$ scheme at 2~GeV. These estimates are in good agreement
with other lattice and phenomenological global fit results as
discussed in Sec.~\ref{sec:results}. The most extensive and precise
results from global fits are for the unpolarized moments of the
nucleons, the momentum fraction $\langle x \rangle_q$, while those for
the helicity fraction, the polarized moment $\langle x \rangle_{\Delta
q}$, have a large spread and our lattice results are consistent with
the smaller error global fit values at the lower end.  Lattice QCD
results for the transversity $\langle x\rangle_{\delta q}$ are a
prediction due to lack of sufficient experimental
data~\cite{Lin:2017snn,Lin:2020rut}.

The paper is organized as follows: In Sec.~\ref{sec:lattice}, we
briefly summarize the lattice parameters and methodology. The
definitions of moments and operators investigated are given in
Sec.~\ref{sec:moments}. The two- and three-point functions calculated,
and their connection to the moments, are specified in
Sec.~\ref{sec:correlators}. The analysis of excited state
contributions and the extraction of the ground state matrix elements is presented in
Sec.~\ref{sec:ESC}.  Results for the moments after the
chiral-continuum-finite-volume (CCFV) extrapolation are given in
Sec.~\ref{sec:results}, and compared with other lattice calculations
and global fit values.  We end with conclusions in
Sec.~\ref{sec:summary}. The data and fits used to remove excited-state
contamination are shown in Appendix~\ref{sec:ratios} and the
calculation of the renormalization factors, $Z_{VD,AD,TD}$, for the
three operators is discussed in Appendix~\ref{sec:renormalization}.

\section{Lattice Methodology}
\label{sec:lattice}

This work follows closely the methodology described in
Ref.~\cite{Mondal:2020cmt}, with two major differences. The first is
the calculation here uses 2+1-flavors of Wilson-clover fermions in a
clover-on-clover unitary formulation of lattice QCD, whereas the
clover-on-HISQ formulation was used in Ref.~\cite{Mondal:2020cmt}. 
The clover action includes one iteration of stout
smearing with weight $\rho = 0.125$ for the staples~\cite{Morningstar:2003gk}. The tadpole
corrected tree-level
Sheikholeslami-Wohlert coefficient 
$c_{SW} = 1/u_0$~\cite{Sheikholeslami:1985ij}, where $u_0$ is the fourth root of the plaquette
expectation value, is very close to the nonperturbative value
determined, a posteriori, using the Schr\"odinger functional
method~\cite{Luscher:1996ug}, a consequence of the stout smearing. The update of
configurations was carried out using the rational hybrid Monte Carlo
(RHMC) algorithm~\cite{Duane:1987de} as described in
Ref.~\cite{Yoon:2016jzj}.  

The parameters of the seven clover ensembles generated by the
JLab/W\&M/LANL/MIT collaborations~\cite{JLAB:2016} and used in the 
analysis are summarized in Table~\ref{tab:ensembles}. The range of
lattice spacings covered is $0.071 \leq a \leq 0.127$ fm and of
lattice size is $3.7 \leq M_\pi L \leq 6.2$.  So far simulations have
been carried out at two pion masses, $M_\pi \approx 270$ and 170~MeV.
These seven data points allow us to perform
chiral-continuum-finite-volume fits to obtain physical results.

The second improvement is higher statistics data that lead to a more 
robust analysis of three strategies for evaluating ESC.
Table~\ref{tab:ensembles} also gives the
number of configurations, the source-sink separations $\tau$, high
precision (HP) and low precision (LP) measurements made to
cost-effectively increase statistics using the bias-corrected 
truncated-solver 
method~\cite{Bali:2009hu,Blum:2012uh}.

The parameters used to construct the Gaussian smeared
sources~\cite{Gusken:1988yi,Yoon:2016dij,Gupta:2018qil,Mondal:2020cmt},
are given in Table~\ref{tab:cloverparams}. To construct the smeared source, 
the gauge links were first smoothened using twenty hits of the stout
algorithm with $\rho=0.08$ and including only the spatial staples. The
root-mean-square size of the Gaussian smearing, $\sqrt{\int dr \, r^4
S^\dag S /\int dr \, r^2 S^\dag S} $ with $S(r)$ the value of the
smeared source at radial distance $r$, was adjusted to be between 
0.72---0.76~fm to reduce ESC.  The quark propagators from these smeared sources
were generated by inverting the Dirac operator (same as what was used
to generate the lattices) using the multigrid
algorithm~\cite{Babich:2010qb,Clark:2009wm,10.5555/3014904.3014995}.
These propagators are then used to construct the two- and three-point
correlation functions.

\begin{table*}[tbhp]  
\centering
\renewcommand{\arraystretch}{1.2}
\begin{tabular}{|c|c|c|c|c|c|c|c|c| }
\hline
      &    &      &           &                        &            &    &   &   \\
Ensemble&$a$&$M_\pi$& $L^3\times T$&$M_\pi L$&$\tau/a$        
&$N_{conf}$&$N_{HP}$ &$N_{LP}$\\
ID &  (fm)&(MeV)&&&&&&\\
\hline
\hline
$a127m285$ &$0.127(2)$&$285(3)$&$32^3\times96$ &$5.85$&$\{8,10,12,14\}$     &$2001$    &$8,004$  &$256,128$   \\
\hline
$a094m270$ &$0.094(1)$&$270(3)$&$32^3\times64$ &$4.11$&$\{10,12,14,16\}$    &$1464$    &$4,392$  &$140,544$    \\
$a094m270L$&$0.094(1)$&$269(3)$&$48^3\times128$&$6.16$&$\{8,10,12,14,16,18\}$ &$4501$  &$18,004$ &$576,128$    \\
\hline
$a091m170$&$0.091(1)$&$169(2)$&$48^3\times96$  &$3.75$&$\{8,10,12,14,16\}$  &$4015$    &$16,060$ &$513,920$   \\
$a091m170L$&$0.091(1)$&$169(2)$&$64^3\times128$&$5.08$&$\{8,10,12,14,16\}$  &$1533$    &$7,665$  &$245,280$   \\
\hline
$a073m270$&$0.0728(8)$&$272(3)$&$48^3\times128$&$4.82$&$\{11,13,15,17,19\}$ &$4477$    &$17,908$ &$573,056$   \\
\hline
$a071m170$&$0.0707(8)$&$167(2)$&$72^3\times192$&$4.26$&$\{13,15,17,19,21\}$ &$1500$    &$9,000$  &$144,000$    \\
\hline
\end{tabular}
\caption{Lattice parameters of the 2+1-flavor clover ensembles generated by the 
JLab/W\&M/LANL/MIT collaboration and analyzed in this study. We specify 
the lattice spacing $a$, pion mass $M_\pi$, lattice
size $L^3\times T$, the values of source-sink separation
$\tau$ simulated, the number of configurations
analyzed, and the total number of high precision (HP) and low
precision (LP) measurements made. }
\label{tab:ensembles}
\end{table*}

\begin{table}[htbp]      
\centering
\begin{ruledtabular}
\setlength{\tabcolsep}{1.5pt}
\begin{tabular}{l|lc|c|c}
\multicolumn{1}{c|}{ID}  & \multicolumn1c{$m_l$} &  $c_{\text{SW}}$ & Smearing    & RMS  \\
                    &                       &                  & Parameters  & smearing       \\
                    &                       &                  & $\{\sigma, N_{\text{KG}}\}$    & radius       \\
\hline
$a127m285$          & $-0.2850$  & 1.24931 & \{5, 50\}    & 5.79(1)  \\
\hline
$a094m270$          & $-0.2390$  & 1.20537 & \{7, 91\}    & 7.72(3)  \\
\hline
$a094m270L$         & $-0.2390$  & 1.20537 & \{7, 91\}    & 7.76(4)  \\
\hline
$a091m170$          & $-0.2416$  & 1.20537 & \{7, 91\}    & 7.64(3)  \\
\hline
$a091m170L$         & $-0.2416$  & 1.20537 & \{7, 91\}    & 7.76(4)  \\
\hline
$a073m270$          & $-0.2070$  & 1.17008 & \{9, 150\}   & 9.84(1)  \\
\hline
$a071m170$          & $-0.2091$  & 1.17008 & \{10, 185\}  & 10.71(2)  \\
\hline
\end{tabular}
\end{ruledtabular}
\caption{The parameters used in the calculation of the clover
  propagators.  The hopping parameter for the light quarks,
  $\kappa_l$, in the clover action is given by $2\kappa_{l} =
  1/(m_{l}+4)$.  $c_{\rm SW} $ is the Sheikholeslami-Wohlert
  improvement coefficient in the clover action. The parameters used to
  construct Gaussian smeared sources~\protect\cite{Gusken:1989ad},
  $\{\sigma, N_{\text{KG}}\}$ in the Chroma
  convention~\cite{Edwards:2004sx}, are given in the fourth column
  where $N_{\text{KG}}$ is the number of applications of the
  Klein-Gordon operator and $\sigma$ controls the width of the smearing.  
  The resulting root-mean-square radius of
  the smearing in lattice units, defined as $\sqrt{\int dr \, r^4
  S^\dag S /\int dr \, r^2 S^\dag S} $ with $S(r)$ the value of the
  smeared source at radial distance $r$, is given in the last column.
  }
\label{tab:cloverparams}
\end{table}


\section{Moments and Matrix elements}
\label{sec:moments}

The first moments of spin independent (or
unpolarized), $q=q_\uparrow+q_\downarrow$, helicity (or polarized),
$\Delta q=q_\uparrow-q_\downarrow$, and transversity, 
$\delta q =q_\top+q_{\perp}$ distributions, are defined as
\be
\langle x \rangle_q &=& \int_0^1~x~[q(x)+\ol{q}(x)]~dx \,, \\
\langle x \rangle_{\Delta q} &=& \int_0^1~x~[\Delta q(x)+\Delta \ol{q}(x)]~dx \,, \\
\langle x \rangle_{\delta q} &=& \int_0^1~x~[\delta q(x)+\delta \ol{q}(x)]~dx \,,
\ee
where $q_{\uparrow(\downarrow)}$ corresponds to quarks with helicity
aligned (anti-aligned) with that of a longitudinally polarized target,
and $q_{\top(\perp)}$ corresponds to quarks with spin aligned
(anti-aligned) with that of a transversely polarized target.

These moments, at leading twist, are extracted from the forward matrix
elements of one-derivative vector, axial-vector and tensor operators
within ground state nucleons at rest.  The complete
set of the relevant twist two operators are
\be
{\cal O}^{\mu \nu}_{V^a}&=&\ol{q} \gamma^{\{\mu}\olra{D}^{\nu\}} \tau^a q\nn \,, \\
{\cal O}^{\mu \nu}_{A^a}&=&\ol{q}\gamma^{\{\mu}  \olra{D}^{\nu\}} \gf \tau^a q\nn \,, \\
{\cal O}^{\mu \nu \rho}_{T^a}&=&\ol{q} \sigma^{[\mu\{\nu]} \olra{D}^{\rho\}} \tau^a q \,, 
\label{operators}
\ee 
where $q=\{u,d\}$ is the isodoublet of light quarks and  
$\sigma^{\mu\nu} = (\gamma^\mu\gamma^\nu - \gamma^\nu\gamma^\mu)/2$. 
The derivative $\olra{D}^{\nu}\equiv\frac{1}{2}(\overrightarrow{D}^\nu-\overleftarrow{D}^\nu)$ 
consists of four terms defined in Ref.~\cite{Mondal:2020cmt}. 
Lorentz indices within $\{ ~\}$ in Eq.~\eqref{operators} are
symmetrized and within $[\, ]$ are antisymmetrized. It is also
implicit that, where relevant, the traceless part of the above
operators is taken.  Their renormalization is carried out
nonperturbatively in the regularization independent RI${}^\prime$-MOM
scheme as discussed in Appendix~\ref{sec:renormalization}. A more
detailed discussion of these twist-2 operators and their
renormalization can be found in Refs.~\cite{Gockeler:1995wg}
and~\cite{Harris:2019bih}.

In our setup to calculate the isovector moments, we work with $\tau^a
= \tau^3$ and fix the spin of the nucleon state to be in the
``3'' direction. With these choices, the explicit operators calculated
are
\be
{\cal O}^{44}_{V^3} &=&  \ol{q} (\gamma^{4}\olra{D}^{4}  -\frac{1}{3}
{\bm \gamma} \cdot \olra{\bf D}) \tau^3 q \,,
\label{eq:finaloperatorV} \\
{\cal O}^{34}_{A^3} &=&\ol{q} \gamma^{\{3}\olra{D}^{4\}} \gf \tau^3 q \,,
\label{eq:finaloperatorA} \\
{\cal O}^{124}_{T^3} &=& \ol{q} \sigma^{[1\{2]}\olra{D}^{4\}} \tau^3 q \,. 
\label{eq:finaloperatorT}
\ee
The forward matrix elements ($ME$) of these operators within the
ground state of the nucleon with mass $M_N$ are related to the moments
as follows:
\be
\la 0 | {\cal O}^{44}_{V^3}| 0 \ra &=&  -  M_N\, \la x \ra_{u-d} \,, 
\label{eq:me2momentV} \\
\la 0 | {\cal O}^{34}_{A^3}| 0 \ra &=&  - \frac{i  M_N}{2} \, \la x \ra_{\Delta u-\Delta d} \,, 
\label{eq:me2momentA} \\
\la 0 | {\cal O}^{124}_{T^3}| 0 \ra &=& - \frac{i M_N}{2} \, \la x \ra_{\delta u-\delta d} \,. 
\label{eq:me2momentT}
\ee 
The moments are, by construction, dimensionless. 

\section{Correlation functions and Moments}
\label{sec:correlators}

To construct the two- and three-point correlation functions needed to calculate 
the matrix elements, the interpolating operator ${\mathcal N}$ used to
create/annihilate the nucleon state is 
\be 
{\mathcal N} = \epsilon^{abc} \Big[  q_1^{aT} (x) C \gf \frac{(1\pm \gamma_4)}{2}q_2^b(x) \Big] q^c_1(x) \,, 
\label{nucop}
\ee 
where $\{a,b,c\}$ are color indices,
$q_1,q_2 \in \{u,d\}$ and $C=\gamma_4 \gamma_2$ is the charge conjugation
matrix in our convention.  The nonrelativistic projection $(1\pm \gamma_4)/2 $
is inserted to improve the signal, with the plus and minus signs
applied to the forward and backward propagation in Euclidean time,
respectively~\cite{Gockeler:1995wg}. At zero momentum, this operator
couples only to the spin-$\frac12$ states.  The zero momentum
two-point and three-point nucleon correlation functions are defined as 
\begin{flalign}
\bold{C}^{\rm 2pt}_{\alpha \beta} (\tau ) &= \sum_{\bm x}\la 0 | {\mathcal N}_\alpha (\tau, {\bm x}) \ol{{\mathcal N}}_\beta(0,{\bm 0})| 0\ra\\ 
\bold{C}^{\rm 3pt}_{\mathcal{O},\alpha \beta}
(\tau,t ) &= \sum_{{\bm x}',{\bm x}} \la 0 | {\mathcal N}_\alpha (\tau, {\bm x}) {\cal O} (t, {\bm x}') \ol{{\mathcal N}}_\beta(0,{\bm 0})| 0\ra 
\end{flalign}
where $\alpha$, $\beta$ are spin indices. The source is placed at time
slice 0, the sink is at $\tau$ and the one-derivative operators,
defined in Sec.~\ref{sec:moments}, are inserted at time slice $t$.
Data have been accumulated for the values of $\tau$ specified in
Table~\ref{tab:ensembles}, and for each $\tau$ for all intermediate
times $0  < t < \tau$.

To isolate the various contributions, projected $2$- and $3$-point
functions are constructed as
\be
C^{\rm 2pt}&=& {\rm Tr} \big( {\cal P}_{\rm 2pt} \bold{C}^{\rm 2pt} \big)\\
C_\mathcal{O}^{\rm 3pt}&=& {\rm Tr} \big( {\cal P}_{\rm 3pt} \bold{C}^{\rm 3pt}_{\mathcal{O}} \big) \,.
\ee
The projector ${\cal P}_{\rm 2pt} = \frac{1}{2}\, (1 + \gamma_4)$ in the nucleon 
correlator gives the positive parity contribution for the nucleon propagating
in the forward direction.
For the connected $3$-point contributions ${\cal P}_{\rm 3pt}= \frac{1}{2}(1 + 
\gamma_4)(1+i \gf \gamma^3)$ is used. With these spin projections, 
the three moments are obtained using
Eqs.~\ref{eq:me2momentV},~\ref{eq:me2momentA} and~\ref{eq:me2momentT}.

To display the data, we construct the ratios
\begin{equation}
R_\mathcal{O}(\tau;t) = C_\mathcal{O}^{3\text{pt}}(\tau;t)/C^{2\text{pt}}(\tau)
\label{eq:Ratio}
\end{equation}
that give the ground state matrix element in the limits $t \to \infty$ and
$(\tau - t) \to \infty$. These ratios are shown in
Figs.~\ref{fig:Ratio-mom-1}--\ref{fig:Ratio-transversity-2} in Appendix~\ref{sec:ratios}. 
We re-emphasize that the ground state matrix element 
$\matrixe{0}{\mathcal{O}}{0}$ used in the analysis is obtained from fits to
$C_\mathcal{O}^{3\text{pt}}(\tau;t)$ with input of spectral quantities from
$C^{2\text{pt}}(\tau)$. These fits are carried out within a
single-elimination jackknife process, which is used to get both the
central values and the errors.

\begin{table*}[tbhp]   
\centering
\renewcommand{\arraystretch}{1.1}
\begin{tabular}{|c|c|c|c|c|c|c|c|c|c| }
\hline
      &     &  &                     &                        &            &    &   &   \\
Ensemble    &$aM_N^{\{4\}}$ &$a M_N^{\{4^{N\pi}\}}$  
& $a \Delta M_1^{\{2\}}$ & $a \Delta M_1^{\{4\}}$ & $a \Delta M_1^{\{4^{N\pi}\}}$ 
& $a \Delta M_1^{\{2^{\rm free}\}}$
& $a \Delta M_1^{\{2^{\rm free}\}}$ 
& $a \Delta M_1^{\{2^{\rm free}\}}$ \\
ID          &                        &                        &            &     & 
& $\langle x \rangle_{u -d}$ 
&$\langle x \rangle_{\Delta u- \Delta d}$ 
&$\langle x \rangle_{\delta u- \delta d}$ \\
\hline
\hline
$a127m285$  & $0.6181(19)$ & $0.6167(14)$ & $0.413(46)$  & $0.376(52)$ &  $0.326(20)$  &  $0.359(35)$ &  $0.706(58)$  &  $0.64(11)$   \\ 
\hline                                                                                                                         
$a094m270$  & $0.4709(35)$ & $0.4706(25)$ & $0.349(91)$  & $0.273(63)$ &  $0.2643(95)$ &  $0.521(64)$ &  $0.647(66)$  &  $0.510(66)$  \\ 
$a094m270L$ & $0.4668(12)$ & $0.4656(9) $ & $0.357(19)$  & $0.303(42)$ &  $0.249(27)$  &  $0.344(26)$ &  $0.400(53)$  &  $0.466(29)$  \\ 
\hline                                                                                                                         
$a091m170$  & $0.4163(23)$ & $0.4119(19)$ & $0.346(22)$  & $0.293(45)$ &  $0.195(19)$  &  $0.311(48)$ &  $0.441(64)$  &  $0.424(61)$  \\ 
$a091m170L$ & $0.4143(25)$ & $0.4093(28)$ & $0.307(25)$  & $0.252(36)$ &  $0.157(11)$  &  $0.297(43)$ &  $0.388(52)$  &  $0.34(11)$   \\ 
\hline                                                                                                                         
$a073m270$  & $0.3719(11)$ & $0.3716(8)$  & $0.321(18)$  & $0.229(41)$ &  $0.217(24)$  & $0.311(16)$ &  $0.457(14)$  &  $0.431(16)$  \\ 
$a071m170$  & $0.3300(17)$ & $0.3266(17)$ & $0.301(28)$  & $0.249(34)$ &  $0.155(12)$ &  $0.404(28)$ &  $0.561(21)$  & $0.491(28)$  \\ 
\hline
\end{tabular}
\caption{Results for the nucleon mass $aM_N^{\{4\}}$ and $a M_N^{\{4^{N\pi}\}}$ obtained from the two 4-state fits 
to the two-point functions. The next six columns give the values of the
mass gap, $a \Delta M_1 \equiv a(M_1-M_0)$, of the first excited state
obtained from different fits studied in this work. The notation used
is $\{2\}$ ($\{4\}$) is a two-state (4-state) fit to the two-point
functions, $\{4^{N\pi}\}$ is a 4-state fit to the two-point functions
with a prior with a narrow width for $a\Delta M_1$ corresponding to the
non-interacting $N \pi$ state. In the three $\{2^{\rm free}\}$ cases, 
the $a\Delta M_1$ are determined from fits to the three-point functions used to
extract the three moments as explained in the text. }
\label{tab:massgap}
\end{table*}

\begin{table*}[htbp]   
\centering
\renewcommand{\arraystretch}{1.2}
\begin{tabular}{ |c|c|c|c|c|c|c|c|c|c| }
 \hline
\multicolumn{1}{|c|}{}&\multicolumn{3}{c|}{$\{4^{N\pi},3^*\}$} &\multicolumn{3}{c|}{$\{4,3^*\}$} &\multicolumn{3}{c|}{$\{4,2^{\rm free}\}$} \\
\hline
moment&$\tau$&${ t_{skip}}$&$\langle x \rangle$&$\tau$&${ t_{skip}}$&$\langle x \rangle$&$\tau$&${ t_{skip}}$&$\langle x \rangle$\\
\hline
\multicolumn{10}{|c|}{$a127m285$} \\
\hline
$\la x\ra_{u-d}$&$\{10,12,14\}$&$2$&$0.181(4)$&$\{10,12,14\}$&$1$&$0.190(6)$&$\{10,12,14\}$&$2$&$0.186(7)$\\
$\la x \ra_{\Delta u-\Delta
d}$&$\{10,12,14\}$&$3$&$0.234(5)$&$\{10,12,14\}$&$2$&$0.233(3)$&$\{10,12,14\}$&$2$&$0.243(2)$\\
$\la x \ra_{\delta u-\delta
d}$&$\{10,12,14\}$&$3$&$0.238(7)$&$\{10,12,14\}$&$2$&$0.234(5)$&$\{10,12,14\}$&$3$&$0.249(5)$\\
\hline
\multicolumn{10}{|c|}{$a094m270$} \\
\hline
$\la x \ra_{u-d}$&$\{12,14,16\}$&$2$&$0.190(7)$&$\{12,14,16\}$&$2$&$0.190(8)$&$\{12,14,16\}$&$2$&$0.203(5)$\\
$\la x \ra_{\Delta u-\Delta
d}$&$\{12,14,16\}$&$2$&$0.222(9)$&$\{12,14,16\}$&$2$&$0.222(9)$&$\{12,14,16\}$&$2$&$0.240(5)$\\
$\la x \ra_{\delta u-\delta
d}$&$\{12,14,16\}$&$2$&$0.222(10)$&$\{12,14,16\}$&$2$&$0.223(11)$&$\{12,14,16\}$&$2$&$0.243(6)$\\
\hline
\multicolumn{10}{|c|}{$a094m270L$} \\
\hline
$\la x\ra_{u-d}$&$\{14,16,18\}$&$3$&$0.173(4)$&$\{14,16,18\}$&$3$&$0.179(5)$&$\{14,16,18\}$&$3$&$0.182(3)$\\
$\la x\ra_{\Delta u-\Delta
d}$&$\{14,16,18\}$&$3$&$0.208(3)$&$\{14,16,18\}$&$3$&$0.211(4)$ &$\{14,16,18\}$&$4$&$0.217(4)$\\
$\la x\ra_{\delta u-\delta
d}$&$\{14,16,18\}$&$3$&$0.217(4)$&$\{14,16,18\}$&$3$&$0.219(4)$&$\{14,16,18\}$&$3$&$0.228(2)$\\
\hline
\multicolumn{10}{|c|}{$a091m170$} \\
\hline
$\la x \ra_{u-d}$&$\{12,14,16\}$&$3$&$0.148(14)$&$\{12,14,16\}$&$3$&$0.167(9)$&$\{12,14,16\}$&$3$&$0.169(12)$\\
$\la x\ra_{\Delta u-\Delta
d}$&$\{12,14,16\}$&$3$&$0.195(13)$&$\{12,14,16\}$&$3$&$0.204(7)$ &$\{12,14,16\}$&$3$&$0.216(7)$\\
$\la x\ra_{\delta u-\delta
d}$&$\{12,14,16\}$&$3$&$0.183(18)$&$\{12,14,16\}$&$3$&$0.200(11)$&$\{12,14,16\}$&$3$&$0.217(9)$\\
\hline
\multicolumn{10}{|c|}{$a091m170L$} \\
\hline
$\la x\ra_{u-d}$&$\{12,14,16\}$&$2$&$0.146(18)$&$\{12,14,16\}$&$3$&$0.156(11)$&$\{12,14,16\}$&$3$&$0.167(12)$\\
$\la x\ra_{\Delta u-\Delta
d}$&$\{12,14,16\}$&$3$&$0.178(21)$&$\{12,14,16\}$&$3$&$0.191(9)$&$\{12,14,16\}$&$3$&$0.209(9)$\\
$\la x\ra_{\delta u-\delta
d}$&$\{12,14,16\}$&$3$&$0.221(35)$&$\{12,14,16\}$&$3$&$0.206(10)$&$\{12,14,16\}$&$4$&$0.211(23)$\\
\hline
\multicolumn{10}{|c|}{$a073m270$} \\
\hline
$\la x\ra_{u-d}$&$\{15,17,19\}$&$3$&$0.168(4)$&$\{15,17,19\}$&$3$&$0.170(6)$&$\{15,17,19\}$&$3$&$0.180(3)$\\
$\la x\ra_{\Delta u-\Delta
d}$&$\{15,17,19\}$&$3$&$0.210(4)$&$\{15,17,19\}$&$3$&$0.210(3)$&$\{15,17,19\}$&$3$&$0.222(2)$\\
$\la x\ra_{\delta u-\delta
d}$&$\{15,17,19\}$&$3$&$0.211(5)$&$\{15,17,19\}$&$3$&$0.211(4)$&$\{15,17,19\}$&$3$&$0.227(2)$\\
\hline
\multicolumn{10}{|c|}{$a071m170$} \\
\hline
$\la x \ra_{u-d}$&$\{15,17,19\}$&$2$&$0.150(9)$&$\{15,17,19\}$&$2$&$0.167(6)$&$\{15,17,19\}$&$2$&$0.187(4)$\\
$\la x \ra_{\Delta u-\Delta
d}$&$\{15,17,19\}$&$2$&$0.211(13)$&$\{15,17,19\}$&$2$&$0.201(5)$&$\{15,17,19\}$&$2$&$0.221(3)$\\
$\la x \ra_{\delta u-\delta
d}$&$\{15,17,19\}$&$2$&$0.195(15)$&$\{15,17,19\}$&$2$&$0.198(7)$&$\{15,17,19\}$&$2$&$0.224(4)$\\
\hline
\hline
\end{tabular}
\caption{
Estimates of the three unrenormalized moments from the three fit
  strategies, $\{4^{N\pi},3^\ast \}$, $\{4,3^\ast \}$ and $\{4,2^{\rm
  free} \}$, used to analyze the two- and three-point functions and
  remove ESC.  For each fit strategy we give the $\tau$ values and the
  number of time slices $t_{skip}$ omitted next to the source and the
  sink in the final fits to the three-point functions. }
\label{tab:ESC-fits}
\end{table*}

\section{Controlling excited state contamination}
\label{sec:ESC}

A major challenge to precision results is removing the contribution of
excited states in the three-point functions. These occur because the
lattice nucleon interpolating operator ${\mathcal N}$, defined in
Eq.~\eqref{nucop}, couples to the nucleon, all its excitations and
multiparticle states with the same quantum numbers. Previous lattice
calculations have shown that these ESC can be
large~\cite{Mondal:2020cmt,Bhattacharya:2013ehc,Bali:2014gha,
Bali:2012av}. The strategy to remove these artifacts in this work is
the same as described in Ref.~\cite{Mondal:2020cmt}: reduce ESC by
using smeared sources in the generation of quark propagators and then
fit the data at multiple source-sink separations $\tau$ using the
spectral decomposition of the correlation functions
(Eqs.~\eqref{eq:2pt} and ~\eqref{eq:3pt}) keeping as many excited
states as possible without overparameterizing the fits.  In this work,
we examine three strategies, $\{4,3^*\}$, $\{4^{N\pi},3^*\}$ and
$\{4,2^{\rm free}\}$, that use different estimates of the excited
state masses in the fits as described below.

The spectral decomposition of the zero-momentum two-point function,
${ C_{\rm 2pt}}$, truncated at four states, is given by
\begin{equation}
C_{\rm 2pt}(\tau) = \sum_{i=0}^3  |{\cal A}_i|^2e^{-M_i \tau} \,. 
\label{eq:2pt}
\end{equation}
We fit the data over the largest time range,
$\{\tau_{min}-\tau_{max}\}$, allowed by statistics, i.e., by the
stability of the covariance matrix, to extract $M_i$ and ${\cal A}_i$,
the masses and the amplitudes for the creation/annihilation of the
four states by the interpolating operator ${\mathcal N}$.  We perform
two types of 4-state fits.  In the fit denoted $\{4\}$, we use the
empirical Bayesian technique described in the Ref.~\cite{Yoon:2016jzj}
to stabilize the three excited-state parameters. In the second fit,
denoted $\{4^{N\pi}\}$, we use a normally distributed prior for $M_1$, with value 
given by 
the lower of the non-interacting energy of $N({-\bm 1}) \pi({\bm 1})$
or the $N({\bm 0})\pi({\bm 0}) \pi({\bm 0})$ state\footnote{When
priors are used, the augmented $\chi^2$ is defined as the standard
correlated $\chi^2$ plus the square of the deviation of the parameter
from the prior mean normalized by the prior width. This quantity is
minimized in the fits. In the following we quote this augmented
$\chi^2$ divided by the degrees of freedom calculated without
reference to the prior, and call it $\chi^2$/dof for brevity.}. The
masses of these two states are roughly equal for the seven ensembles
and lower than the $M_1$ obtained from the $\{4\}$ fit. The lower
energy $N({-\bm 1}) \pi({\bm 1})$ state has been shown to contribute
in the axial channel~\cite{Jang:2019vkm}, whereas for the vector
channel the $N({\bm 0}) \pi({\bm 0}) \pi({\bm 0})$ state is expected
to be the relevant one. Since the two states have roughly the same
mass, which is all that matters in the fits, we do not distinguish
between them and use the common label $\{4^{N\pi}\}$. We also
emphasize that even though we use a Bayesian procedure for stabilizing
the fits, the errors are calculated using the jackknife method and are
thus the usual frequentist standard errors.

In the fits to the two-point functions, the $\{4\}$ and $\{4^{N\pi}\}$
strategies cannot be distinguished on the basis of the $\chi^2/$dof.
In fact, the full range of $M_1$ values between the two estimates,
from $\{4\}$ and $\{4^{N\pi}\}$, are viable on the basis of
$\chi^2/$dof alone. The same is true of the values for $M_2$,
indicating a large flat region in parameter space.  Because of this
large region of possible values for the excited-state masses, $M_i$,
we carry out the full analysis with three strategies that use
different estimates of $M_i$ and investigate the sensitivity of the
results on them.  The ground-state nucleon mass obtained from the two
fits is denoted by the common symbol $M_N \equiv M_0$ and the
successive mass gaps by $\Delta M_i \equiv M_i - M_{i-1}$. These are
given in Table~\ref{tab:massgap}.

The analysis of the three-point functions,
$C_\mathcal{O}^{3\text{pt}}$, with insertion of the operators with
zero momentum defined in
Eqs.~\eqref{eq:finaloperatorV},~\eqref{eq:finaloperatorA}
and~\eqref{eq:finaloperatorT}, is performed retaining up to three
states $|i\rangle$ in the spectral decomposition:
\begin{equation}
  C_\mathcal{O}^{3\text{pt}}(\tau;t) = 
   \sum_{i,j=0}^2 \abs{\mathcal{A}_i} \abs{\mathcal{A}_j}\matrixe{i}{\mathcal{O}}{j} e^{-M_i t - M_j(\tau-t)}\,.
   \label{eq:3pt}
\end{equation}
To get the forward matrix element, we also fix the momentum at the
sink to zero. To remove the ESC and extract the desired ground-state
matrix element, $\matrixe{0}{\mathcal{O}}{0}$, we make a simultaneous
fit in $t$ and $\tau$. The full set of values of $\tau$ investigated
are given in Table \ref{tab:ensembles}. In choosing the set of points,
$\{t,\tau\}$, to include in the final fit, we attempt to balance
statistical and systematic errors. First, we neglect $t_{\rm skip}$
points next to the source and sink in the fits as these have the
largest ESC. Next, noting that the data at smaller $\tau$ have exponentially smaller
errors but larger ESC, we pick the largest three values of $\tau$ that
have statistically precise data.  Since errors in the data grow with
$\tau$, we partially compensate for the larger weight given to smaller
$\tau$ data by choosing $t_{\rm skip}$ to be the same for all $\tau$,
i.e., by including increasingly more $t$ points with larger $\tau$,
the weight of the larger $\tau$ data points is increased.  Most of our
analysis uses a $3^\ast$-fit, which is a three-state fit with the term
involving $\langle 2 | {\mathcal O} | 2\rangle$ set to zero, as it is
undetermined and its inclusion results in an overparameterization
based on the Akaike information criteria~\cite{1100705}.

The key challenge to $3$-state fits using Eq.~\eqref{eq:3pt} is
determining the relevant $M_i$ to use because fits to the two-point
function show a large flat region in the space of the $M_i$ with
roughly the same $\chi^2/$dof. Theoretically, there are many
candidate intermediate states, and their contribution to the
three-point functions with the insertion of operator ${\mathcal O}$ is not known 
{\it a priori}. To investigate the sensitivity of
$\matrixe{0}{\mathcal{O}}{0}$ to possible values of $M_1$ and $M_2$,
we carry out the full analysis with the following three strategies:
\begin{itemize}
\item
$\{4,3^*\}$: The spectrum is taken from a $\{4\}$ state
fit to the two-point function using Eq.~\eqref{eq:2pt} and then a
$\{3^\ast\}$ fit is made to the three-point function using
Eq.~\eqref{eq:3pt}. Both fits are made within a single jackknife loop.
This is the standard strategy, which assumes that the same set of
states are dominant in the two- and three-point functions.
\item
$\{4^{N\pi},3^*\}$: The excited state spectrum is taken from a 4-state
fit to the two-point function but with a narrow prior for the first
excited state mass taken to be the energy of a non-interacting $N
({\bm p}=1) \pi({\bm p}=-1)$ state (or $N (0) \pi(0) \pi(0)$ that has
roughly the same energy).  This spectrum is then used in a
$\{3^\ast\}$ fit to the three-point function. This variant of the
$\{4,3^*\}$ strategy assumes that the lowest of the theoretically
allowed tower of $N \pi$ (or $N \pi \pi $) states contribute.
\item
$\{4,2^{\rm free}\}$: The only parameters taken from the $\{4\}$ state
fit are the ground state amplitude ${\cal A}_0$ and mass $M_0$. In the
two-state fit to the three-point function, the mass of the first
excited state, $M_1$, is left as a free parameter, ie, the most
important determinant of ESC, $M_1$, is obtained from the fit to the
three-point function.
\end{itemize}
The mnemonic $\{m,n\}$ denotes an $m$-state fit to the
two-point function and an $n$-state fit to the three-point function.

The data for the ratios $R_\mathcal{O}(\tau;t)$ are plotted in 
Figs.~\ref{fig:Ratio-mom-1}--\ref{fig:Ratio-transversity-2} in 
Appendix~\ref{sec:ratios} for the three operators, the three
strategies, and all seven ensembles. We note the following features:
\begin{itemize}
\item
The fractional statistical errors are less than $2\%$ for all three
operators and on all seven ensembles.  The only exceptions are the
$\tau =16$ ($\tau = 21$) data on the $a094m270$ and $a091m170L$  ($a071m170$)
ensemble.
\item
The errors grow, on average, by a factor between 1.3--1.5 for every 
two units increase in $\tau/a$. This is smaller than the 
asymptotic factor, $e^{(M_N - 3M_\pi/2) \tau}$, expected for nucleon correlation
functions. There is also a small increase in this factor between
$\langle x \rangle_{u-d} \to \langle x \rangle_{\Delta u- \Delta
d} \to \langle x \rangle_{\delta u- \delta d} $.
\item
The data for all three operators is symmetric about $t=\tau/2$ as
predicted by the spectral decomposition. Only on three ensembles,
$a094m270$, $a091m170L$ and $a071m170$, the symmetry about $t=\tau/2$ is not
manifest in the largest $\tau$ data. As stated above, these data have the largest
errors, and the deviations are within errors.
\item
In all cases (operators and ensembles), the convergence of the data
towards the $\tau \to \infty$ value is monotonic and from above. Thus,
ESC causes all three moments to be overestimated.
\end{itemize}
With the data satisfing the expected conditions, we are able to make
$3^\ast$-state (three-state fit neglecting the term with $\matrixe{2}{\mathcal{O}}{2} $)
fits in almost all cases to data with the largest three values of
$\tau$. The one exceptions is the $a071m170$ ensemble where we neglect
the largest $\tau=21$ data as the statistics are still inadequate. Including it in 
the fits does not change the results. We have also checked that the
results from fits keeping the largest four values of $\tau$ overlap
with these within $1\sigma$. The data and the result of the fit evaluated for 
various values of $\tau$ are shown in each panel in
Figs.~\ref{fig:Ratio-mom-1}--\ref{fig:Ratio-transversity-2} along with
the $\tau=\infty$ value as the blue band. The figure labels also give
the ensemble ID, the value of the moment obtained using
Eqs.~\eqref{eq:me2momentV}, or~\eqref{eq:me2momentA},
or~\eqref{eq:me2momentT}, the $\chi^2/$dof of the fit, and the values
of $\tau$ at which data have been collected and displayed.

The three panels in each row of Figs.~\ref{fig:Ratio-mom-1}--\ref{fig:Ratio-transversity-2} 
have the same data but show fits with the
three strategies that are being compared.  The scale for the y-axis is
chosen to be the same for all the plots to facilitate this comparison.  The
values of $\Delta M_1$ entering/determined by the various fits are
given in Table~\ref{tab:massgap}.

As mentioned above, the key parameter needed to control ESC is the
mass gap $\Delta M_1$ of the first excited state that provides the
dominant contribution. Theoretically, the lightest possible state with
positive parity contributing to the forward matrix elements is either
$N ({\bm p}=1) \pi({\bm p}=-1)$ or $N (0) \pi(0) \pi(0)$ depending
on the value of $M_\pi$ and the lowest momenta, which is larger than
200~MeV on all seven ensembles. For our ensembles, the non-interacting
energies for these two states are roughly equal. Since the fits do not
rely on knowing the identity of the state but only on $\Delta M_1$, we
regard these two possible states as operationally the same and
label them $N\pi$. Thus in the strategy $\{4^{N\pi},3^\ast \}$, 
$\Delta M_1$ is  approximately the lowest possible value, and accounts for the 
possibility that one (or both) of these states gives the dominant ESC. 

The strategy $\{4,3^*\}$ assumes that the relevant states are the same
in two- and three-point functions. The strategy $\{4,2^{\rm free}\}$
only takes ${\cal A}_0$ and $M_0$ from the two-point fit, whose
determination is robust--the variation in their values between
$\{4,3^*\}$ and $\{4^{N\pi},3^*\}$ is less than a percent as shown in
Table~\ref{tab:massgap}.  The value of $\Delta M_1$ is an output in
this case. The relative limitation of the $\{4,2^{\rm free}\}$
strategy is that, with the current data, we can only make two-state
fits to the three-point functions, ie, include only one excited state.

The data for $\Delta M_1$ summarized in Table~\ref{tab:massgap}
displays the following qualitative features:
\begin{itemize}
\item 
The values $a \Delta M_1^{\{4\}} \approx 0.6 aM_N^{\{4\}}$. This
suggests that the lowest excited state in the $\{4\}$ fit to the
two-point function is close to the $N(1440)$.
\item
$ a \Delta M_1^{\{4^{N\pi}\}}$ is significantly smaller than $a \Delta
M_1^{\{4\}}$ as mentioned above.
\item
On five ensembles, $a \Delta M_1^{\{2^{\rm free}\}}$ from fits to the
momentum-fraction data are consistent with $a \Delta M_1^{\{4\}}$. (We
have also given the two-state fit value $a \Delta M_1^{\{2\}}$ in
Table~\ref{tab:massgap} to show how much $a \Delta M_1$ can vary
between a two- and four-state fit.)  To check whether this rough
agreement is a possibility for the remaining two ensembles, $a094m270$
and $a071m170$, we made fits with a range of priors but did not find a
flat direction with respect to $a \Delta M_1^{\{2^{\rm
free}\}}$. Thus, the large values of $a \Delta M_1^{\{2^{\rm free}\}}$
from these two ensembles are unexplained, however, as noted
previously, the statistical errors in these two ensembles are the
largest.
\item
The $a \Delta M_1^{\{2^{\rm free}\}}$ for helicity and transversity
moments are roughly the same and much larger than even $a \Delta
M_1^{\{2\}}$.
\end{itemize}

The unrenormalized results for the three moments obtained using
Eqs.~\eqref{eq:me2momentV}, or~\eqref{eq:me2momentA},
or~\eqref{eq:me2momentT} are given in Table~\ref{tab:ESC-fits} along
with the values of $\{t,\tau\}$ used.  The parameters and the
$\chi^2/$dof of the fits for the various strategies are given in
Tables~\ref{tab:5strategy-fits-momfrac},~\ref{tab:5strategy-fits-helfrac},
and~\ref{tab:5strategy-fits-transvmom}. In these tables, we include
results with $\{4,2\}$ and $\{4^{N\pi},2\}$ in addition to the
$\{4,3^\ast\}$, $\{4^{N\pi},3^\ast\}$ and $\{4,2^{\rm free}\}$
strategies to show that the variation on including the second excited
state is small, ie, $\Delta M_1$ is the key parameter in controlling ESC.

We draw the following conclusions from the results presented in
Tables~\ref{tab:massgap}--~\ref{tab:5strategy-fits-transvmom} and the
fits shown in
Figs.~\ref{fig:Ratio-mom-1}--\ref{fig:Ratio-transversity-2}:
\begin{itemize}
\item
The statistics on the $a091m170L$ and $a071m170$ ensembles need to be
increased to make the largest $\tau$ data useful.
\item
The $\chi^2/$dof of most fits are reasonable. 
\item 
The $\{4,2^{\rm free}\}$ fits have reasonable $\chi^2/$dof but do not
indicate a preference for the small $\Delta M_1^{N\pi}$ given in
Table~\ref{tab:massgap}. Their $\Delta M_1$ lie closer to or higher
than $\Delta M_1^{(2)}$.
\item
The $\Delta M_1$ from a two-state fit is expected to be larger since
it is an effective combination of the mass gaps of the full tower of
excited states. This is illustrated by the difference between $\Delta
M_1^{\{2\}}$ and $\Delta M_1^{\{4\}}$.  Thus we take the values
$\Delta M_1^{\{4^{N\pi}\}}$ and $a \Delta M_1^{\{2^{\rm free}\}}$ to bracket
possible values of $\Delta M_1$ in each case.
\end{itemize}
Based on the above arguments, we will choose the $\{4,3^\ast\}$
results obtained after performing the CCFV fits for the final central
value.  We will also take half the spread in results between the
$\{4^{N\pi},3^\ast\}$ and $\{4,2^{\rm free}\}$ strategies, which is
$\approx 0.02$ in most cases, as a second  uncertainty to
account for possible unresolved bias from ESC.

The renormalization of the matrix elements is carried out using
estimates of $Z_{VD},~Z_{AD}$, and $ Z_{TD}$ calculated on the lattice
in the \ripmom scheme and then converted to the $\MSbar$ scheme at
2~GeV. Two methods to control discretization errors are described in
the Appendix~\ref{sec:renormalization}. The final values of
$Z_{VD},~Z_{AD}$, and $ Z_{TD}$ used in the analysis are given in
Table \ref{tab:Z-fac}.  The values of the three renormalized moments
from the seven ensembles and with the three strategies are summarized
in Tables~\ref{tab:renormalized-moments-A} for renomalization method A
and in Table~\ref{tab:renormalized-moments-B} for method B. These data
are used to perform the CCFV fits discussed next.



\begin{table*}[htbp]    
\centering
\setlength{\tabcolsep}{0.6pt}
\renewcommand{\arraystretch}{1.1}
\begin{tabular}{|c|c|c|c|c|c|c|c|c|c| }
\hline

\multicolumn{10}{|c|}{$\langle x \rangle_{u-d}$} \\
\hline
Ensemble & fit-type&$a \Delta M_1$&$a \Delta M_2$
&$\la 0|{\cal O}|0 \ra$
&$\frac{\la 1|{\cal O}| 1\ra |A_1|^2 }{\la 0|{\cal O}|0 \ra |A_0|^2}$
&$\frac{\la 1|{\cal O}| 0\ra |A_1|}{\la 0|{\cal O}|0 \ra |A_0|}$ 
&$\frac{\la 2|{\cal O}| 0\ra |A_2|}{\la 0|{\cal O}|0 \ra |A_0|}$ 
&$\frac{\la 2|{\cal O}| 1\ra |A_2||A_1|}{\la 0|{\cal O}|0 \ra |A_0|^2}$
&$\chi^2$/dof\\
\hline
$a127m285$& $\{4,2\}$&$0.376(52)$ && $0.1166(54)$& $0.24(62)$&$0.629(63)$ & &
&$1.74$\\
$a127m285$& $\{4^{N\pi},2\}$&$0.326(20)$ && $0.1114(28)$& $-0.46(31)$&$0.708(46)$ & &
&$1.72$\\
$a127m285$& $\{4,3^*\}$&$0.376(52)$ &$0.776(59)$& $0.1176(41)$& $0.29(60)$&$0.584(31)$ &$0.048(45)$ &$1.17(78)$
&$1.69$\\
$a127m285$& $\{4^{N\pi},3^*\}$&$0.326(20)$ &$0.688(61)$& $0.1118(28)$& $-0.59(42)$&$0.696(68)$ &$-0.12(12)$ &$2.6(1.2)$
&$1.73$\\
$a127m285$& $\{4,2^{\rm free}\}$&$0.359(35)$ && $0.1148(41)$& $-0.01(59)$&$0.651(52)$ & &
&$1.81$\\
\hline
\hline
$a094m270$& $\{4,2\}$&$0.273(63)$ && $0.0796(88)$& $-0.19(52)$&$0.77(18)$ & &
&$1.22$\\
$a094m270$& $\{4^{N\pi},2\}$&$0.2643(95)$ && $0.0784(27)$& $-0.24(37)$&$0.80(10)$ & &
&$1.25$\\
$a094m270$& $\{4,3^*\}$&$0.273(63)$ &$0.567(83)$& $0.0855(53)$& $0.44(49)$&$0.473(98)$ &$0.28(13)$ &$1.02(89)$
&$1.23$\\
$a094m270$& $\{4^{N\pi},3^*\}$&$0.2643(95)$ &$0.56(15)$& $0.0850(33)$& $0.40(38)$&$0.47(13)$ &$0.29(12)$ &$1.00(88)$
&$1.23$\\
$a094m270$& $\{4,2\}$&$0.485(42)$ && $0.0951(22)$& $6.8(3.6)$&$0.586(20)$ & &
&$1.06$\\
\hline
\hline
$a094m270L$& $\{4,2\}$&$0.303(42)$ && $0.0822(35)$& $0.74(95)$&$0.680(47)$ & &
&$1.08$\\
$a094m270L$& $\{4^{N\pi},2\}$&$0.249(27)$ && $0.0767(36)$& $-0.18(45)$&$0.775(75)$ & &
&$1.57$\\
$a094m270L$& $\{4,3^*\}$&$0.303(42)$ &$0.561(83)$& $0.0834(23)$& $1.29(99)$&$0.588(53)$ &$0.23(13)$ &$-1.2(3.4)$
&$0.95$\\
$a094m270L$& $\{4^{N\pi},3^*\}$&$0.249(27)$ &$0.440(61)$& $0.0806(21)$& $0.55(36)$&$0.534(61)$ &$0.316(93)$ &$-0.001213(3)$
&$0.93$\\
$a094m270L$& $\{4,2^{\rm free}\}$&$0.344(26)$ && $0.0849(16)$& $2.4(1.6)$&$0.654(16)$ & &
&$1.02$\\
\hline
\hline
$a091m170$ & $\{4,2\}$&$0.293(45)$ & &$0.0682(57)$&$0.89(95)$&$0.94(14)$& &
&$0.98$\\

$a091m170$ & $\{4^{N\pi},2\}$&$0.195(19)$&
&$0.0473(71)$&$-0.92(67)$&$1.78(43)$& &
&$1.34$\\

$a091m170$ &
$\{4,3^*\}$&$0.293(45)$&$0.578(81)$&$0.0696(39)$&$1.4(1.2)$&$0.81(15)$&$0.28(41)$&$-1.2(4.9)$&$1.04$\\

$a091m170$ &
$\{4^{N\pi},3^*\}$&$0.195(19)$&$0.413(66)$&$0.0609(59)$&$0.7(1.7)$&$0.76(43)$&$0.59(69)$&$0.05(4)$&$1.02$\\

$a091m170$ & $\{4,2^{\rm free}\}$         &$0.320(51)$&
&$0.0702(50)$&$1.5(1.7)$&$0.89(11)$& & &$1.02$\\
\hline
\hline

$a091m170L$ & $\{4,2\}$&$0.252(36)$ & &$0.0630(65)$&$0.73(69)$&$1.03(20)$& &
&$1.15$\\

$a091m170L$ & $\{4^{N\pi},2\}$&$0.157(11)$&
&$0.0337(72)$&$-1.44(99)$&$2.77(86)$& &
&$1.63$\\

$a091m170L$ &
$\{4,3^*\}$&$0.252(36)$&$0.530(38)$&$0.0647(49)$&$1.00(66)$&$0.90(17)$&$0.16(28)$&$-0.02(1)$&$1.23$\\

$a091m170L$ &
$\{4^{N\pi},3^*\}$&$0.157(11)$&$0.386(73)$&$0.0596(75)$&$1.13(59)$&$0.54(37)$&$0.89(18)$&$-0.06(1)$&$1.27$\\

$a091m170L$ & $\{4,2^{\rm free}\}$         &$0.297(43)$& &$0.0694(51)$&$2.1(1.7)$&$0.87(11)$& & &$1.25$\\
\hline
\hline
$a073m270$& $\{4,2\}$&$0.229(41)$ && $0.0612(45)$& $0.17(77)$&$0.76(11)$ & &
&$0.98$\\
$a073m270$& $\{4^{N\pi},2\}$&$0.217(24)$ && $0.0598(31)$& $-0.05(47)$&$0.792(82)$ & &
&$1.02$\\
$a073m270$& $\{4,3^*\}$&$0.229(41)$ &$0.386(79)$& $0.0633(25)$& $0.97(71)$&$0.563(64)$ &$0.35(11)$ &$-0.9(1.2)$
&$1.03$\\
$a073m270$& $\{4^{N\pi},3^*\}$&$0.217(24)$ &$0.363(79)$& $0.0626(16)$& $0.83(44)$&$0.545(75)$ &$0.383(86)$ &$-0.75(87)$
&$1.03$\\
$a073m270$& $\{4,2^{\rm free}\}$&$0.311(16)$ && $0.0670(10)$& $3.3(1.3)$&$0.695(15)$ & &
&$1.56$\\
\hline
\hline
$a071m170$& $\{4,2\}$&$0.249(34)$ && $0.0530(36)$& $1.1(1.1)$&$0.88(11)$ & &
&$0.75$\\
$a071m170$& $\{4^{N\pi},2\}$&$0.155(12)$ && $0.0420(41)$& $-0.75(77)$&$1.23(31)$ & &
&$0.34$\\
$a071m170$& $\{4,3^*\}$&$0.249(34)$ &$0.471(60)$& $0.0553(23)$& $2.3(1.2)$&$0.636(78)$ &$0.53(11)$ &$-4.2(3.2)$
&$0.84$\\
$a071m170$& $\{4^{N\pi},3^*\}$&$0.155(12)$ &$0.377(40)$& $0.0491(29)$& $1.04(45)$&$0.54(17)$ &$0.86(13)$ &$-1.28(91)$
&$0.87$\\
$a071m170$& $\{4,2^{\rm free}\}$&$0.404(28)$ && $0.0617(12)$& $18.0(9.7)$&$0.781(20)$ & &
&$1.15$\\

\hline

\end{tabular}
\caption{Comparison of results of the fits to remove the excited-state contamination 
for the momentum fraction $\langle x \rangle_{ u- d}$ using the five strategies, $\{4,2\}$,
  $\{4^{N\pi},2\}$, $\{4,3^*\}$, $\{4^{N\pi},3^*\}$ and $\{4,2^{\rm
    free}\}$. The fit parameters, defined in Eq.~\protect\eqref{eq:3pt}, are given for all seven ensembles
 along with the $\chi^2/$dof of the fit.
}
\label{tab:5strategy-fits-momfrac}
\end{table*}

\begin{table*}[htbp]   
\centering
\setlength{\tabcolsep}{0.6pt}
\renewcommand{\arraystretch}{1.1}
\begin{tabular}{|c|c|c|c|c|c|c|c|c|c| }
\hline

\multicolumn{10}{|c|}{$\langle x \rangle_{\Delta u-\Delta d}$} \\
\hline
Ensemble & fit-type&$a \Delta M_1$&$a \Delta M_2$
&$\la 0|{\cal O}|0 \ra$
&$\frac{\la 1|{\cal O}| 1\ra |A_1|^2 }{\la 0|{\cal O}|0 \ra |A_0|^2}$
&$\frac{\la 1|{\cal O}| 0\ra |A_1|}{\la 0|{\cal O}|0 \ra |A_0|}$ 
&$\frac{\la 2|{\cal O}| 0\ra |A_2|}{\la 0|{\cal O}|0 \ra |A_0|}$ 
&$\frac{\la 2|{\cal O}| 1\ra |A_2||A_1|}{\la 0|{\cal O}|0 \ra |A_0|^2}$
&$\chi^2$/dof\\
\hline
\hline
$a127m285$& $\{4,2\}$&$0.376(52)$ && $0.1404(38)$& $-0.19(35)$&$0.404(30)$ & &
&$1.10$\\
$a127m285$& $\{4^{N\pi},2\}$&$0.326(20)$ && $0.1363(24)$& $-0.25(25)$&$0.431(30)$ & &
&$1.30$\\
$a127m285$& $\{4,3^*\}$&$0.376(52)$ &$0.776(59)$& $0.1443(24)$& $0.40(41)$&$0.252(51)$ &$0.529(94)$ &$-0.6(1.7)$
&$1.07$\\
$a127m285$& $\{4^{N\pi},3^*\}$&$0.326(20)$ &$0.688(61)$& $0.1445(33)$& $0.59(49)$&$0.162(96)$ &$0.62(30)$ &$-0.04(2.14)$
&$0.83$\\
$a127m285$& $\{4,2^{\rm free }\}$&$0.706(58)$ && $0.1503(14)$& $15(12)$&$0.532(28)$ & &
&$1.20$\\
\hline
\hline
$a094m270$& $\{4,2\}$&$0.273(63)$ && $0.0960(80)$& $-0.01(41)$&$0.56(12)$ & &
&$1.30$\\
$a094m270$& $\{4^{N\pi},2\}$&$0.2643(95)$ && $0.0949(27)$& $-0.05(31)$&$0.574(74)$ & &
&$1.33$\\
$a094m270$& $\{4,3^*\}$&$0.273(63)$ &$0.567(83)$& $0.1048(40)$& $0.80(43)$&$0.20(11)$ &$0.69(11)$ &$-0.29(80)$
&$1.09$\\
$a094m270$& $\{4^{N\pi},3^*\}$&$0.2643(95)$ &$0.56(15)$& $0.1048(52)$& $0.80(48)$&$0.19(18)$ &$0.703(96)$ &$-0.30(85)$
&$1.09$\\
$a094m270$& $\{4,2^{\rm free}\}$&$0.689(50)$ && $0.1136(18)$& $39(22)$&$0.692(28)$ & &
&$0.94$\\

\hline
\hline
$a094m270L$& $\{4,2\}$&$0.303(42)$ && $0.0971(30)$& $0.19(59)$&$0.543(25)$ & & &$1.26$\\
$a094m270L$& $\{4^{N\pi},2\}$&$0.249(27)$ && $0.0940(26)$& $0.18(39)$&$0.499(39)$ & & &$1.30$\\
$a094m270L$& $\{4,3^*\}$&$0.303(42)$ &$0.561(83)$& $0.0986(19)$& $1.14(65)$&$0.418(63)$ &$0.497(96)$ &$-3.1(3.5)$ &$0.86$\\
$a094m270L$& $\{4^{N\pi},3^*\}$&$0.249(27)$ &$0.440(61)$& $0.0970(17)$& $0.83(41)$&$0.324(68)$ &$0.540(87)$ &$-1.4(1.3)$ &$0.85$\\
$a094m270L$& $\{4,2^{\rm free}\}$&$0.400(53)$ && $0.1015(18)$& $4.4(4.5)$&$0.559(35)$ & & &$1.11$\\
\hline
\hline
$a091m170$ & $\{4,2\}$&$0.293(45)$ & &$0.0794(57)$&$-0.12(56)$&$0.83(12)$& &
&$1.36$\\

$a091m170$ & $\{4^{N\pi},2\}$&$0.195(19)$&
&$0.0646(59)$&$-0.73(43)$&$1.14(21)$& &
&$1.39$\\

$a091m170$ &
$\{4,3^*\}$&$0.293(45)$&$0.578(81)$&$0.0848(32)$&$0.69(78)$&$0.52(11)$&$0.48(27)$&$1.3(3.2)$&$1.02$\\

$a091m170$ &
$\{4^{N\pi},3^*\}$&$0.195(19)$&$0.413(66)$&$0.0802(53)$&$0.58(71)$&$0.36(23)$&$0.65(27)$&$0.3(1.5)$&$1.03$\\

$a091m170$ & $\{4,2^{\rm free}\}$         &$0.441(64)$&
&$0.0899(28)$&$6.6(6.4)$&$0.72(3)$& & &$1.09$\\
\hline
\hline
$a091m170L$ & $\{4,2\}$&$0.252(36)$ & &$0.0721(69)$&$-0.25(48)$&$1.00(19)$& &
&$0.81$\\

$a091m170L$ & $\{4^{N\pi},2\}$&$0.157(11)$&
&$0.0453(67)$&$-1.54(68)$&$2.11(49)$& &
&$0.89$\\

$a091m170L$ &
$\{4,3^*\}$&$0.252(36)$&$0.530(38)$&$0.0789(40)$&$0.72(92)$&$0.60(13)$&$0.52(39)$&$-0.04(4.2)$&$0.36$\\

$a091m170L$ &
$\{4^{N\pi},3^*\}$&$0.157(11)$&$0.386(73)$&$0.0728(87)$&$0.71(87)$&$0.42(36)$&$0.80(36)$&$-0.02(1.6)$&$0.35$\\

$a091m170L$ & $\{4,2^{\rm free}\}$         &$0.388(52)$&
&$0.0865(33)$&$4.6(3.7)$&$0.76(3)$& & &$0.43$\\
\hline
\hline
$a073m270$& $\{4,2\}$&$0.229(41)$ && $0.0747(33)$& $-0.13(42)$&$0.542(50)$ & &
&$1.56$\\
$a073m270$& $\{4^{N\pi},2\}$&$0.217(24)$ && $0.0737(23)$& $-0.23(31)$&$0.555(47)$ & &
&$1.63$\\
$a073m270$& $\{4,3^*\}$&$0.229(41)$ &$0.386(79)$& $0.0780(13)$& $0.57(30)$&$0.30(14)$ &$0.630(96)$ &$0.61(97)$
&$1.12$\\
$a073m270$& $\{4^{N\pi},3^*\}$&$0.217(24)$ &$0.363(79)$& $0.0780(17)$& $0.59(33)$&$0.25(14)$ &$0.660(84)$ &$0.39(87)$
&$1.12$\\
$a073m270$& $\{4,2^{\rm free}\}$&$0.457(14)$ && $0.0827(5)$& $24.0(7.1)$&$0.758(13)$ & &
&$1.28$\\
\hline
\hline
$a071m170$& $\{4,2\}$&$0.249(34)$ && $0.0618(35)$& $-0.33(70)$&$0.809(86)$ & &
&$0.97$\\
$a071m170$& $\{4^{N\pi},2\}$&$0.155(12)$ && $0.0517(41)$& $-0.76(61)$&$0.98(23)$ & &
&$1.18$\\
$a071m170$& $\{4,3^*\}$&$0.249(34)$ &$0.471(60)$& $0.0665(18)$& $1.49(73)$&$0.38(12)$ &$0.97(11)$ &$-2.0(2.0)$
&$1.09$\\
$a071m170$& $\{4^{N\pi},3^*\}$&$0.155(12)$ &$0.377(40)$& $0.0690(45)$& $1.56(40)$&$-0.05(19)$ &$1.255(92)$ &$-1.30(70)$
&$1.24$\\
$a071m170$& $\{4,2^{\rm free}\}$&$0.561(21)$ && $0.0729(8)$& $122(57)$&$1.034(23)$ & &
&$1.41$\\

\hline
\end{tabular}
\caption{Comparison of results of the fits to remove the excited-state contamination for the helicity  moment  
$\langle x \rangle_{\Delta u - \Delta d}$ using the five strategies,
  $\{4,2\}$, $\{4^{N\pi},2\}$, $\{4,3^*\}$, $\{4^{N\pi},3^*\}$ and
  $\{4,2^{\rm free}\}$. The fit parameters, defined in Eq.~\protect\eqref{eq:3pt}, are given for all seven ensembles 
 along with the $\chi^2/$dof of the fit.
}
\label{tab:5strategy-fits-helfrac}
\end{table*}


\begin{table*}[htbp]   
\centering
\setlength{\tabcolsep}{0.4pt}
\renewcommand{\arraystretch}{1.1}
\begin{tabular}{|c|c|c|c|c|c|c|c|c|c| }
\hline

\multicolumn{10}{|c|}{$\langle x \rangle_{\delta u-\delta d}$} \\
\hline
Ensemble & fit-type&$a \Delta M_1$&$a \Delta M_2$
&$\la 0|{\cal O}|0 \ra$
&$\frac{\la 1|{\cal O}| 1\ra |A_1|^2 }{\la 0|{\cal O}|0 \ra |A_0|^2}$
&$\frac{\la 1|{\cal O}| 0\ra |A_1|}{\la 0|{\cal O}|0 \ra |A_0|}$ 
&$\frac{\la 2|{\cal O}| 0\ra |A_2|}{\la 0|{\cal O}|0 \ra |A_0|}$ 
&$\frac{\la 2|{\cal O}| 1\ra |A_2||A_1|}{\la 0|{\cal O}|0 \ra |A_0|^2}$
&$\chi^2$/dof\\
\hline
$a127m285$& $\{4,2\}$&$0.376(52)$ && $0.1406(55)$& $0.10(57)$&$0.61(5)$ & &
&$1.28$\\
$a127m285$& $\{4^{N\pi},2\}$&$0.326(20)$ && $0.1344(33)$& $-0.22(32)$&$0.66(5)$ & &
&$1.59$\\
$a127m285$& $\{4,3^*\}$&$0.376(52)$ &$0.776(59)$& $0.1448(36)$& $0.51(57)$&$0.45(6)$ &$0.36(14)$ &$2.5(2.2)$
&$0.82$\\
$a127m285$& $\{4^{N\pi},3^*\}$&$0.326(20)$ &$0.688(61)$& $0.1466(44)$& $0.30(89)$&$0.30(15)$ &$0.42(55)$ &$5.4(4.5)$
&$0.57$\\
$a127m285$& $\{4,2^{\rm free}\}$&$0.64(11)$ && $0.1539(32)$& $17(21)$&$0.68(10)$ & &
&$0.81$\\
\hline
\hline
$a094m270$& $\{4,2\}$&$0.273(63)$ && $0.096(10)$& $-0.19(53)$&$0.74(17)$ & &
&$1.63$\\
$a094m270$& $\{4^{N\pi},2\}$&$0.2643(95)$ && $0.0944(32)$& $-0.24(42)$&$0.76(10)$ & &
&$1.66$\\
$a094m270$& $\{4,3^*\}$&$0.273(63)$ &$0.567(83)$& $0.1056(53)$& $0.54(51)$&$0.34(13)$ &$0.66(15)$ &$1.1(1.2)$
&$1.08$\\
$a094m270$& $\{4^{N\pi},3^*\}$&$0.2643(95)$ &$0.56(15)$& $0.1054(58)$& $0.54(51)$&$0.32(20)$ &$0.67(14)$ &$1.1(1.2)$
&$1.08$\\
$a094m270$& $\{4,2^{\rm free}\}$&$0.633(45)$ && $0.1166(20)$& $29(15)$&$0.80(3)$ & &
&$0.86$\\
\hline
\hline
$a094m270L$& $\{4,2\}$&$0.303(42)$ && $0.0997(35)$& $-0.30(54)$&$0.644(31)$ & & &$0.80$\\
$a094m270L$& $\{4^{N\pi},2\}$&$0.249(27)$ && $0.0943(36)$& $-0.69(33)$&$0.700(55)$ & & &$1.06$\\
$a094m270L$& $\{4,3^*\}$&$0.303(42)$ &$0.561(83)$& $0.1024(20)$& $0.39(59)$&$0.460(77)$ &$0.53(13)$ &$3.3(3.3)$ &$0.62$\\
$a094m270L$& $\{4^{N\pi},3^*\}$&$0.249(27)$ &$0.440(61)$& $0.1009(20)$& $0.31(42)$&$0.348(85)$ &$0.59(11)$ &$1.4(1.6)$ &$0.64$\\
$a094m270L$& $\{4^{\rm free},2\}$&$0.466(29)$ && $0.1066(12)$& $12.3(7.0)$&$0.746(25)$ & & &$0.82$\\

\hline
\hline

$a091m170$ & $\{4,2\}$&$0.293(45)$ & &$0.0780(72)$&$-0.59(93)$&$1.05(17)$& &
&$1.17$\\

$a091m170$ & $\{4^{N\pi},2\}$&$0.195(19)$&
&$0.0517(88)$&$-1.29(78)$&$2.07(54)$& &
&$1.89$\\

$a091m170$ &
$\{4,3^*\}$&$0.293(45)$&$0.578(81)$&$0.0834(47)$&$-0.20(1.07)$&$0.80(16)$&$-0.06(39)$&$10(6)$&$0.87$\\

$a091m170$ &
$\{4^{N\pi},3^*\}$&$0.195(19)$&$0.413(66)$&$0.0754(77)$&$-0.83(1.2)$&$0.76(36)$&$0.14(47)$&$4.8(2.8)$&$0.87$\\

$a091m170$ & $\{4,2^{\rm free}\}$         &$0.424(61)$&
&$0.0904(39)$&$8.4(7.0)$&$0.89(4)$& & &$0.95$\\
\hline
\hline
$a091m170L$ & $\{4,2\}$&$0.252(36)$ & &$0.0766(70)$&$0.19(62)$&$0.97(17)$& &
&$1.27$\\

$a091m170L$ & $\{4^{N\pi},2\}$&$0.157(11)$&
&$0.0430(80)$&$-1.65(86)$&$2.5(7)$& &
&$1.48$\\

$a091m170L$ &
$\{4,3^*\}$&$0.252(36)$&$0.530(38)$&$0.0855(43)$&$1.64(87)$&$0.43(17)$&$1.00(36)$&$0.3(3)$&$1.25$\\

$a091m170L$ &
$\{4^{N\pi},3^*\}$&$0.157(11)$&$0.386(73)$&$0.0907(148)$&$1.91(90)$&$-0.11(43)$&$1.29(33)$&$-0.85(2)$&$1.22$\\

$a091m170L$ & $\{4,2^{\rm free}\}$         &$0.339(114)$&
&$0.0872(96)$&$2.8(5.3)$&$0.82(8)$& & &$1.30$\\
\hline
\hline
$a073m270$& $\{4,2\}$&$0.229(41)$ && $0.0736(49)$& $-0.65(46)$&$0.774(92)$ & &
&$1.01$\\
$a073m270$& $\{4^{N\pi},2\}$&$0.217(24)$ && $0.0738(29)$& $-0.43(38)$&$0.695(68)$ & &
&$0.73$\\
$a073m270$& $\{4,3^*\}$&$0.229(41)$ &$0.386(79)$& $0.0786(17)$& $0.38(36)$&$0.41(14)$ &$0.66(12)$ &$1.3(1.4)$
&$0.79$\\
$a073m270$& $\{4^{N\pi},3^*\}$&$0.217(24)$ &$0.363(79)$& $0.0785(18)$& $0.40(37)$&$0.36(15)$ &$0.70(10)$ &$1.0(1.2)$
&$0.80$\\
$a073m270$& $\{4,2^{\rm free}\}$&$0.431(16)$ && $0.0843(7)$& $17.7(5.6)$&$0.868(14)$ & &
&$1.11$\\

\hline
\hline

$a071m170$& $\{4,2\}$&$0.249(34)$ && $0.0615(43)$& $0.5(1.1)$&$0.98(12)$ & &
&$1.59$\\
$a071m170$& $\{4^{N\pi},2\}$&$0.155(12)$ && $0.0449(61)$& $-1.9(1.2)$&$1.70(51)$ & &
&$1.02$\\
$a071m170$& $\{4,3^*\}$&$0.249(34)$ &$0.471(60)$& $0.0655(26)$& $2.2(1.2)$&$0.58(13)$ &$0.90(14)$ &$-1.8(3.0)$
&$1.41$\\
$a071m170$& $\{4^{N\pi},3^*\}$&$0.155(12)$ &$0.377(40)$& $0.0637(50)$& $1.44(53)$&$0.23(23)$ &$1.25(14)$ &$-0.8(1.1)$
&$1.49$\\
$a071m170$& $\{4,2^{\rm free}\}$&$0.491(28)$ && $0.0738(12)$& $72(36)$&$1.07(3)$ & &
&$1.54$\\
\hline

\end{tabular}
\caption{Comparison of results of the fits to remove the excited-state contamination 
for the transversity moment $\langle x \rangle_{\delta u -
\delta d}$ using the five strategies, $\{4,2\}$,
  $\{4^{N\pi},2\}$, $\{4,3^*\}$, $\{4^{N\pi},3^*\}$ and $\{4,2^{\rm
    free}\}$. The fit parameters, defined in Eq.~\protect\eqref{eq:3pt}, are given for all seven ensembles 
 along with the $\chi^2/$dof of the fit.
}
\label{tab:5strategy-fits-transvmom}
\end{table*}

\begin{table*}[htbp]   
\centering
\renewcommand{\arraystretch}{1.2}
\begin{tabular}{ |c|c|c|c|c|c|c|c|c| }
 \hline moment & strategy & $a127m285$ & $a094m270$ & $a094m270L$ &
$a091m170$ & $a091m170L$ & $a073m270$ & $a071m170$ \\
\hline
$\la x\ra_{u-d}$                     & $\{4^{N\pi},3^*\}$         & $0.179(5)$ & $0.197(8)$  & $0.177(5)$ & $0.150(15)$  & $0.151(18)$  & $0.181(5)$ & $0.158(9)$ \\
\hline 
$\la x\ra_{u-d}$                     & $\{4,3^*\}$                & $0.188(7)$ &  $0.197(9)$ & $0.183(5)$ & $0.170(9)$ & $0.162(12)$  & $0.183(7)$ & $0.177(7)$ \\
\hline
$\la x\ra_{u-d}$                     & $\{4,2^{\rm free}\}$       & $0.184(7)$ & $0.211(6)$  & $0.187(4)$ & $0.171(12)$  & $0.174(13)$  & $0.193(4)$ & $0.197(4)$ \\
\hline
$\la x\ra_{\Delta u-\Delta d}$       &  $\{4^{N\pi},3^*\}$        & $0.237(7)$ & $0.235(10)$  & $0.217(5)$ & $0.200(13)$  & $0.188(22)$  & $0.228(6)$ & $0.228(14)$  \\
\hline
$\la x\ra_{\Delta u-\Delta d}$       &  $\{4,3^*\}$               & $0.236(5)$ & $0.235(10)$  & $0.220(5)$ & $0.210(8)$ & $0.202(10)$ & $0.227(5)$ & $0.217(6)$ \\
\hline
$\la x\ra_{\Delta u-\Delta d}$       &  $\{4,2^{\rm free}\}$      & $0.246(5)$ & $0.255(6)$  & $0.226(5)$ & $0.222(8)$ & $0.221(9)$ & $0.241(4)$ & $0.238(4)$ \\
\hline
$\la x\ra_{\delta u-\delta d}$       &  $\{4^{N\pi},3^*\}$        & $0.244(8)$ & $0.241(11)$   & $0.232(6)$ & $0.194(20)$  & $0.241(39)$  & $0.237(7)$ & $0.217(17)$  \\
\hline
$\la x\ra_{\delta u-\delta d}$       &  $\{4,3^*\}$               & $0.240(7)$ & $0.242(13)$   & $0.235(5)$ & $0.213(12)$  & $0.225(11)$  & $0.237(6)$ & $0.221(8)$ \\
\hline
$\la x\ra_{\delta u-\delta d}$       &  $\{4,2^{\rm free}\}$      & $0.256(7)$ & $0.263(8)$  & $0.245(5)$ & $0.231(10)$  & $0.229(25)$  & $0.254(5)$ & $0.249(5)$ \\
\hline
\hline
\end{tabular}
\caption{Renormalized moments for all three fit-strategies with $Z$ factors
obtained using Method A defined in appendix~\protect\ref{sec:renormalization}.}
\label{tab:renormalized-moments-A}
\end{table*}

\begin{table*}[htbp]   
\centering
\renewcommand{\arraystretch}{1.2}
\begin{tabular}{ |c|c|c|c|c|c|c|c|c| }
 \hline
moment                               & strategy                   & $a127m285$   & $a094m270$   & $a094m270L$  & $a091m170$   & $a091m170L$  & $a073m270$   & $a071m170$   \\  
\hline
\hline
$\la x\ra_{u-d}$                     & $\{4^{N\pi},3^*\}$         & $0.171(5)$ & $0.190(8)$ & $0.171(5)$ & $0.144(14)$  & $0.146(18)$  & $0.177(6)$ & $0.150(9)$  \\ 
\hline
$\la x\ra_{u-d}$                     & $\{4,3^*\}$                & $0.179(6)$ & $0.190(9)$ & $0.177(5)$ & $0.163(9)$ & $0.156(11)$  & $0.179(7)$ & $0.167(6)$ \\
\hline
$\la x\ra_{u-d}$                     & $\{4,2^{\rm free}\}$       & $0.175(7)$ & $0.204(6)$ & $0.180(4)$ & $0.165(12)$  & $0.167(13)$  & $0.189(5)$ & $0.186(4)$ \\
\hline
$\la x\ra_{\Delta u-\Delta d}$       & $\{4^{N\pi},3^*\}$         & $0.225(7)$ & $0.228(10)$ & $0.208(5)$ & $0.192(13)$  & $0.182(21)$  & $0.222(6)$ & $0.216(14)$  \\
\hline
$\la x\ra_{\Delta u-\Delta d}$       & $\{4,3^*\}$                & $0.225(5)$ & $0.229(10)$ & $0.211(5)$ & $0.201(8)$ & $0.195(10)$ & $0.221(5)$ & $0.206(5)$ \\
\hline
$\la x\ra_{\Delta u-\Delta d}$       & $\{4,2^{\rm free}\}$       & $0.234(5)$ & $0.247(7)$ & $0.217(5)$ & $0.213(8)$ & $0.213(9)$ & $0.235(4)$ & $0.226(4)$ \\
\hline
$\la x\ra_{\delta u-\delta d}$       & $\{4^{N\pi},3^*\}$         & $0.233(8)$ & $0.236(11)$  & $0.226(6)$ & $0.189(19)$  & $0.234(38)$  & $0.233(7)$ & $0.209(16)$  \\
\hline
$\la x\ra_{\delta u-\delta d}$       & $\{4,3^*\}$                & $0.230(7)$ & $0.237(13)$  & $0.229(6)$ & $0.207(11)$  & $0.218(11)$  & $0.233(6)$ & $0.213(8)$ \\
\hline
$\la x\ra_{\delta u-\delta d}$       & $\{4,2^{\rm free}\}$       & $0.244(7)$ & $0.258(8)$ & $0.238(5)$ & $0.224(10)$  & $0.222(25)$  & $0.250(5)$ & $0.240(5)$ \\
\hline
\hline
\end{tabular}
\caption{Renormalized moments for all three fit-strategies with $Z$ factors
obtained using Method B defined in appendix~\protect\ref{sec:renormalization}.}
\label{tab:renormalized-moments-B}
\end{table*}

\section{Chiral, continuum and infinite volume extrapolation}
\label{sec:results}

\begin{figure*}[htbp]   
\begin{subfigure}
\centering
\includegraphics[angle=0,width=0.32\textwidth]{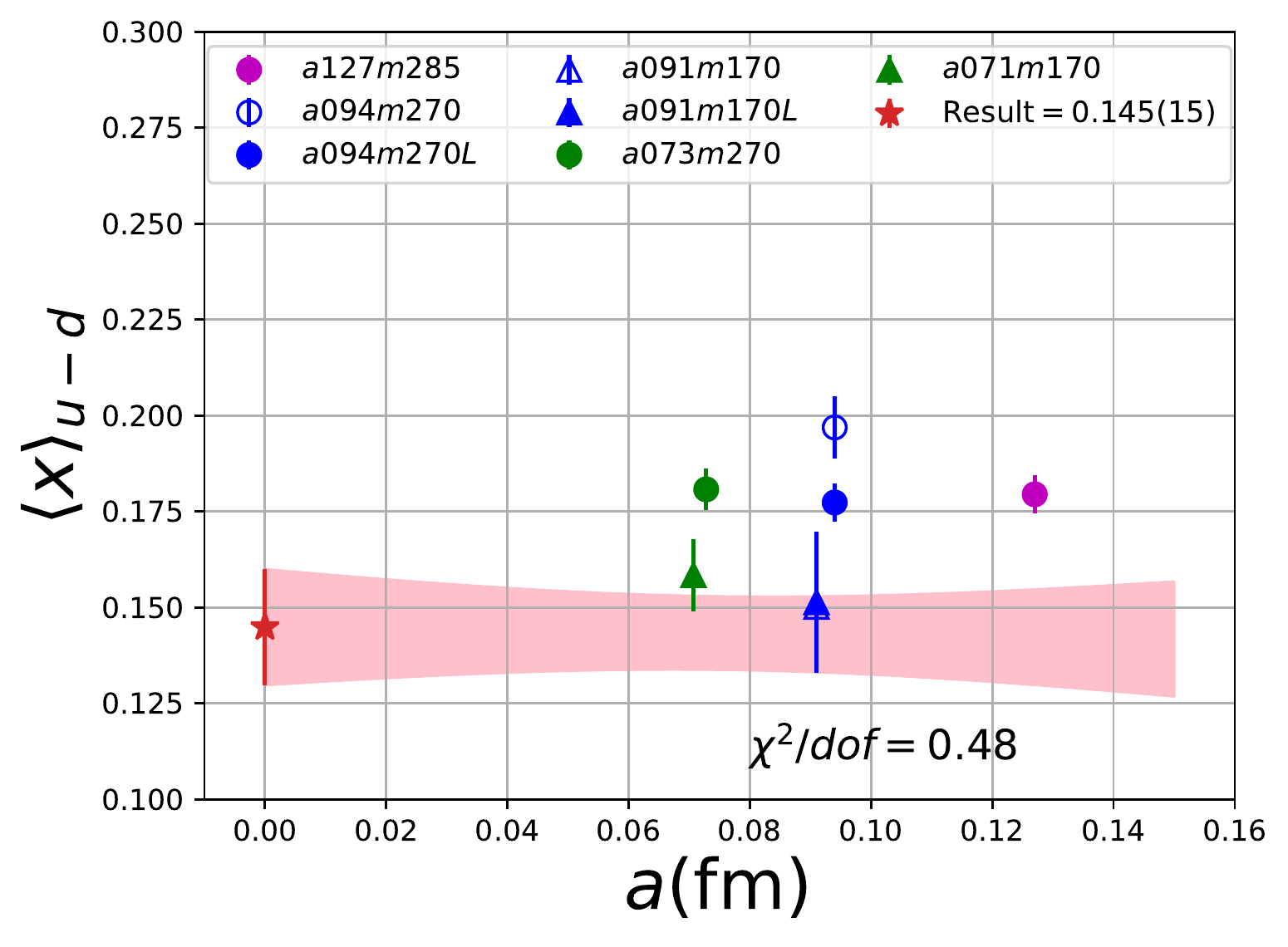}
\includegraphics[angle=0,width=0.32\textwidth]{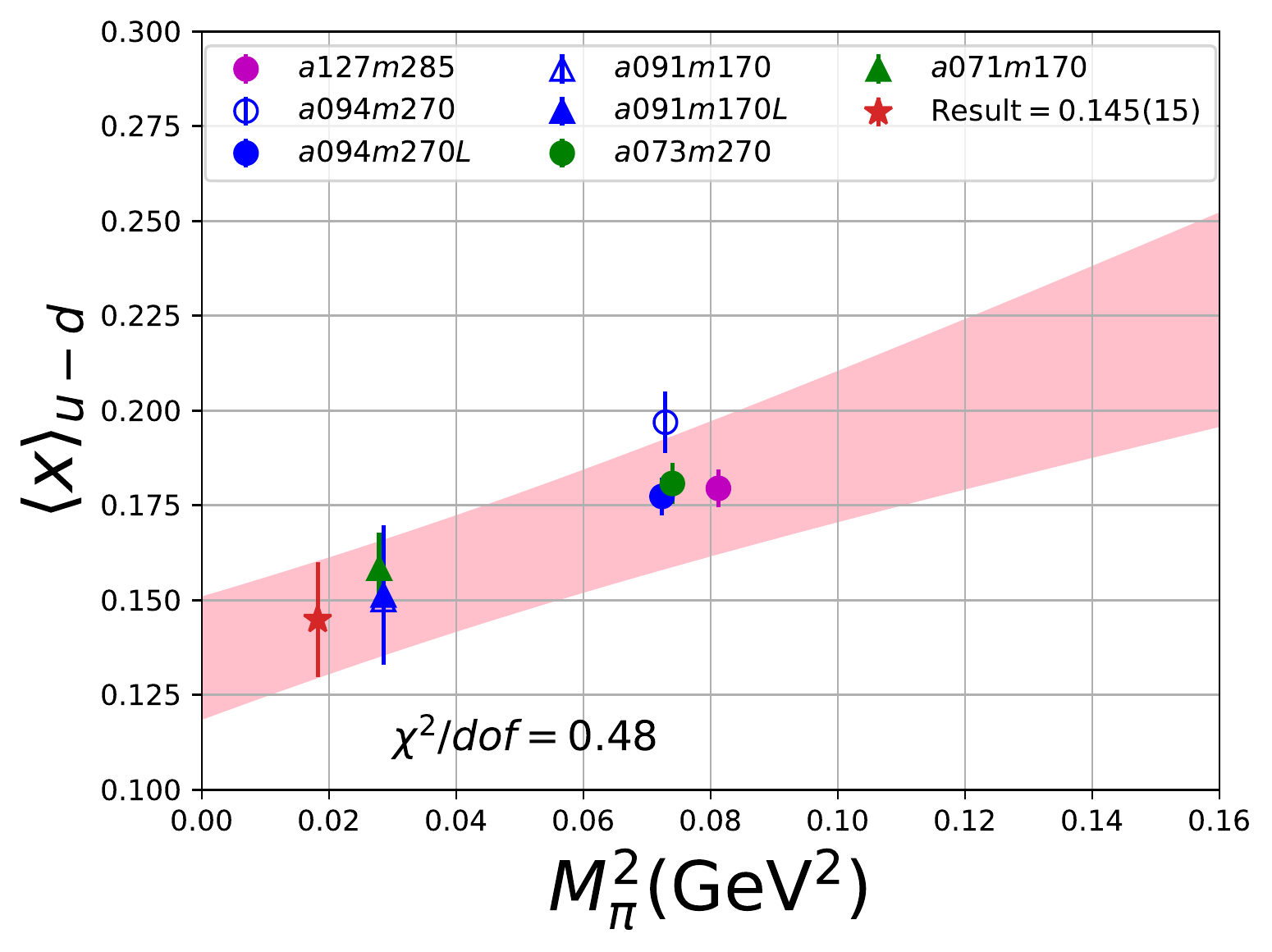}
\includegraphics[angle=0,width=0.32\textwidth]{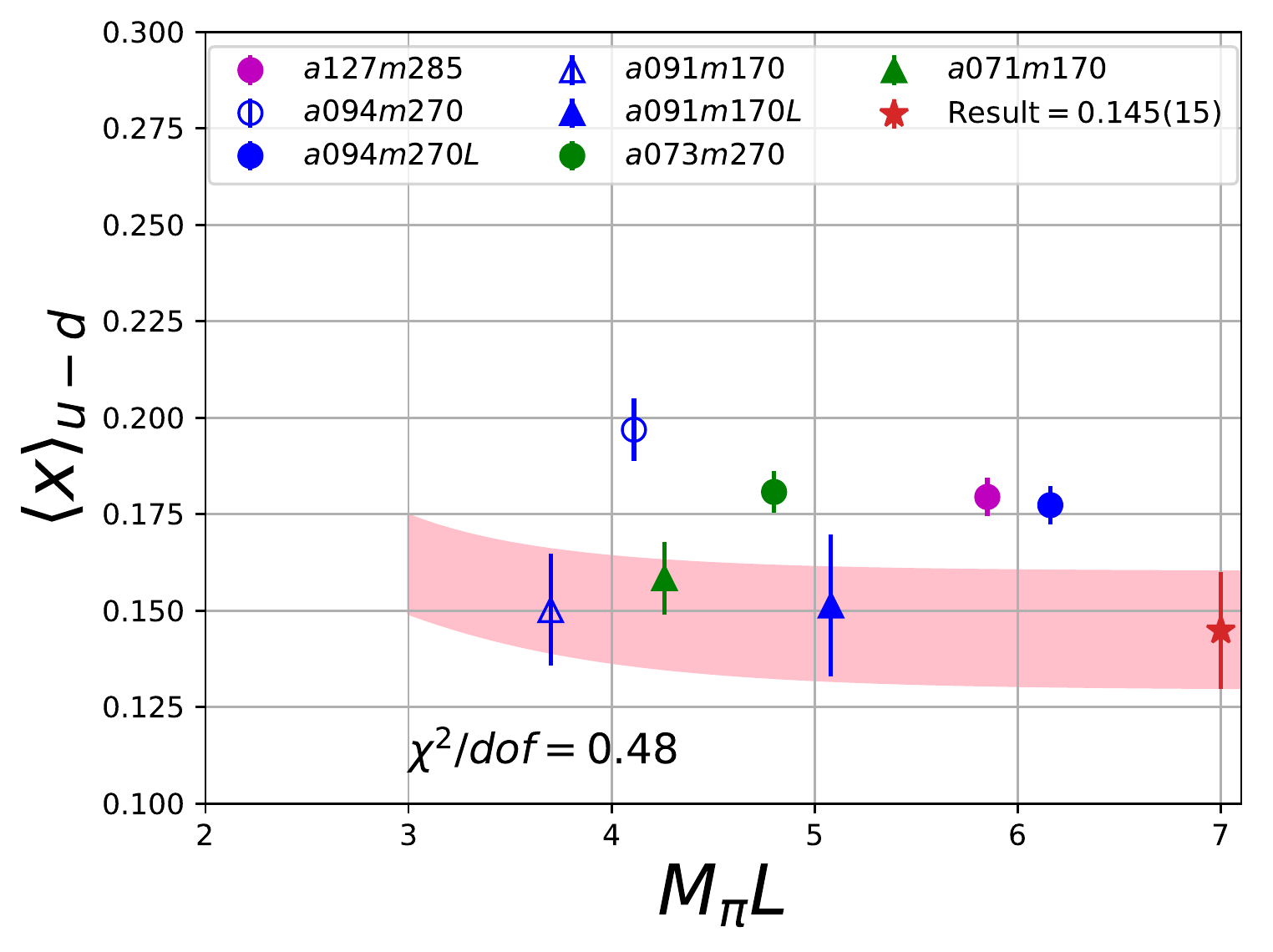}
\end{subfigure}
\begin{subfigure}
\centering
\includegraphics[angle=0,width=0.32\textwidth]{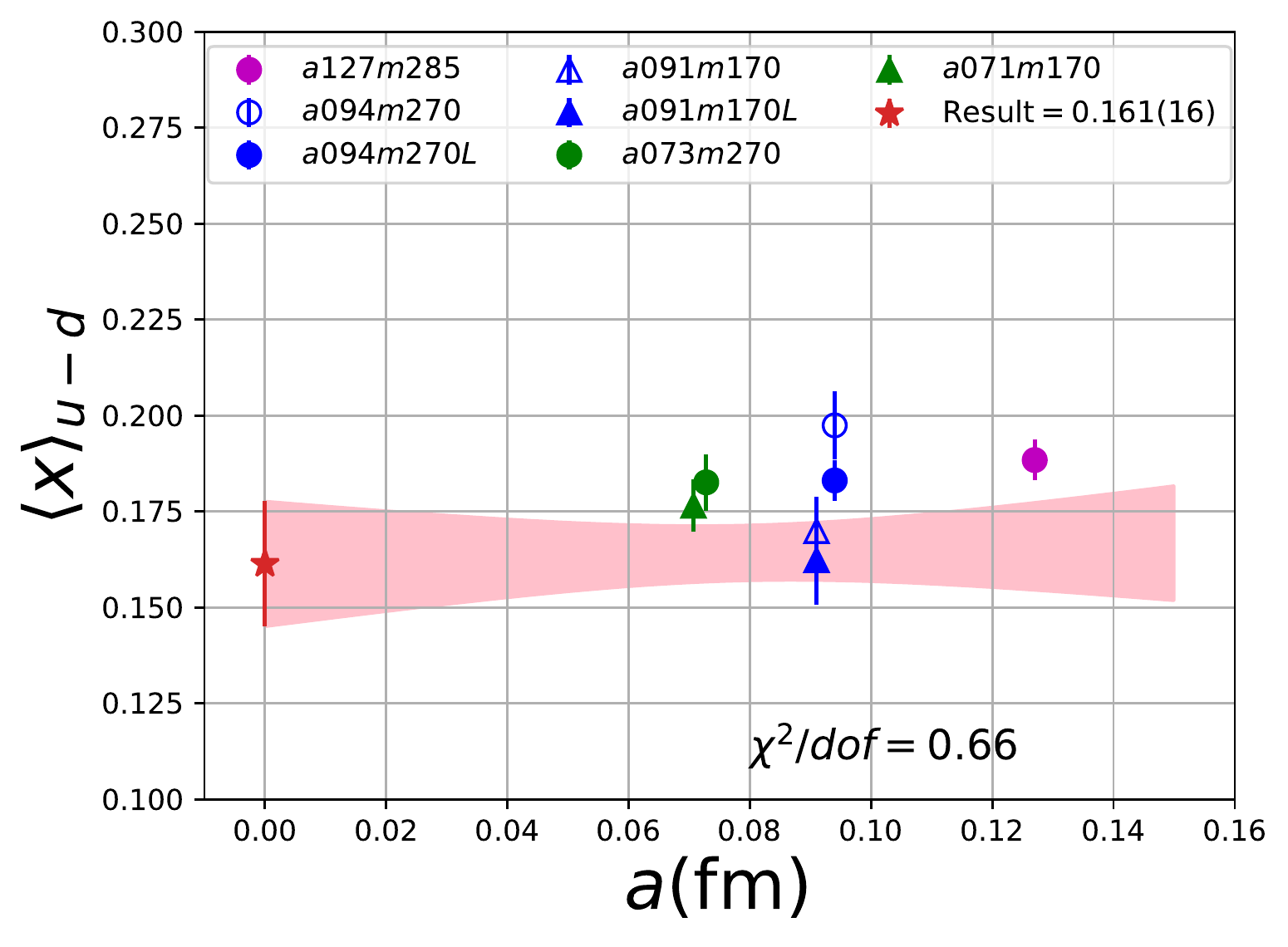}
\includegraphics[angle=0,width=0.32\textwidth]{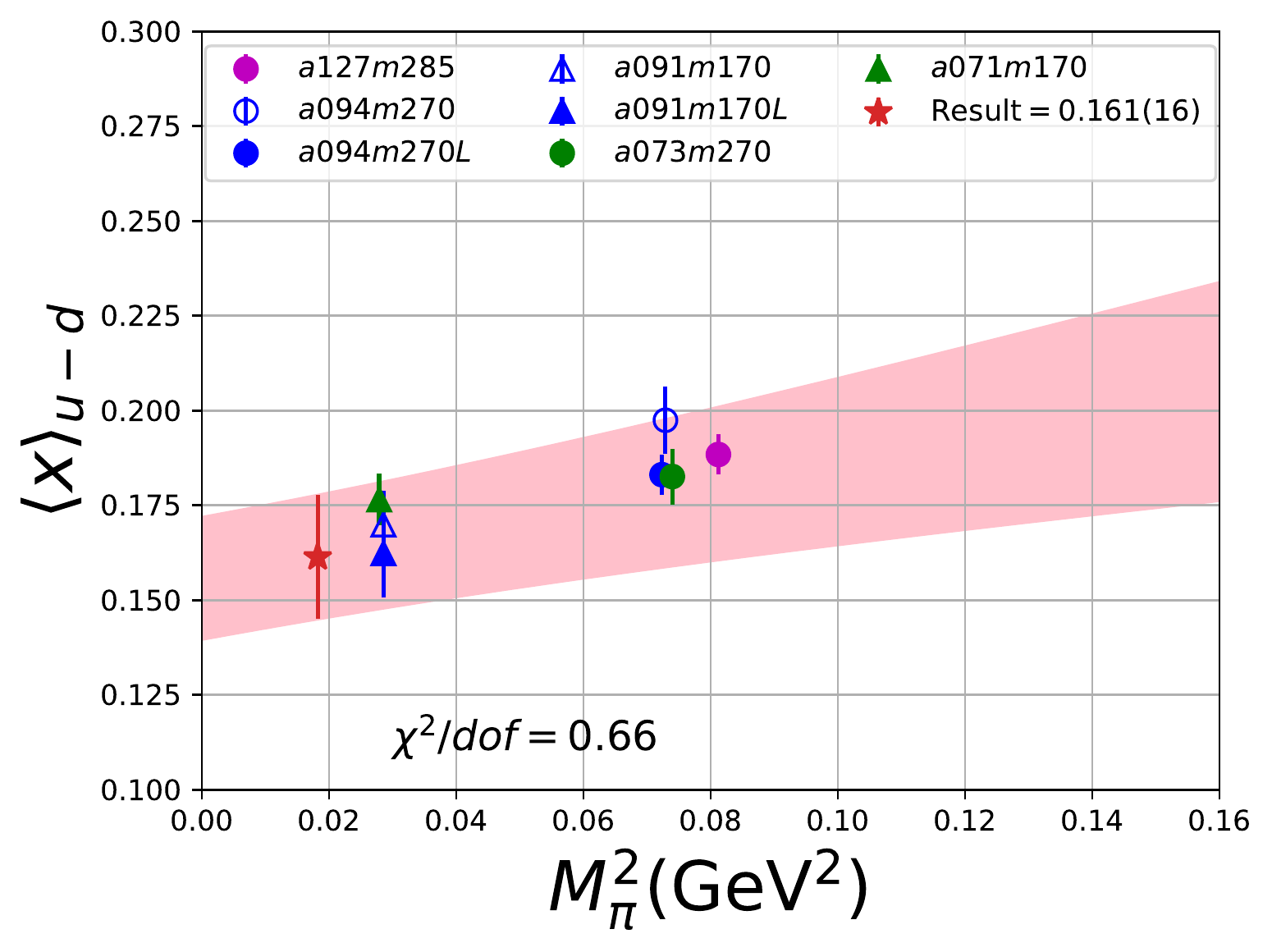}
\includegraphics[angle=0,width=0.32\textwidth]{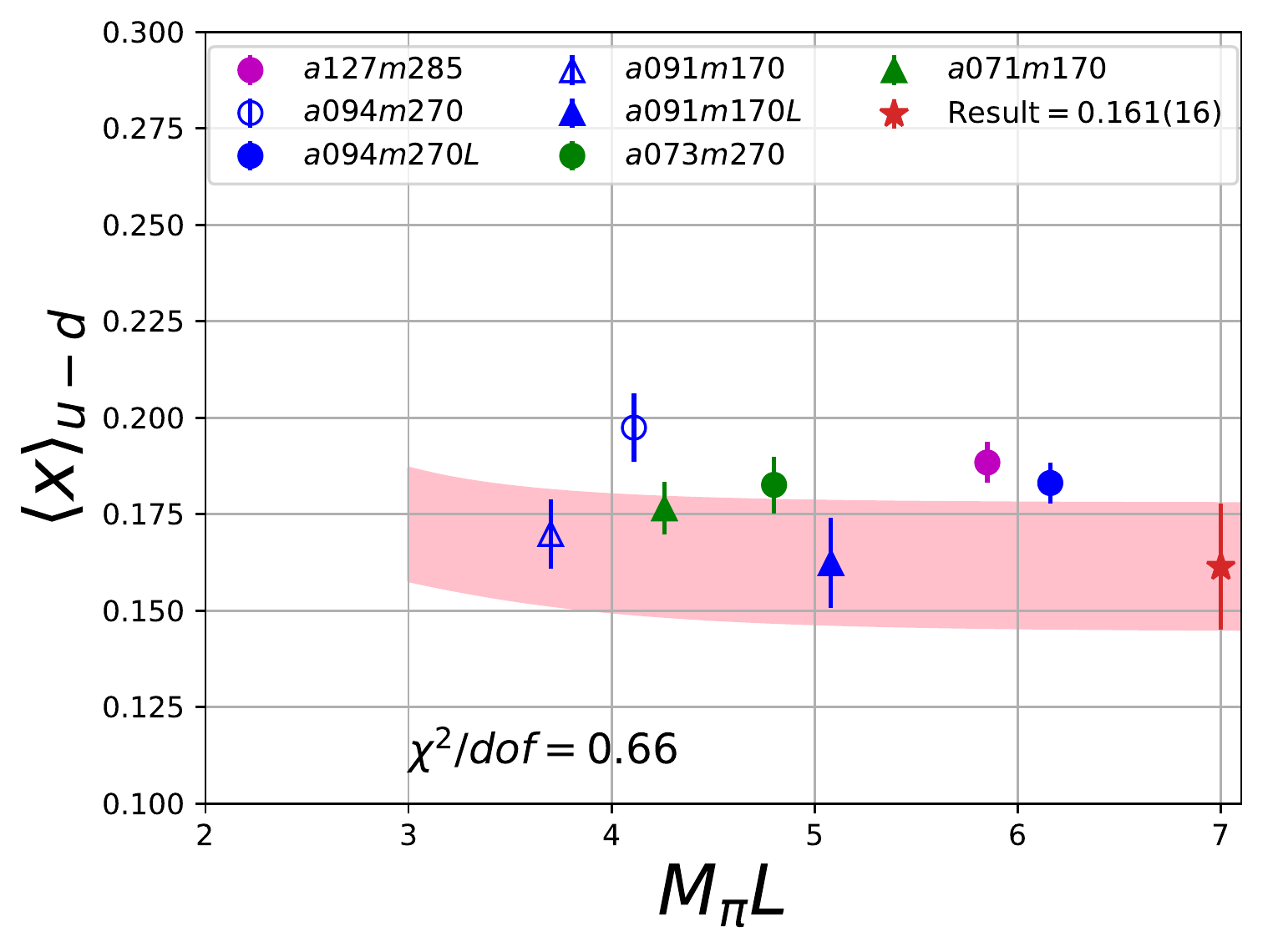}
\end{subfigure}
\begin{subfigure}
\centering
\includegraphics[angle=0,width=0.32\textwidth]{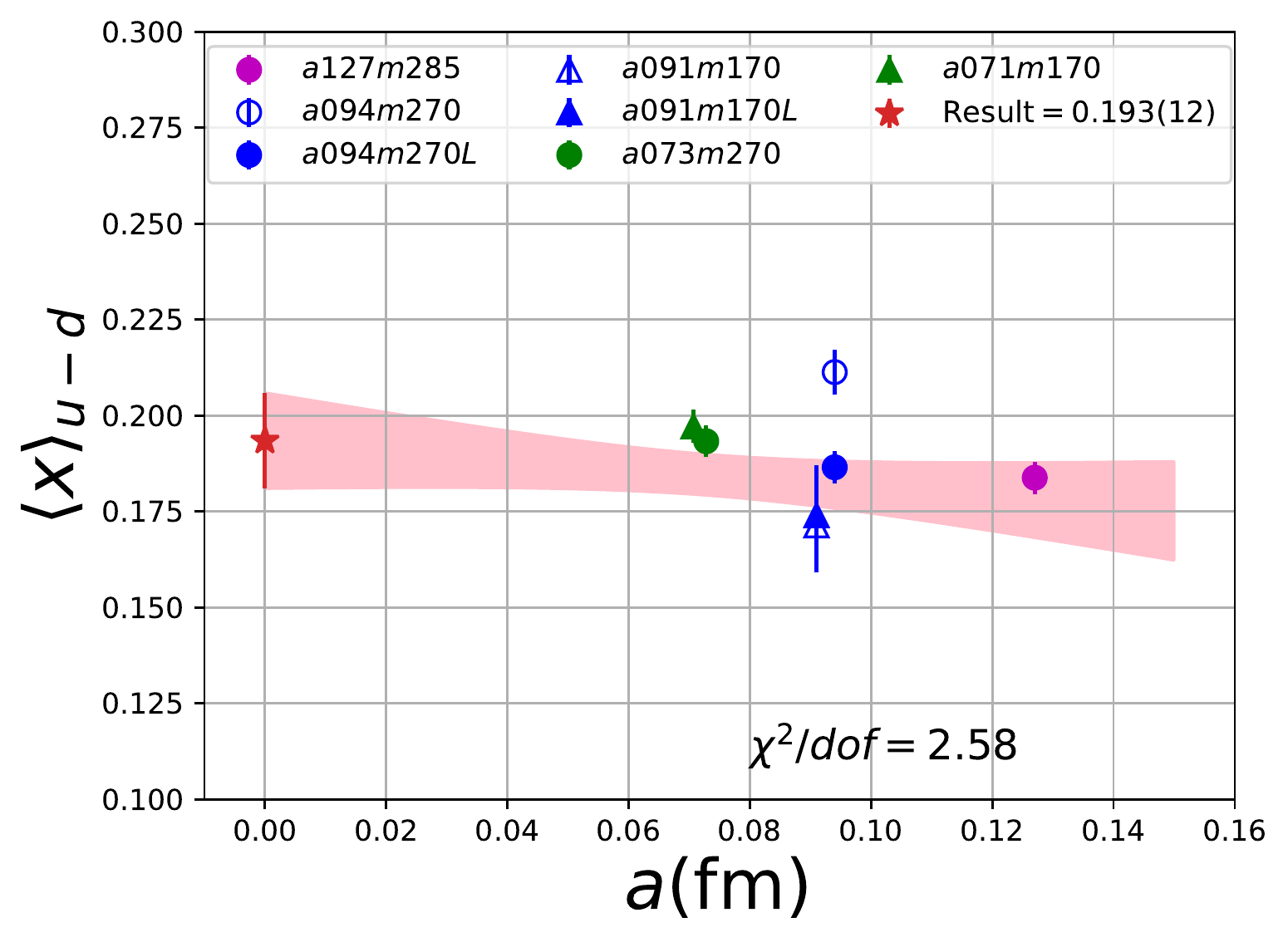}
\includegraphics[angle=0,width=0.32\textwidth]{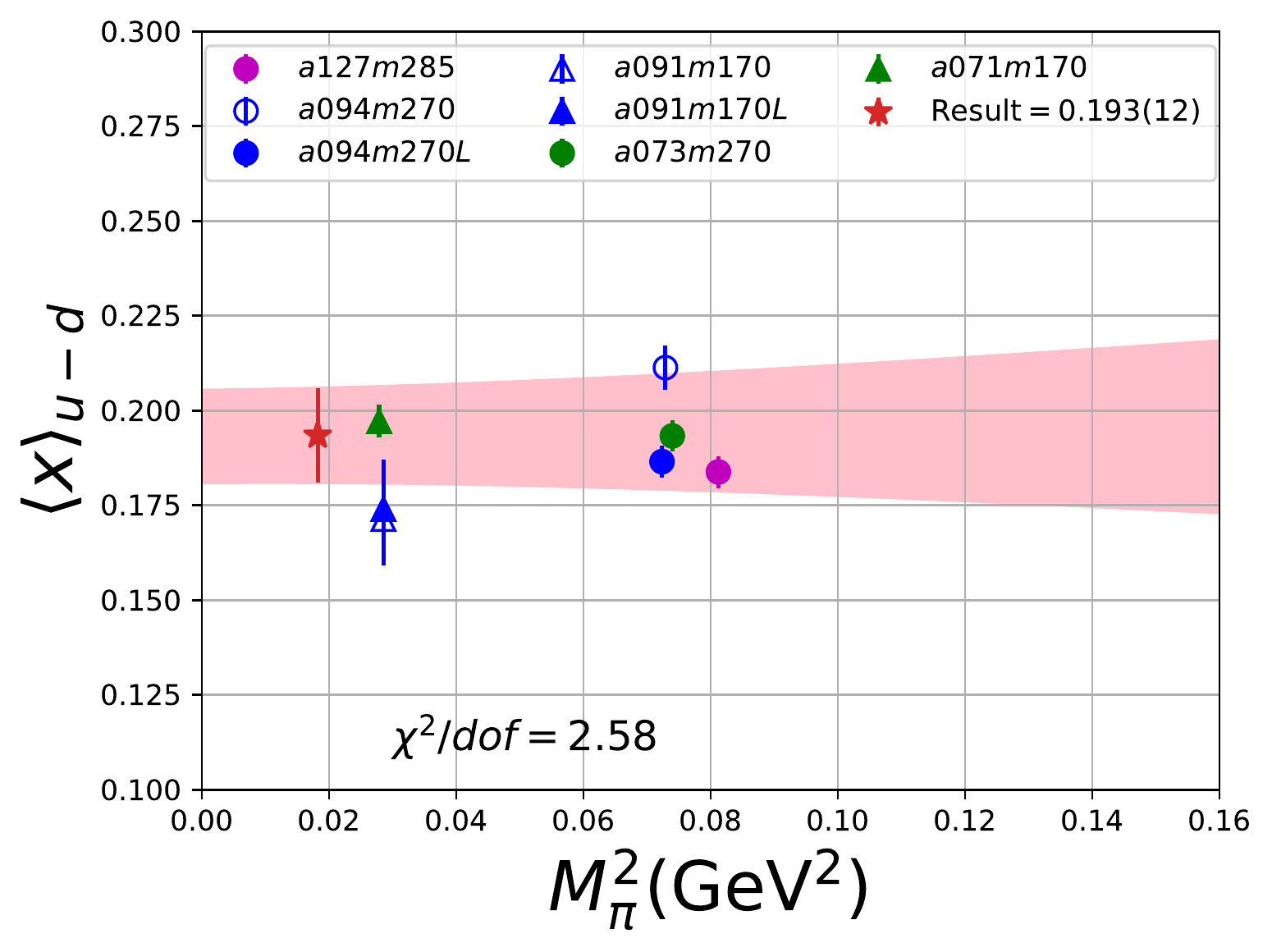}
\includegraphics[angle=0,width=0.32\textwidth]{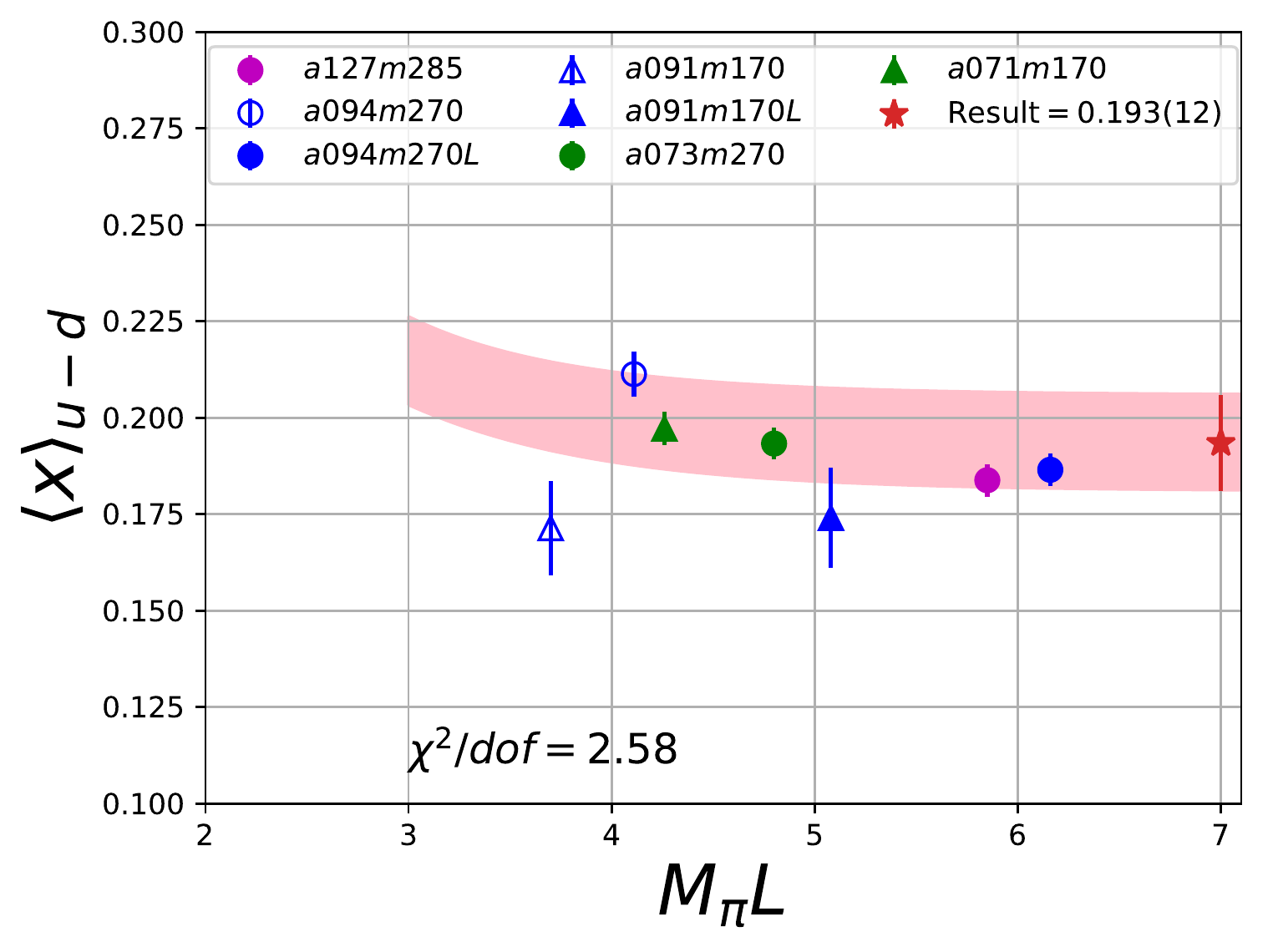}
\end{subfigure}
\vspace{-0.08in}
\caption{Data for the momentum fraction $\la x \ra_{u-d}$ from the seven ensembles 
  renormalized in the $\MSbar$ scheme at $\mu=2$ GeV. The top row
shows data obtained using the $\{4^{N\pi},3^*\}$ fits strategy, middle
from $\{4,3^*\}$ and bottom from $\{4,2^{\rm free}\}$.  The pink band
shows the result of the CCFV fit plotted versus $a$ (left panel),
versus $M_\pi^2$ (middle panel) and versus $M_\pi L$ (right panel)
with the other two variables set to their physical values in each
case.}
\label{fig:momfrac-CCFV-Z-via-averaging}
\end{figure*}

\begin{figure*}[htbp]  
\begin{subfigure}
\centering
\includegraphics[angle=0,width=0.32\textwidth]{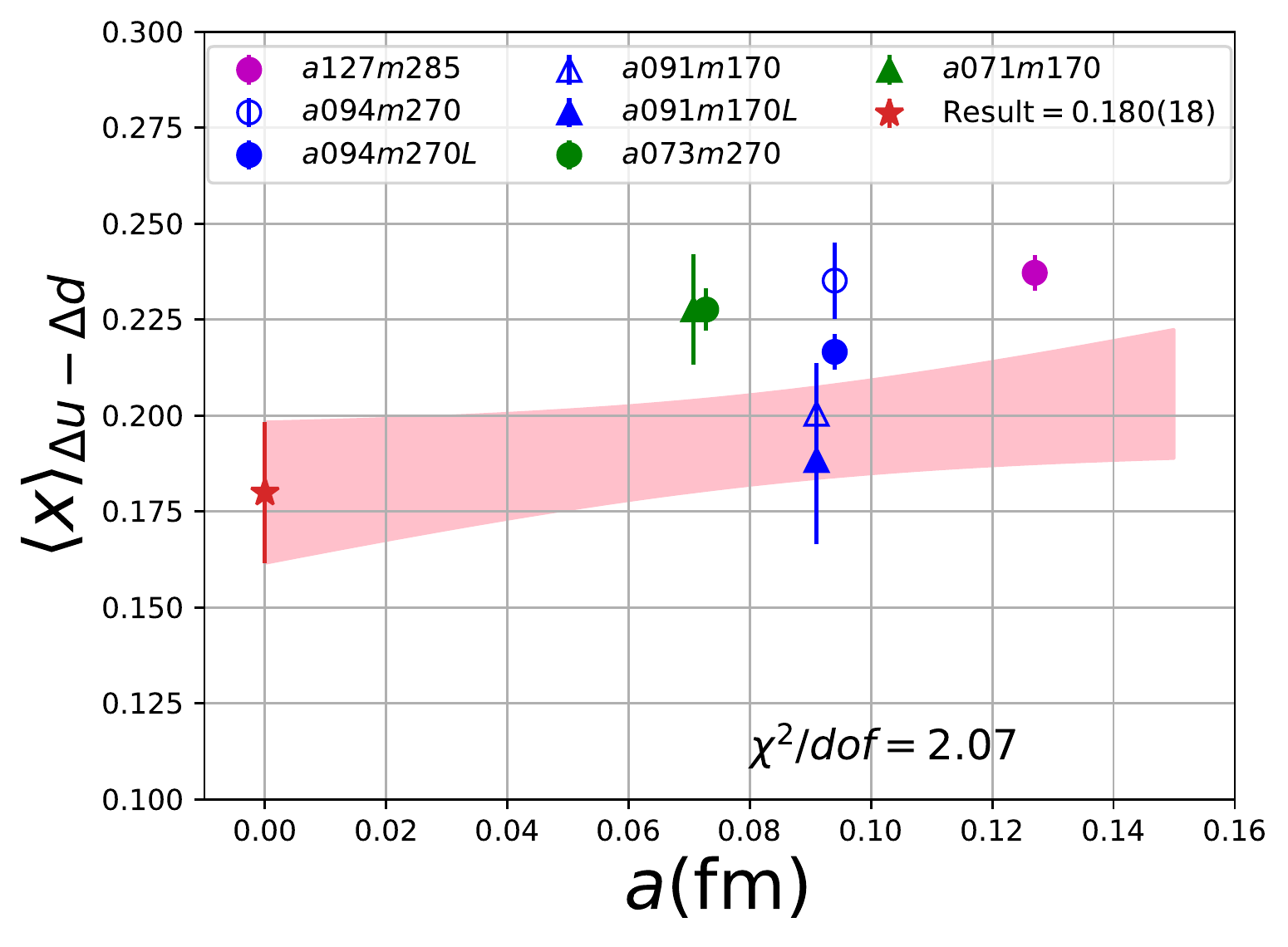}
\includegraphics[angle=0,width=0.32\textwidth]{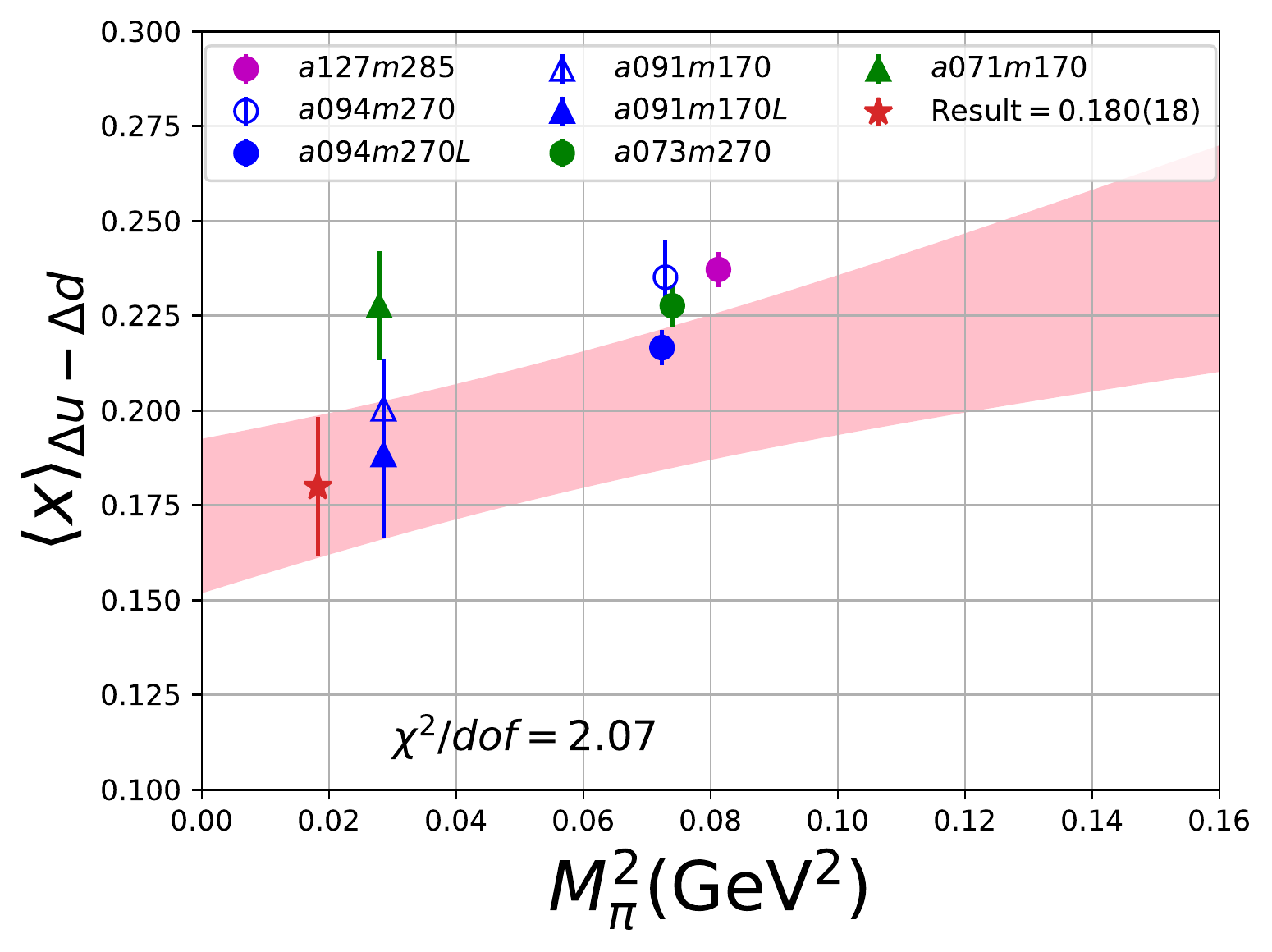}
\includegraphics[angle=0,width=0.32\textwidth]{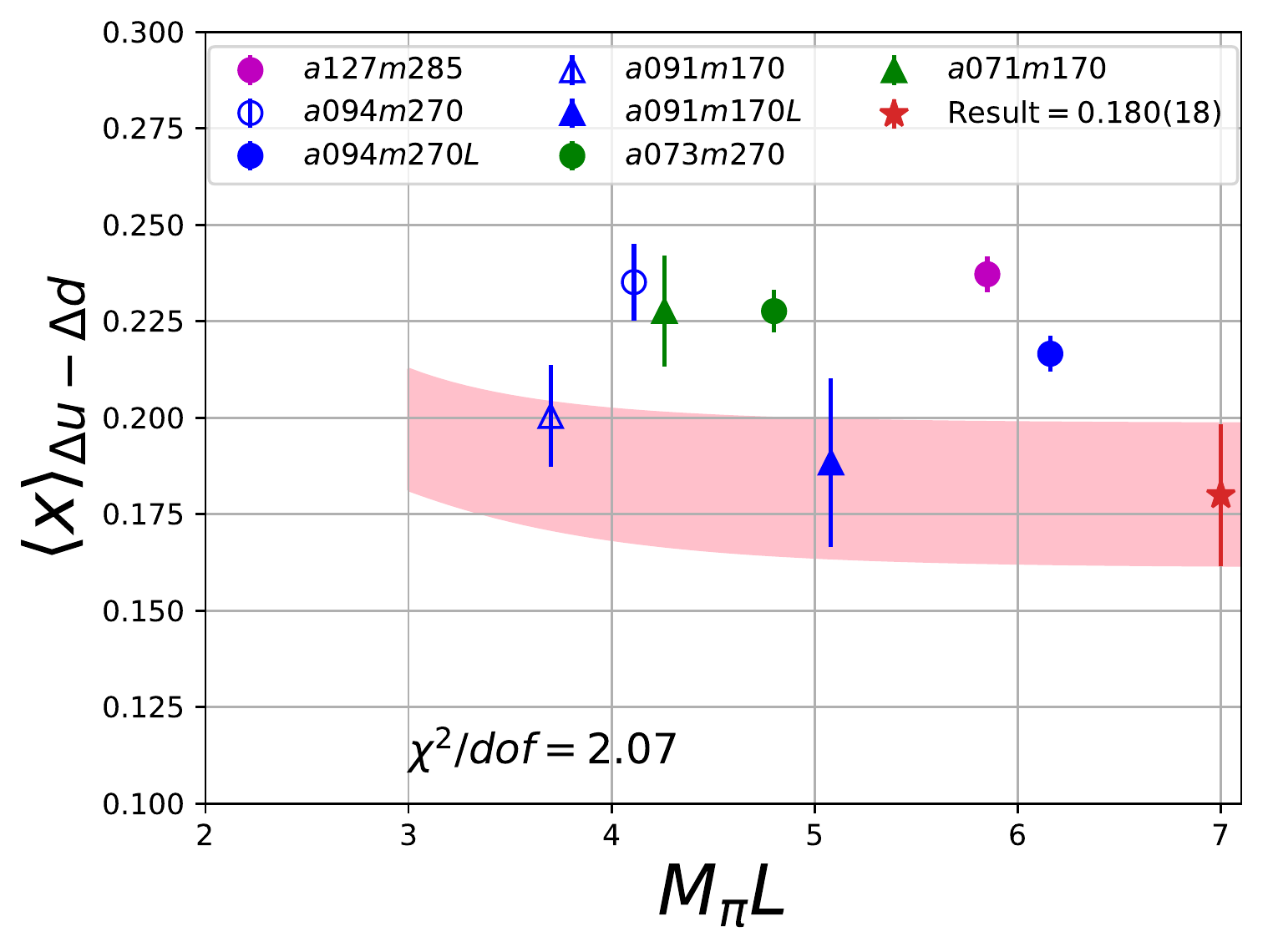}
\end{subfigure}
\begin{subfigure}
\centering
\includegraphics[angle=0,width=0.32\textwidth]{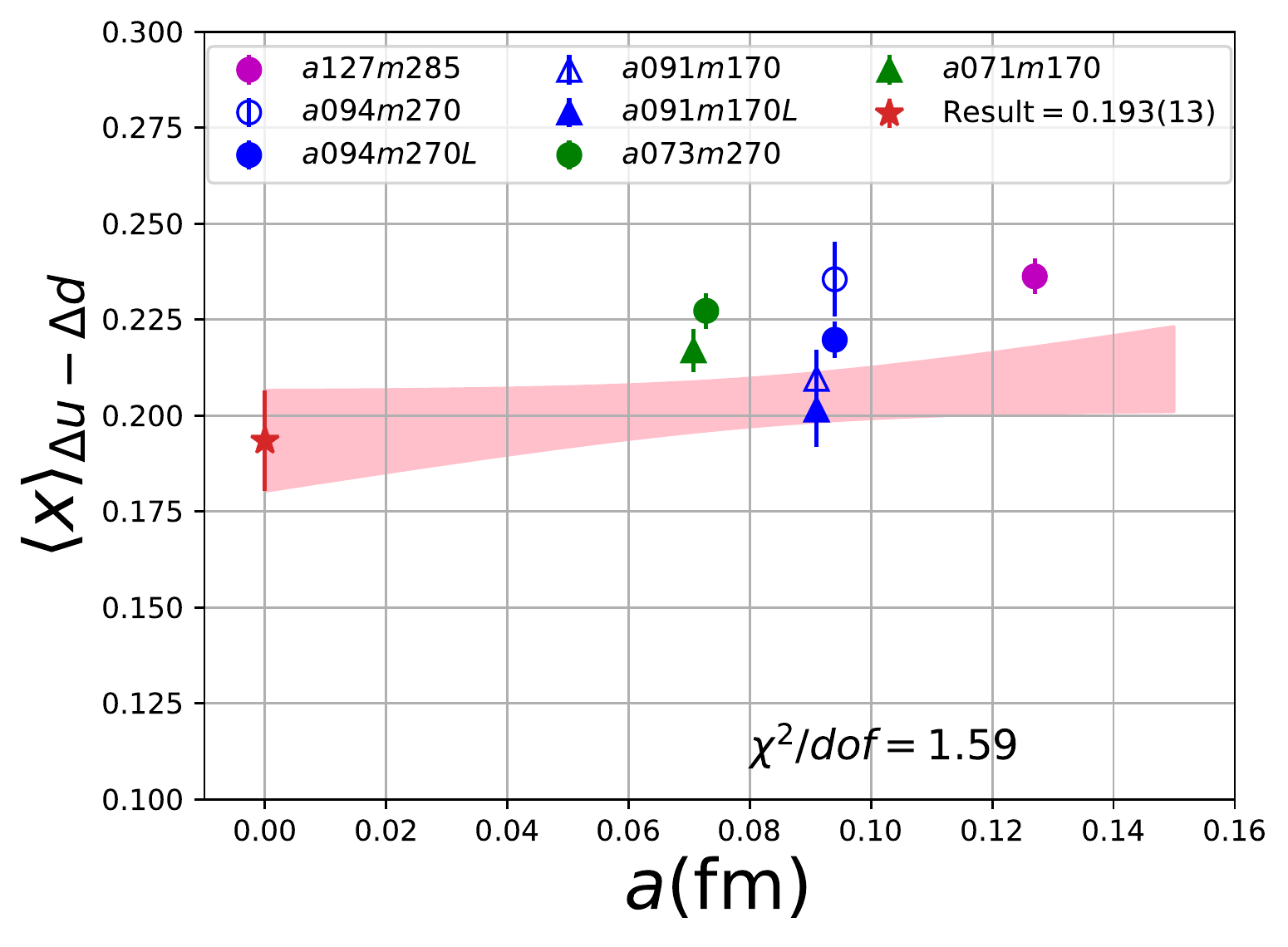}
\includegraphics[angle=0,width=0.32\textwidth]{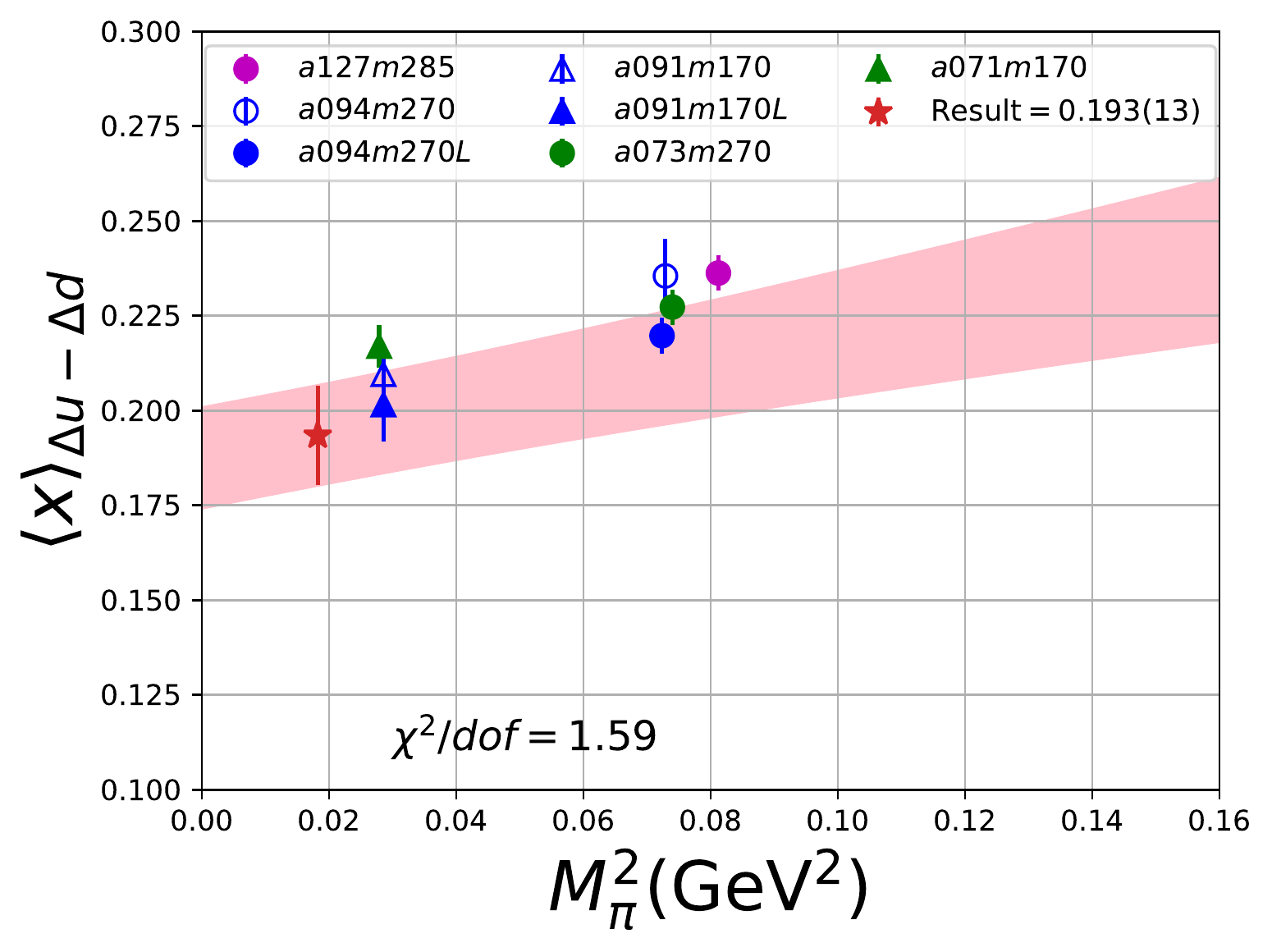}
\includegraphics[angle=0,width=0.32\textwidth]{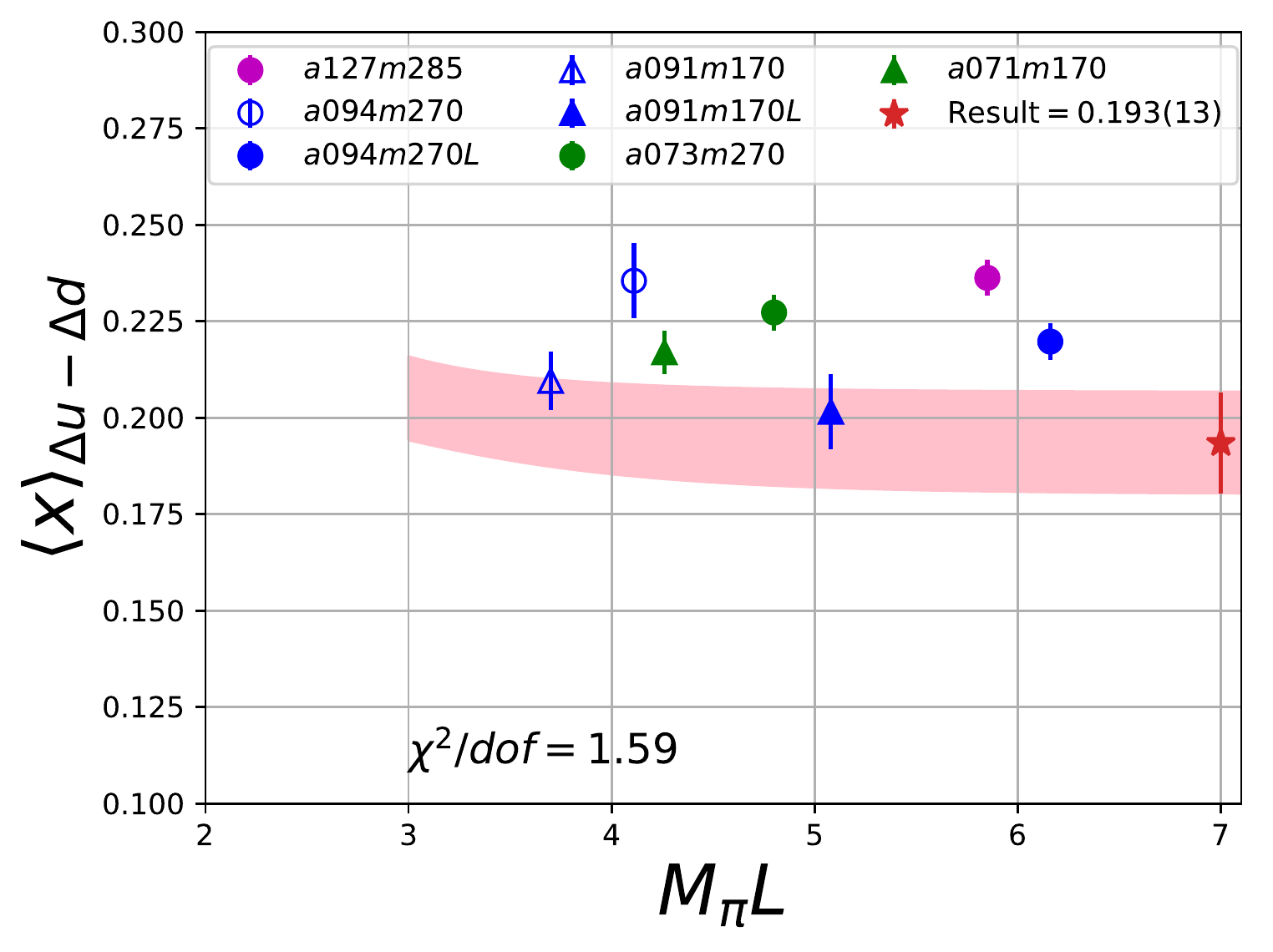}
\end{subfigure}
\begin{subfigure}
\centering
\includegraphics[angle=0,width=0.32\textwidth]{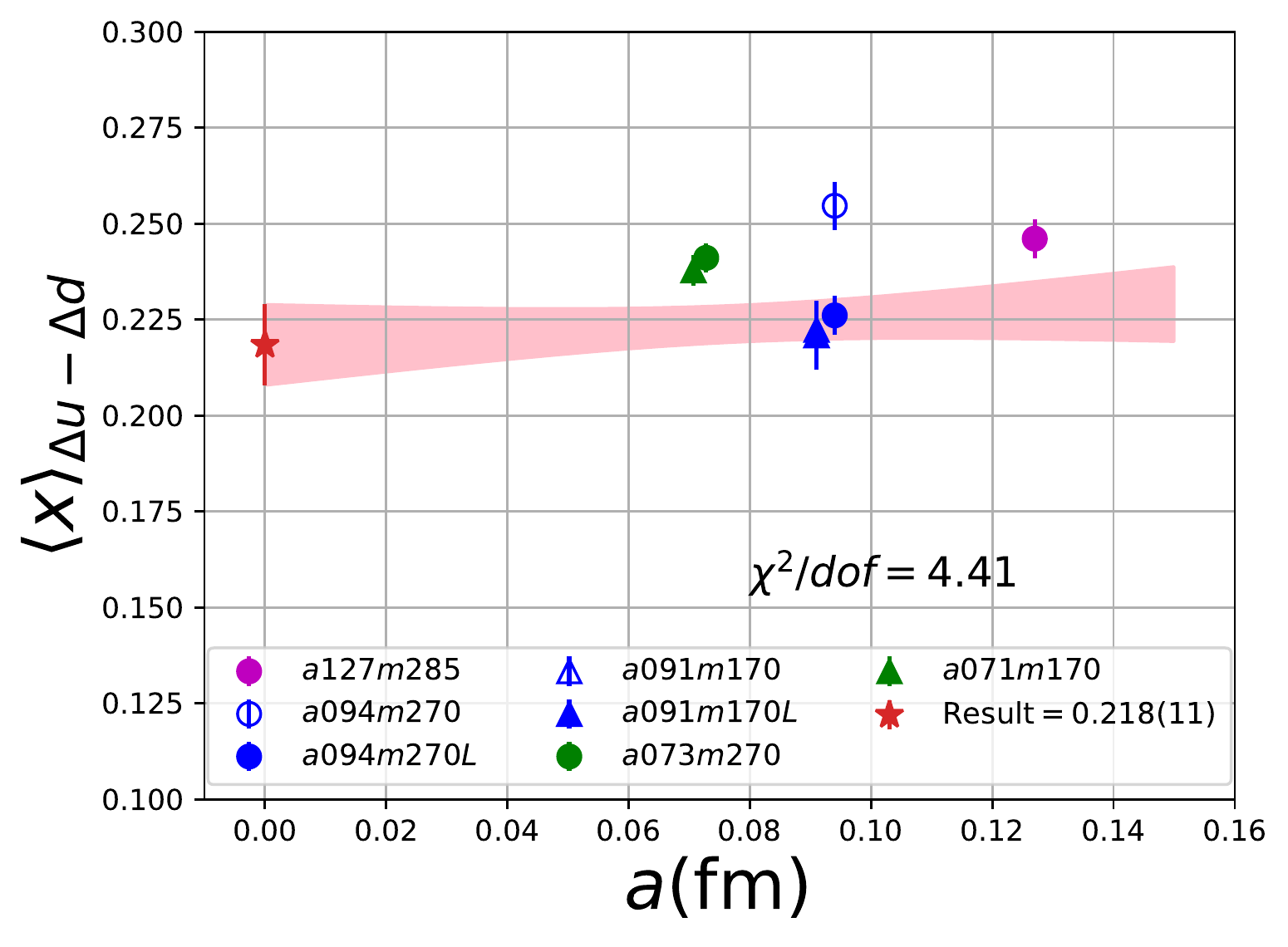}
\includegraphics[angle=0,width=0.32\textwidth]{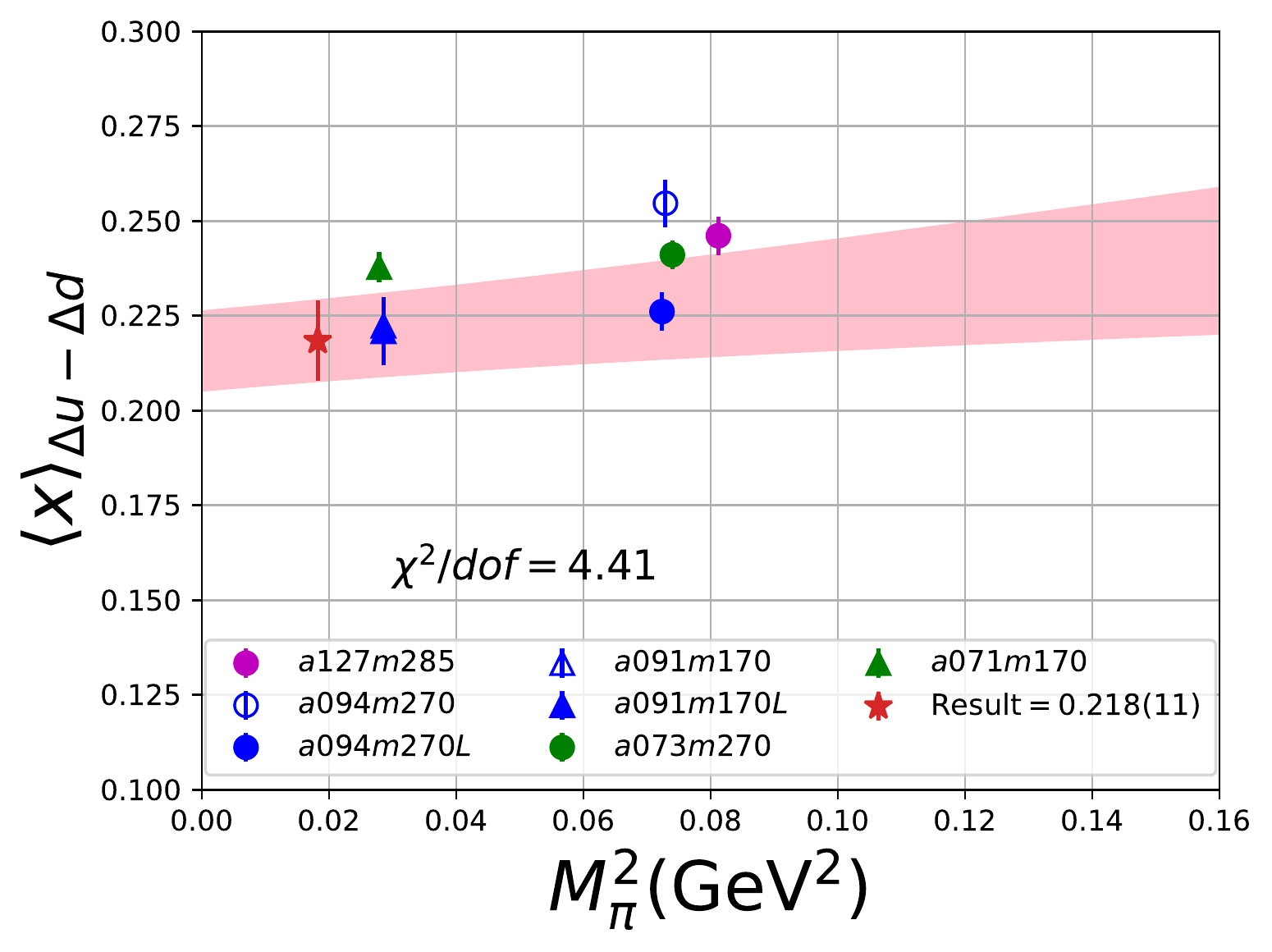}
\includegraphics[angle=0,width=0.32\textwidth]{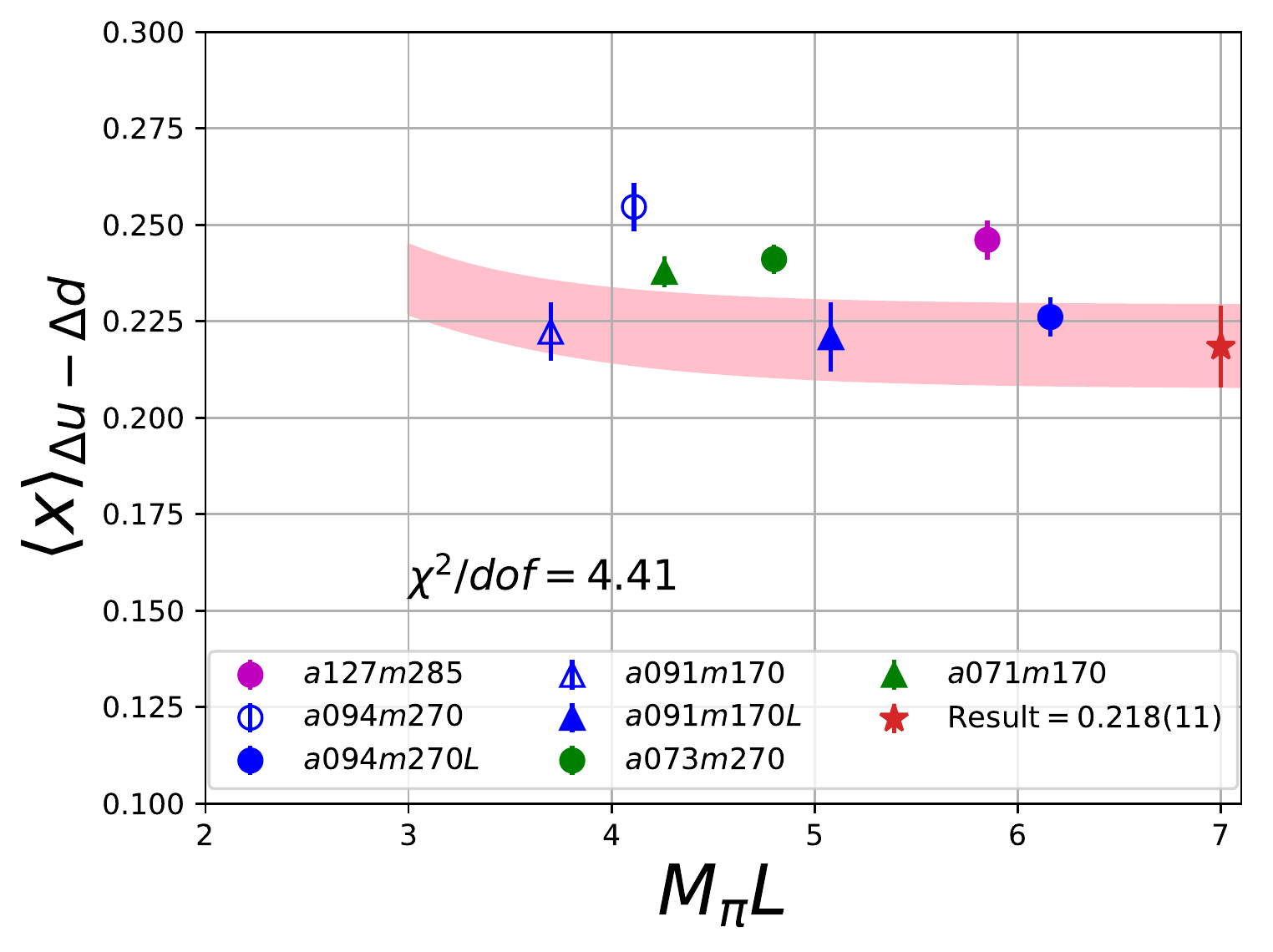}
\end{subfigure}
\vspace{-0.08in}
\caption{Data for the helicity moment $\la x \ra_{\Delta u- \Delta d}$
from the seven ensembles renormalized using method A (Averaging) in
  the $\MSbar$ scheme at $\mu=2$ GeV. The top row shows data obtained
  using the $\{4^{N\pi},3^*\}$ fits strategy, middle from $\{4,3^*\}$
  and bottom from $\{4,2^{\rm free}\}$.  The pink band shows the
  result of the CCFV fit plotted versus $a$ (left panel), versus
  $M_\pi^2$ (middle panel) and versus $M_\pi L$ (right panel) with the
  other two variables set to their physical values in each case.}
\label{fig:helfrac-CCFV-Z-via-averaging}
\end{figure*}

\begin{figure*}[htbp]   
\begin{subfigure}
\centering
\includegraphics[angle=0,width=0.32\textwidth]{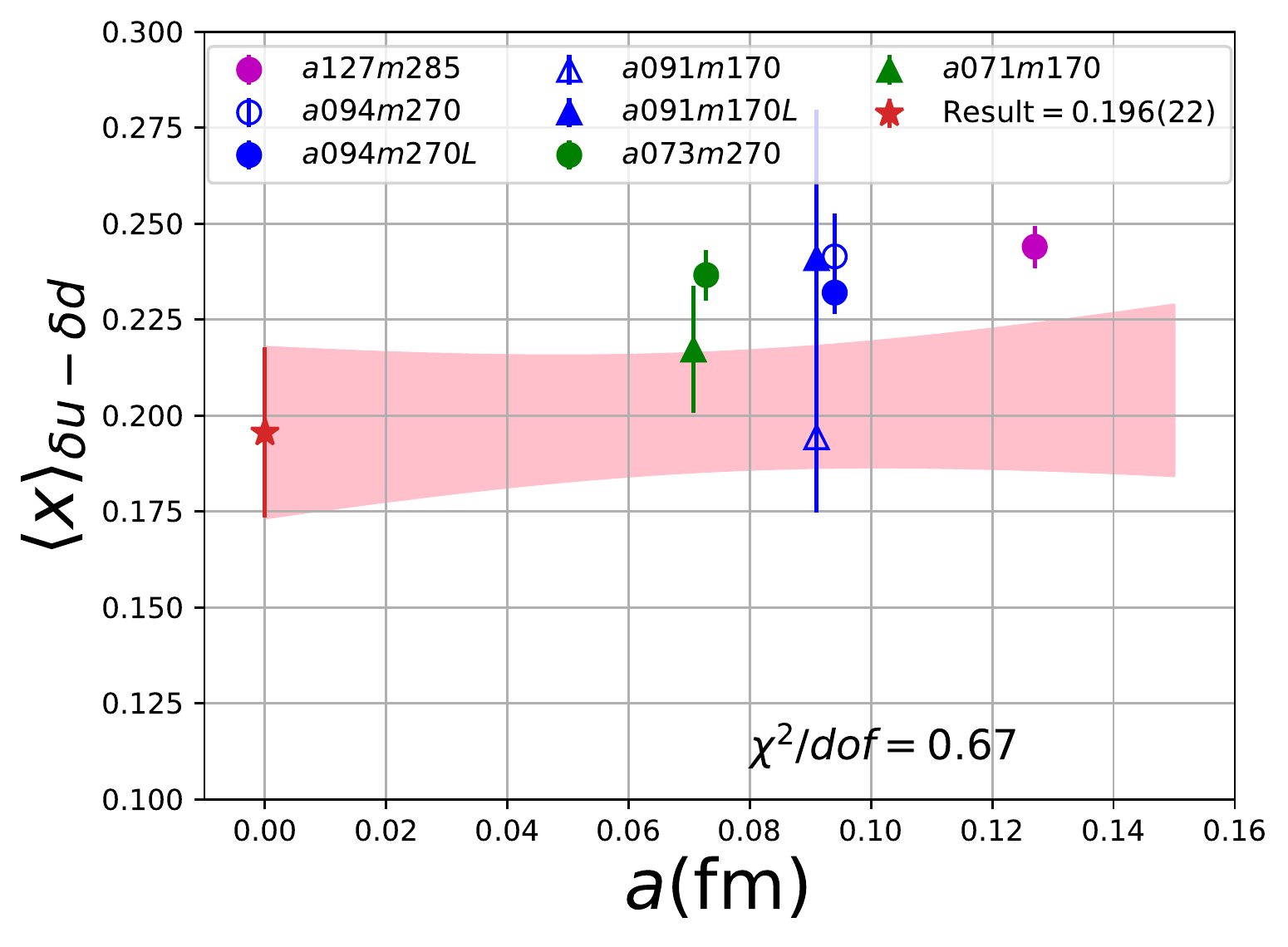}
\includegraphics[angle=0,width=0.32\textwidth]{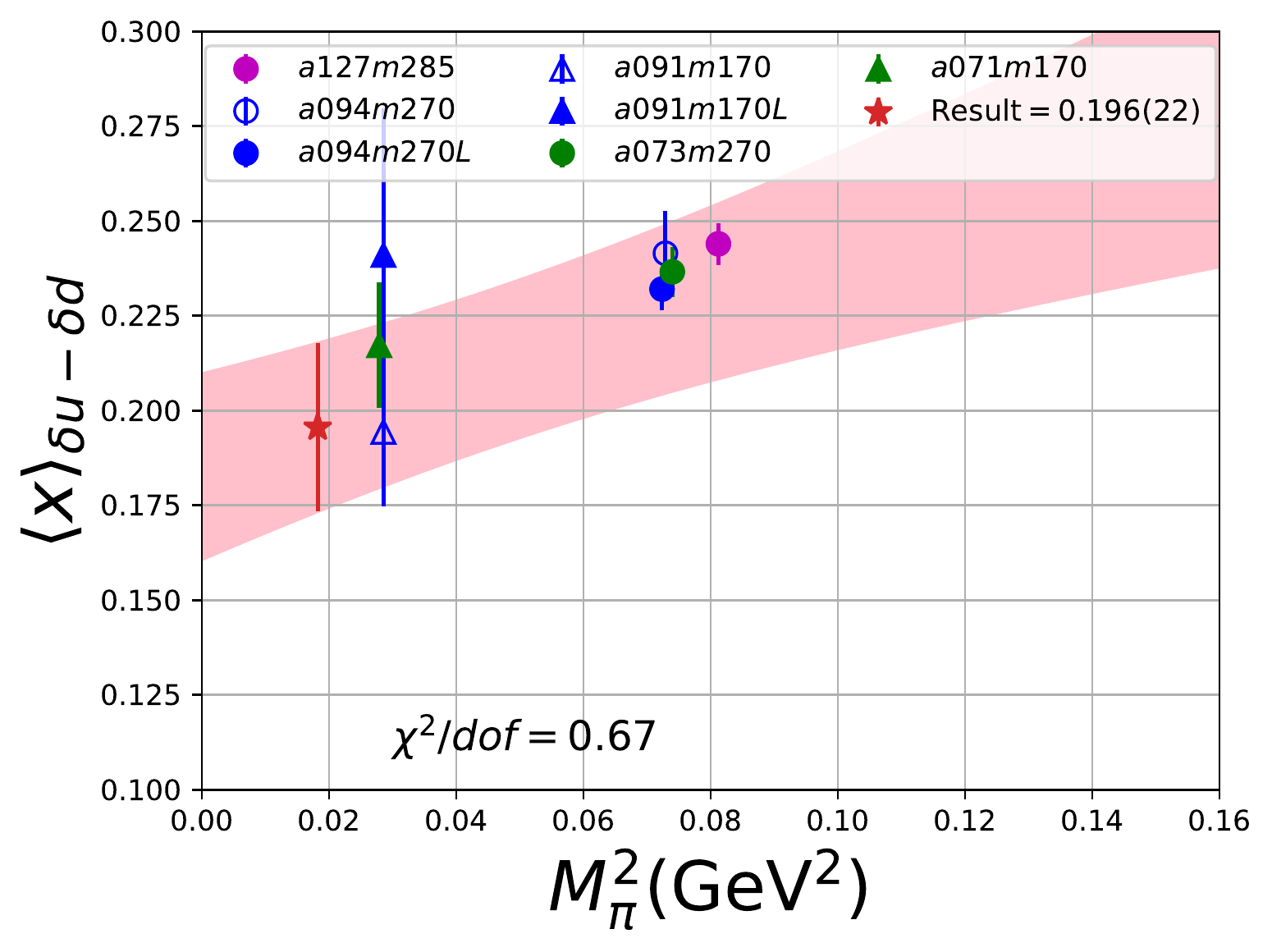}
\includegraphics[angle=0,width=0.32\textwidth]{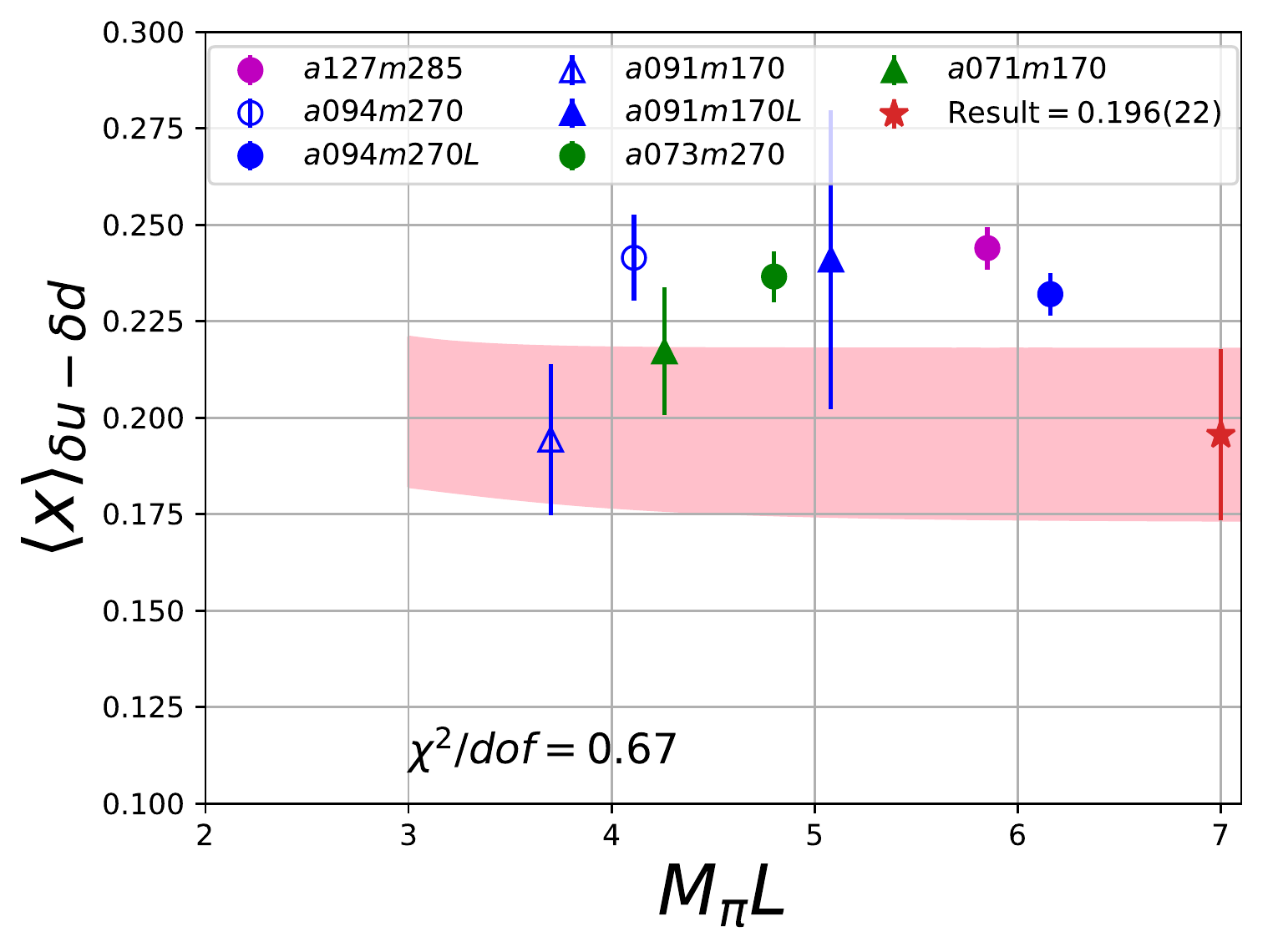}
\end{subfigure}
\begin{subfigure}
\centering
\includegraphics[angle=0,width=0.32\textwidth]{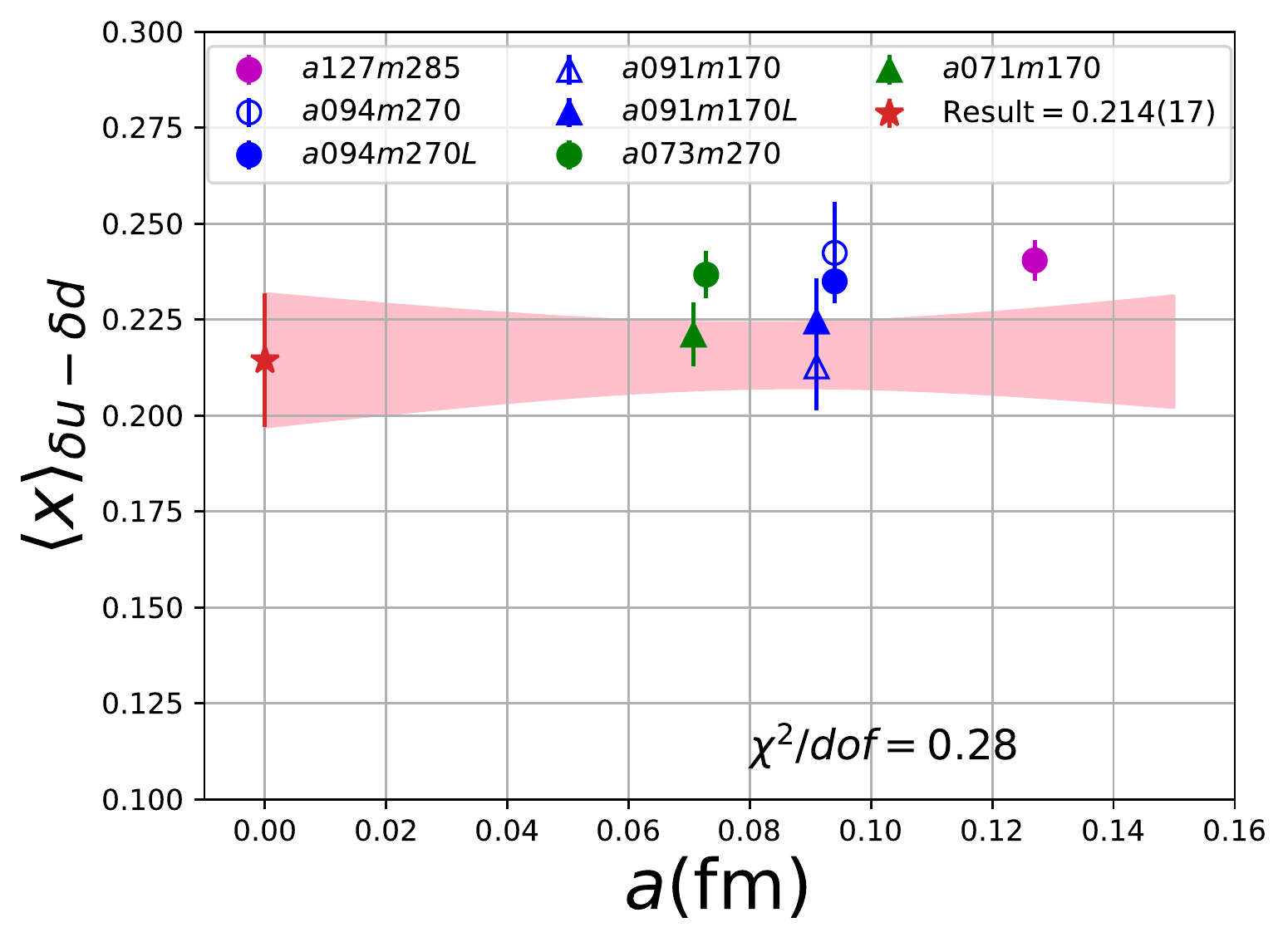}
\includegraphics[angle=0,width=0.32\textwidth]{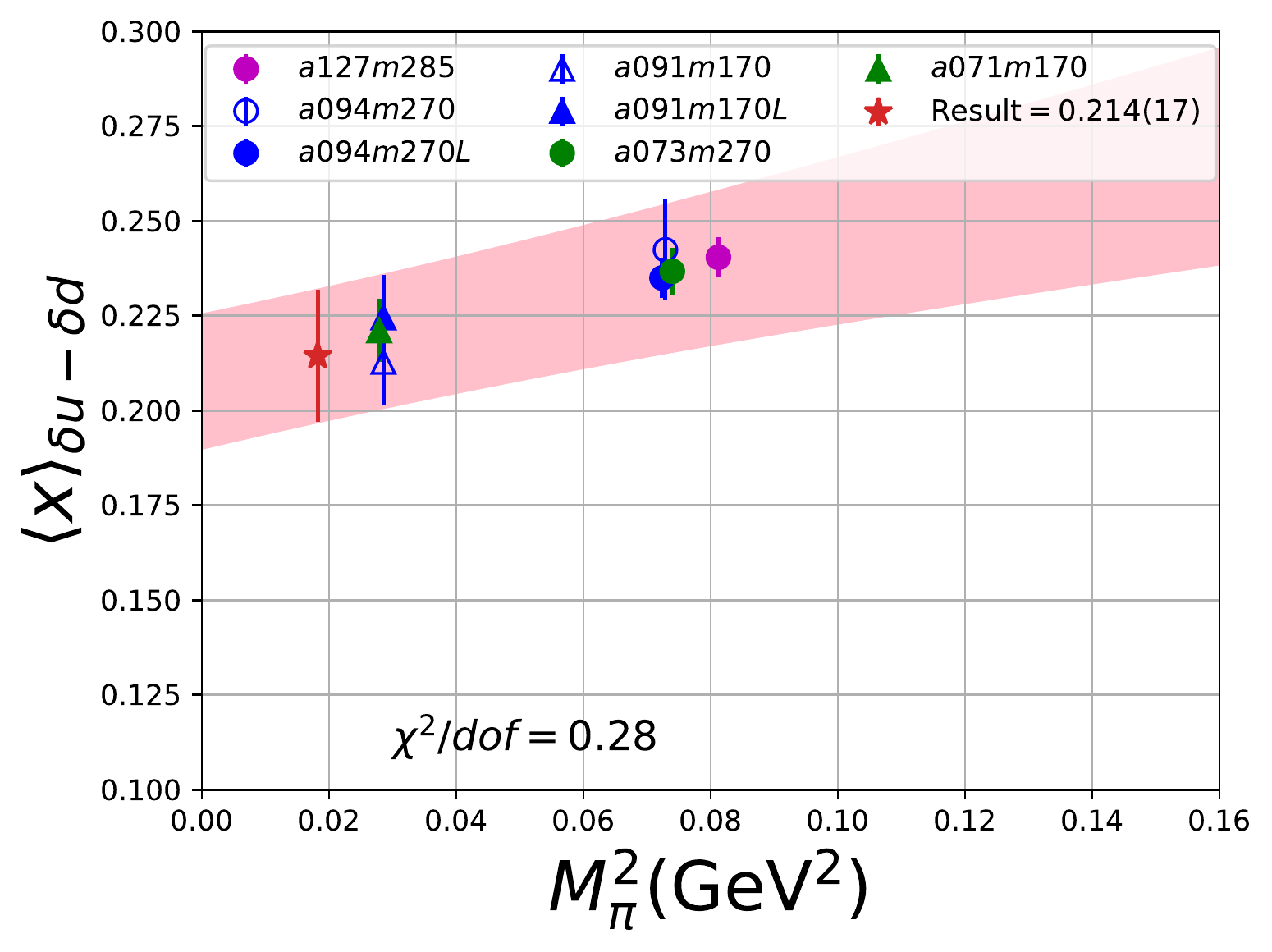}
\includegraphics[angle=0,width=0.32\textwidth]{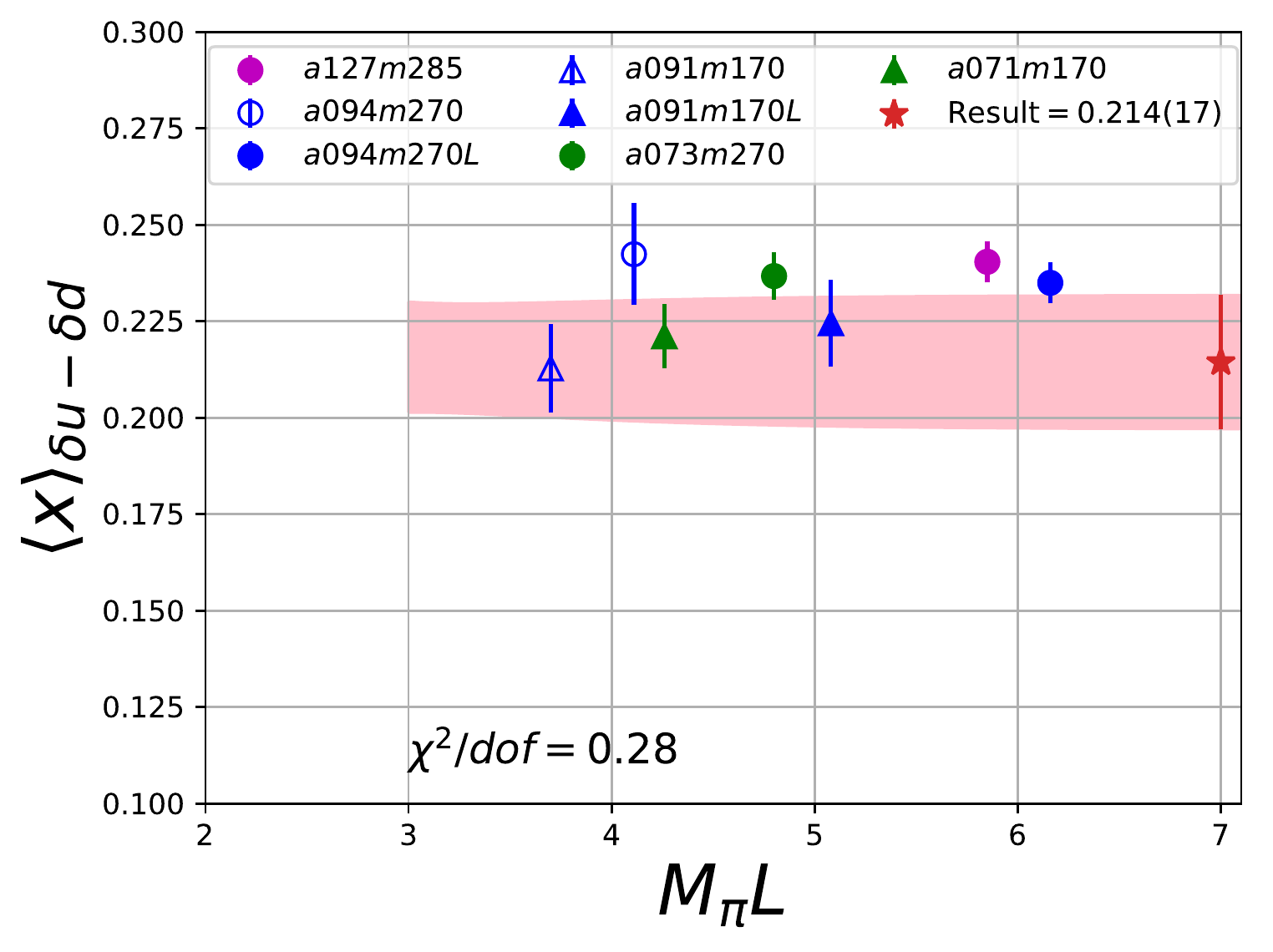}
\end{subfigure}
\begin{subfigure}
\centering
\includegraphics[angle=0,width=0.32\textwidth]{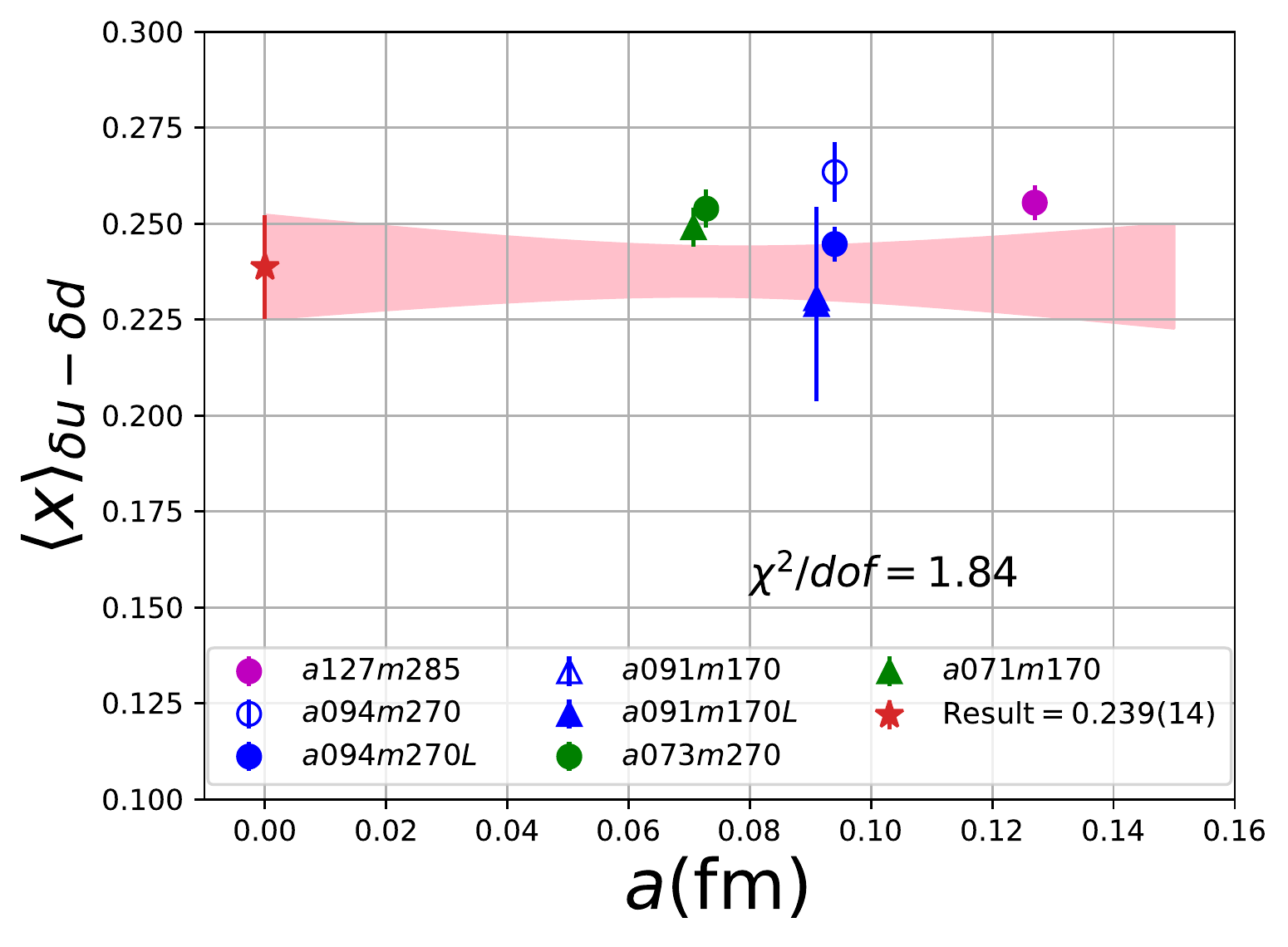}
\includegraphics[angle=0,width=0.32\textwidth]{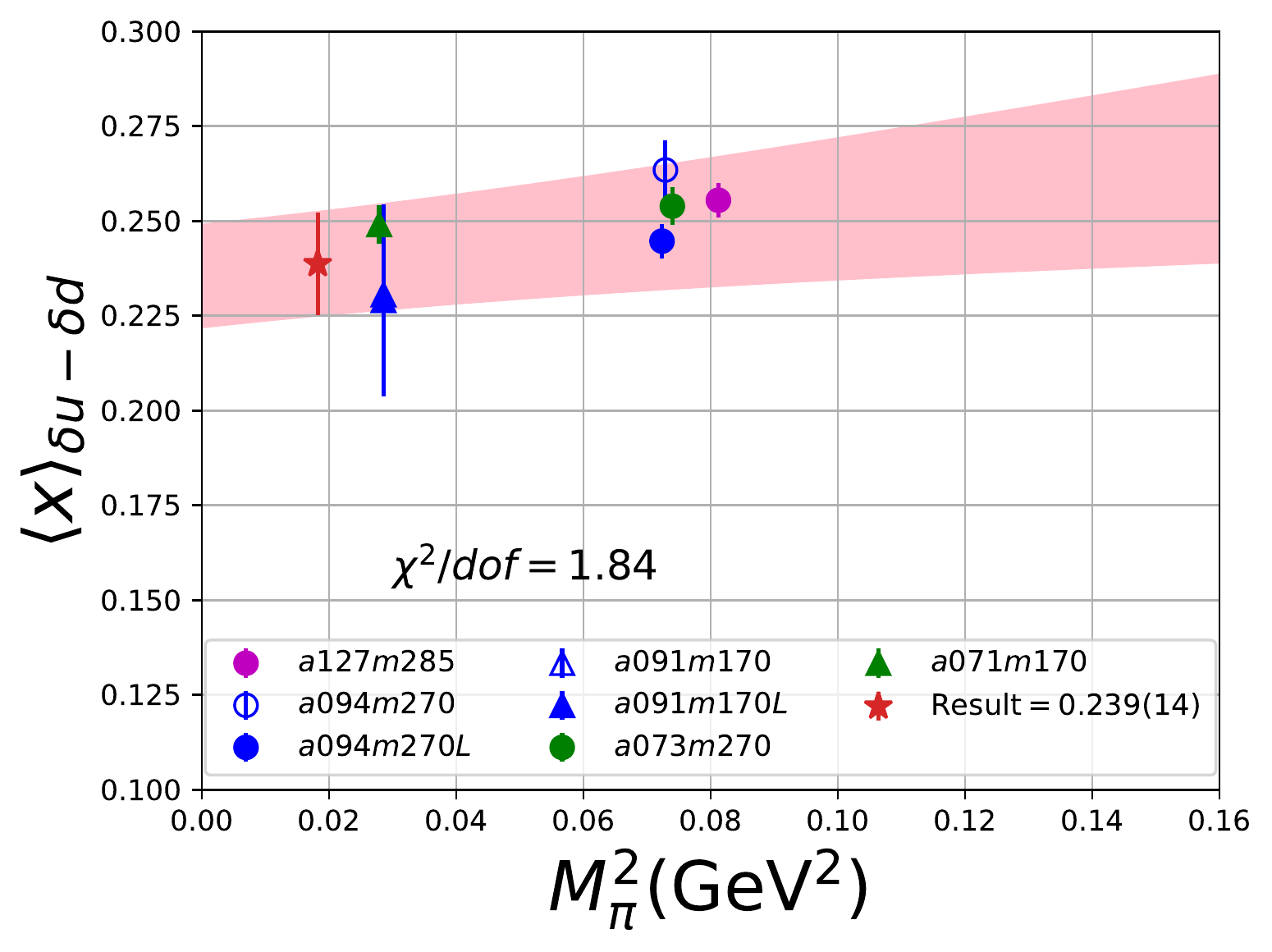}
\includegraphics[angle=0,width=0.32\textwidth]{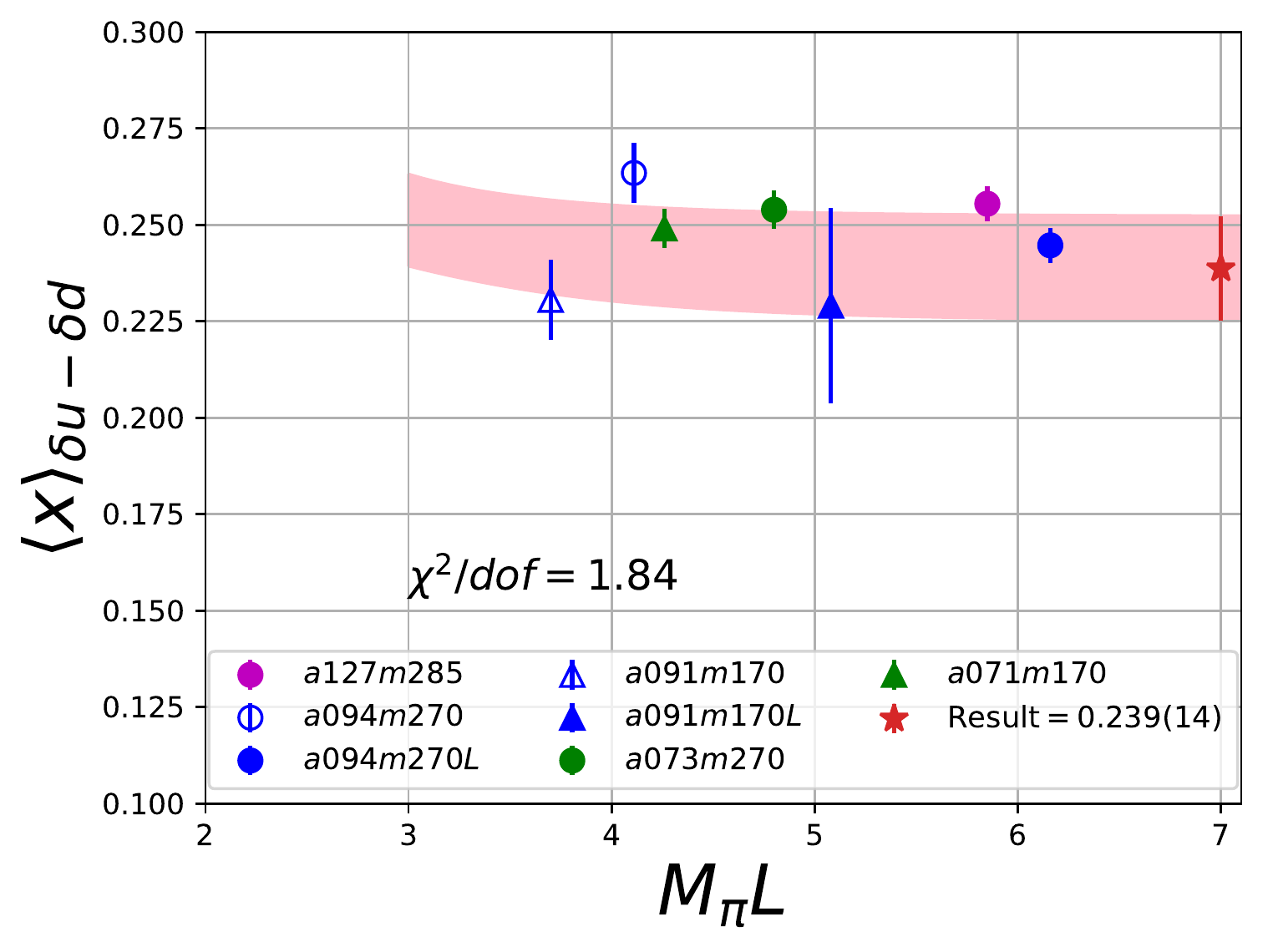}
\end{subfigure}
\vspace{-0.08in}
\caption{Data for the transversity  moment $\la x \ra_{\delta u- \delta d}$
from the seven ensembles renormalized in the $\MSbar$ scheme at $\mu=2$ GeV. The top row
shows data obtained using the $\{4^{N\pi},3^*\}$ fits strategy, middle
from $\{4,3^*\}$ and bottom from $\{4,2^{\rm free}\}$.  The pink band
shows the result of the CCFV fit plotted versus $a$ (left panel),
versus $M_\pi^2$ (middle panel) and versus $M_\pi L$ (right panel)
with the other two variables set to their physical values in each
case.}
\label{fig:transvmom-CCFV-Z-via-averaging}
\end{figure*}

To obtain the final, physical results at $M_\pi=135$~MeV, $M_\pi L \to
\infty$ and $a=0$, we make a simultaneous CCFV fit keeping only the
leading correction term in each variable:
\begin{flalign}
\la x \ra (M_\pi; a;L) &= c_1 + c_2 a +c_3 M_\pi^2  
+ c_4 \frac{M_\pi^2~ e^{-M_\pi L}}{\sqrt{M_\pi L}} \,. 
\label{eq:CCFV}
\end{flalign}
Note that, since the operators are not $O(a)$ improved in our
clover-on-clover formulation, we take the discretization errors to
start with a term linear in $a$. The fits to the data renormalized
using method A for the three strategies are shown in
Figs.~\ref{fig:momfrac-CCFV-Z-via-averaging},~\ref{fig:helfrac-CCFV-Z-via-averaging}
and~\ref{fig:transvmom-CCFV-Z-via-averaging} and the results are
summarized in Table~\ref{tab:CCFVresults}.  

The dependence on $a$ is found to be small. The significant variation
is with $M_\pi^2$, and this is the main discriminant between the three
strategies. The smaller the $\Delta M_1$, the larger is the
extrapolation in the ESC fits (difference between the data at the largest $\tau$
and $\tau = \infty$ extrapolated value) and a larger slope versus $M_\pi^2$ in the CCFV fits.  The
overall consequence for all three moments is that estimates increase
by about 0.02 between
$\{4^{N\pi},3^\ast\} \to \{4,3^\ast\} \to \{4,2^{\rm free}\}$. Based
on the observation that the $\{4,2^{\rm free}\}$ fits do not prefer
the small $\Delta M_1^{4^{N\pi}}$, but are closer to $\Delta M_1^{4}$
(momentum fraction) or larger (helicity and transversity), we take the $ \{4,3^\ast\}$ 
results for our best.  However, to
account for possible bias due to not having resolved which excited
state makes the dominant contribution, we add a second, systematic,
error of $0.02$ to the final results based on the observed differences
in estimates between the three strategies.

The results from the two renormalization methods summarized in
Table~\ref{tab:CCFVresults} overlap--the differences are a fraction of
the errors from the rest of the analysis. Also, note that these
differences are much smaller than the differences between the $Z$'s
from the two methods.  This pattern is expected provided the
differences in the $Z$'s are largely due to discretization errors that
are removed on taking the continuum limit.

Lastly, a comparison between the chiral-continuum (CC) and CCFV fit
results summarized in Table~\ref{tab:CCFVresults} indicate up to 10\%
decrease due to the finite volume correction term, however, this is
comparable to the size of the final errors.  Also, this effect is
clear only between the $a094m270$ and $a094m270L$ data as shown in
Tables~\ref{tab:renormalized-moments-A}
and~\ref{tab:renormalized-moments-B}.  Consequently, most of the
variation in the CCFV fit occurs for $M_\pi L < 4$. The result of a CC
fit to five larger volume ensembles (excluding $a094m270$ and
$a091m170$) lie in between the CC and CCFV data shown in
Tables~\ref{tab:renormalized-moments-A}
and~\ref{tab:renormalized-moments-B}.  With these caveats, for
present, we choose to present final results from the full data set
using CCFV fits.

\begin{figure*}[htbp]  
\begin{subfigure}
\centering
\includegraphics[angle=0,width=0.32\textwidth]{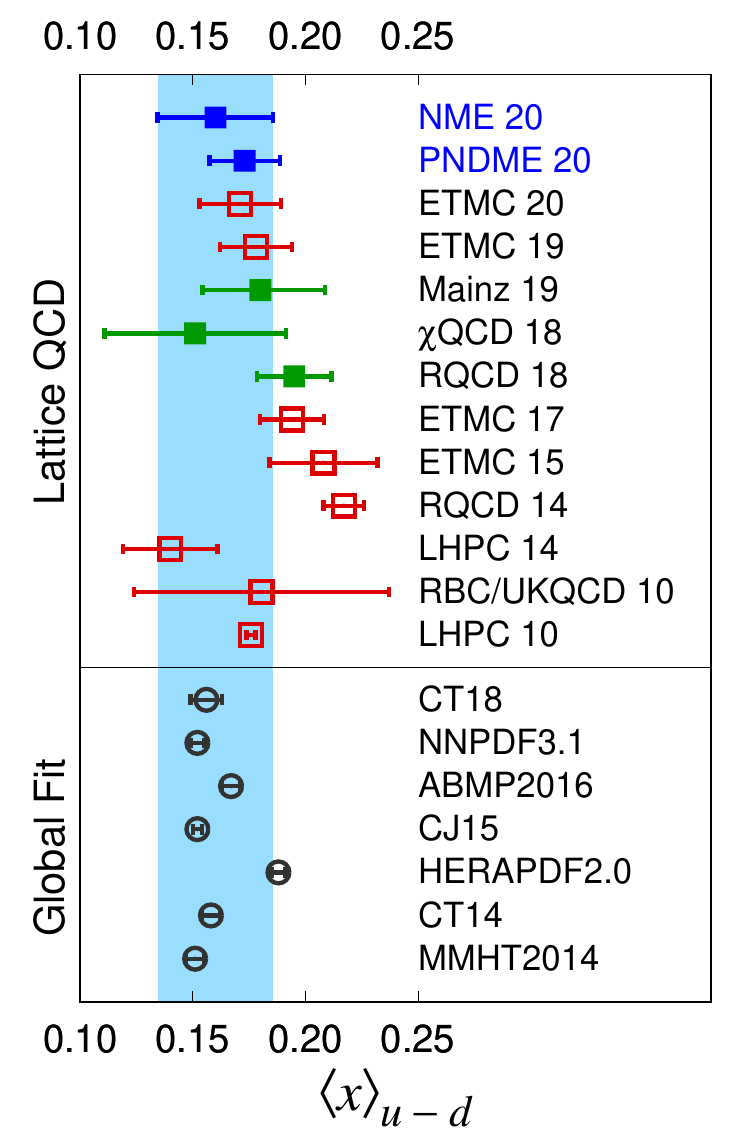}
\includegraphics[angle=0,width=0.32\textwidth]{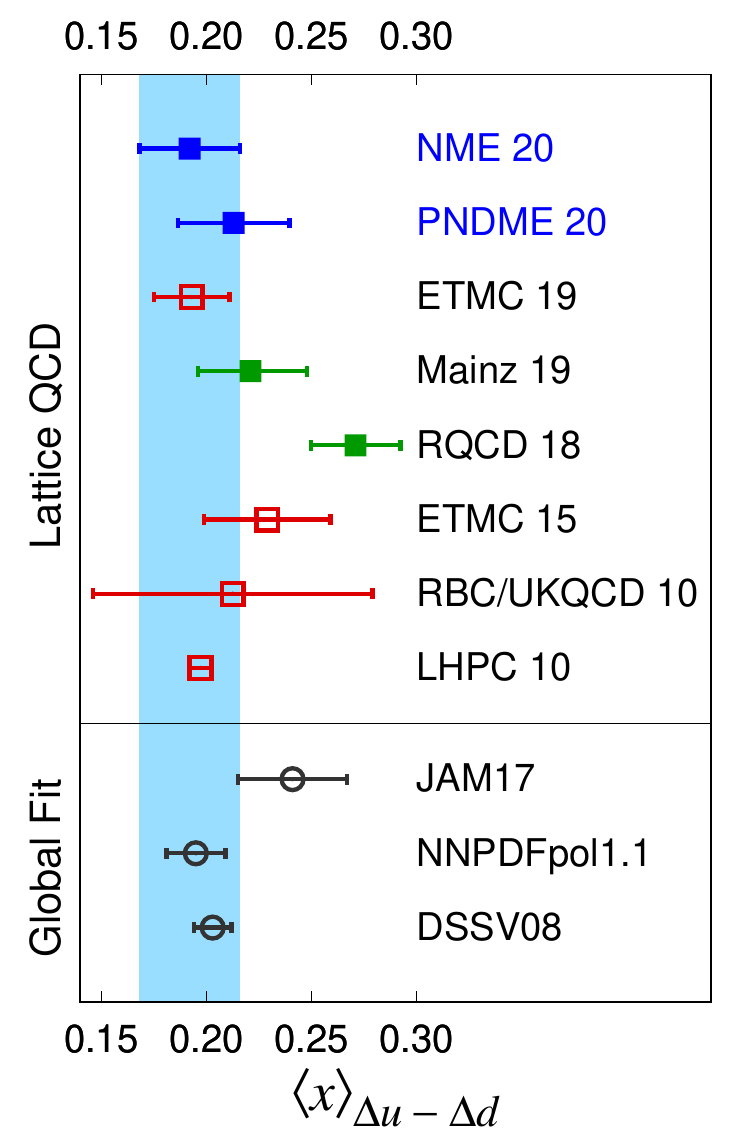}
\includegraphics[angle=0,width=0.32\textwidth]{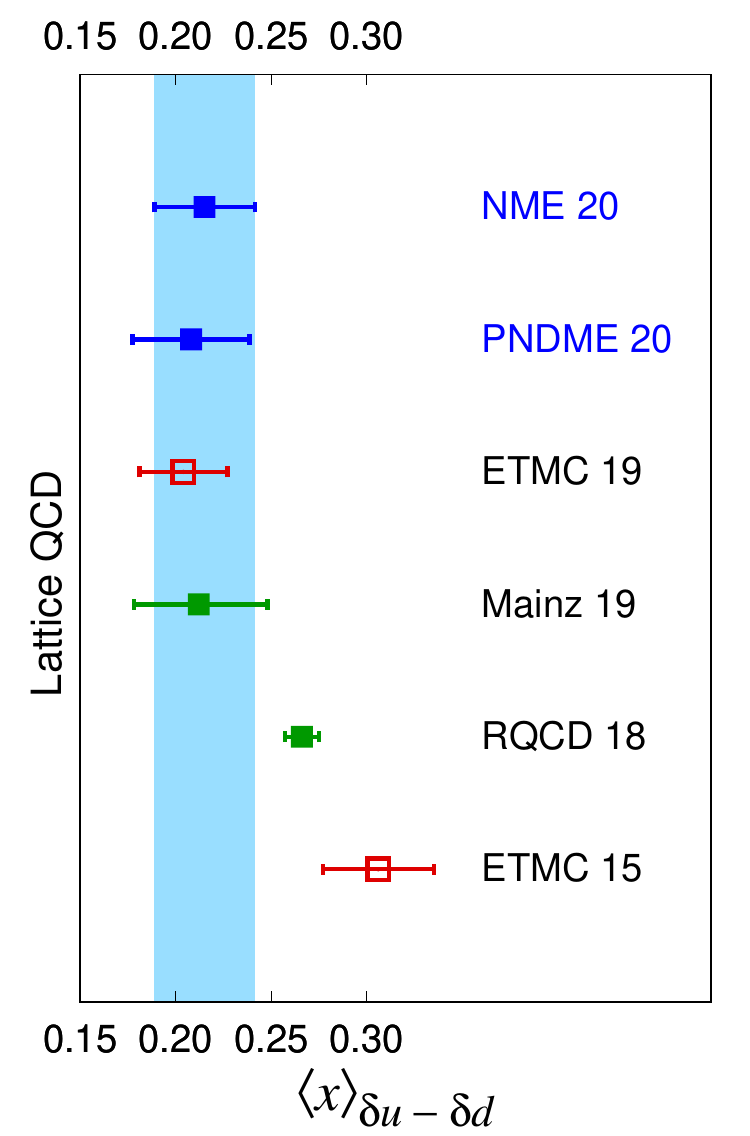}
\end{subfigure}
\vspace{-0.08in}
\caption{A comparison of results from lattice QCD calculations with dynamical fermions and
  global fits (below the black line) summarized in
  Table~\protect\ref{tab:Compare}. The left panel compares results for
  the momentum fraction, the middle for the helicity moment, and the
  right for the transversity moment. Our NME 20 result is also shown
  as the blue band to facilitate comparison.}
\label{fig:summary}
\end{figure*}

With the above choices, our final results are
\begin{align}
\langle x \rangle_{u -d}                &=  0.160(16)(20) \,, \nonumber \\
\langle x \rangle_{\Delta u- \Delta d}  &=  0.192(13)(20) \,, \nonumber \\
\langle x \rangle_{\delta u- \delta d}  &=  0.215(17)(20) \,.
\label{eq:finalresults}
\end{align}
An update of the comparison of lattice QCD calculations on ensembles
with dynamical fermions presented in Ref.~\cite{Mondal:2020cmt} is
shown in the top half of Table~\ref{tab:Compare} and in
Fig.~\ref{fig:summary}. Our new results, Eq.~\eqref{eq:finalresults},
are consistent with the PNDME~20 values published in
Ref.~\cite{Mondal:2020cmt}. This is a valuable check since PNDME~20
calculation used the nonunitary clover-on-HISQ lattice formulation.
Also, the current calculation provides weak evidence for a
finite-volume effect, whereas the PNDME~20 results were obtained using
just CC fits.  On the other hand, the range of lattice spacings and
pion masses simulated is somewhat smaller than in the PNDME~20
calculation.

Our result for the momentum fraction is in very good agreement with
estimates from phenomenological global fits reviewed in
Ref.~\cite{Lin:2017snn}, summarized in the bottom half of
Table~\ref{tab:Compare}, and shown in Fig~\ref{fig:summary}. The
helicity moment is consistent with the smaller error global fit
values, and our transversity moment is a prediction.


\begin{table*}[htbp]      
\centering
\renewcommand{\arraystretch}{1.2}
\begin{tabular}{ |c|c|c|c|c|c|c|c| }
\hline
        &  Renorm  & \multicolumn{2}{|c|}{$\langle x \rangle_{u -d}$} 
                   & \multicolumn{2}{|c|}{$\langle x \rangle_{\Delta u- \Delta d}$} 
                   & \multicolumn{2}{|c|}{$\langle x \rangle_{\delta u- \delta d}$}  \\
strategy& Method   & CC   & CCFV  & CC    &  CCFV   &  CC    &CCFV   \\
\hline
\hline
$\{4^{N\pi},3^*\}$& A     & $0.160(13)$&$0.145(15)$&$0.196(16)$&$0.180(18)$&$0.201(19)$&$0.196(22)$\\
\hline
$\{4^{N\pi},3^*\}$& B     & $0.160(12)$&$0.145(15)$&$0.196(15)$&$0.178(18)$&$0.205(19)$&$0.198(22)$\\
\hline
$\{4^{N\pi},3^*\}$& Final & $0.160(13)$&$0.145(15)$&$0.196(16)$&$0.179(18)$&$0.203(19)$&$0.197(22)$\\
\hline
\hline
$\{4,3^*\}$& A &$0.170(15)$&$0.161(16)$&$0.203(11)$&$0.193(13)$&$0.215(14)$&$0.214(17)$\\
\hline             
$\{4,3^*\}$& B &$0.168(14)$&$0.159(16)$&$0.201(11)$&$0.190(13)$&$0.218(14)$&$0.215(17)$\\
\hline             
$\{4,3^*\}$& Final &$0.169(15)$&$0.160(16)$&$0.202(11)$&$0.192(13)$&$0.217(14)$&$0.215(17)$\\
\hline
\hline
$\{4,2^{\rm free}\}$& A &$0.209(11)$&$0.193(12)$&$0.2324(89)$&$0.218(11)$&$0.249(12)$&$0.239(14)$\\
\hline                      
$\{4,2^{\rm free}\}$& B &$0.206(11)$&$0.190(12)$&$0.2283(92)$&$0.213(11)$&$0.248(12)$&$0.237(14)$\\
\hline                      
$\{4,2^{\rm free}\}$& Final &$0.208(11)$&$0.192(12)$&$0.2304(92)$&$0.216(11)$&$0.249(12)$&$0.238(14)$\\
\hline
\hline
\end{tabular}
\caption{The final results of the chiral-continuum-finite-volume fits for the three moments and the three 
strategies used to remove excited state contamination. Results are
given for the two methods of renormalization (A and B) discussed in
appendix~\ref{sec:renormalization}. The final value for each strategy
is taken to be the average of the two estimates along with the larger of the two
errors.}
\label{tab:CCFVresults}
\end{table*}

\begin{table*}[htbp]  
\setlength{\tabcolsep}{4pt}
\renewcommand{\arraystretch}{1.2}
\centering
\begin{tabular}{|c|c|c|c|c|c| }
\hline
Collaboration&Ref.&$\langle x \rangle_{u-d}$&$\langle x \rangle_{\Delta
u-\Delta d}$& $\langle x \rangle_{\delta u-\delta d}$& Remarks\\
\hline
\hline
NME 20   & &$0.160(16)(20)$&$0.192(13)(20)$&$0.215(17)(20)$& $N_f=2+1$\\
(this work)&&&&& clover-on-clover\\
\hline
PNDME~20 &\cite{Mondal:2020cmt} &$0.173(14)(07)$&$0.213(15)(22)$&$0.208(19)(24)$& $N_f=2+1+1$\\
&&&&& clover-on-HISQ\\
\hline
ETMC 20 &\cite{Alexandrou:2020sml} &$0.171(18)$& & &$N_f=2+1+1$ Twisted Mass\\
&&&&& N-DIS, N-FV\\
\hline
ETMC 19 &\cite{Alexandrou:2019ali} &$0.178(16)$&$0.193(18)$ & $0.204(23)$&$N_f=2+1+1$ Twisted Mass\\
&&&&& N-DIS, N-FV\\
\hline
Mainz 19&\cite{Harris:2019bih}&$0.180(25)_{\rm stat}$&$0.221(25)_{\rm stat}$&$0.212(32)_{\rm stat}$&$N_f=2+1$ Clover\\
        &                     &$(+14,-6)_{\rm sys}$  &$(+10,-0)_{\rm sys}$  &$(+16,-10)_{\rm sys}$ &  \\
\hline
$\chi$QCD 18&\cite{Yang:2018nqn} &$0.151(28)(29)$&        &          &$N_f=2+1$ \\
&&&&& Overlap on Domain Wall\\
\hline
RQCD 18&\cite{Bali:2018zgl} &$0.195(07)(15)$&$0.271(14)(16)$&$0.266(08)(04)$ &$N_f=2$ Clover\\
&&&&& \\
\hline
ETMC 17 &\cite{Alexandrou:2017oeh} &$0.194(9)(11)$& & &$N_f=2$ Twisted Mass\\
&&&&& N-DIS, N-FV\\ 
\hline
ETMC 15
&\cite{Abdel-Rehim:2015owa}&$0.208(24)$&$0.229(30)$&$0.306(29)$&$N_f=2$ Twisted Mass\\
&&&&&N-DIS, N-FV\\
\hline
RQCD 14&\cite{Bali:2014gha} &$0.217(9)$& & &$N_f=2$ Clover\\
&&&&& N-DIS, N-CE, N-FV\\
\hline
LHPC 14 &\cite{Green:2012ud}&$0.140(21)$& & &$N_f=2+1$ Clover\\
&&&&&N-DIS ($a \sim 0.12$ fm)\\
\hline
RBC/&\cite{Aoki:2010xg} &0.124--0.237&0.146--0.279& &$N_f=2+1$ Domain Wall\\
UKQCD 10&&&&&N-DIS, N-CE, N-ES\\
\hline  
LHPC 10 &\cite{Bratt:2010jn} &$0.1758(20)$&$0.1972(55)$& &$N_f=2+1$\\
&&&&& Domain Wall-on-Asqtad\\
&&&&& N-DIS, N-CE, N-NR, N-ES\\
\hline
\hline
\hline
CT18&\cite{Hou:2019efy}&$0.156(7)$& & &\\
\hline
JAM17${}^\dagger$&\protect\cite{Ethier:2017zbq,Lin:2017snn}& &0.241(26) & &\\
\hline
NNPDF3.1&\cite{Ball:2017nwa}&$0.152(3)$& & &\\
\hline
ABMP2016&\cite{Alekhin:2017kpj}&$0.167(4)$& & &\\
\hline
CJ15&\cite{Accardi:2016qay}&$0.152(2)$& & &\\
\hline
HERAPDF2.0&\cite{Abramowicz:2015mha}&$0.188(3)$& & &\\
\hline
CT14&\cite{Dulat:2015mca}&$0.158(4)$& & &\\
\hline
MMHT2014&\cite{Harland-Lang:2014zoa}&$0.151(4)$& & &\\
\hline
NNPDFpol1.1&\cite{Nocera:2014gqa}& &$0.195(14)$& &\\
\hline
DSSV08&\cite{deFlorian:2009vb,deFlorian:2008mr}& &$0.203(9)$& &\\
\hline
\end{tabular}
\caption{Our Lattice QCD results are compared with other lattice
  calculations with $N_f$ flavors of dynamical fermions in rows 2--9, and with results from
  phenomenological global fits in the remainder of the table. In both
  cases, the results are arranged in reverse chronological order.  All
  results are in the $\MSbar$ scheme at scale $2$~GeV. For a
  discussion and comparison of lattice and global fit results up to
  2020, see Ref.~\protect\cite{Lin:2020rut} and also the 
  comparison in~\cite{Hou:2019efy} for $\langle x \rangle_{u-d}$. The
  JAM17${}^\dagger$ estimate for $\langle x \rangle_{\Delta u-\Delta
    d}$ is obtained from~\cite{Lin:2017snn}, where, as part of the
  review, an analysis was carried out using the data in
  \cite{Ethier:2017zbq}. The following abbreviations are used in the
  remarks column for various sources of systematic uncertainties in
  lattice calculations---DIS: Discretization effects, CE: Chiral
  extrapolation, FV: Finite volume effects, NR: Nonperturbative
  renormalization, ES: Excited state contaminations. A prefix "N-"
  means that the systematic uncertainty was neither adequately controlled
  nor estimated.  }
\label{tab:Compare}
\end{table*}

\section{Conclusions}
\label{sec:summary}

We have presented results for the isovector quark momentum fraction,
$\la x \ra_{u-d}^{{\rm \ol{MS}}}$, helicity moment, $\la
x \ra^{{\rm \ol{MS}}}_{\Delta u- \Delta d}$, and transversity moment,
$\la x \ra^{{\rm \ol{MS}}}_{\delta u- \delta d}$ on seven ensembles
with 2+1-flavor Wilson-clover fermions. Using high statistics data we
confirmed the behavior of the correlation functions predicted by the
spectral decomposition as shown in
Figs.~\ref{fig:Ratio-mom-1}--\ref{fig:Ratio-transversity-2}. These
higher precision data allowed us to investigate the systematic
uncertainty associated with excited-state contamination when
extracting the three moments. We carried out the full analysis with
three different estimates of the mass gap of the first excited state
that cover a large range of possible values.  We use the spread in
results to assign a systematic uncertainty to account for possible
residual excited-state bias.

To obtain the final result in the continuum limit, we fit the seven
points using the ansatz in Eq.~\eqref{eq:CCFV} that includes the
leading order terms in $M_\pi$, the lattice spacing $a$ and the finite
volume parameter $M_\pi L$.  Having two pairs of points,
$\{a094m270,a094m270L\}$ and $\{a091m170,a091m170L\}$, that differ
only in the lattice volume, allowed us to quantify finite volume
corrections in all three moments as shown in
Figs.~\ref{fig:momfrac-CCFV-Z-via-averaging},~\ref{fig:helfrac-CCFV-Z-via-averaging}
and~\ref{fig:transvmom-CCFV-Z-via-averaging}. A comparison of the
results with and without the finite-volume correction (CCFV versus CC)
are shown in Table~\ref{tab:CCFVresults}.  Based on this analysis, we
present final results from the CCFV fits that are about 5\% smaller
than the CC-fit values for the momentum fraction and the helicity
moment. These results are consistent with the
PNDME~20~\cite{Mondal:2020cmt} calculation, which used the
clover-on-HISQ lattice formulation and were obtained using just CC
fits.

In appendix~\ref{sec:renormalization}, we describe two methods for
removing the $p^2$ dependent artifacts in the renormalization
constants.  The results for the moments from these two methods are
given in Table~\ref{tab:CCFVresults}. The data show that after
the continuum extrapolation (CCFV or CC fits), the two estimates
overlap even though the renormalization constants themselves differ by
$\approx 5\%$ as shown in Table~\ref{tab:Z-fac}. The better agreement
after the continuum extrapolation suggests that the main difference
between the two methods are indeed discretization artifacts.

The data at three values of the lattice spacing shown in
Figs.~\ref{fig:momfrac-CCFV-Z-via-averaging},~\ref{fig:helfrac-CCFV-Z-via-averaging}
and~\ref{fig:transvmom-CCFV-Z-via-averaging} do not exhibit any
significant dependence on the lattice spacing $a$. The main variation
is with $M_\pi^2$, and its magnitude depends on the mass gap of the
first excited state used in the analysis of the ESC. Since the mass
gaps obtained from fits to the three-point functions ($\{4,2^{\rm
free}\}$ strategy) do not prefer values corresponding to the lowest possible
excitations ($N\pi$ states used in the $\{4^{N\pi},3^*\}$ strategy) but are closer to 
two-state fits to the two-point function (see Table~\ref{tab:massgap}),
we quote final results from the $\{4,3^*\}$ strategy. We add a
systematic error of $0.02$, based on the observed spread (see
Table~\ref{tab:CCFVresults}), to account for possible unresolved
excited-state effects.

Our final results, taken from Table~\ref{tab:CCFVresults}, are given
in Eq.~\eqref{eq:finalresults}. These are compared with other lattice
calculations and phenomenological global fit estimates in
Table~\ref{tab:Compare} and Fig.~\ref{fig:summary}.  They are in good
agreement with other recent lattice results from the 
PNDME~\cite{Mondal:2020cmt},
ETMC~\cite{Alexandrou:2020sml,Alexandrou:2019ali},
Mainz~\cite{Harris:2019bih} and $\chi$QCD~\cite{Yang:2018nqn}
collaborations. Our estimate for the momentum fraction is in good
agreement with most global fit estimates but has much larger
error. The three estimates for the helicity moment from global fits
have a large spread, and our estimate is consistent with the smaller error estimates. Lattice
estimates for the transversity moment are a prediction.

Having established the efficacy of the lattice QCD approach to
reliably calculate these isovector moments, we expect to make steady
progress in reducing the errors by simulating at additional values of
$\{a,M_\pi\}$ and by increasing the statistics.

\begin{acknowledgments}
The calculations used the Chroma software
suite~\cite{Edwards:2004sx}. This research used resources at (i) the
National Energy Research Scientific Computing Center, a DOE Office of
Science User Facility supported by the Office of Science of the
U.S. Department of Energy under Contract No. DE-AC02-05CH11231; (ii)
the Oak Ridge Leadership Computing Facility at the Oak Ridge National
Laboratory, which is supported by the Office of Science of the
U.S. Department of Energy under Contract No. DE-AC05-00OR22725; (iii)
the USQCD Collaboration, which are funded by the Office of Science of
the U.S. Department of Energy, and (iv) Institutional Computing at Los
Alamos National Laboratory.  T. Bhattacharya and R. Gupta were partly
supported by the U.S. Department of Energy, Office of Science, Office
of High Energy Physics under Contract No.~DE-AC52-06NA25396.
F. Winter is supported by the U.S. Department of Energy, Office of
Science, Office of Nuclear Physics under contract DE-AC05-06OR23177.
B. Jo\'o is supported by the U.S. Department of Energy, Office of
Science, under contract DE-AC05-06OR22725. We acknowledge support from
the U.S. Department of Energy, Office of Science, Office of Advanced
Scientific Computing Research and Office of Nuclear Physics,
Scientific Discovery through Advanced Computing (SciDAC) program, and
of the U.S. Department of Energy Exascale Computing Project.
T. Bhattacharya, R. Gupta, S. Mondal, S. Park and B.Yoon were partly
supported by the LANL LDRD program, and S. Park by the Center for
Nonlinear Studies.
\end{acknowledgments}

\appendix
\section{Plots of the Ratio \texorpdfstring{$C_{\mathcal{O}}^{3\text{pt}}(\tau;t)/C^{2\text{pt}}(\tau)$}{C(3pt)/C(2pt)}}
\label{sec:ratios}
\begin{figure*}[p]  
\centering
\begin{subfigure}
\centering
\includegraphics[angle=0,width=0.32\textwidth]{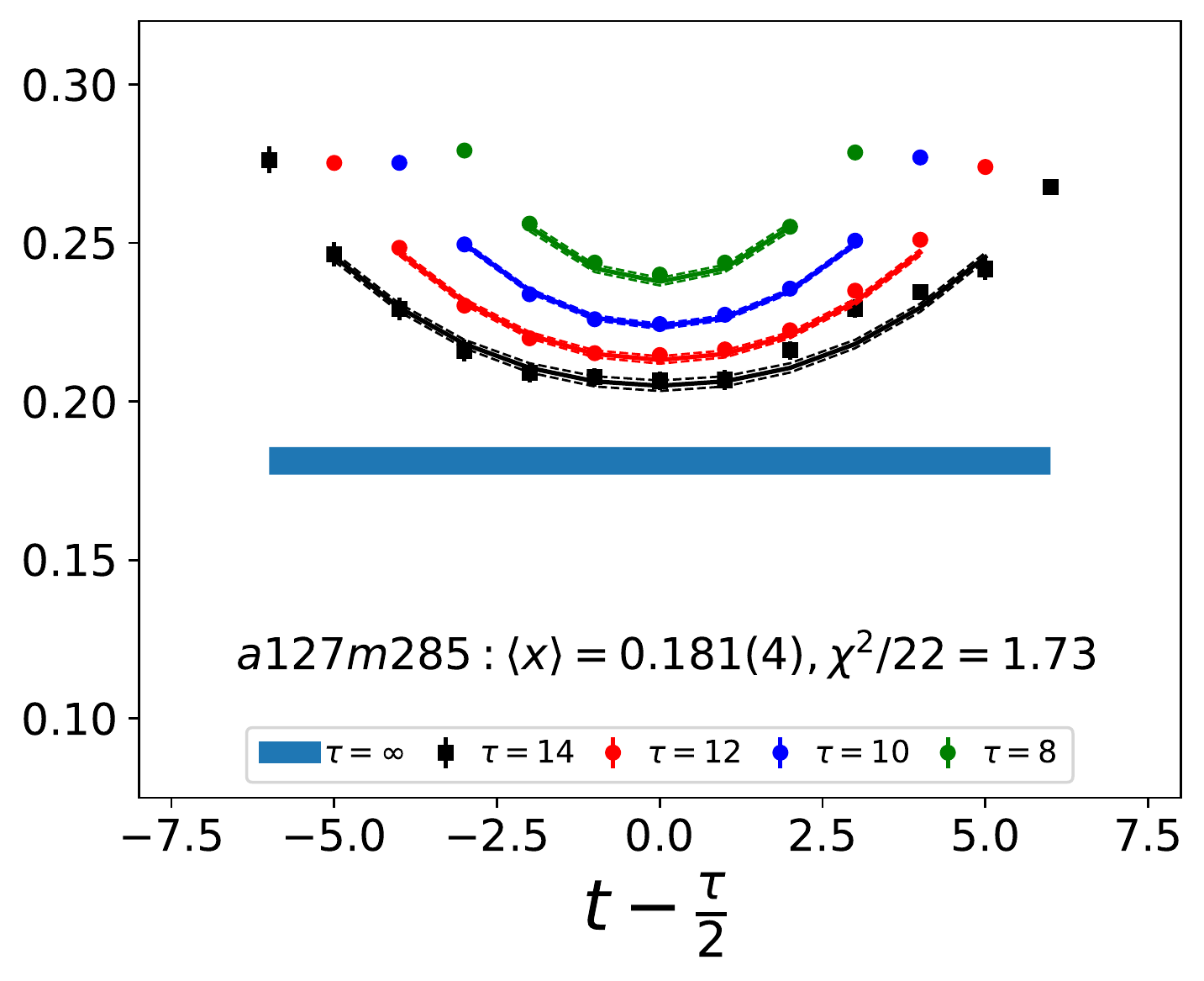}
\includegraphics[angle=0,width=0.32\textwidth]{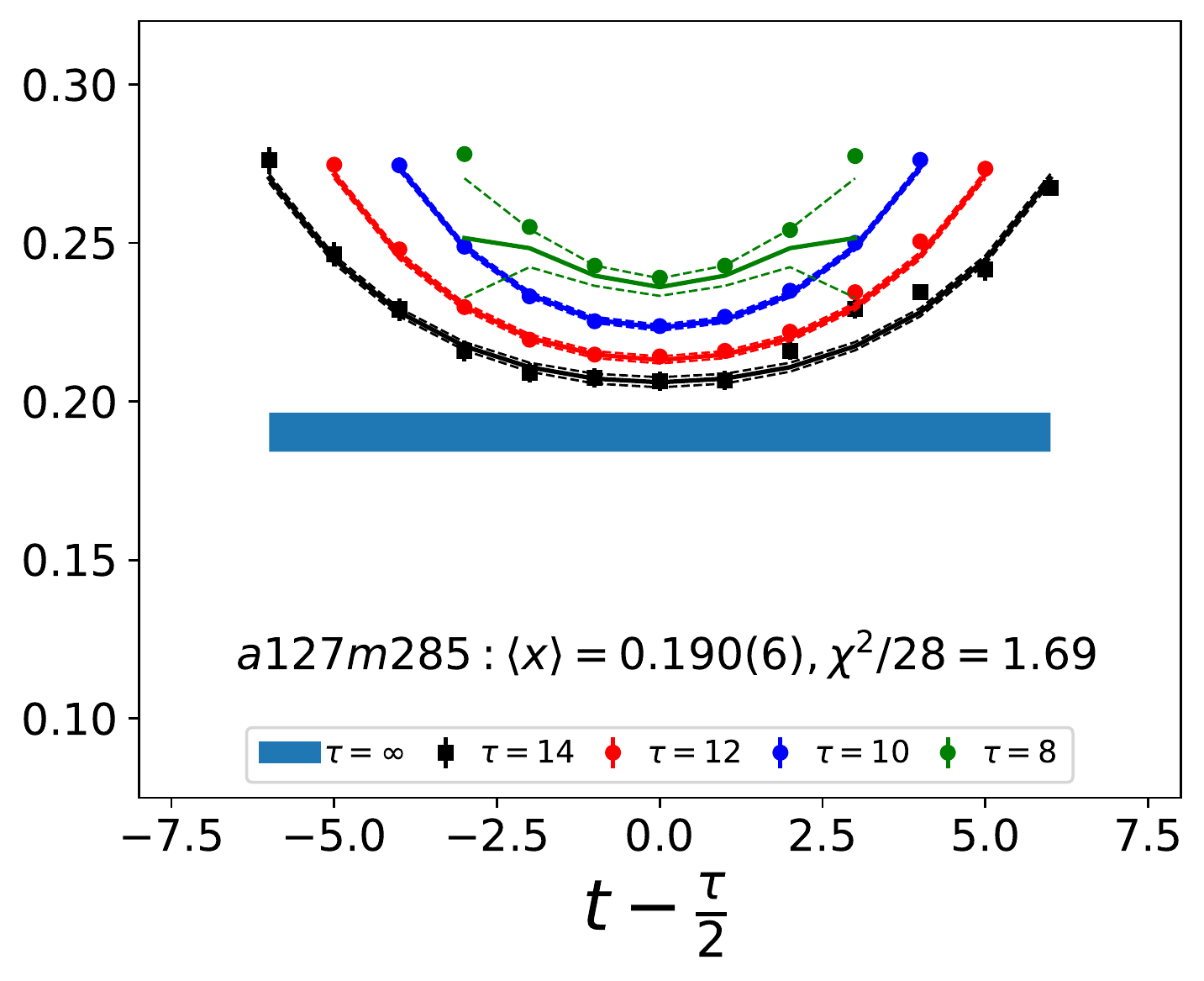}
\includegraphics[angle=0,width=0.32\textwidth]{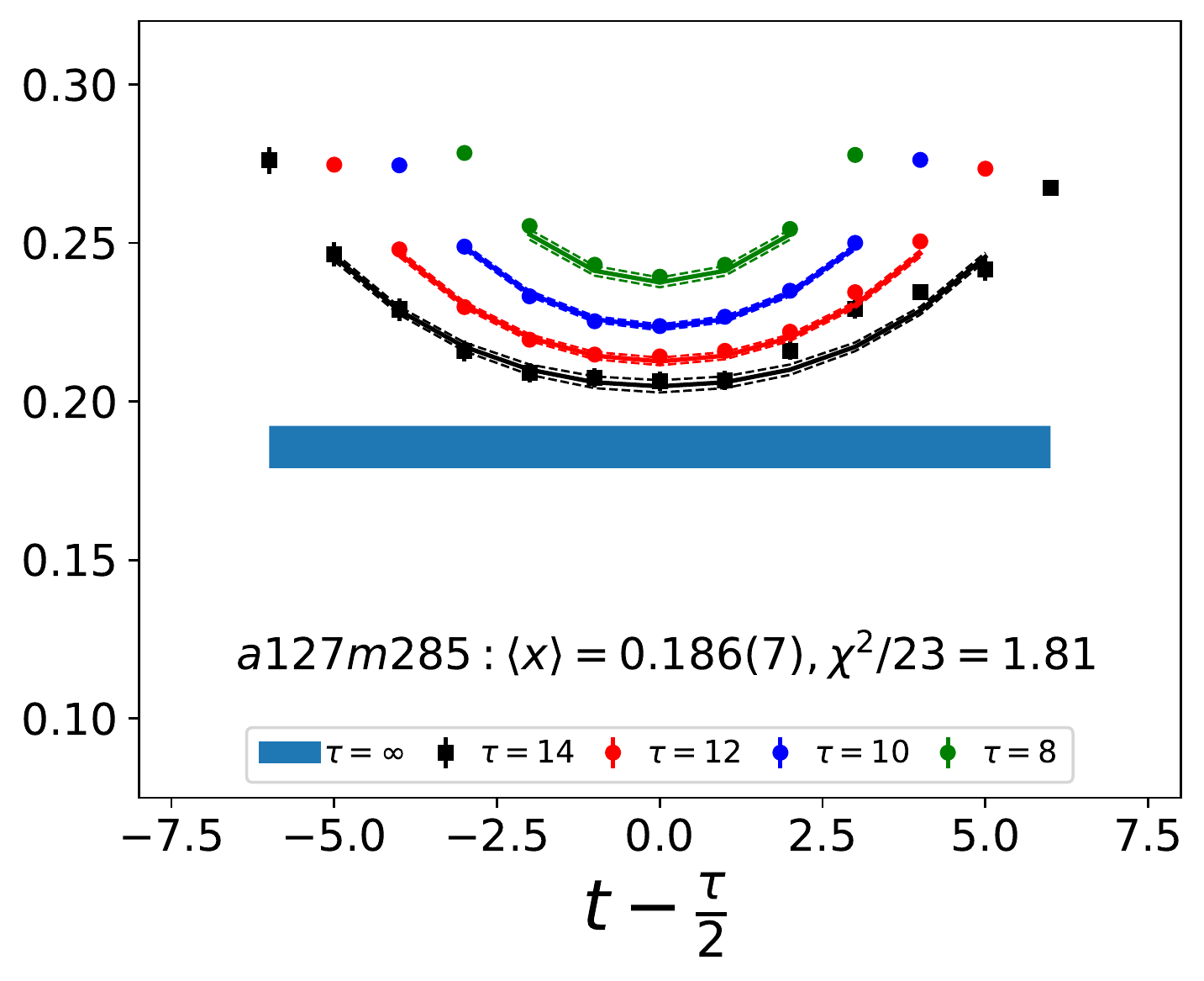}

\includegraphics[angle=0,width=0.32\textwidth]{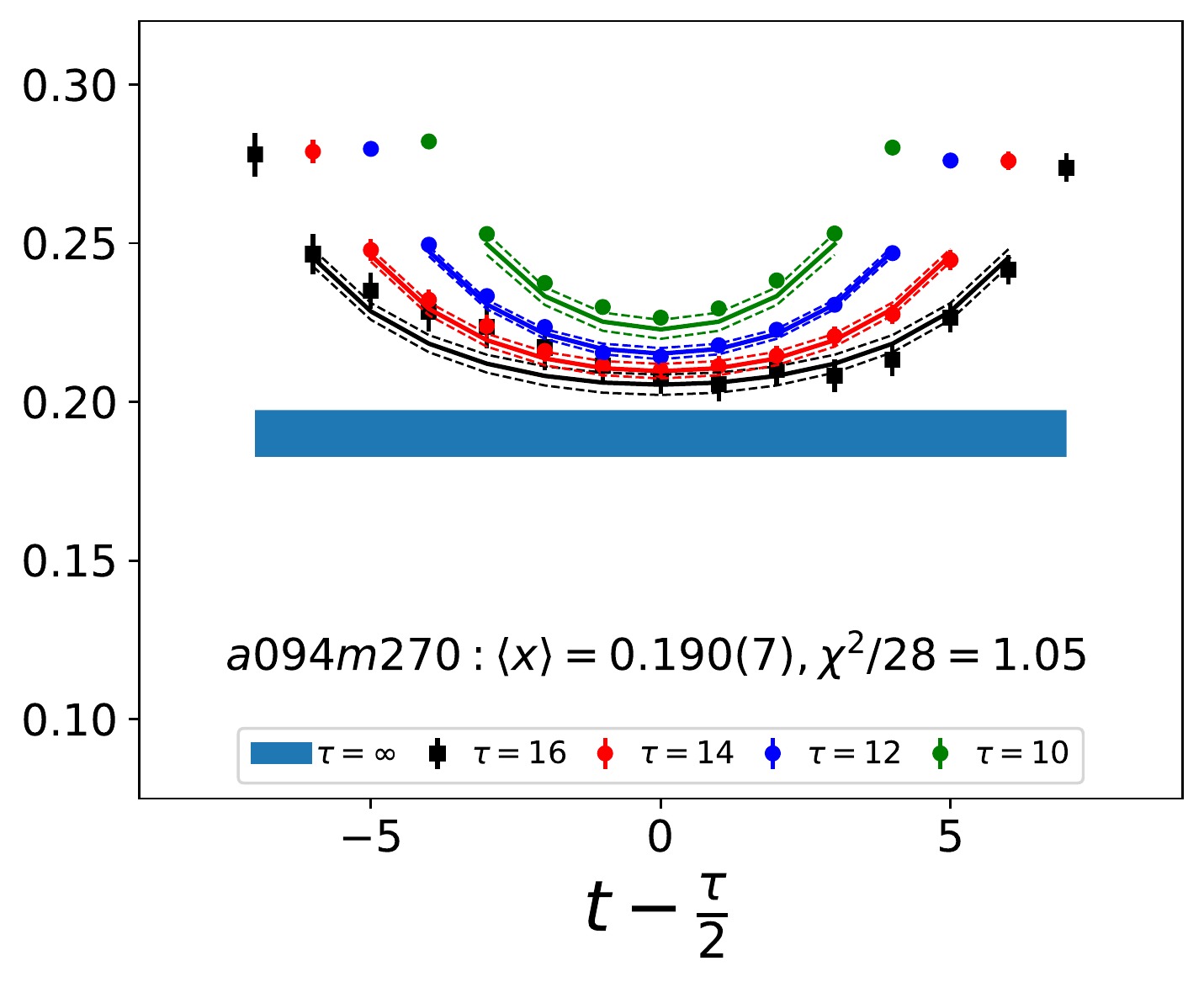}
\includegraphics[angle=0,width=0.32\textwidth]{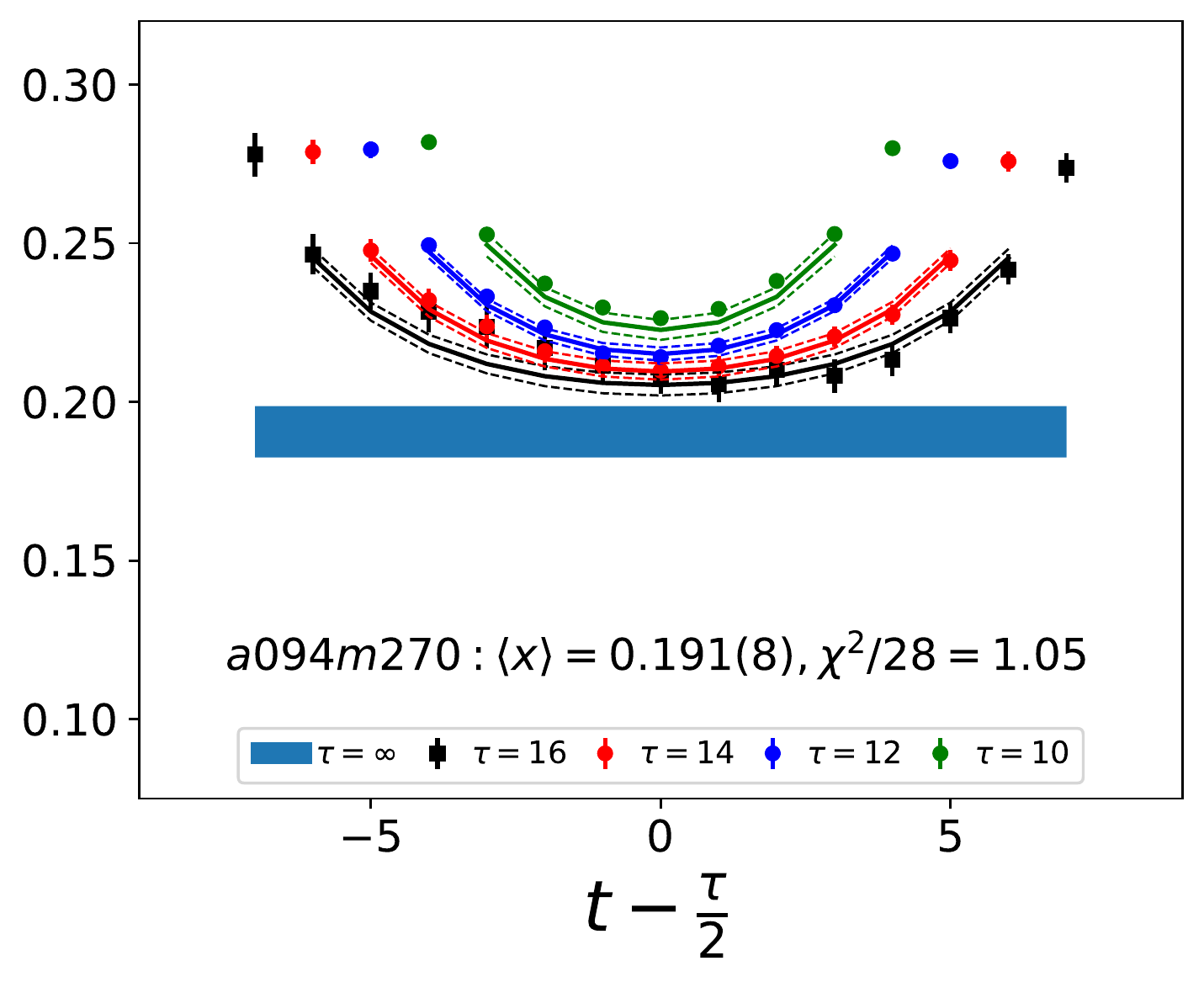}
\includegraphics[angle=0,width=0.32\textwidth]{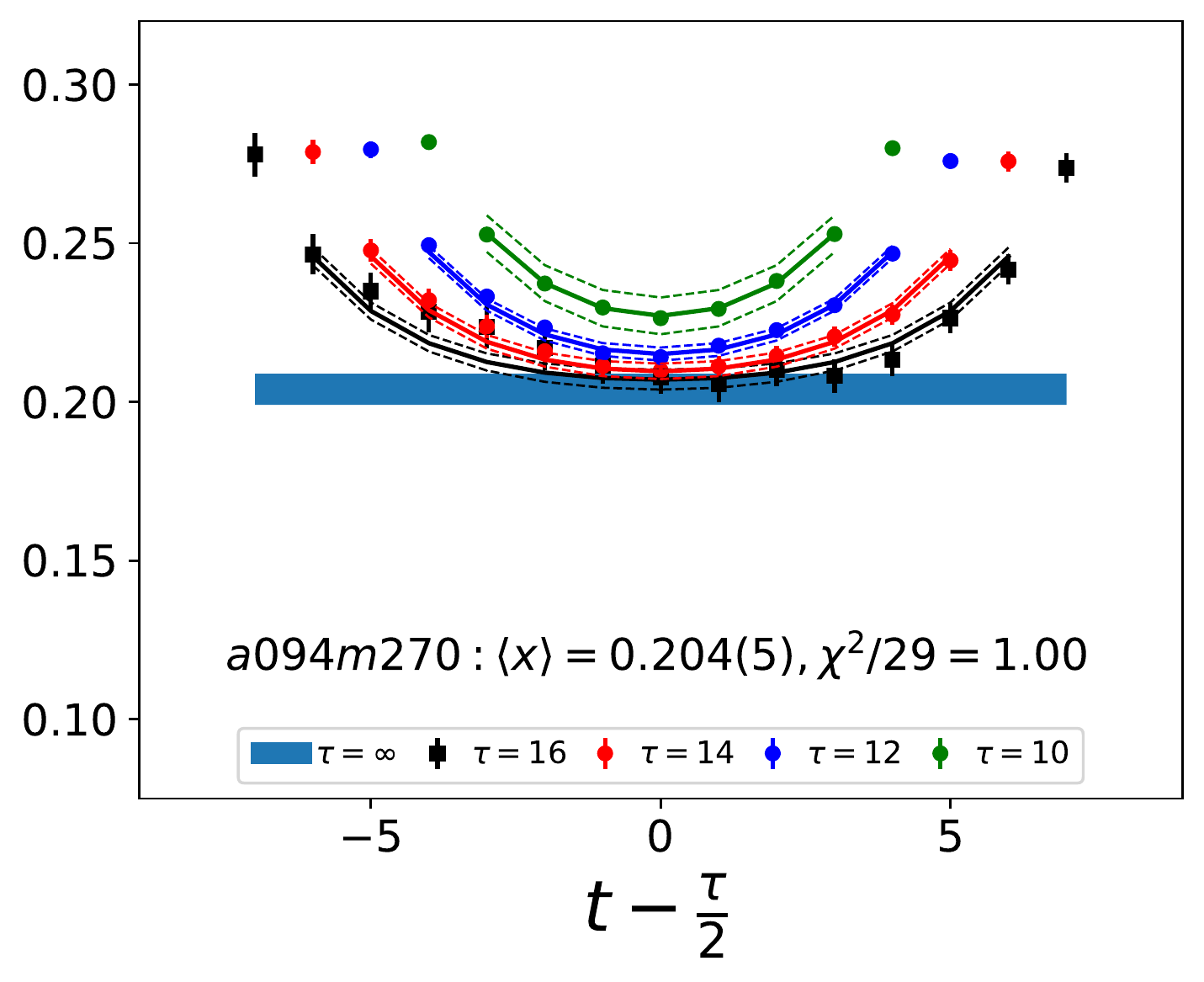}

\includegraphics[angle=0,width=0.32\textwidth]{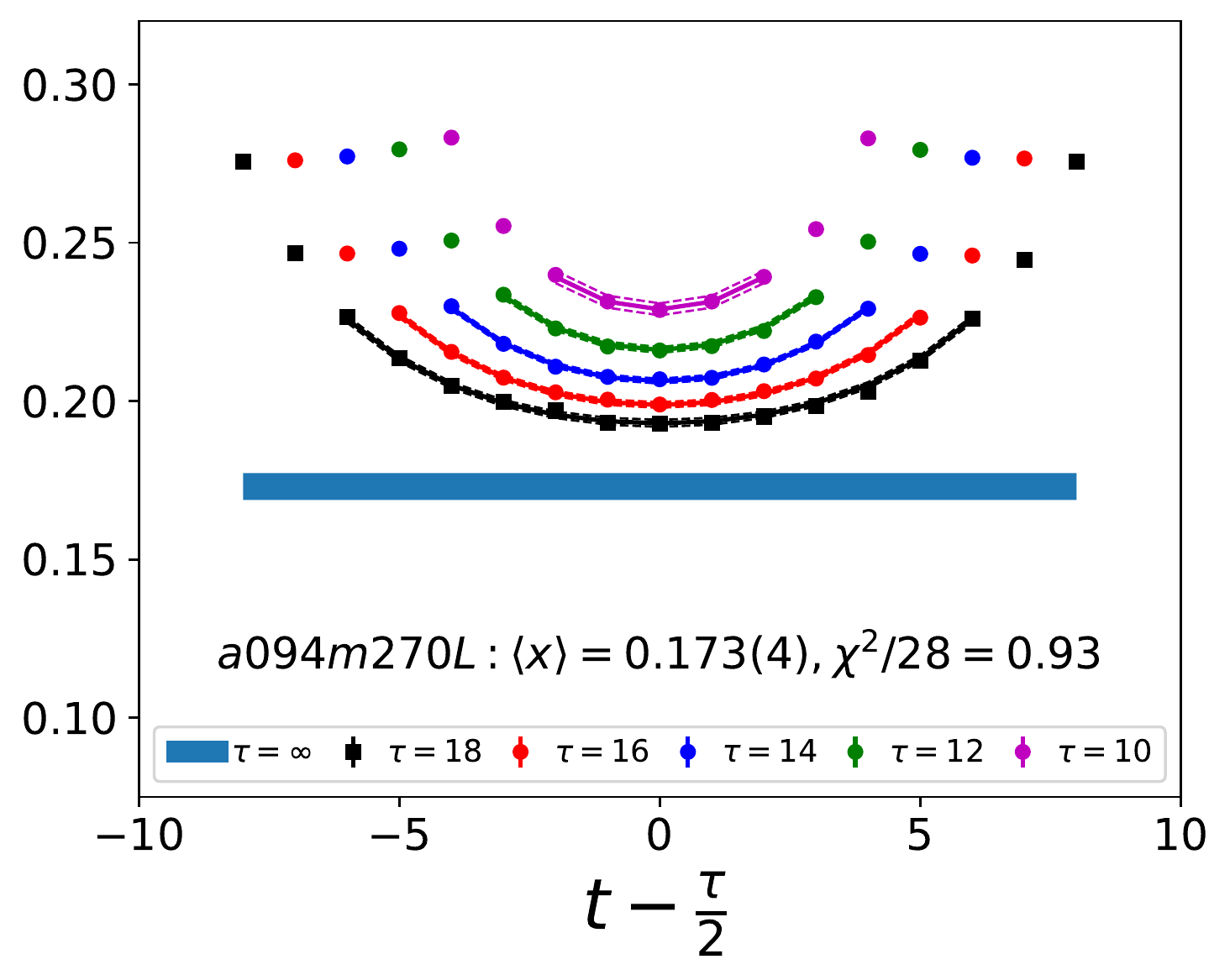}
\includegraphics[angle=0,width=0.32\textwidth]{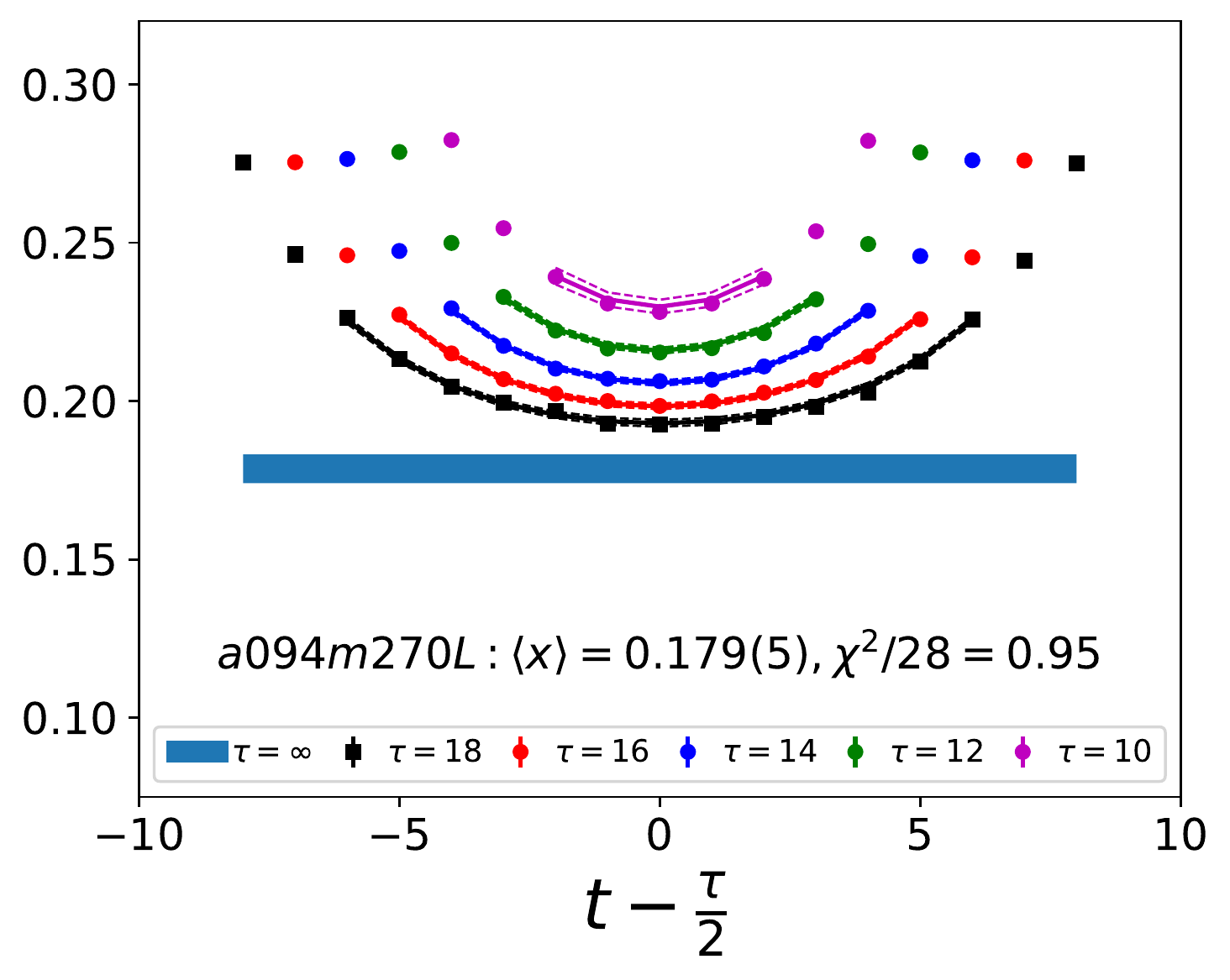}
\includegraphics[angle=0,width=0.32\textwidth]{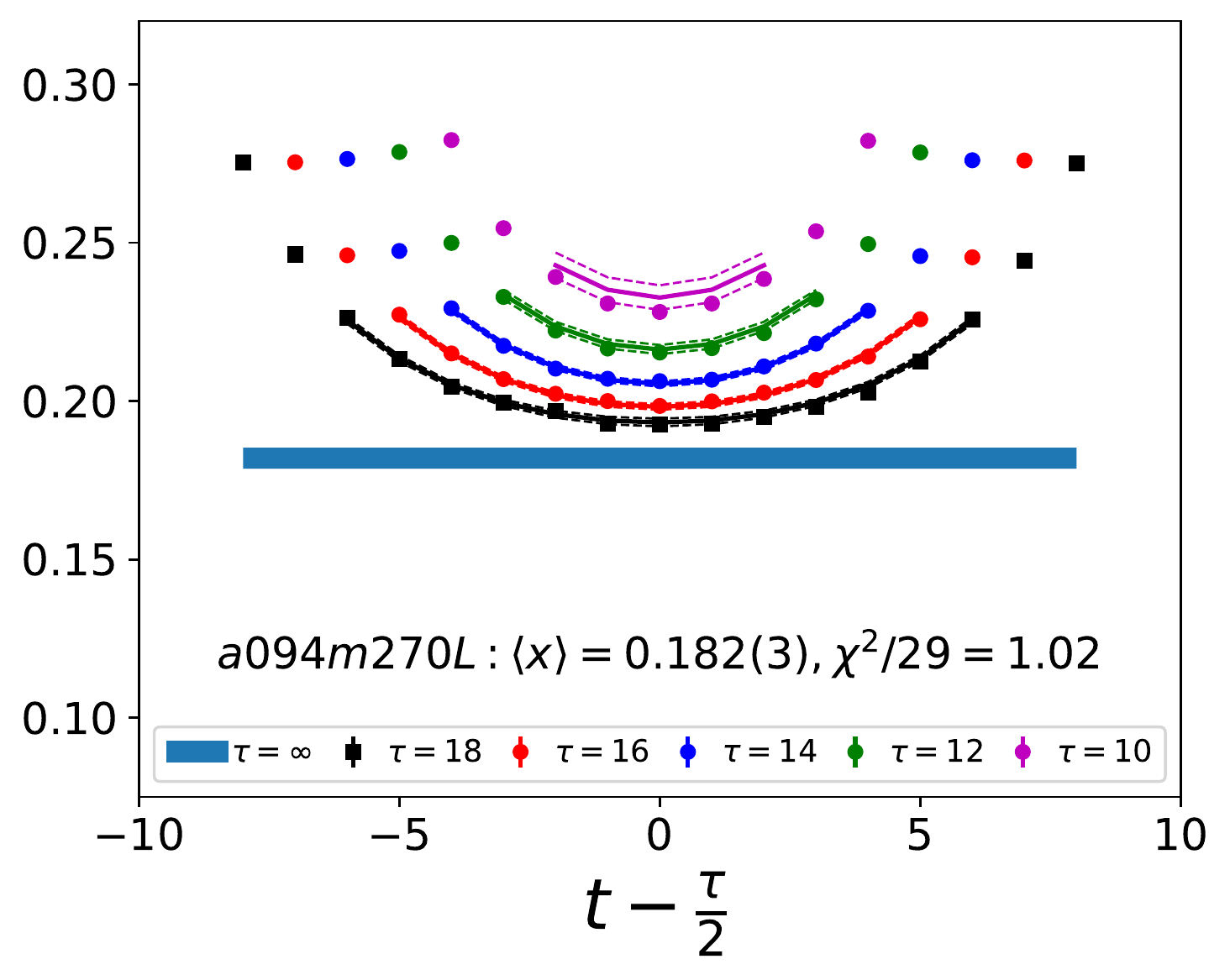}
\end{subfigure}

\caption{Data and fits to remove excited-state contamination in the extraction 
  of the momentum fraction $\langle x \rangle_{u-d}$ for $a127m285$
  (top row), $a094m270$ (second row), and $a094m270L$ (bottom row) ensembles. The data for the ratio
  $C_\mathcal{O}^{3\text{pt}}(\tau;t)/C^{2\text{pt}}(\tau)$ is scaled
  using Eq.~\protect\eqref{eq:me2momentV} to give $\langle
  x \rangle_{u-d}$, and the fit parameters are listed in
  Tables~\protect\ref{tab:5strategy-fits-momfrac}.  In each row, the
  three panels shows data for the three strategies: $\{4^{N\pi},3^*\}$
  (left), $\{4^{},3^*\}$ (middle) and $\{4^{},2^{\rm free}\}$ (right).
  For each $\tau$, the line in the same color as the data points is
  the result of the fit used (see Sec.~\protect\ref{sec:ESC}) to
  obtain the ground state matrix element. The result for the
  unrenormalized ground-state value of the moment is shown by the blue
  band and summarized in Table~\protect\ref{tab:ESC-fits} along with
  the values of $\tau$ and $t_{\rm skip}$ used in the fit. The
  y-interval is selected to be the same for all the panels to
  facilitate comparison. }
\label{fig:Ratio-mom-1}
\end{figure*}

\begin{figure*}[t] 

\begin{subfigure}
\centering
\includegraphics[angle=0,width=0.32\textwidth]{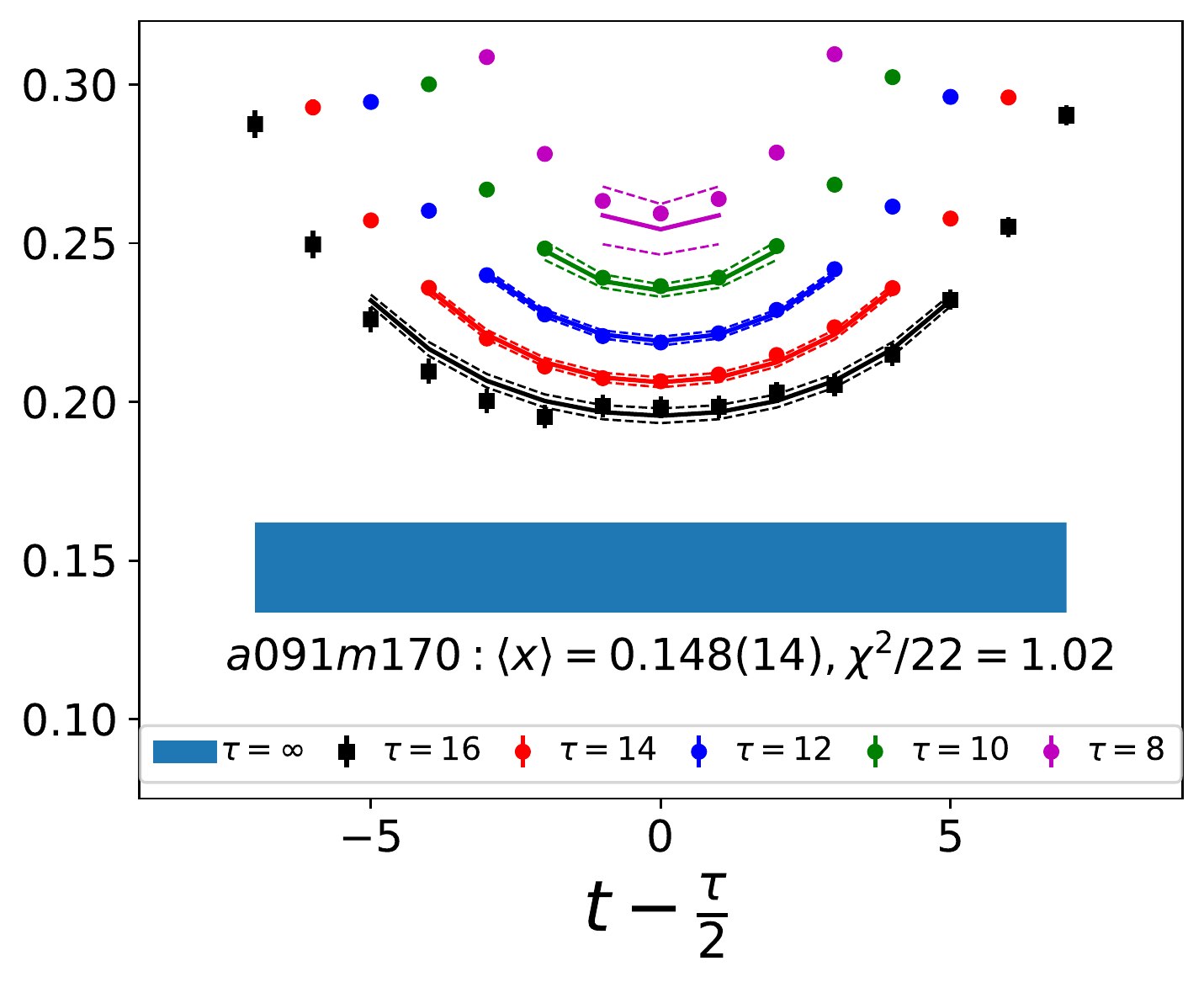}
\includegraphics[angle=0,width=0.32\textwidth]{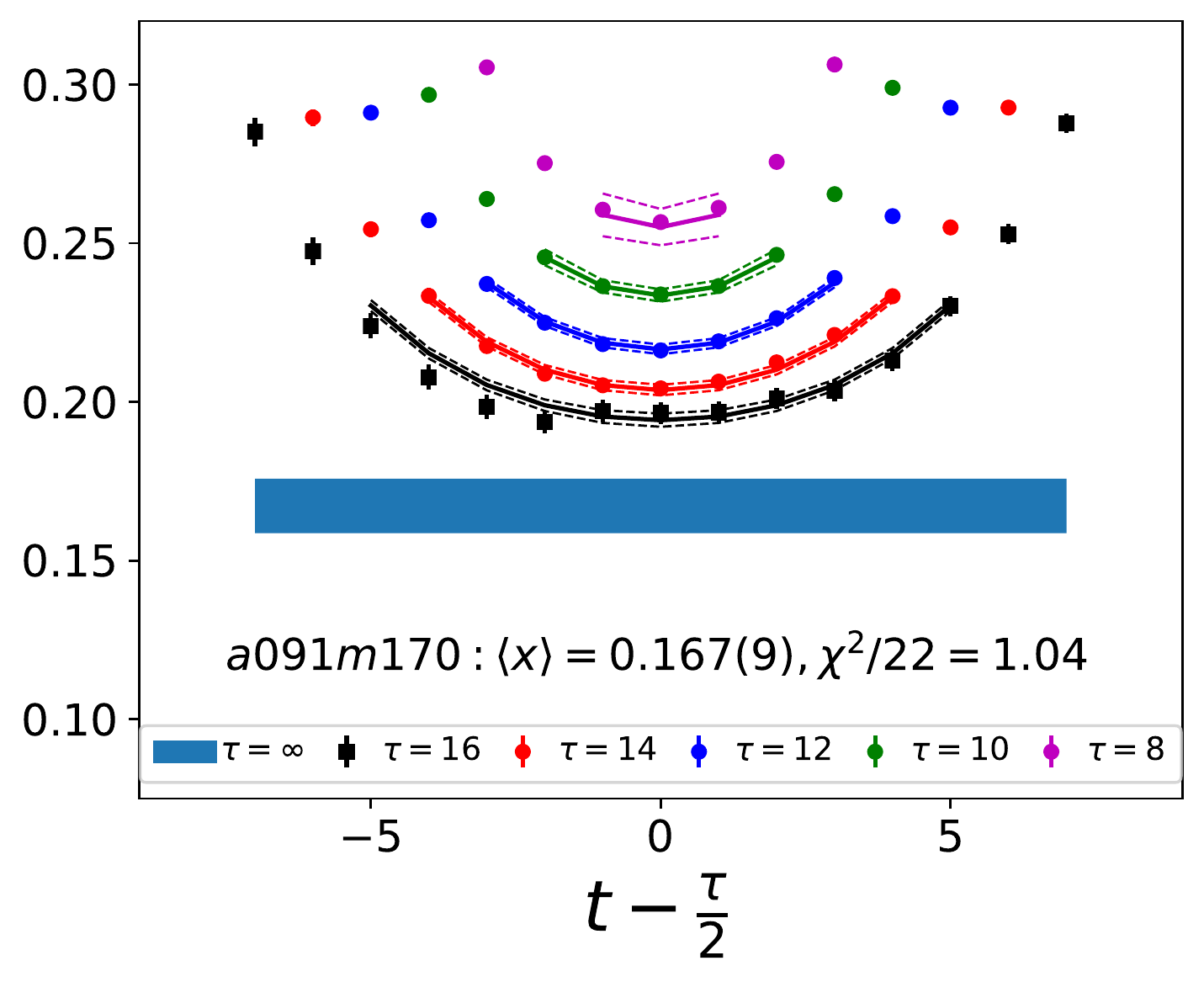}
\includegraphics[angle=0,width=0.32\textwidth]{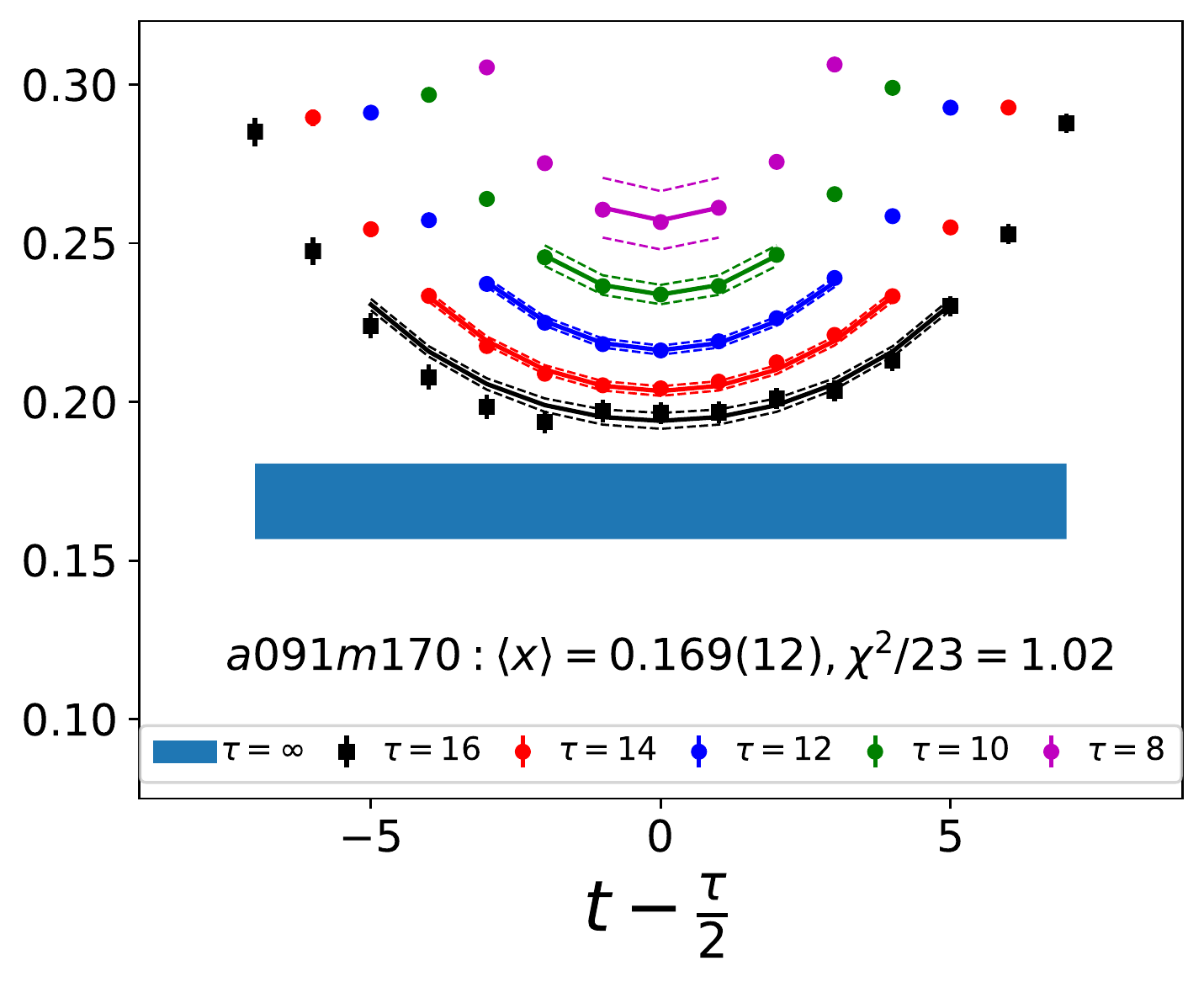}
\end{subfigure}

\begin{subfigure}
\centering
\includegraphics[angle=0,width=0.32\textwidth]{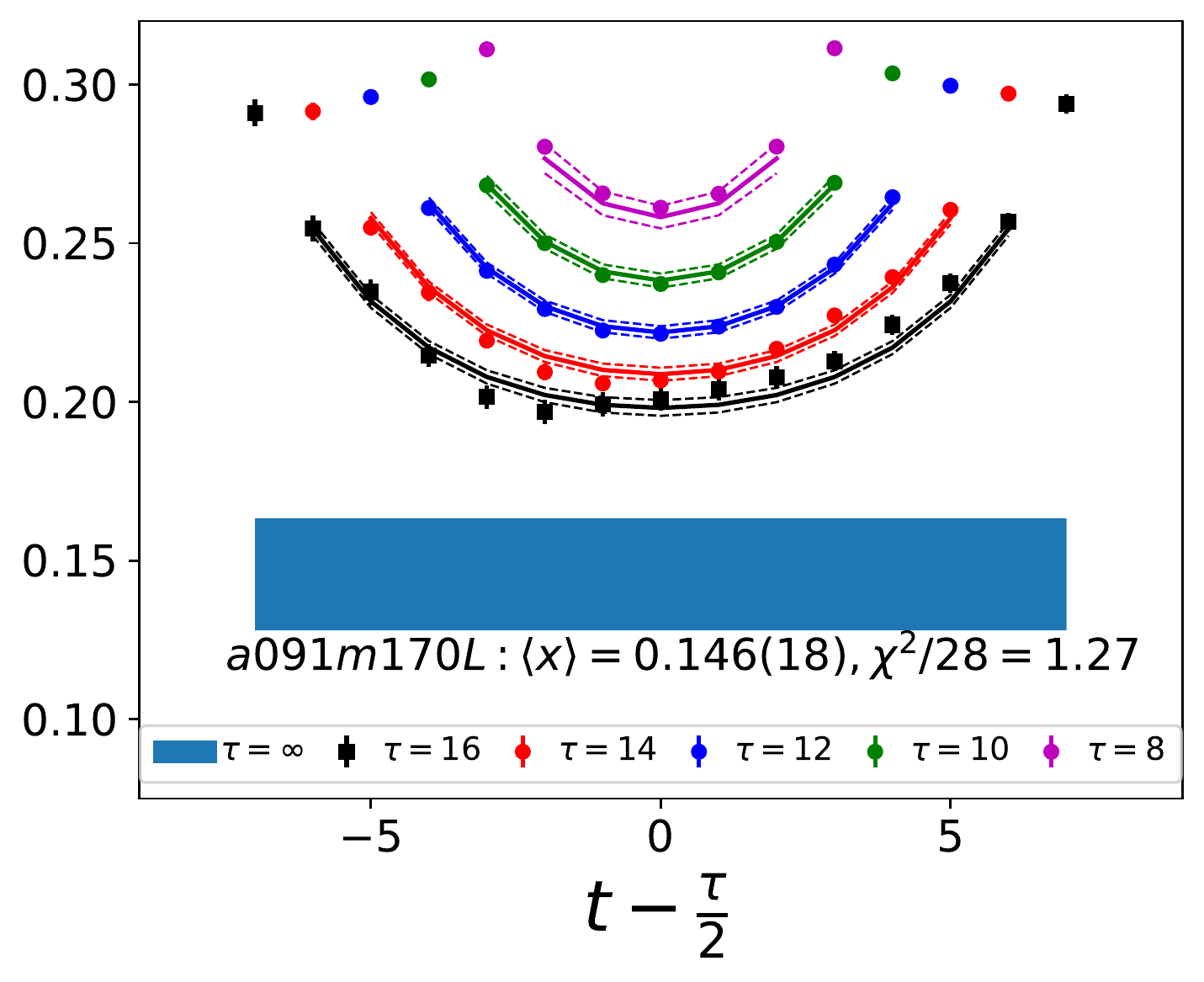}
\includegraphics[angle=0,width=0.32\textwidth]{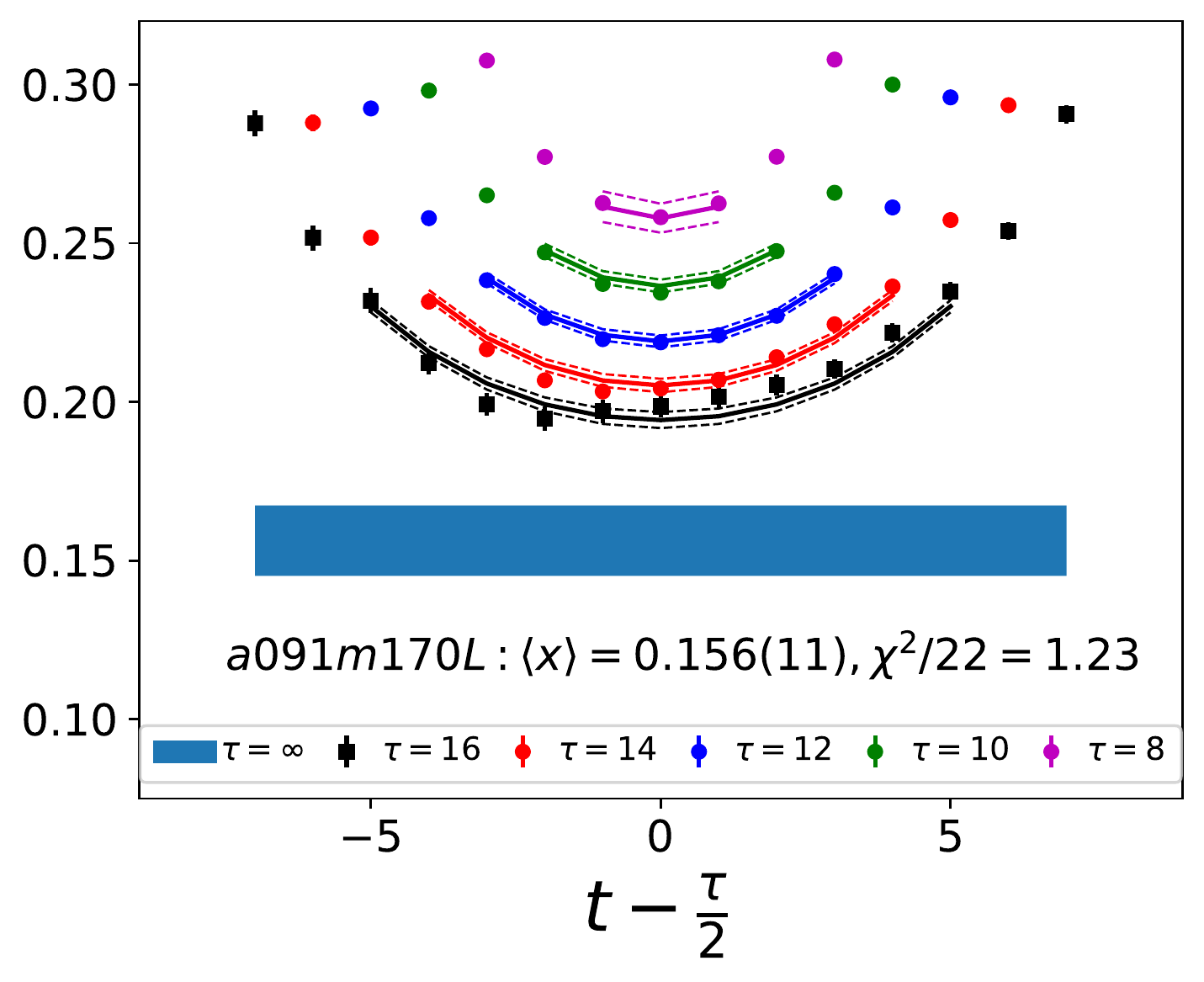}
\includegraphics[angle=0,width=0.32\textwidth]{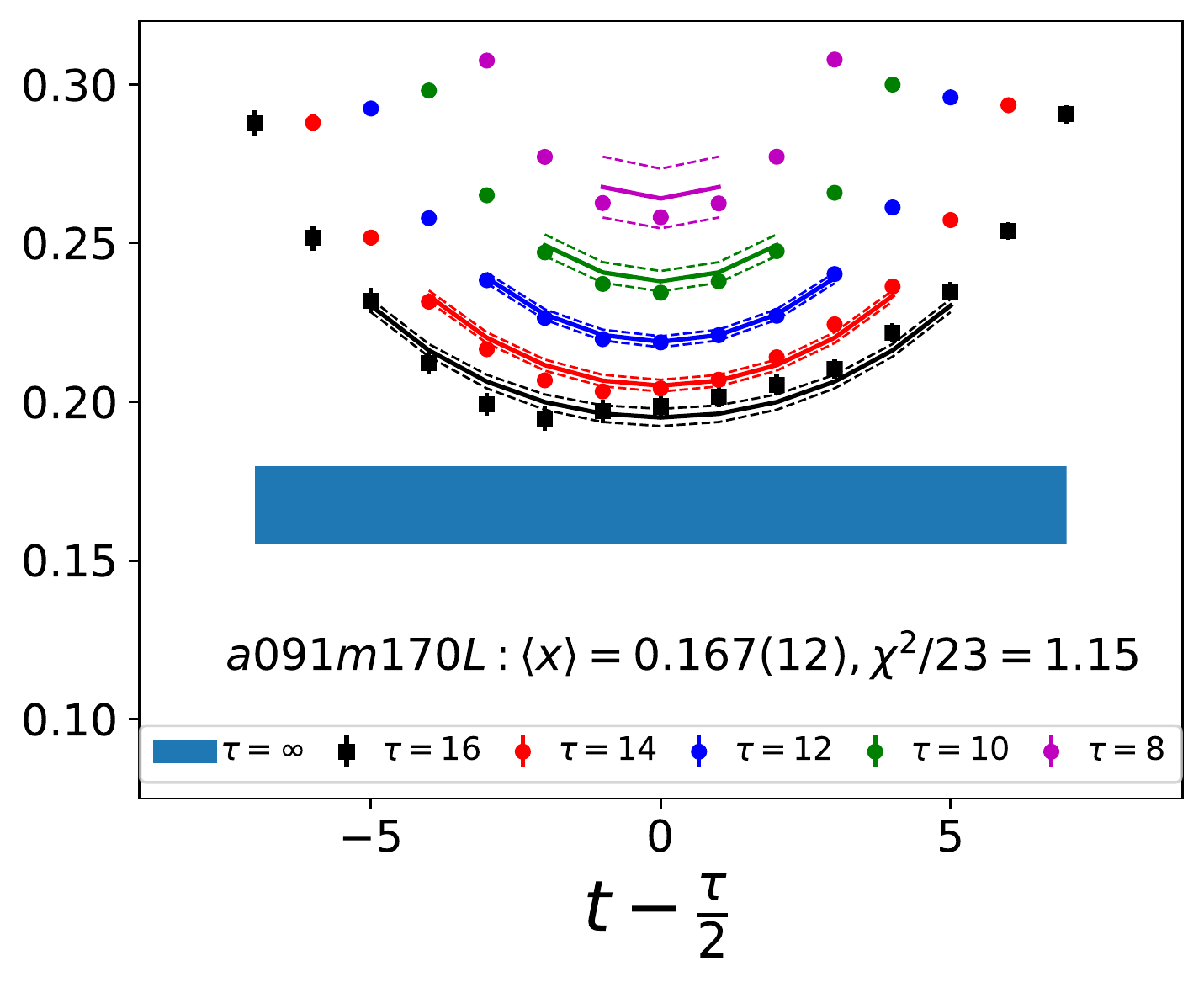}
\end{subfigure}

\begin{subfigure}
\centering
\includegraphics[angle=0,width=0.32\textwidth]{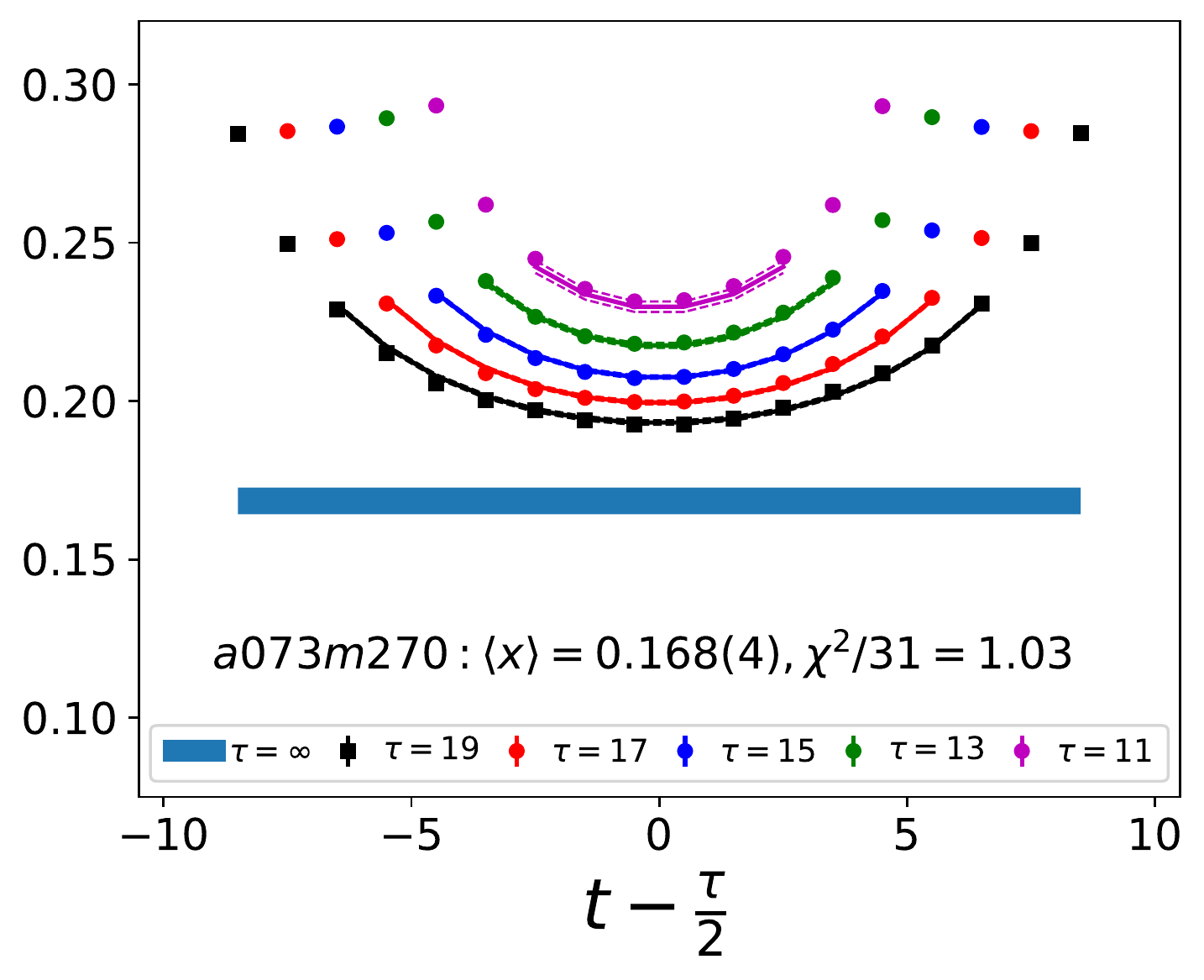}
\includegraphics[angle=0,width=0.32\textwidth]{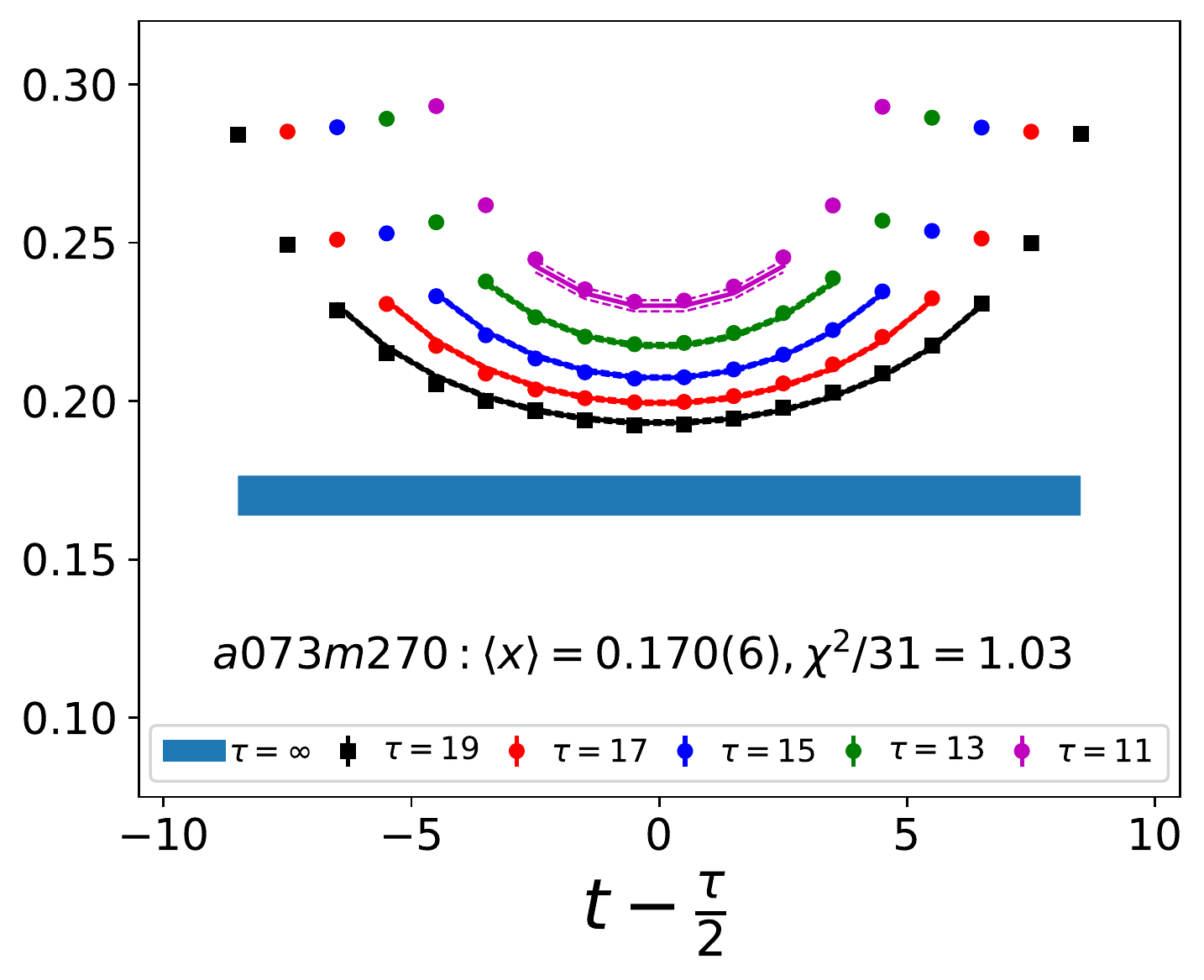}
\includegraphics[angle=0,width=0.32\textwidth]{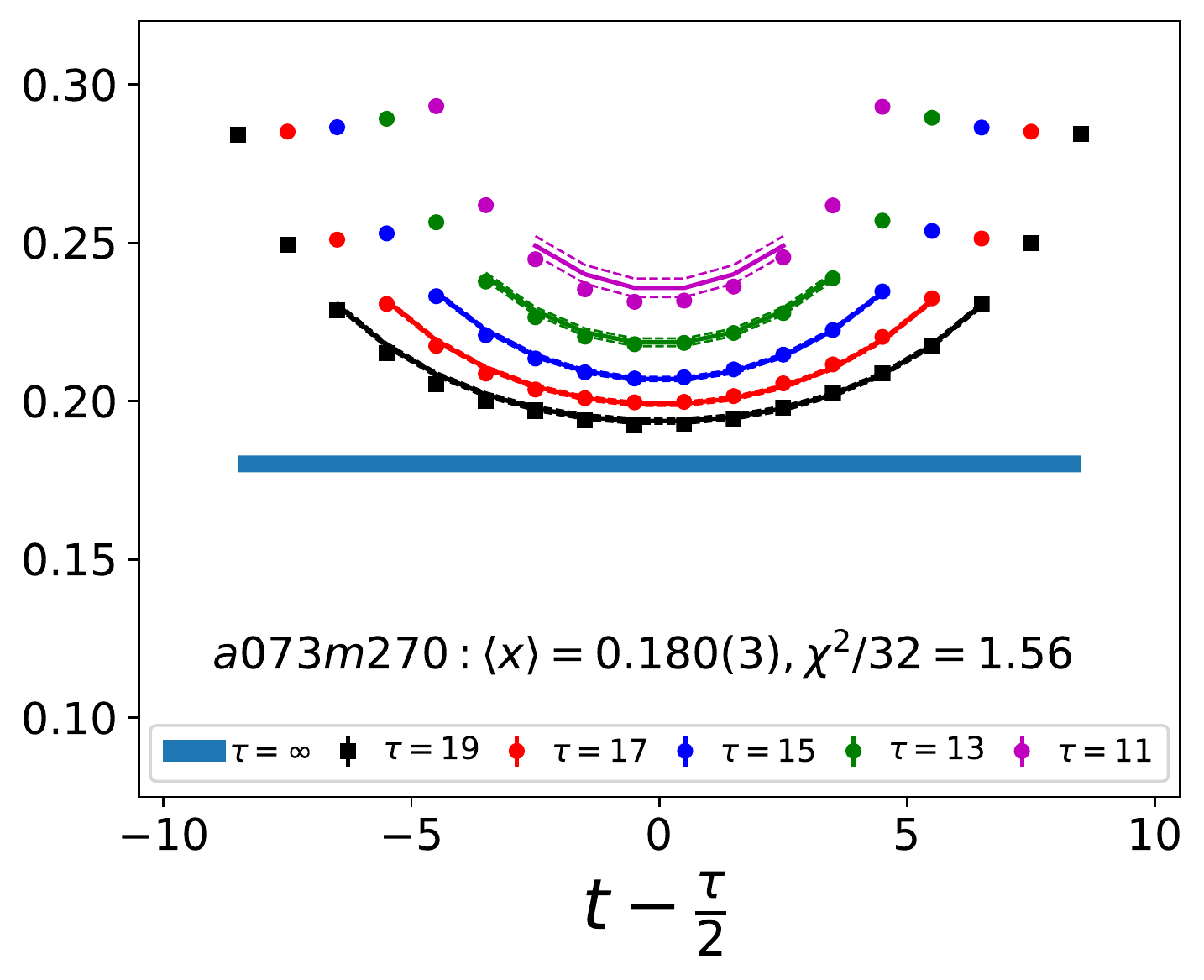}
\end{subfigure}

\begin{subfigure}
\centering
\includegraphics[angle=0,width=0.32\textwidth]{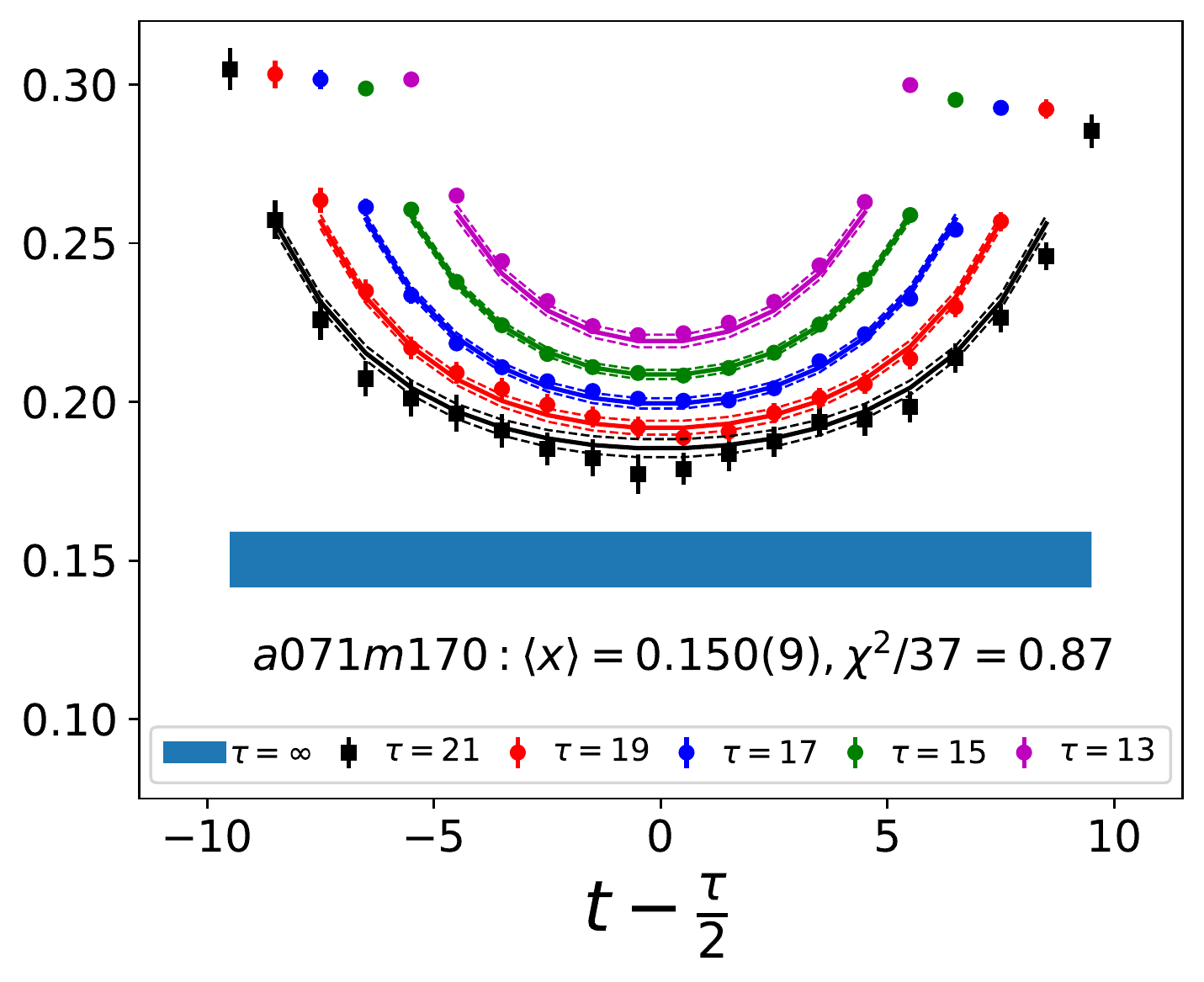}
\includegraphics[angle=0,width=0.32\textwidth]{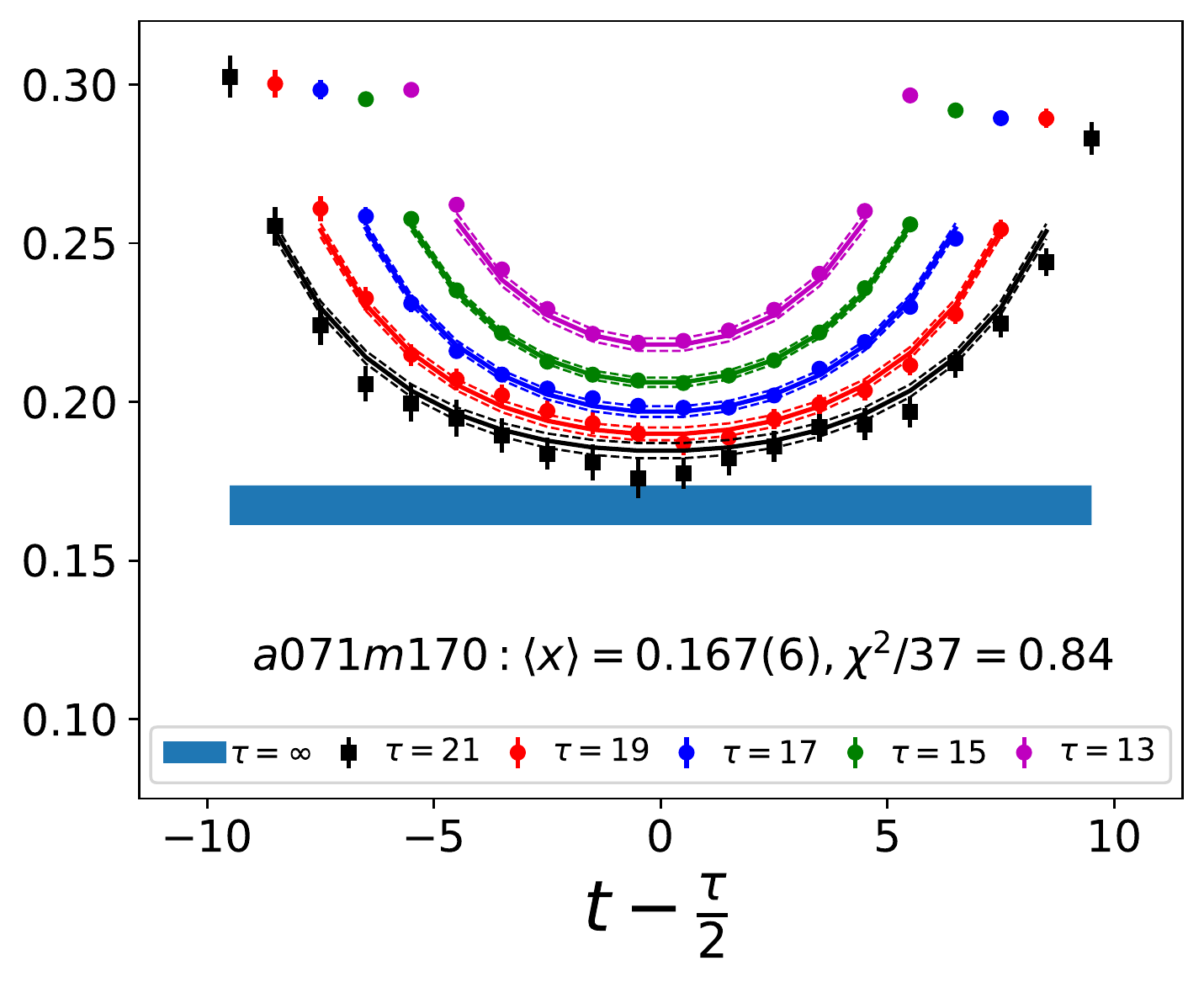}
\includegraphics[angle=0,width=0.32\textwidth]{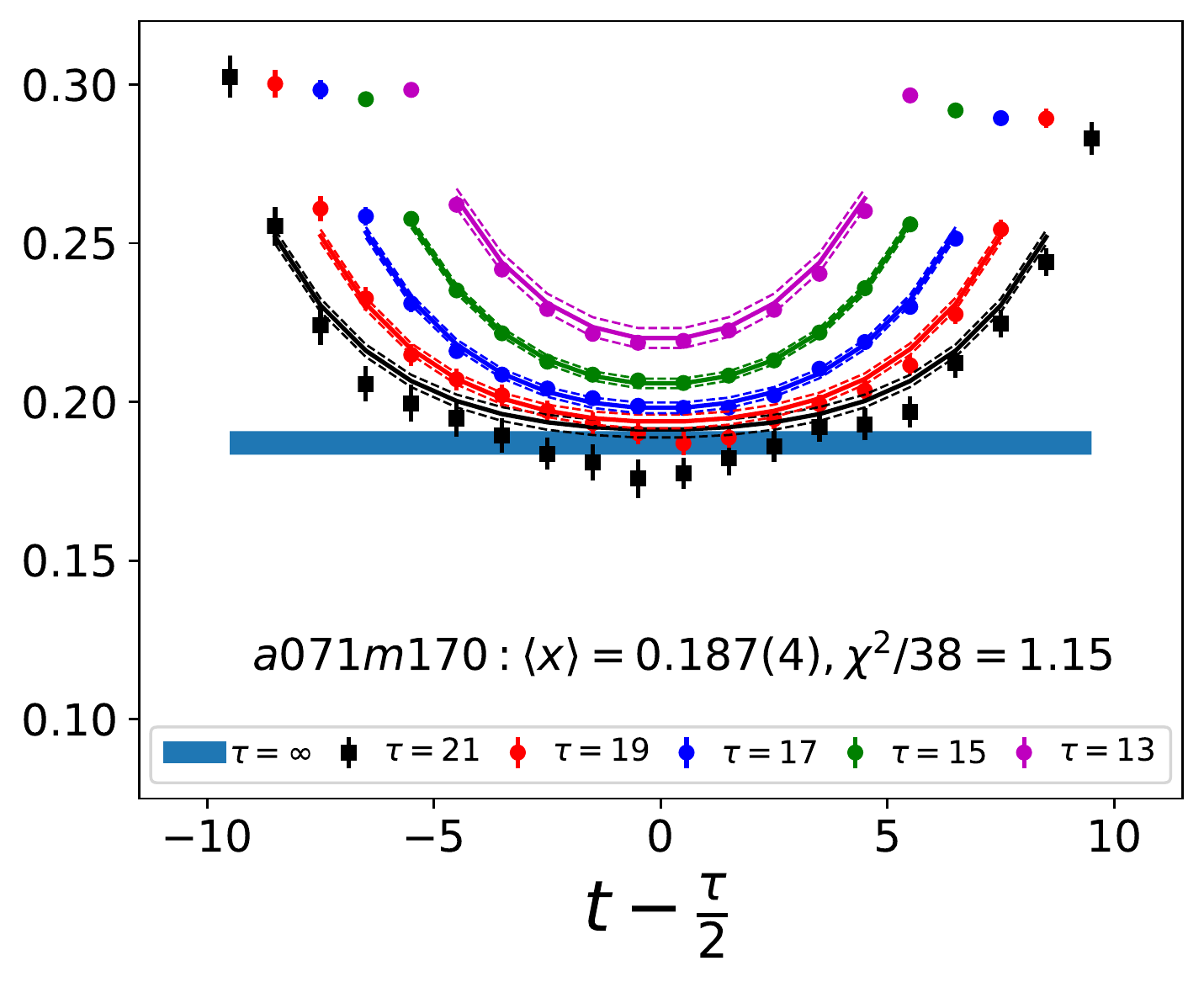}
\end{subfigure}

\caption{Continuation of the data and fits to remove excited-state contamination in the extraction 
  of the momentum fraction $\langle   x \rangle_{u-d}$ for $a091m170$ (top row), 
  $a091m170L$ (middle row), $a073m270$ (third row), and $a071m170$ (bottom row).  The data for the ratio
  $C_\mathcal{O}^{3\text{pt}}(\tau;t)/C^{2\text{pt}}(\tau)$ is scaled
  using Eq.~\protect\eqref{eq:me2momentV} to give $\langle x \rangle_{u-d}$, and 
  the fit parameters are listed in
  Table~\protect\ref{tab:5strategy-fits-momfrac}.  The rest is the same as in
  Fig.~\protect\ref{fig:Ratio-mom-1}.}
\label{fig:Ratio-mom-2} 
\end{figure*}

\begin{figure*}[t]  
\begin{subfigure}
\centering
\includegraphics[angle=0,width=0.32\textwidth]{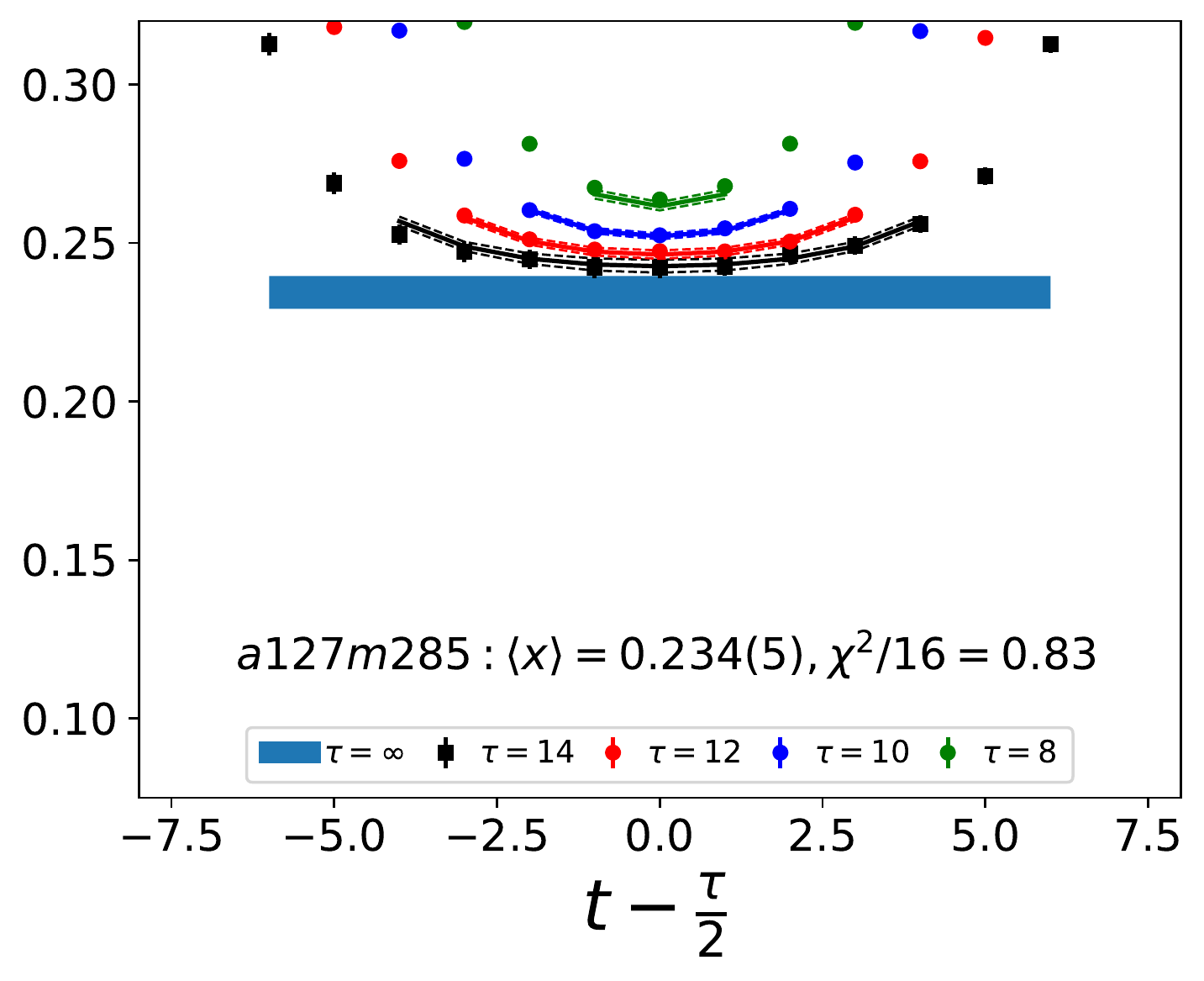}
\includegraphics[angle=0,width=0.32\textwidth]{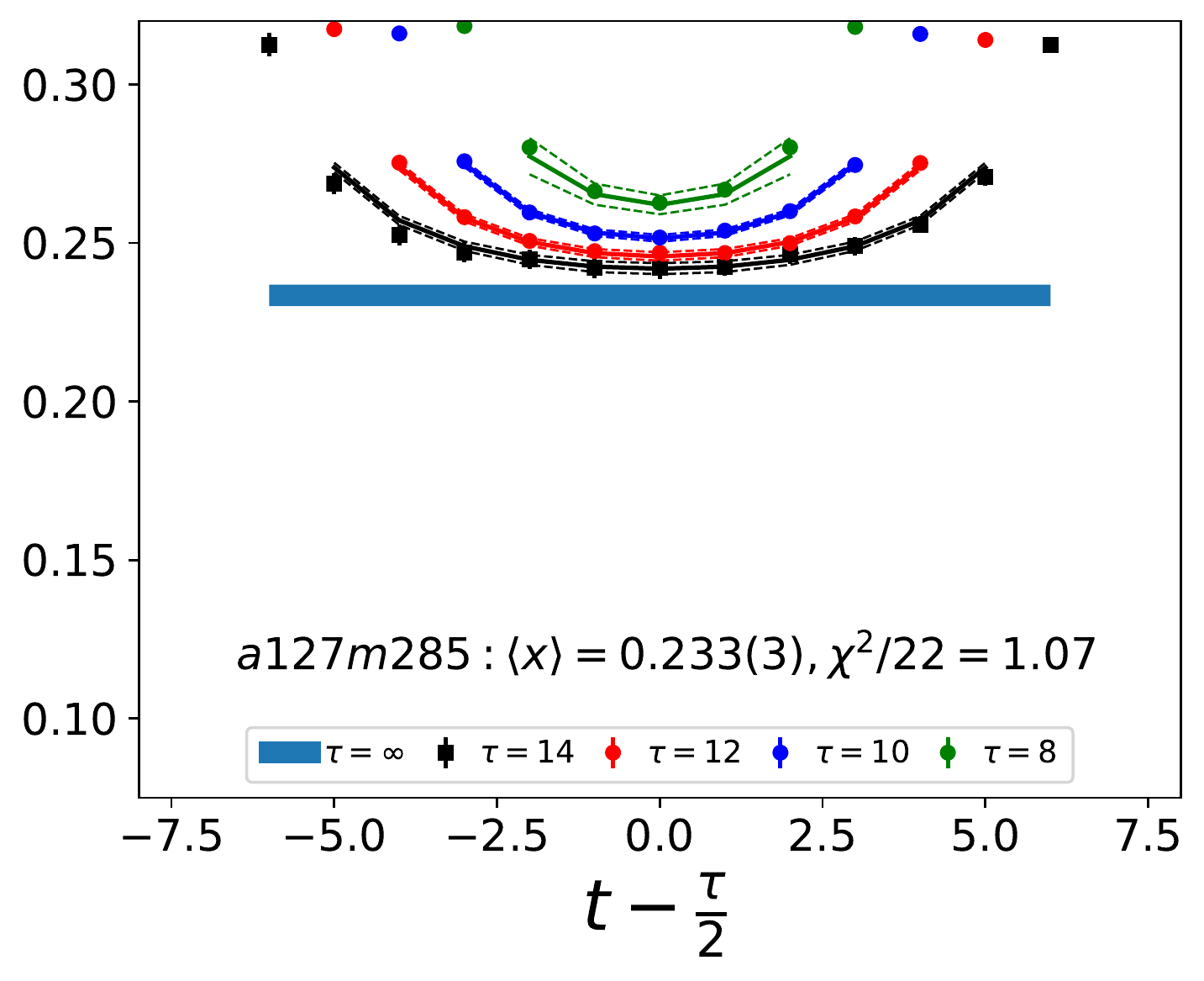}
\includegraphics[angle=0,width=0.32\textwidth]{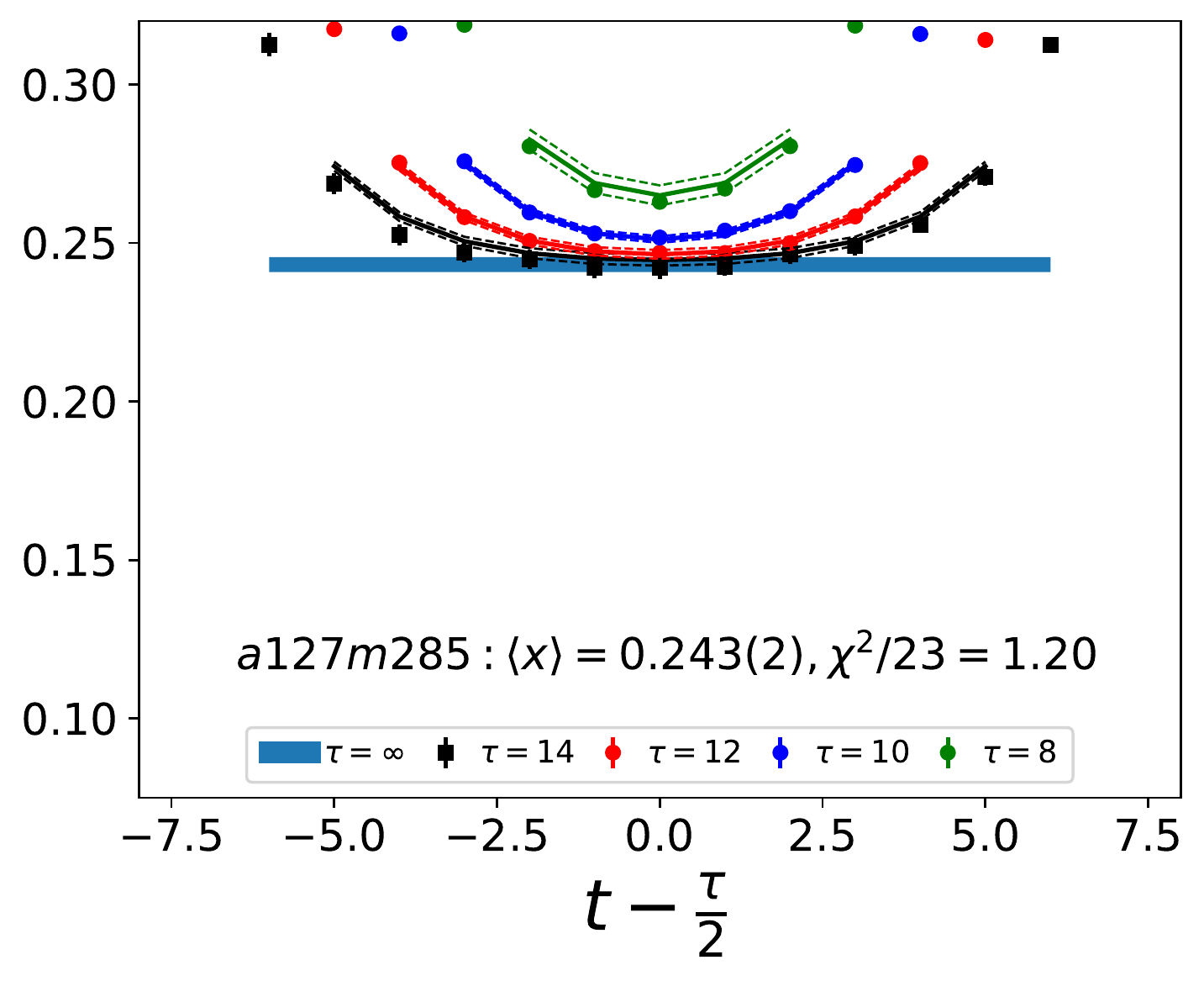}

\includegraphics[angle=0,width=0.32\textwidth]{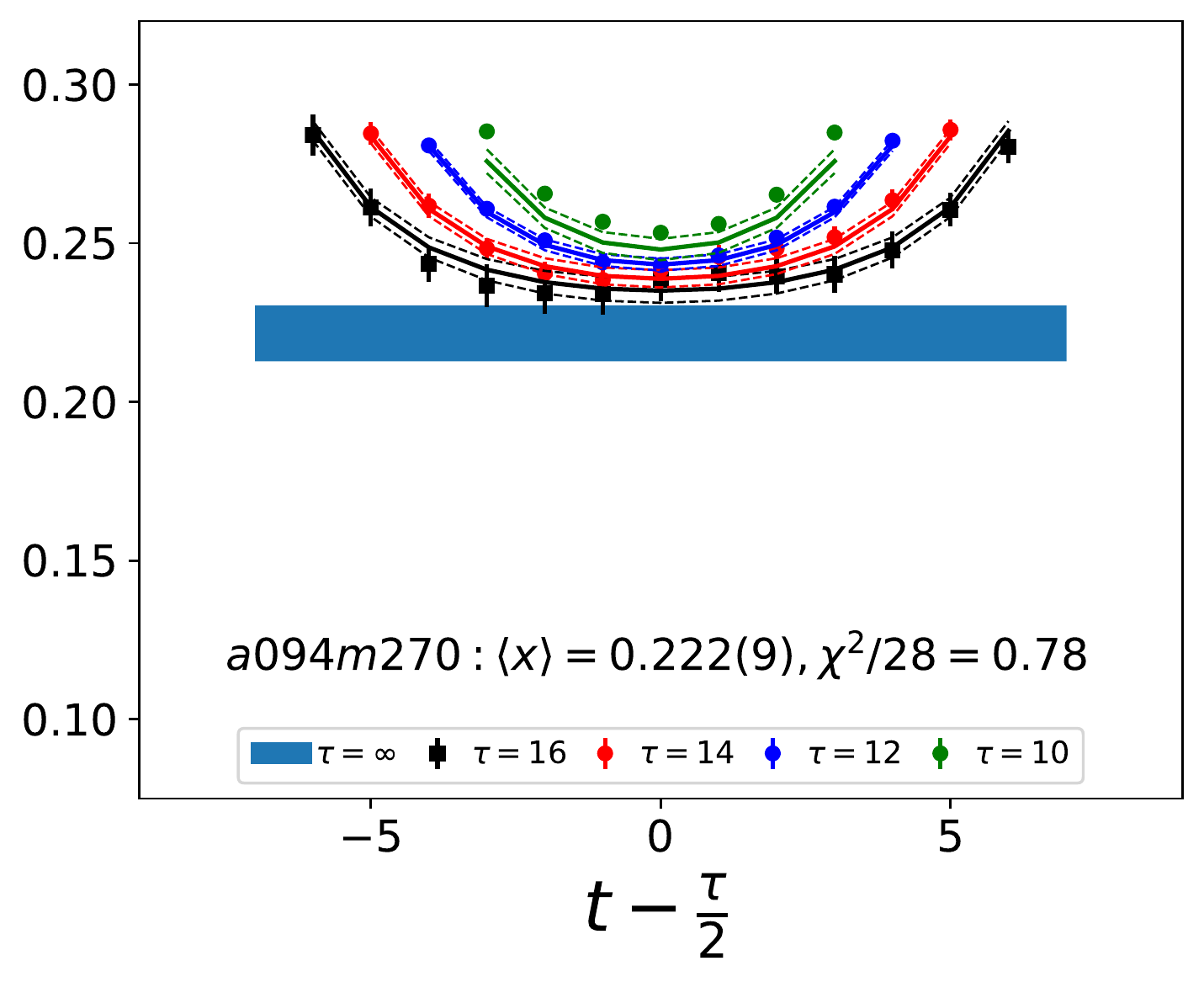}
\includegraphics[angle=0,width=0.32\textwidth]{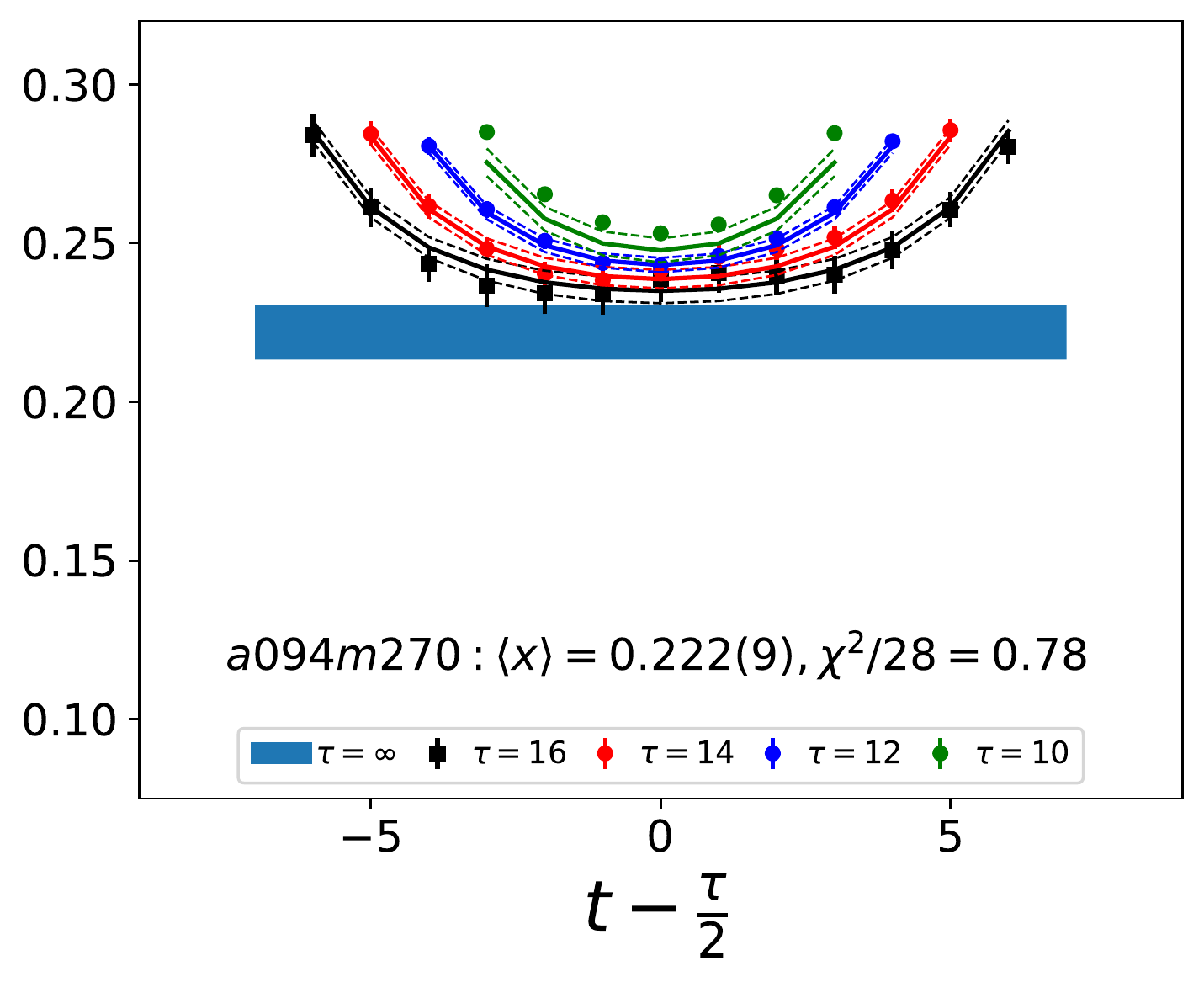}
\includegraphics[angle=0,width=0.32\textwidth]{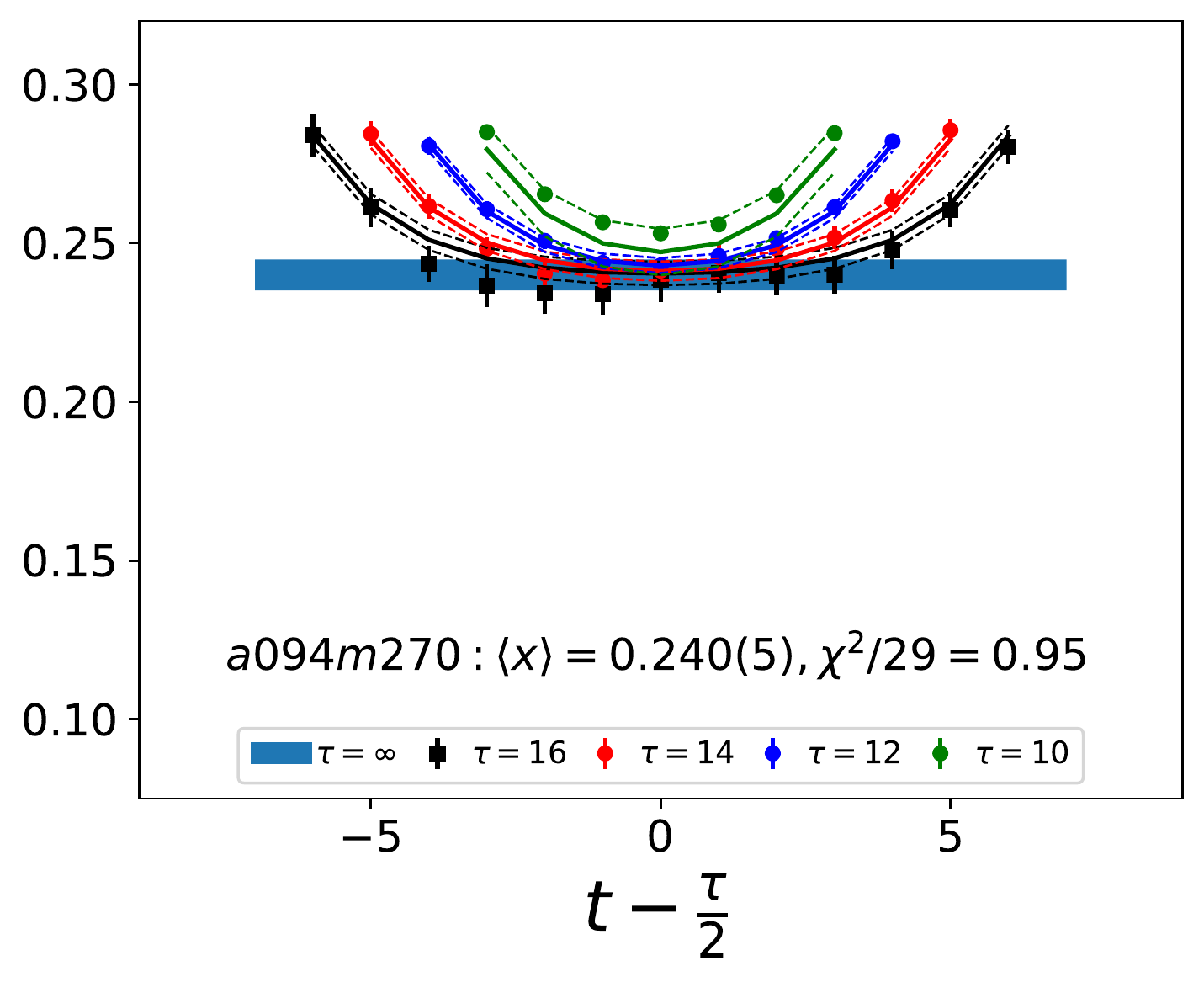}

\includegraphics[angle=0,width=0.32\textwidth]{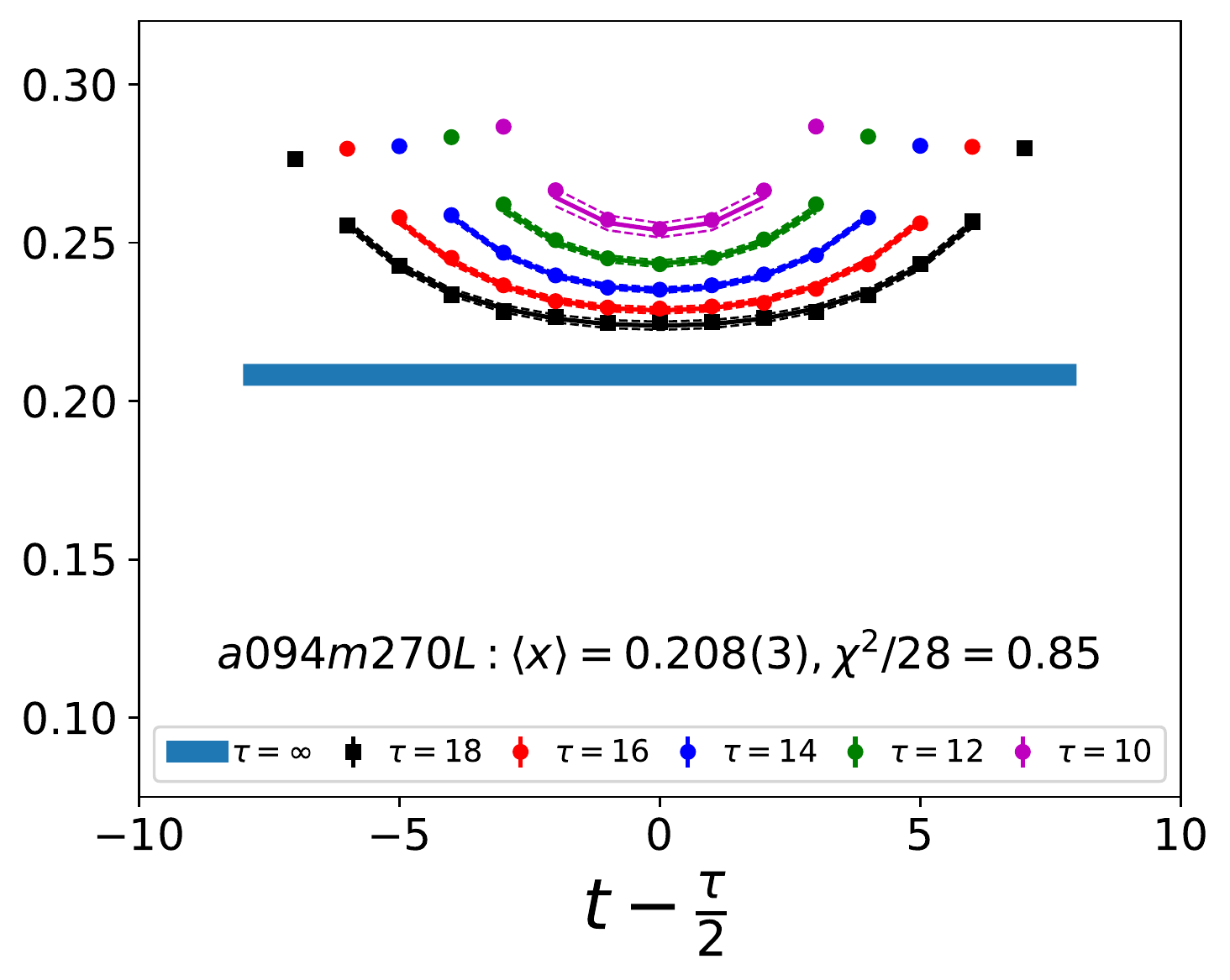}
\includegraphics[angle=0,width=0.32\textwidth]{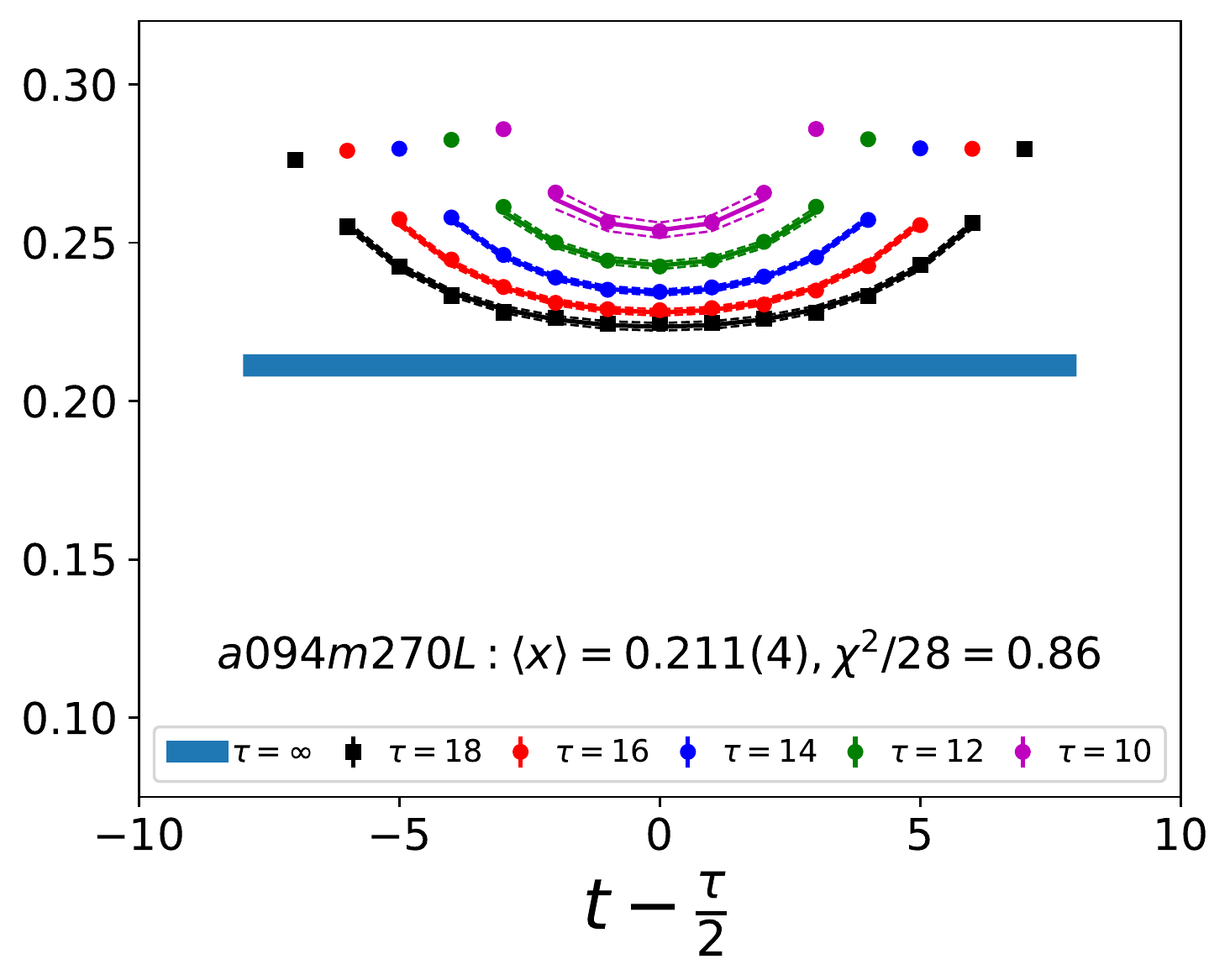}
\includegraphics[angle=0,width=0.32\textwidth]{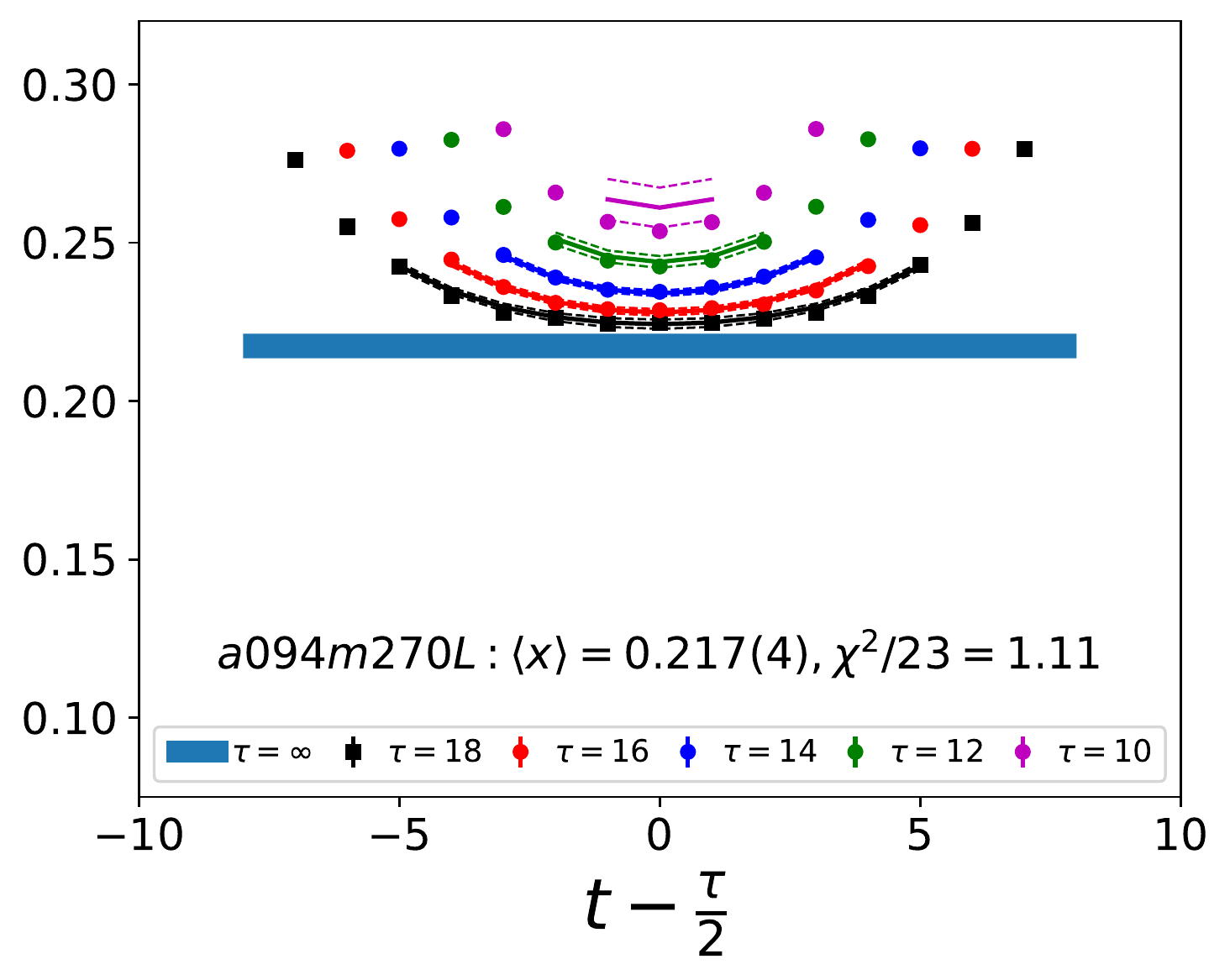}
\end{subfigure}

\caption{Data and fits to remove excited-state contamination in the extraction 
  of the helicity moment $\langle x \rangle_{\Delta u- \Delta d}$ for
  for $a127m285$ (top row), $a094m270$ (second row), and $a094m270L$ (bottom row) ensembles. 
  The data for the ratio
  $C_\mathcal{O}^{3\text{pt}}(\tau;t)/C^{2\text{pt}}(\tau)$ is scaled
  using Eq.~\protect\eqref{eq:me2momentA} to give $\langle
  x \rangle_{\Delta u-\Delta d}$, and the fit parameters are listed in
  Table~\protect\ref{tab:5strategy-fits-helfrac}.  The rest is the
  same as in Fig.~\protect\ref{fig:Ratio-mom-1}.}
\label{fig:Ratio-helicity-1}
\end{figure*}

\begin{figure*}[t]   
\centering

\begin{subfigure}
\centering
\includegraphics[angle=0,width=0.32\textwidth]{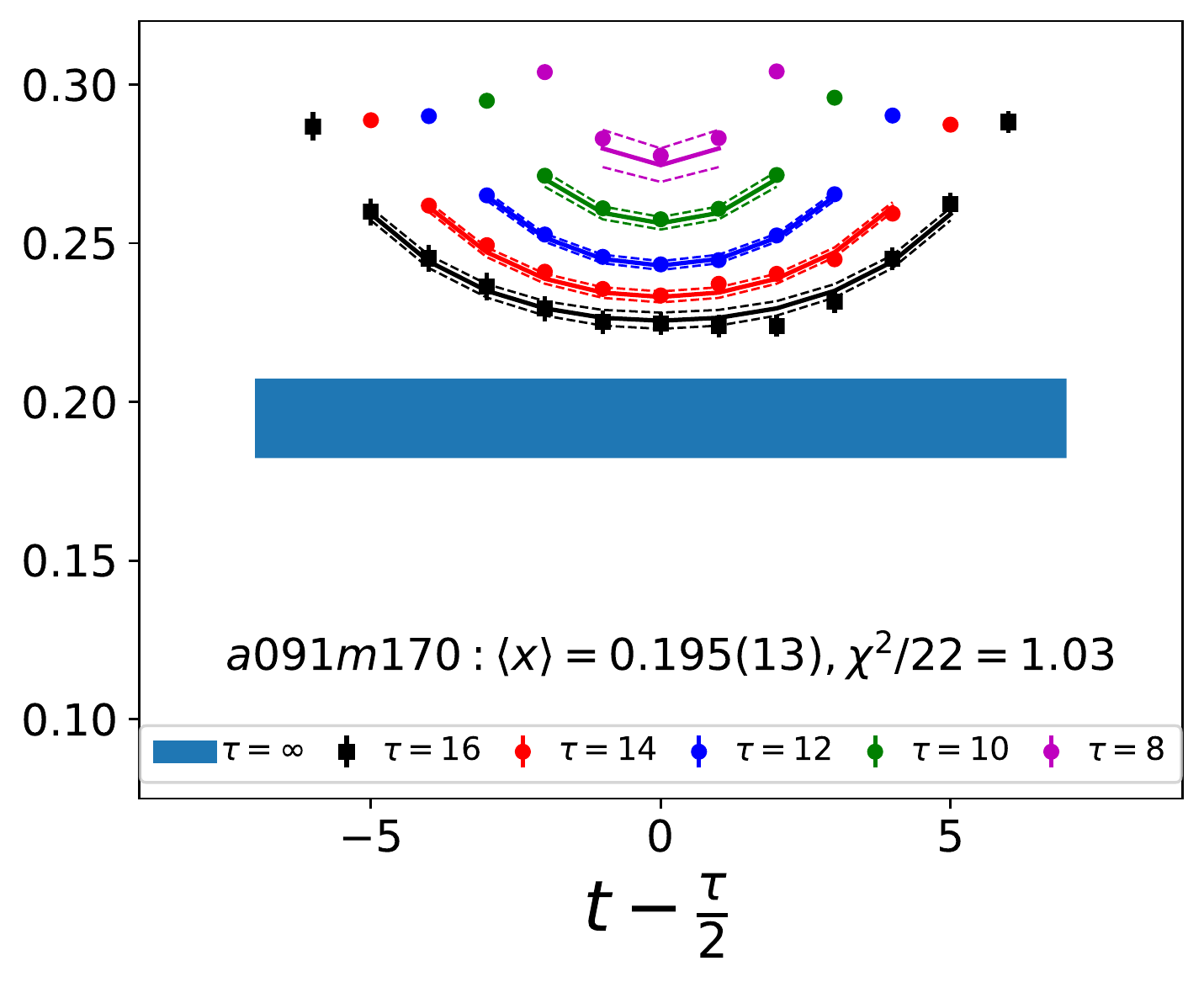}
\includegraphics[angle=0,width=0.32\textwidth]{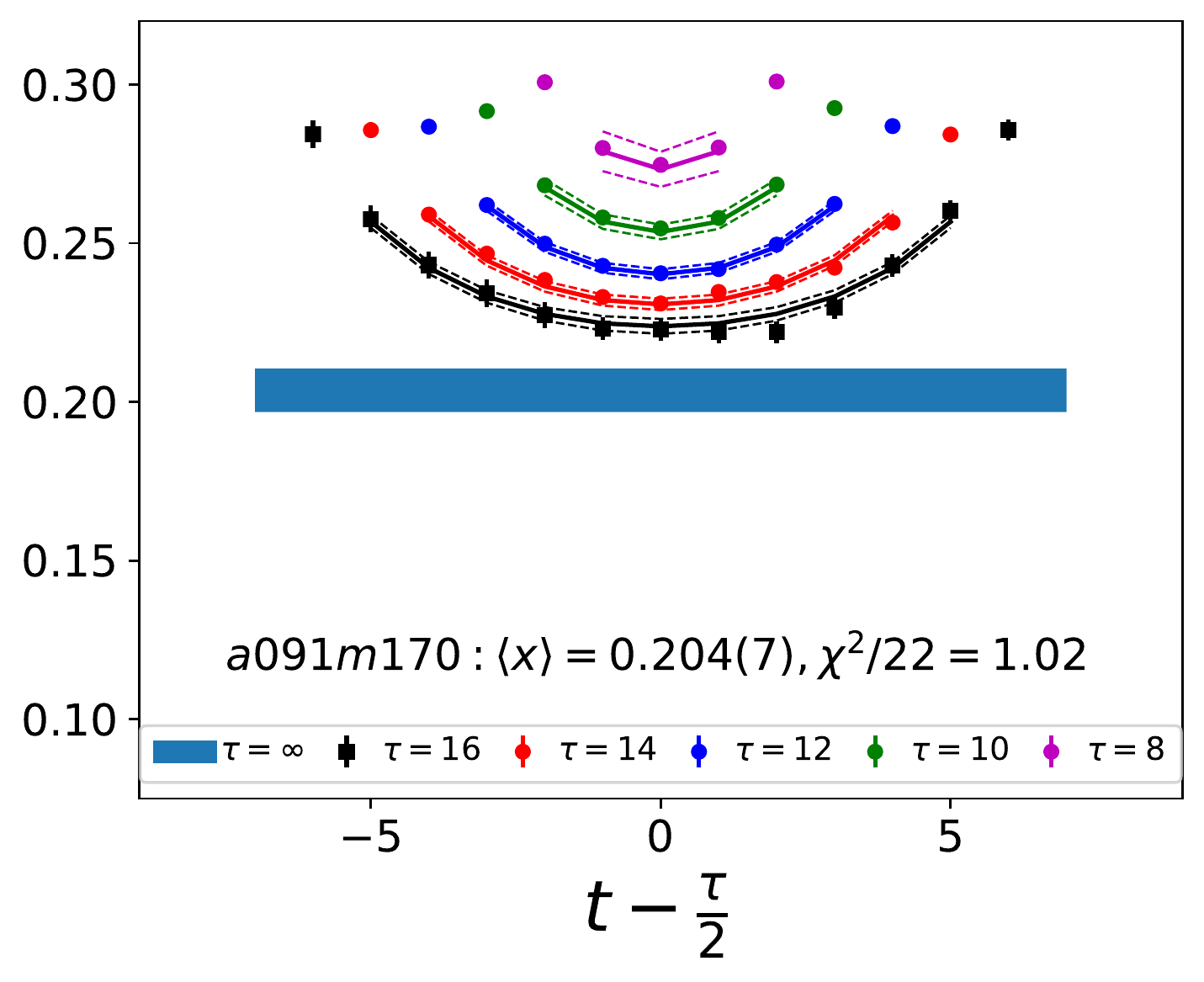}
\includegraphics[angle=0,width=0.32\textwidth]{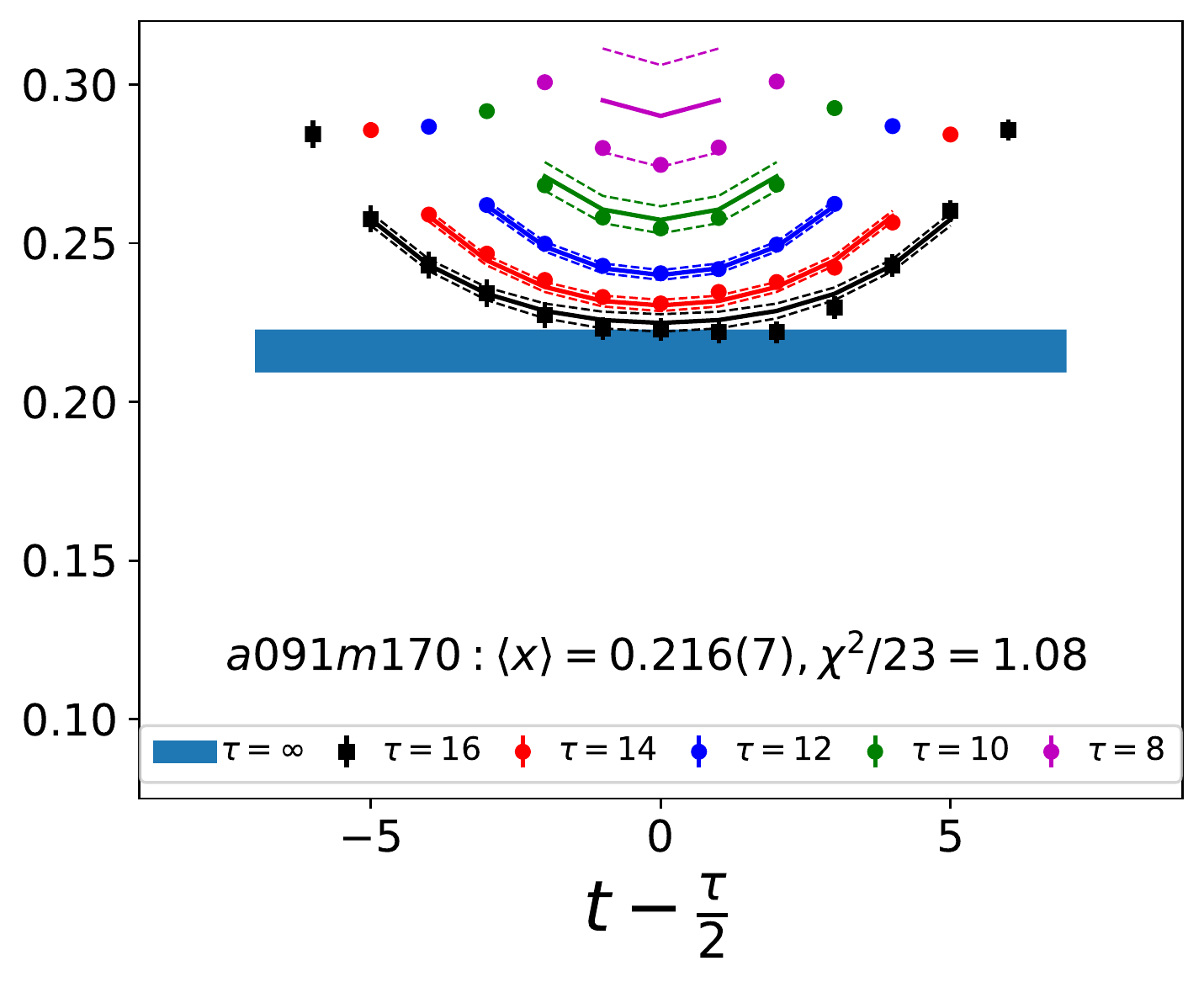}
\end{subfigure}

\begin{subfigure}
\centering
\includegraphics[angle=0,width=0.32\textwidth]{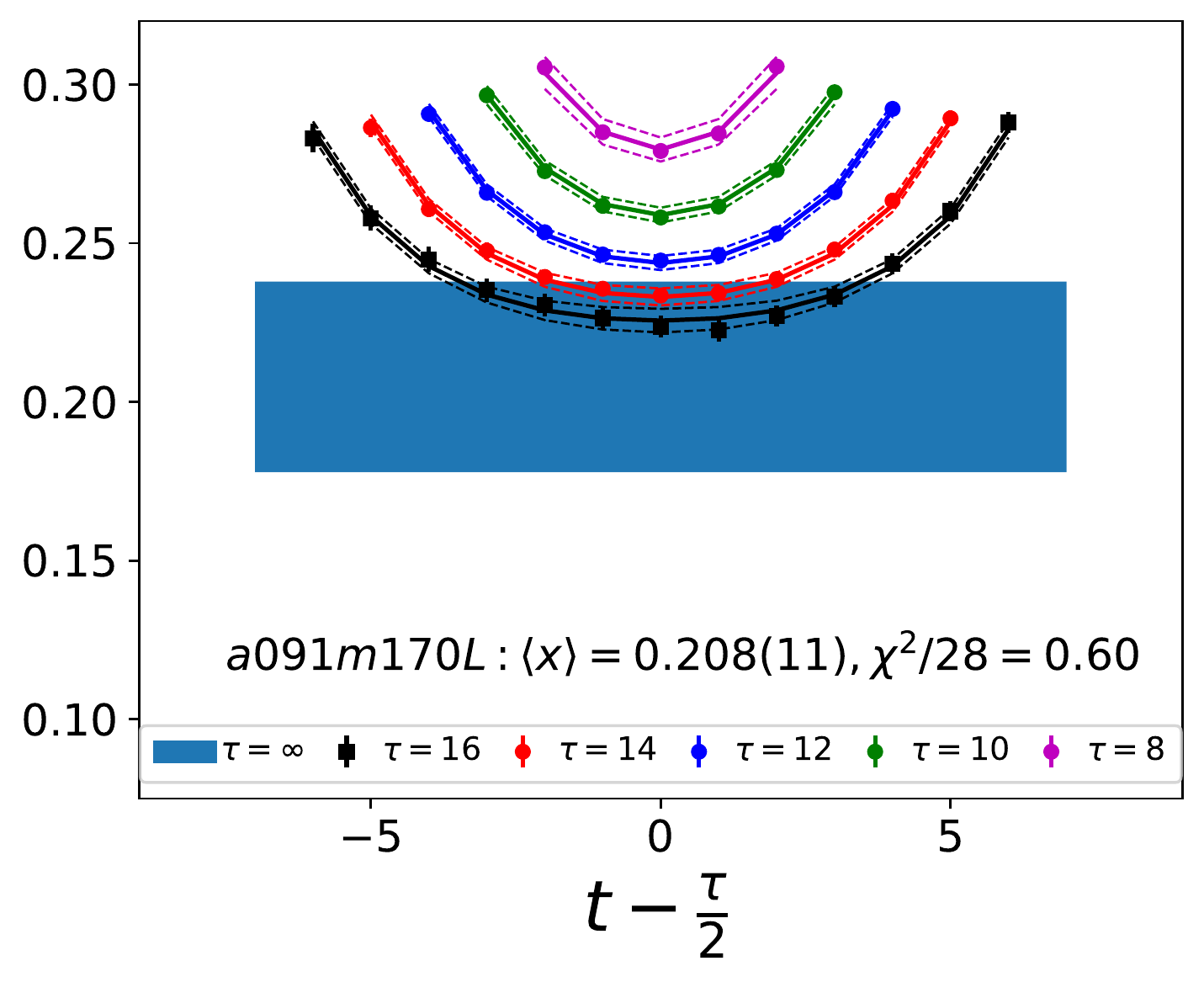}
\includegraphics[angle=0,width=0.32\textwidth]{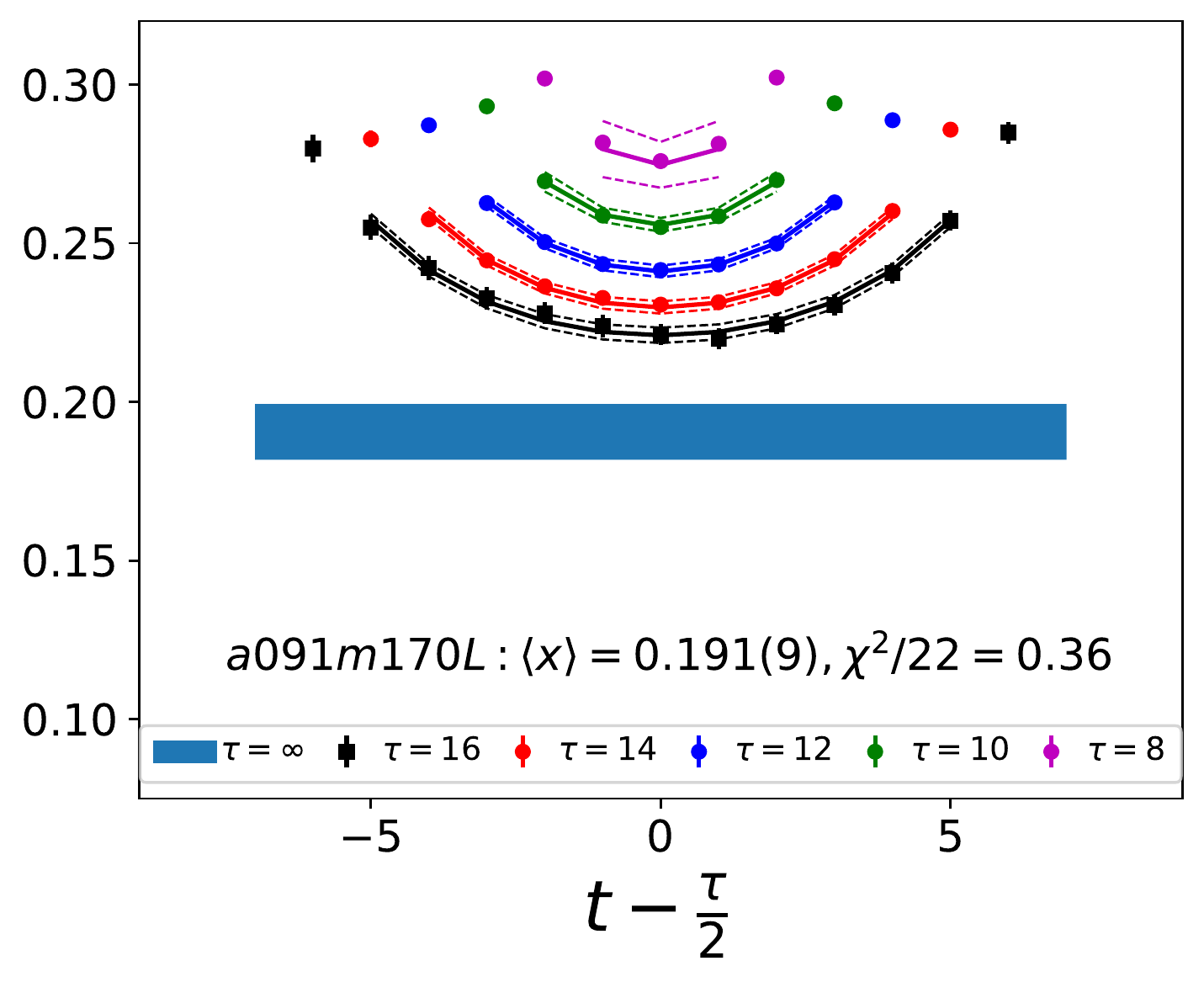}
\includegraphics[angle=0,width=0.32\textwidth]{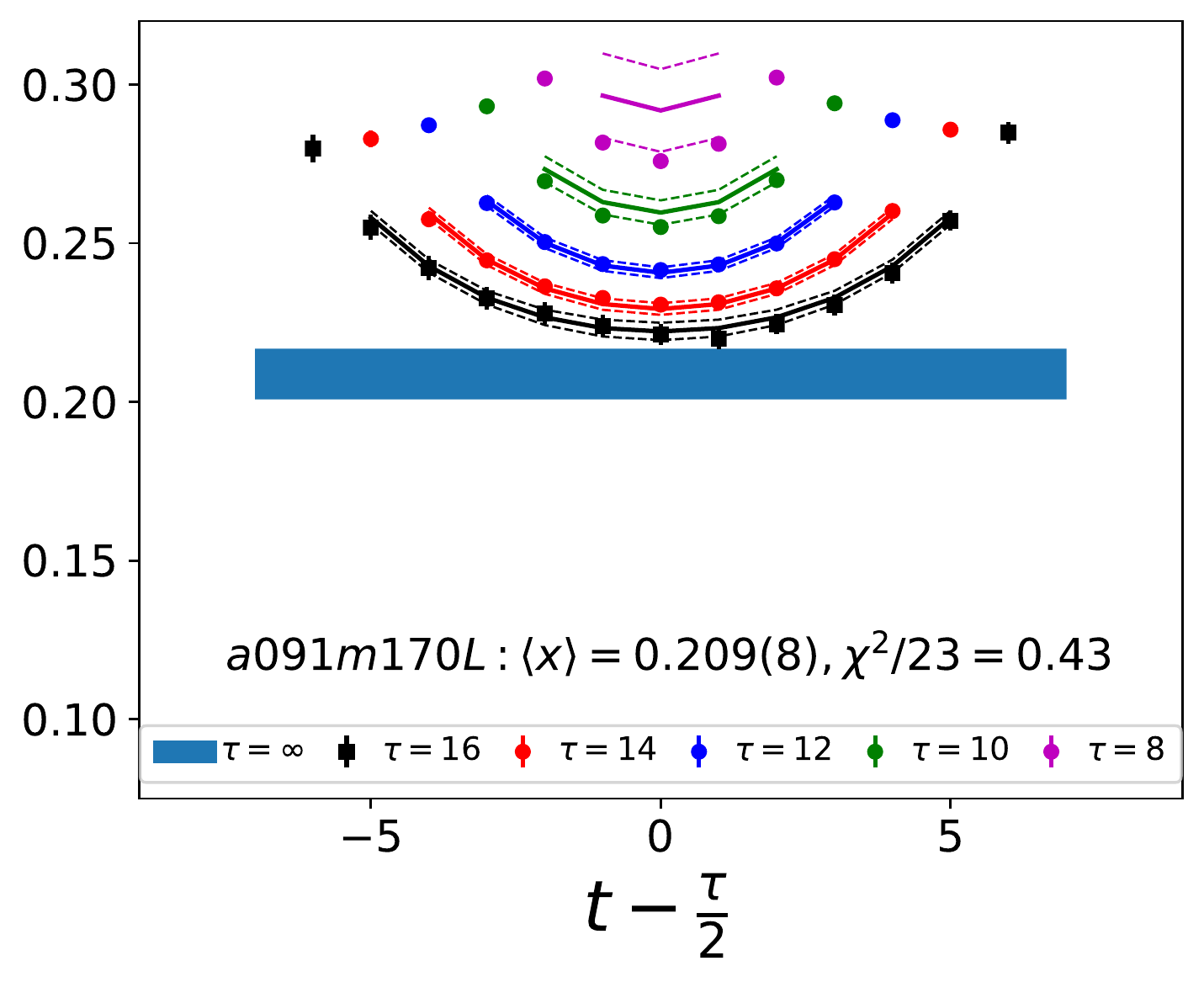}
\end{subfigure}

\begin{subfigure}
\centering
\includegraphics[angle=0,width=0.32\textwidth]{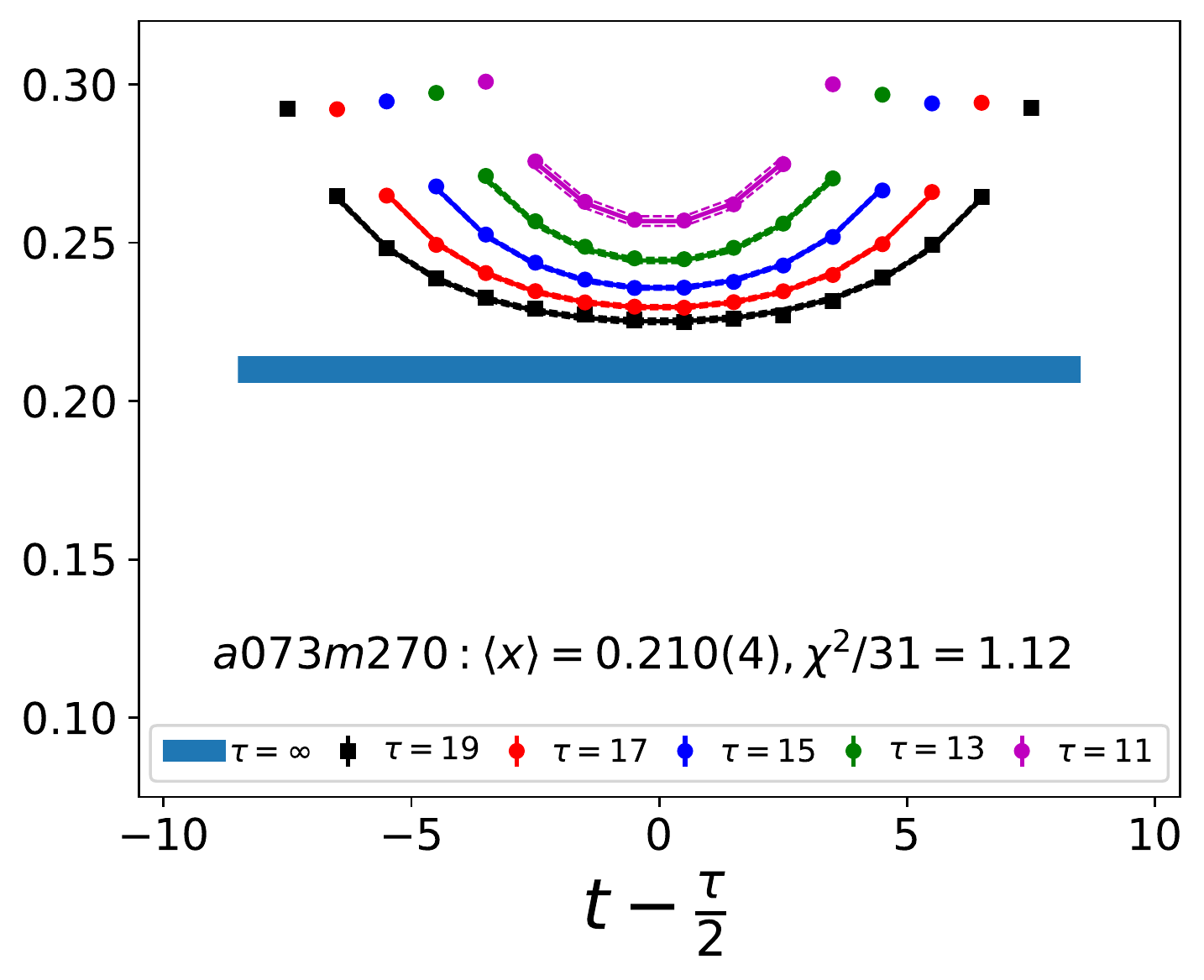}
\includegraphics[angle=0,width=0.32\textwidth]{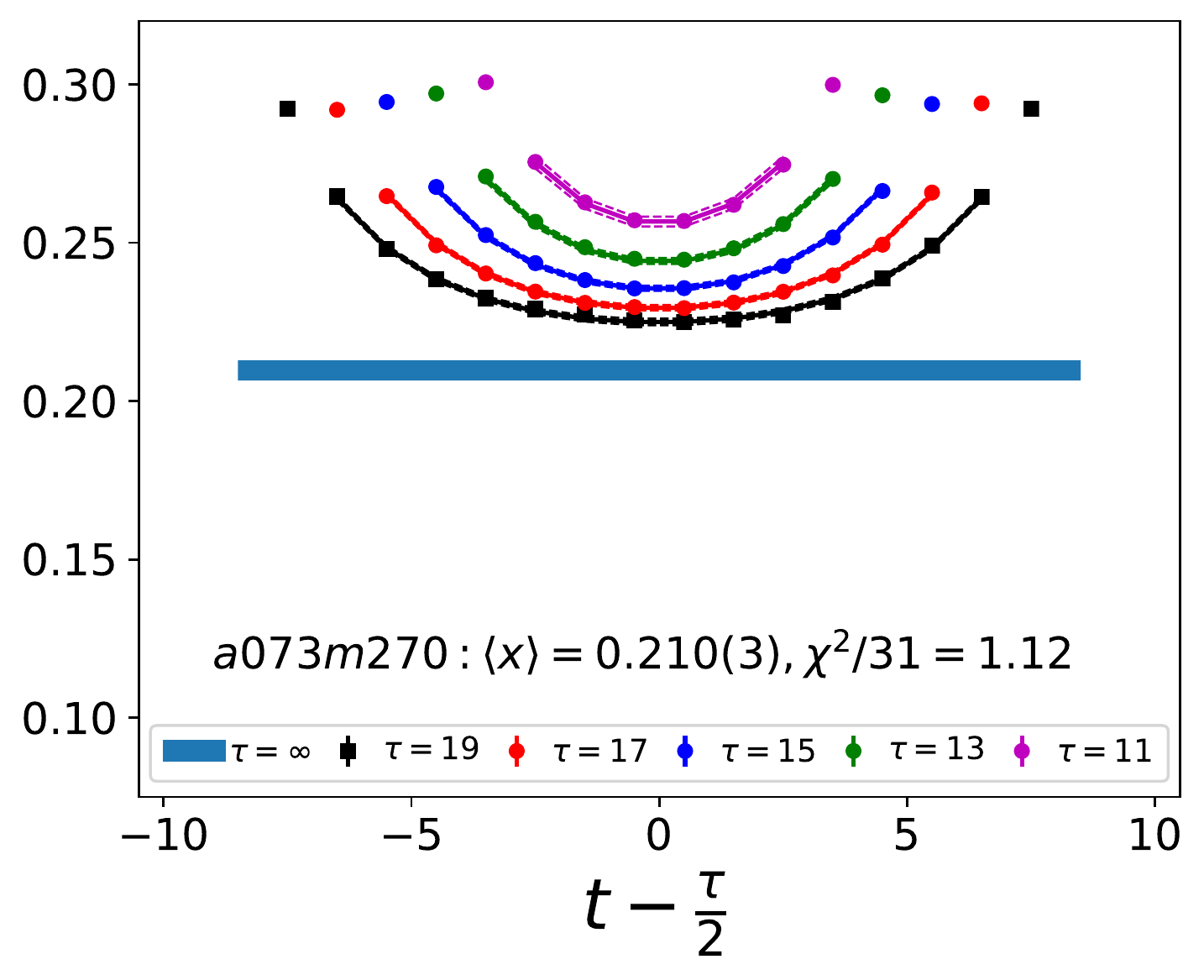}
\includegraphics[angle=0,width=0.32\textwidth]{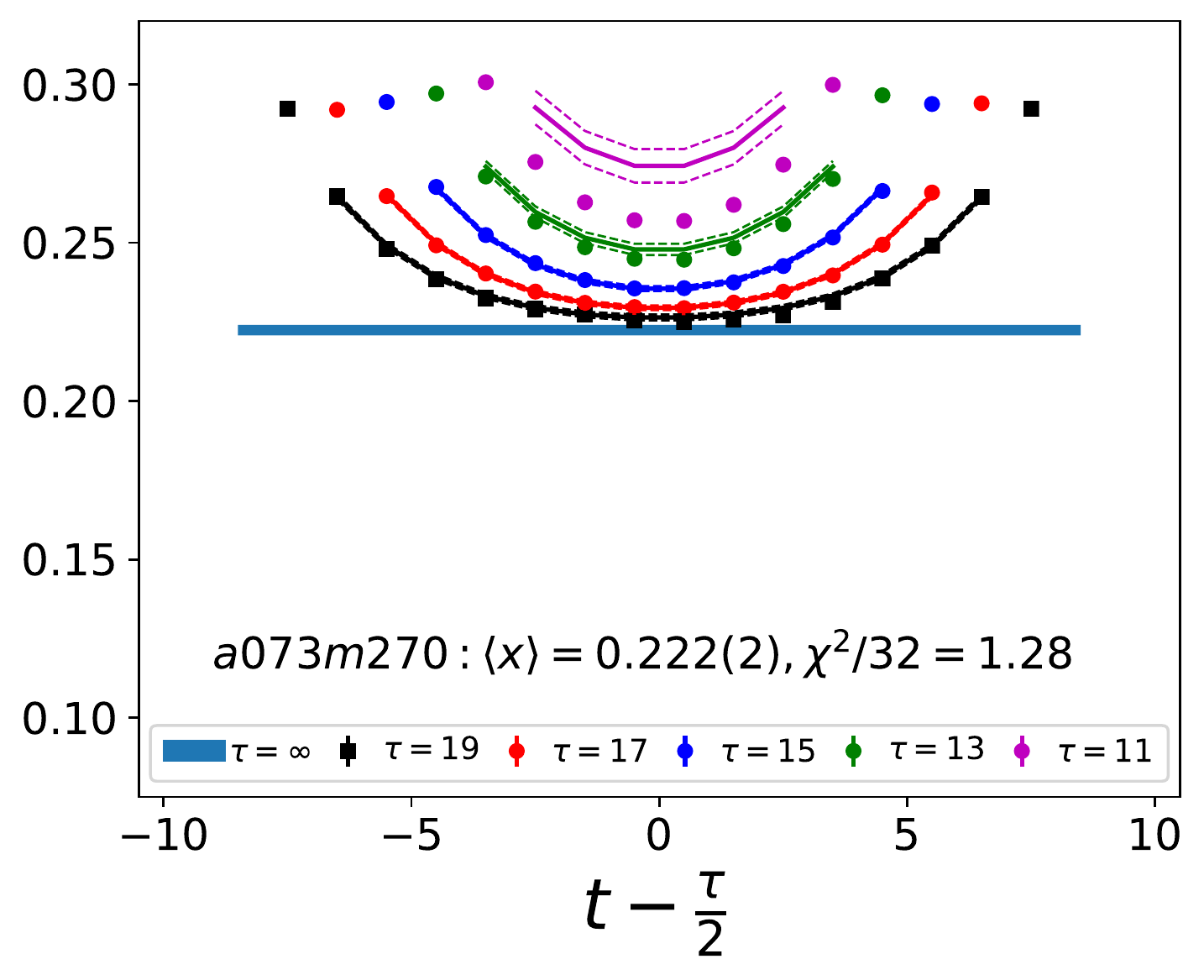}
\end{subfigure}

\begin{subfigure}
\centering
\includegraphics[angle=0,width=0.32\textwidth]{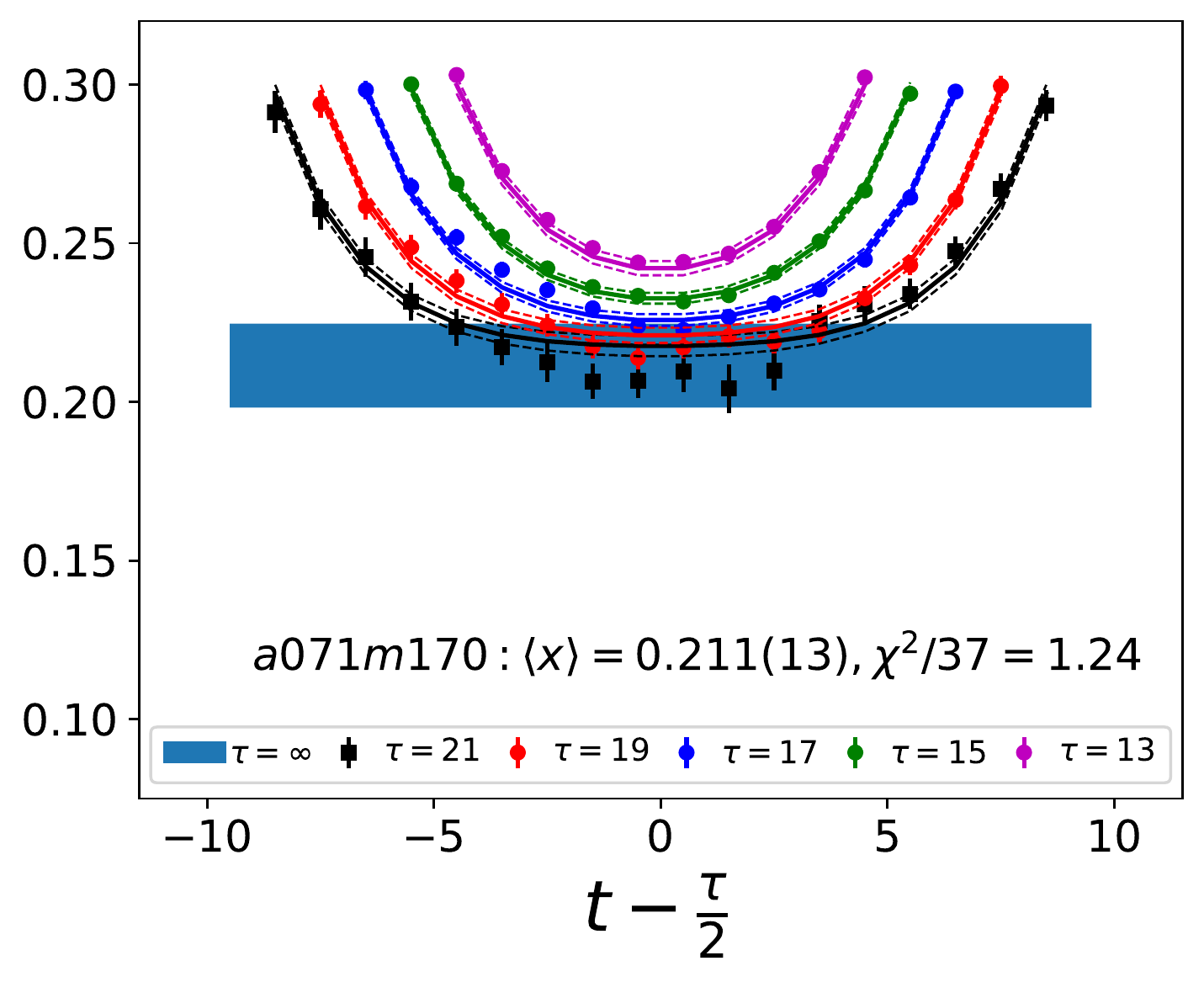}
\includegraphics[angle=0,width=0.32\textwidth]{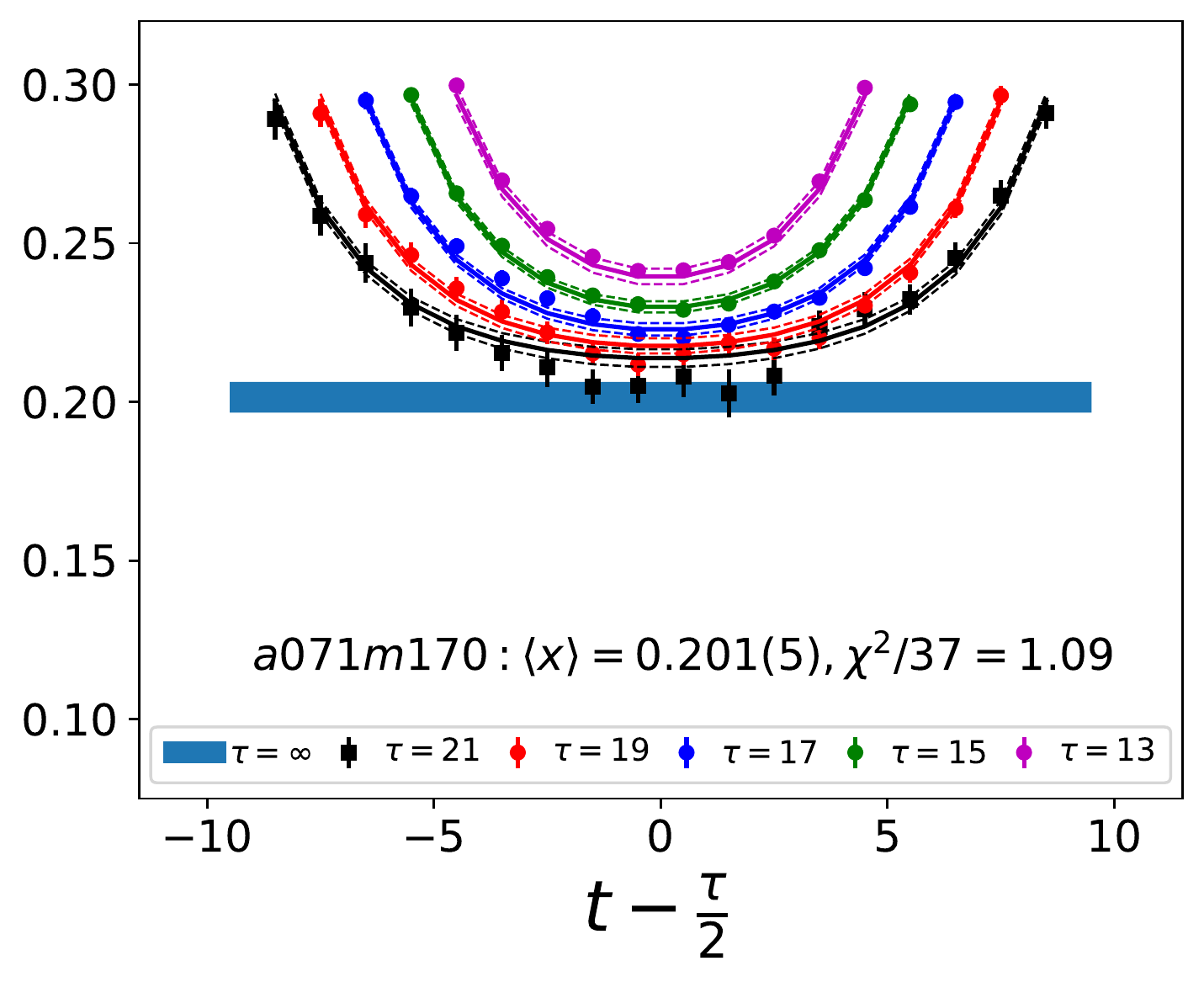}
\includegraphics[angle=0,width=0.32\textwidth]{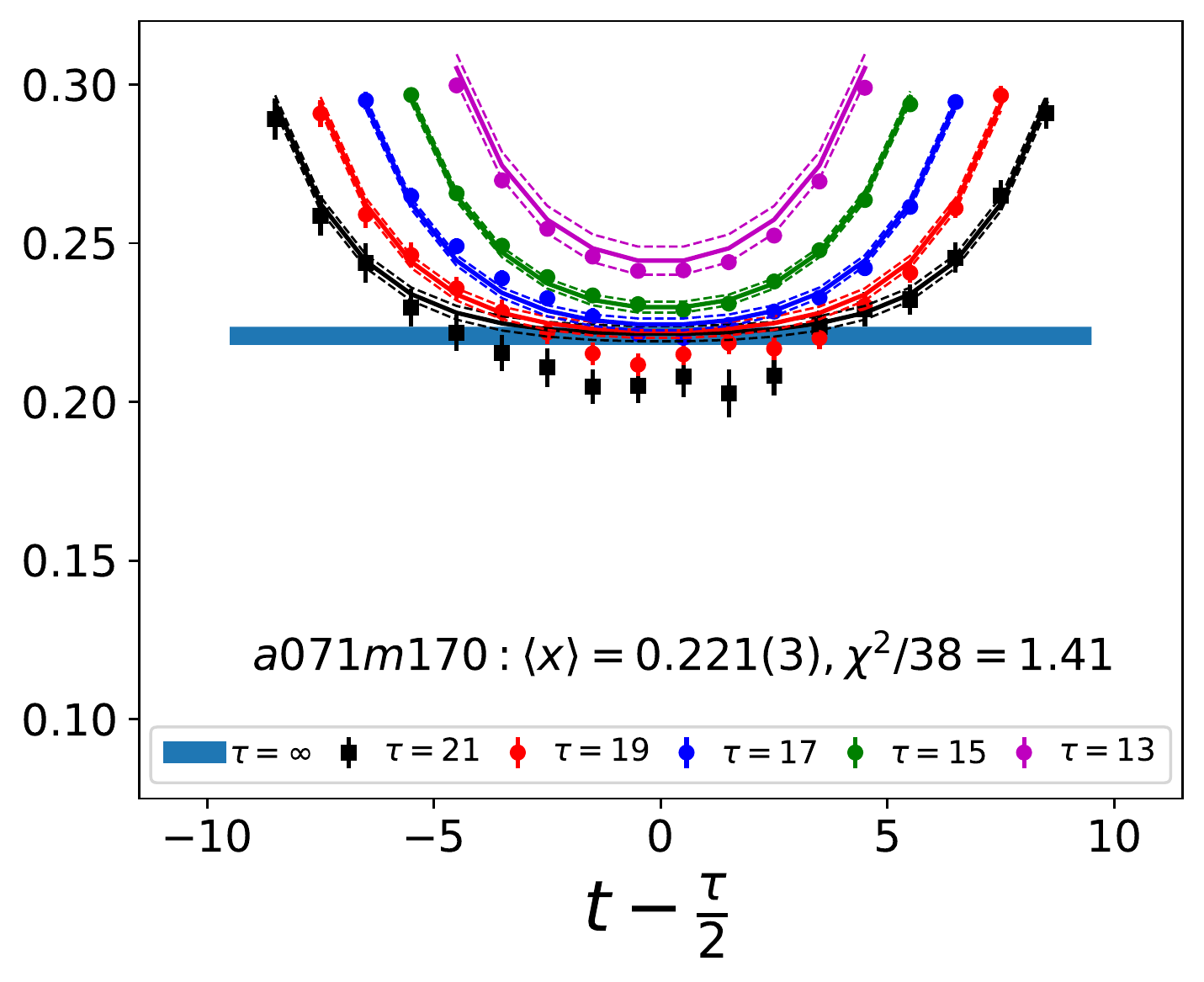}
\end{subfigure}

\caption{Continuation of the data and fits to remove excited-state contamination in the extraction 
  of the helicity moment $\langle   x \rangle_{\Delta u - \Delta d}$ for $a091m170$ (top row),
  $a091m170L$ (middle row), $a073m270$ (third row), and $a071m170$ (bottom row).  The data for the ratio
  $C_\mathcal{O}^{3\text{pt}}(\tau;t)/C^{2\text{pt}}(\tau)$ is scaled
  using Eq.~\protect\eqref{eq:me2momentA} to give $\langle
  x \rangle_{\Delta u-\Delta d}$, and the fit parameters are listed in
  Table~\protect\ref{tab:5strategy-fits-helfrac}.  The rest is the same as in
  Fig.~\protect\ref{fig:Ratio-mom-1}. }
\label{fig:Ratio-helicity-2} 
\end{figure*}

\begin{figure*}[t]  
\begin{subfigure}
\centering

\includegraphics[angle=0,width=0.32\textwidth]{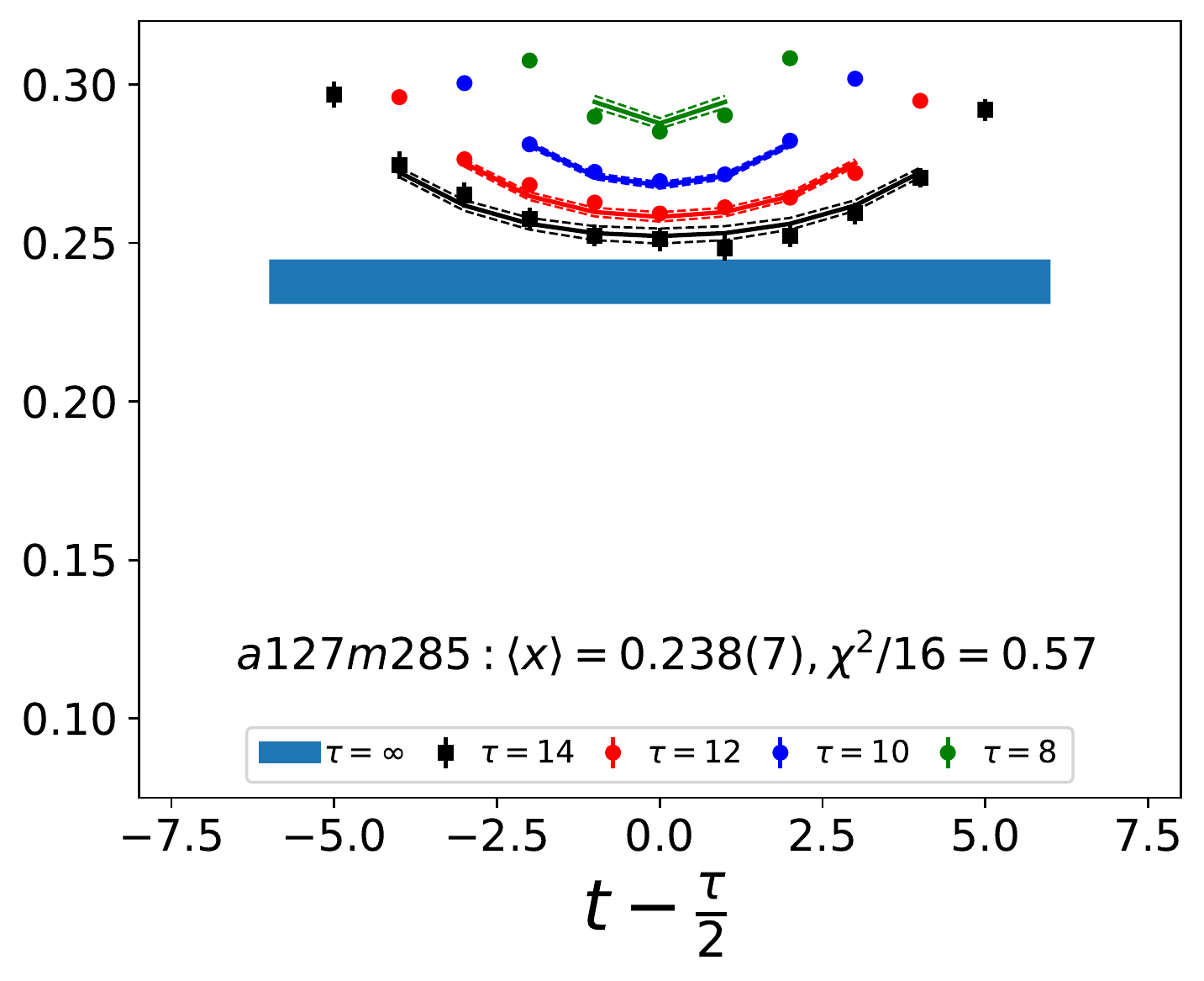}
\includegraphics[angle=0,width=0.32\textwidth]{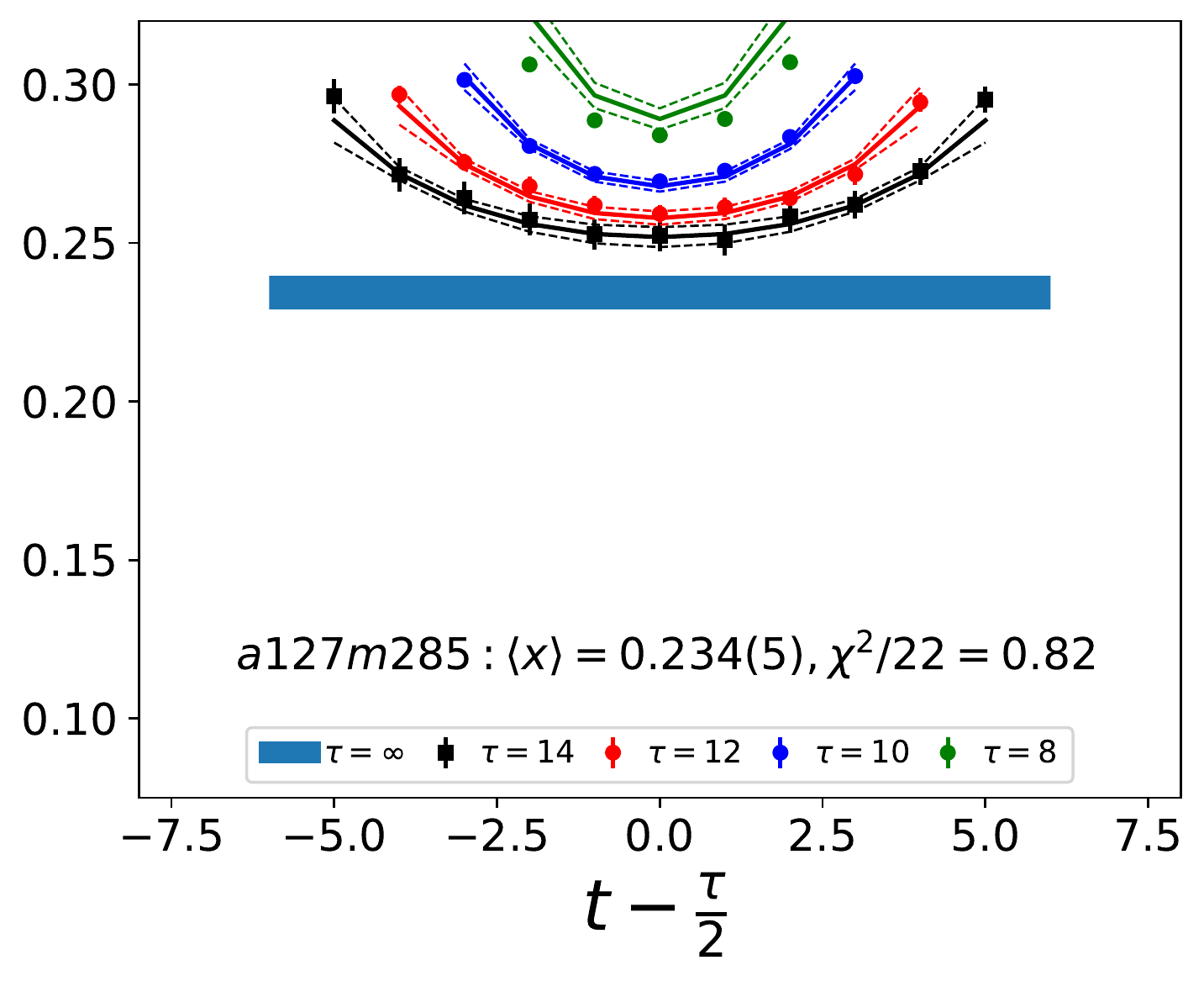}
\includegraphics[angle=0,width=0.32\textwidth]{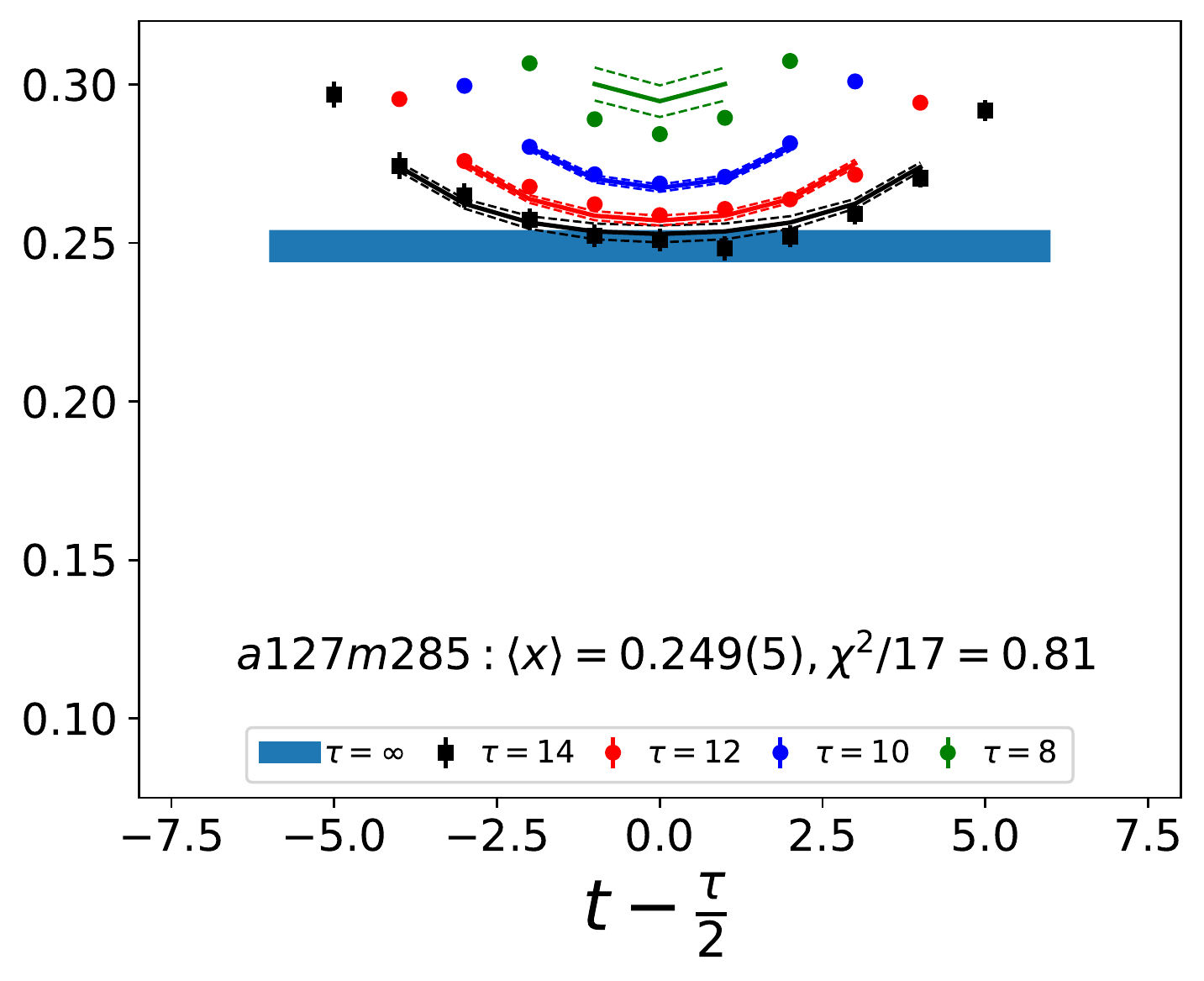}

\includegraphics[angle=0,width=0.32\textwidth]{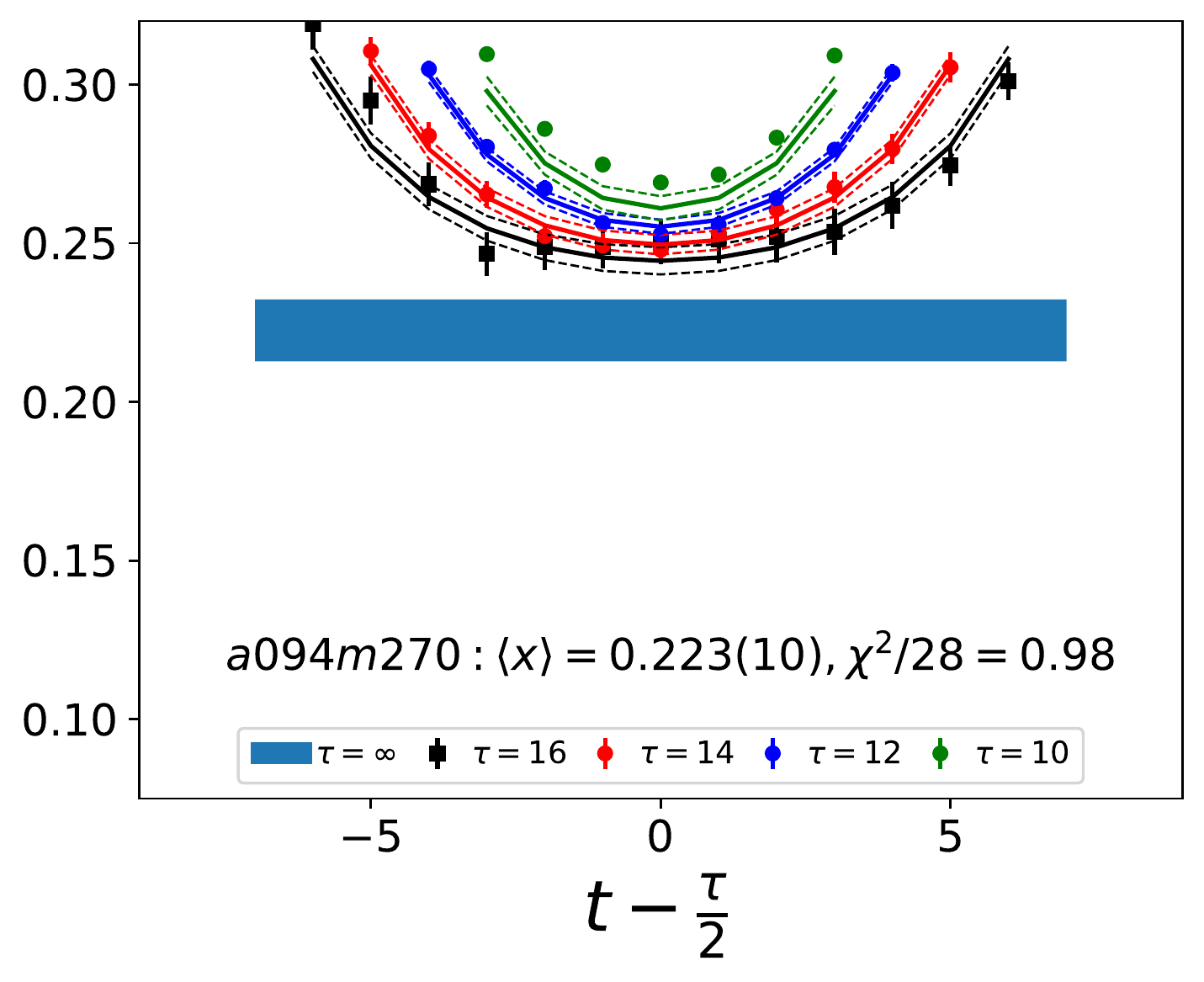}
\includegraphics[angle=0,width=0.32\textwidth]{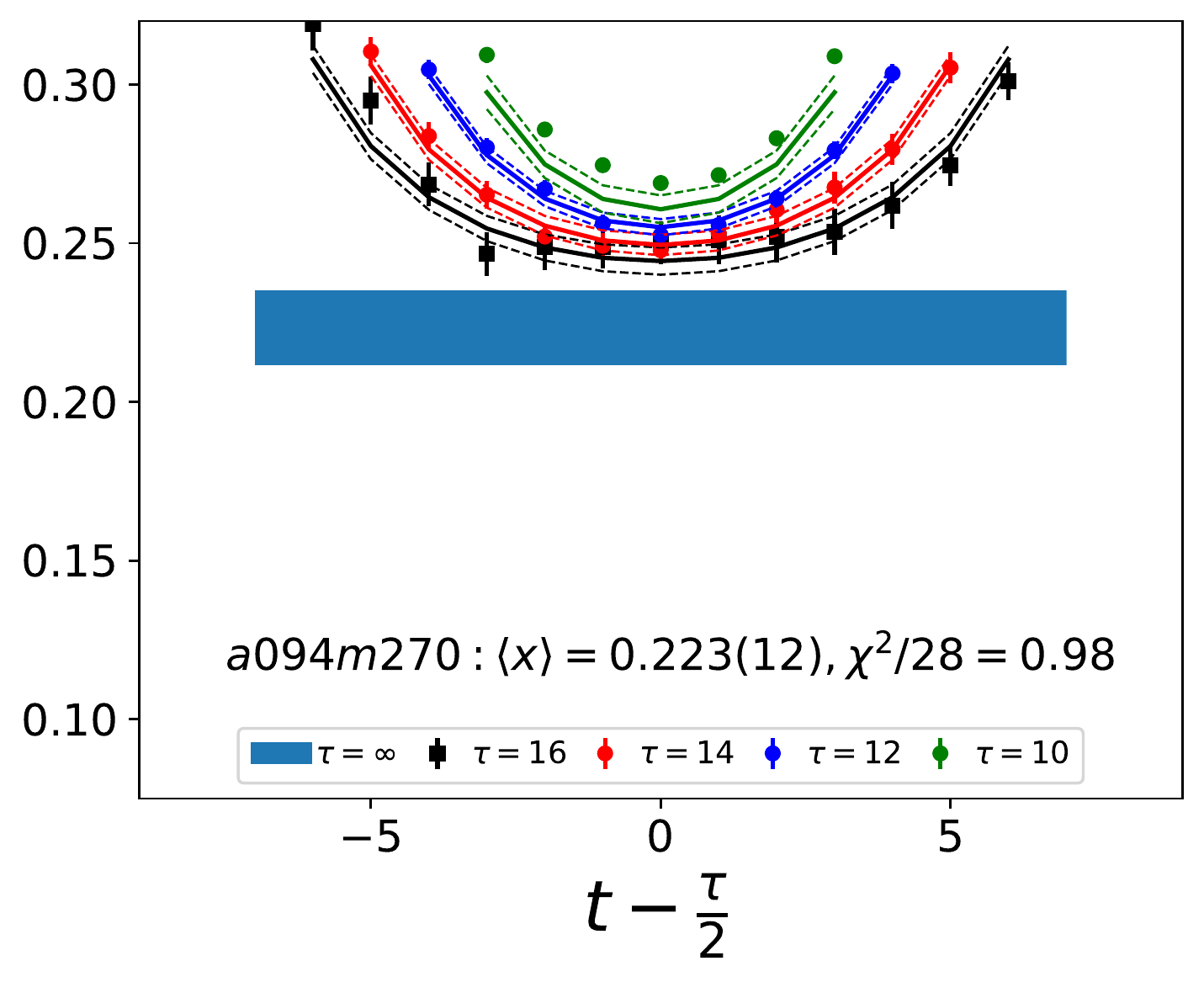}
\includegraphics[angle=0,width=0.32\textwidth]{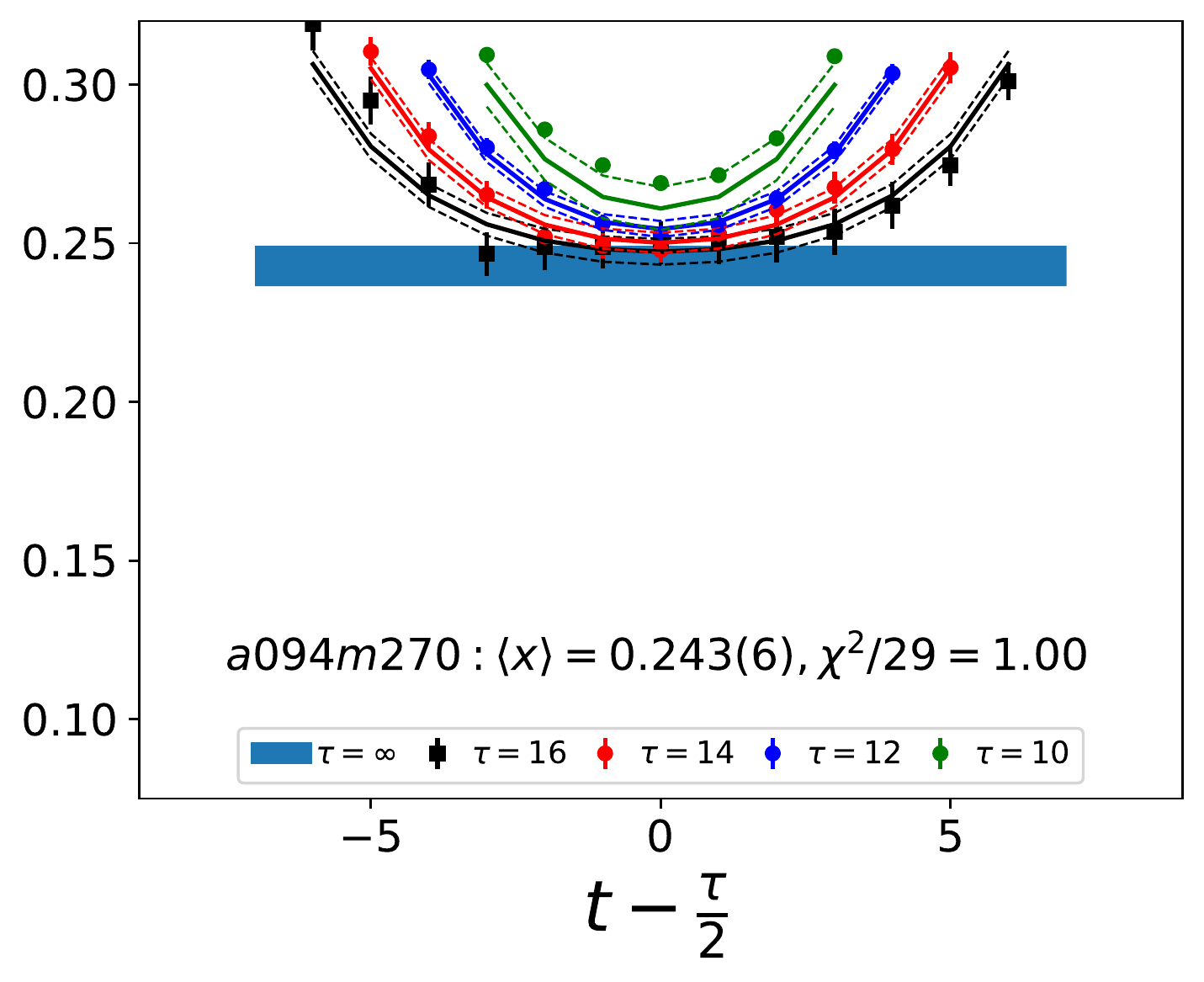}

\includegraphics[angle=0,width=0.32\textwidth]{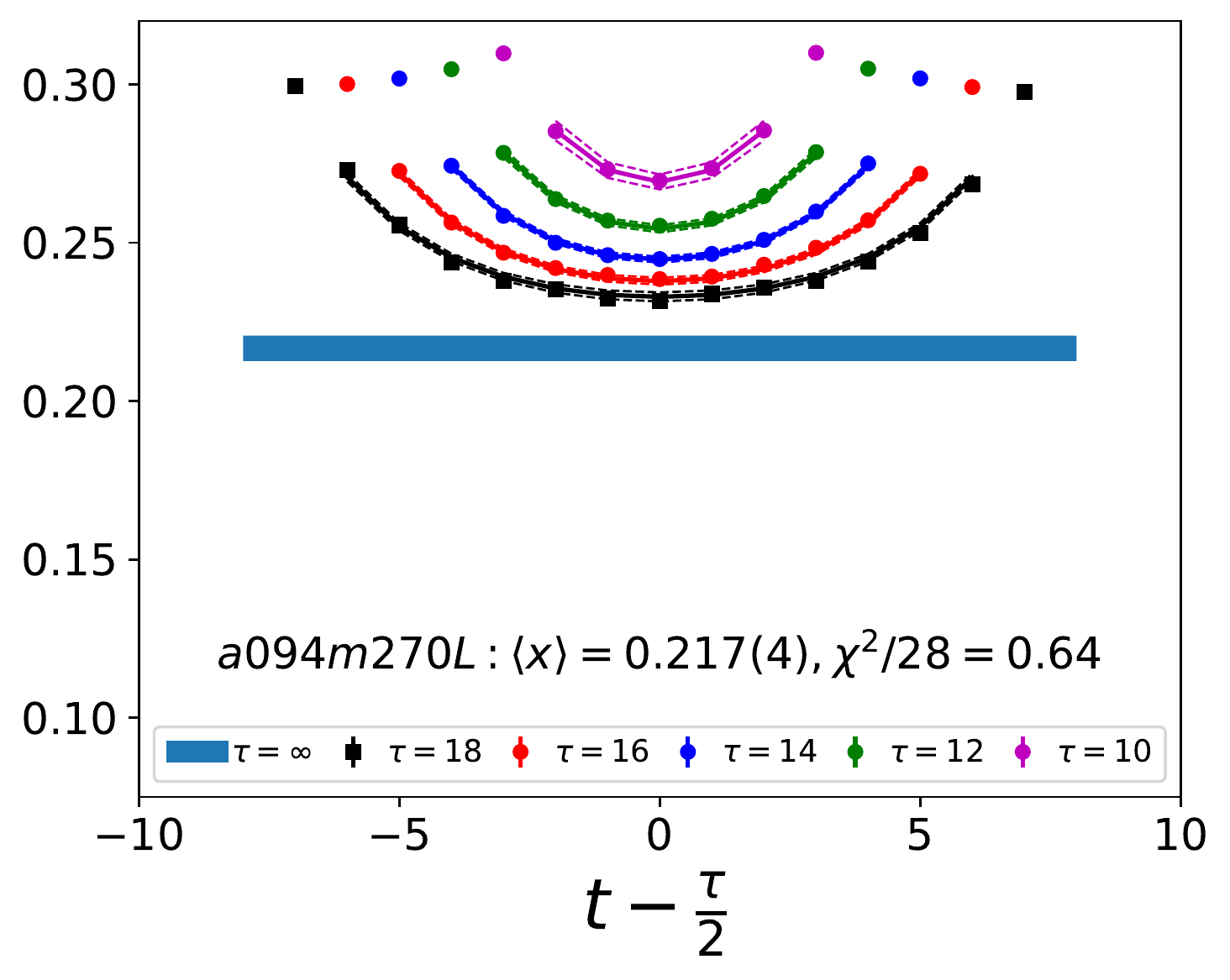}
\includegraphics[angle=0,width=0.32\textwidth]{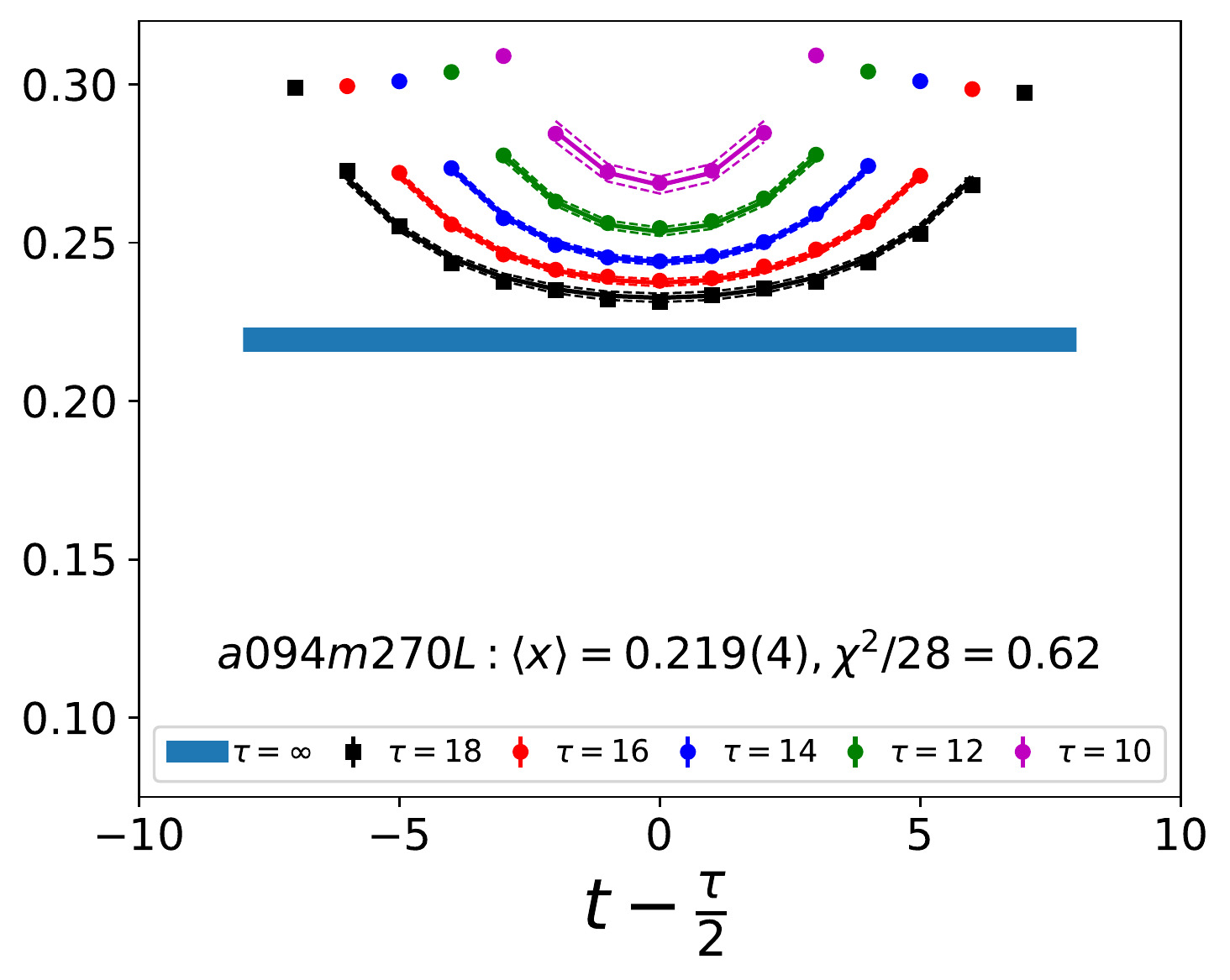}
\includegraphics[angle=0,width=0.32\textwidth]{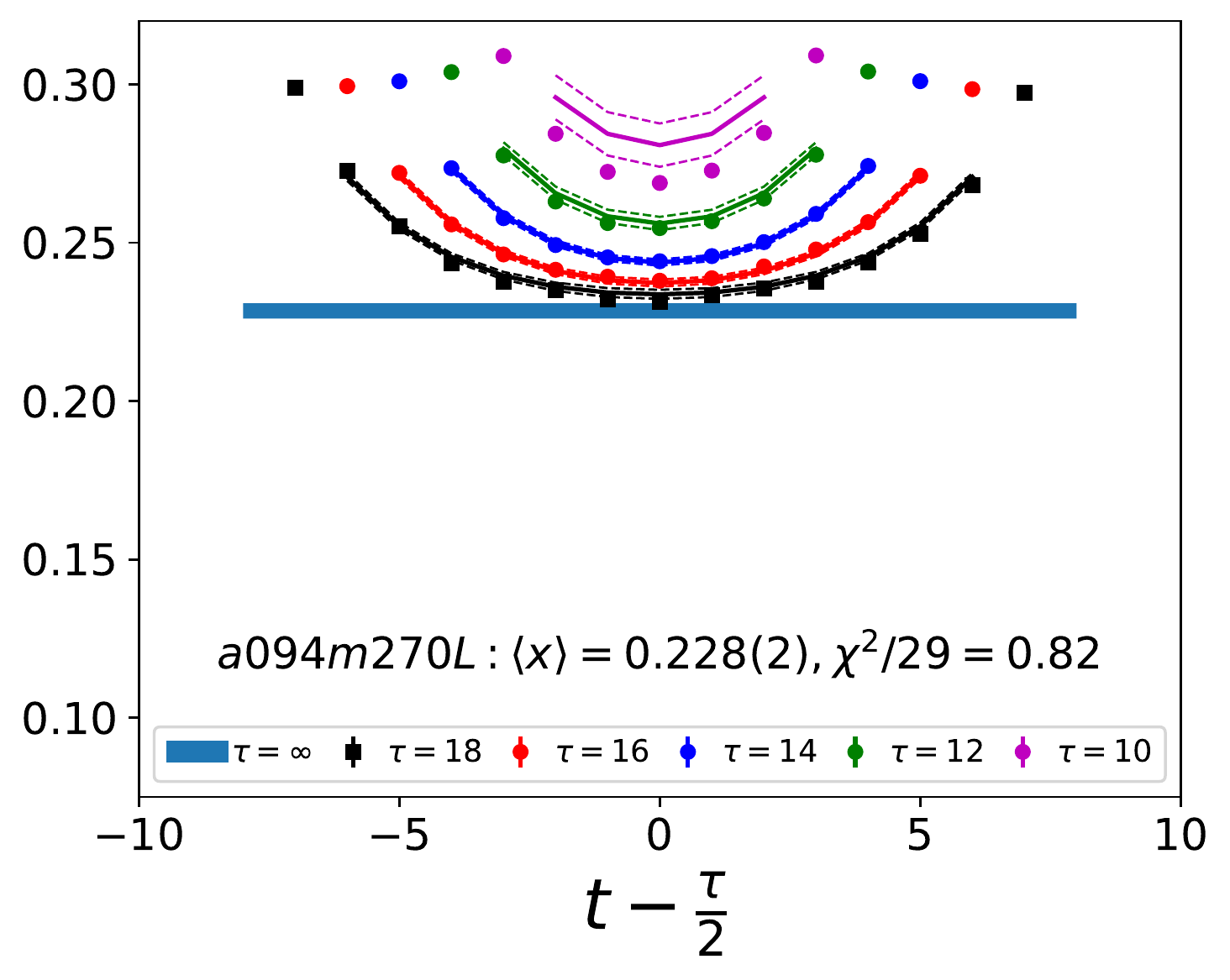}
\end{subfigure}

\caption{Data and fits to remove excited-state contamination in the extraction 
  of the transversity moment $\langle x \rangle_{\delta u- \delta d}$
  for $a127m285$ (top row), $a094m270$ (second row), and $a094m270L$ (bottom row) ensembles. 
  The data for the ratio
  $C_\mathcal{O}^{3\text{pt}}(\tau;t)/C^{2\text{pt}}(\tau)$ is scaled
  using Eq.~\protect\eqref{eq:me2momentT} to give $\langle
  x \rangle_{\delta u-\delta d}$, and the fit parameters are listed in
  Table~\protect\ref{tab:5strategy-fits-transvmom}. The rest is the same as in
  Fig.~\protect\ref{fig:Ratio-mom-1}. }
\label{fig:Ratio1-transversity-1}
\end{figure*}

\begin{figure*}[tp]  

\centering

\begin{subfigure}
\centering
\includegraphics[angle=0,width=0.32\textwidth]{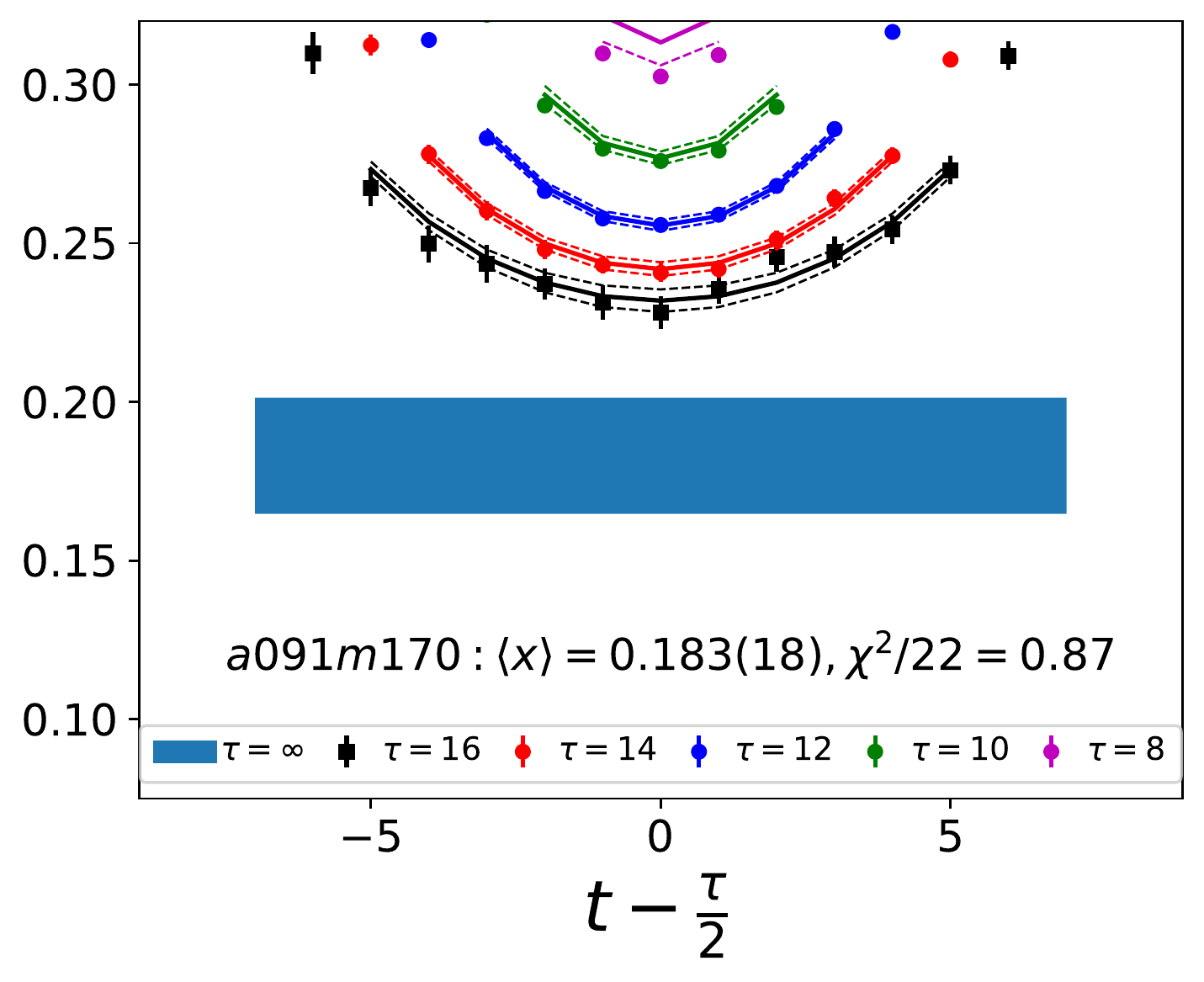}
\includegraphics[angle=0,width=0.32\textwidth]{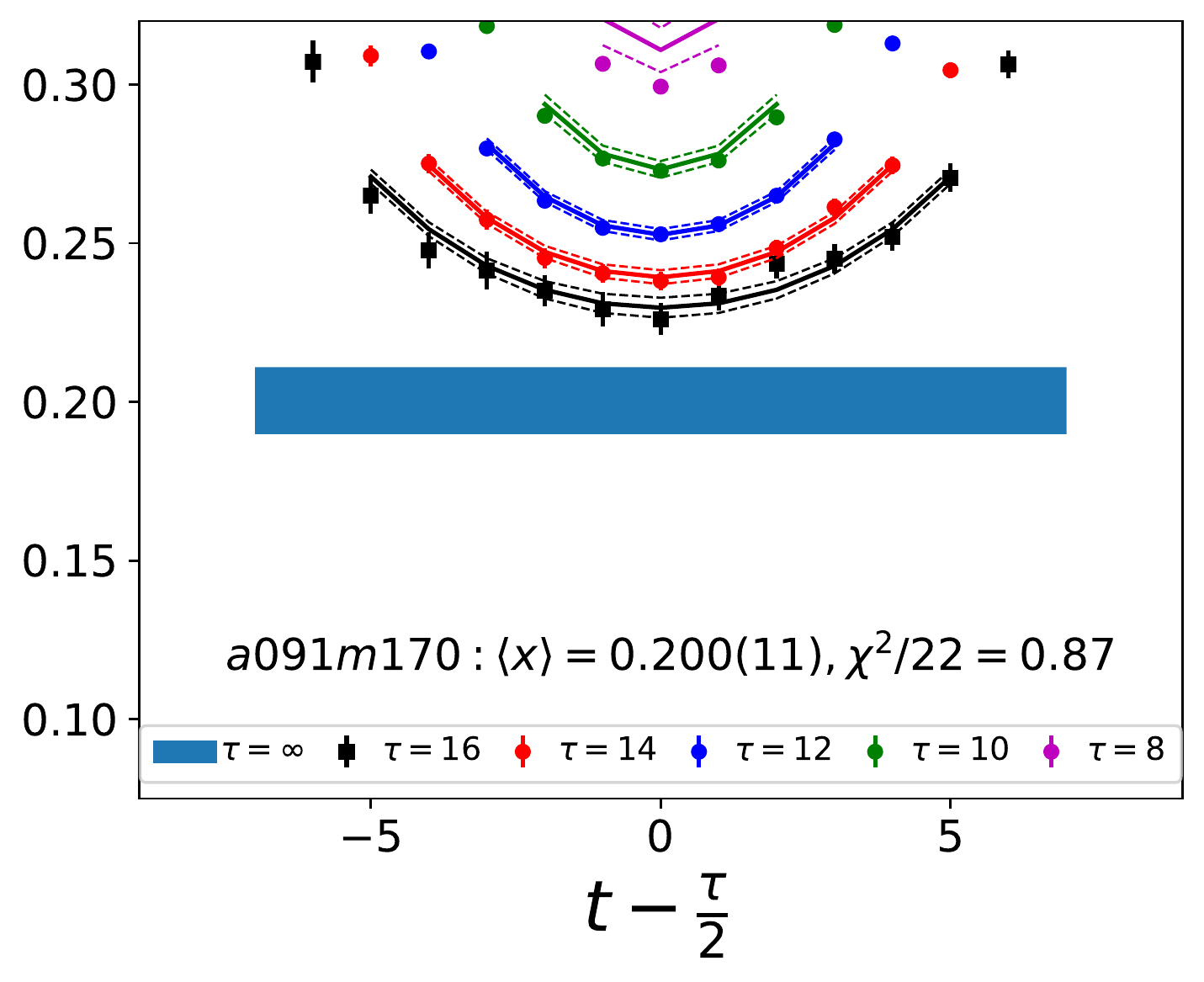}
\includegraphics[angle=0,width=0.32\textwidth]{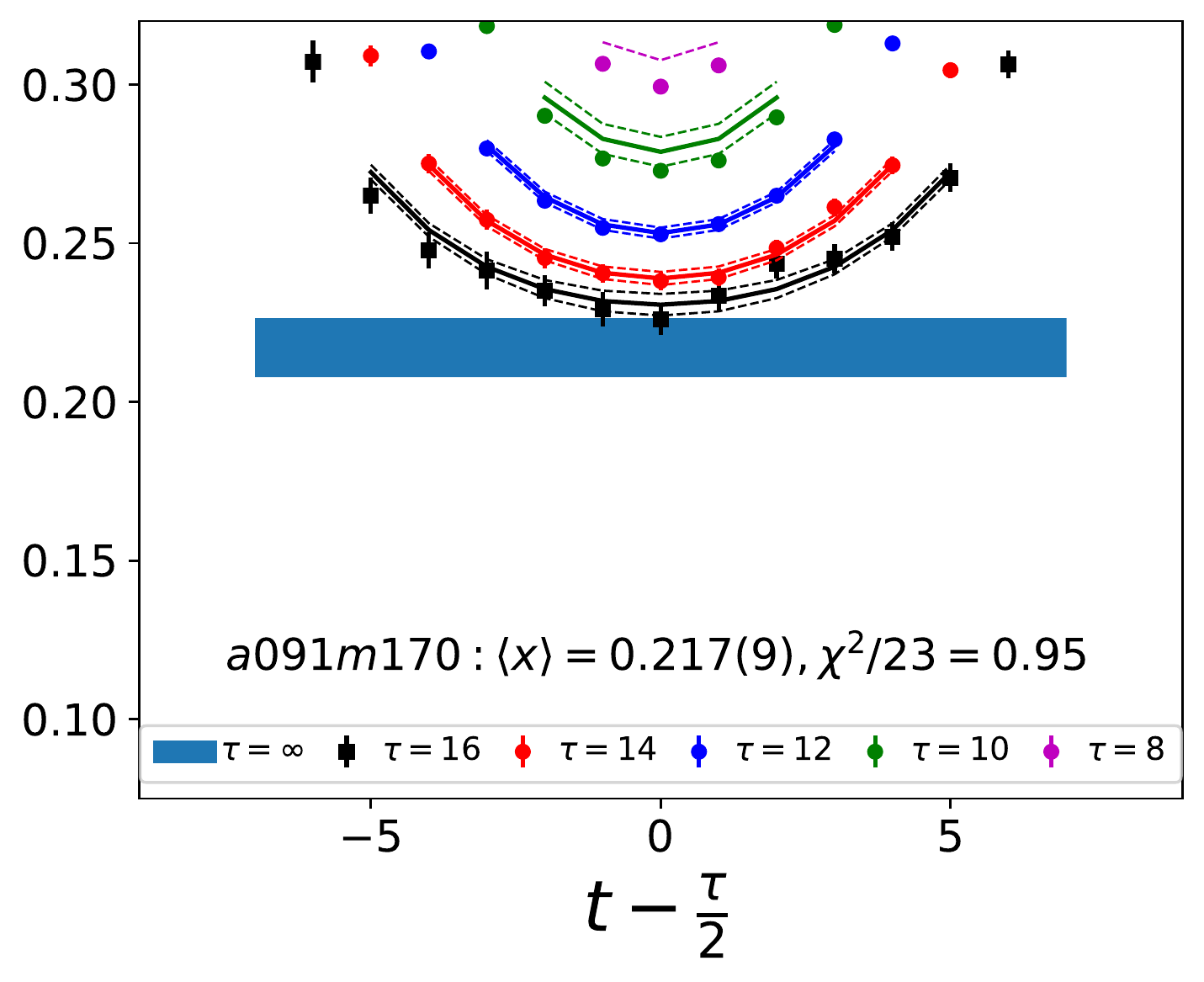}
\end{subfigure}

\begin{subfigure}
\centering
\includegraphics[angle=0,width=0.32\textwidth]{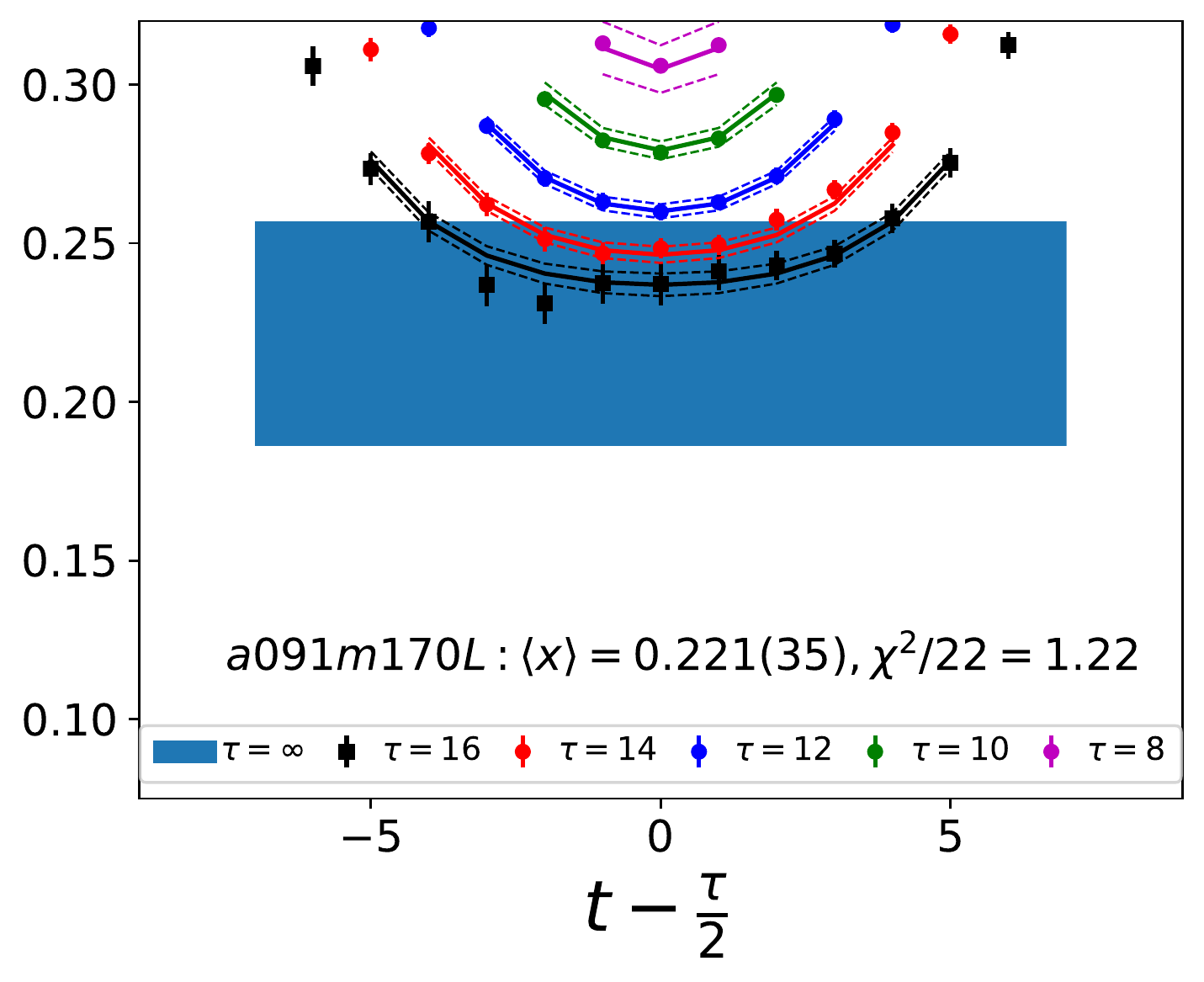}
\includegraphics[angle=0,width=0.32\textwidth]{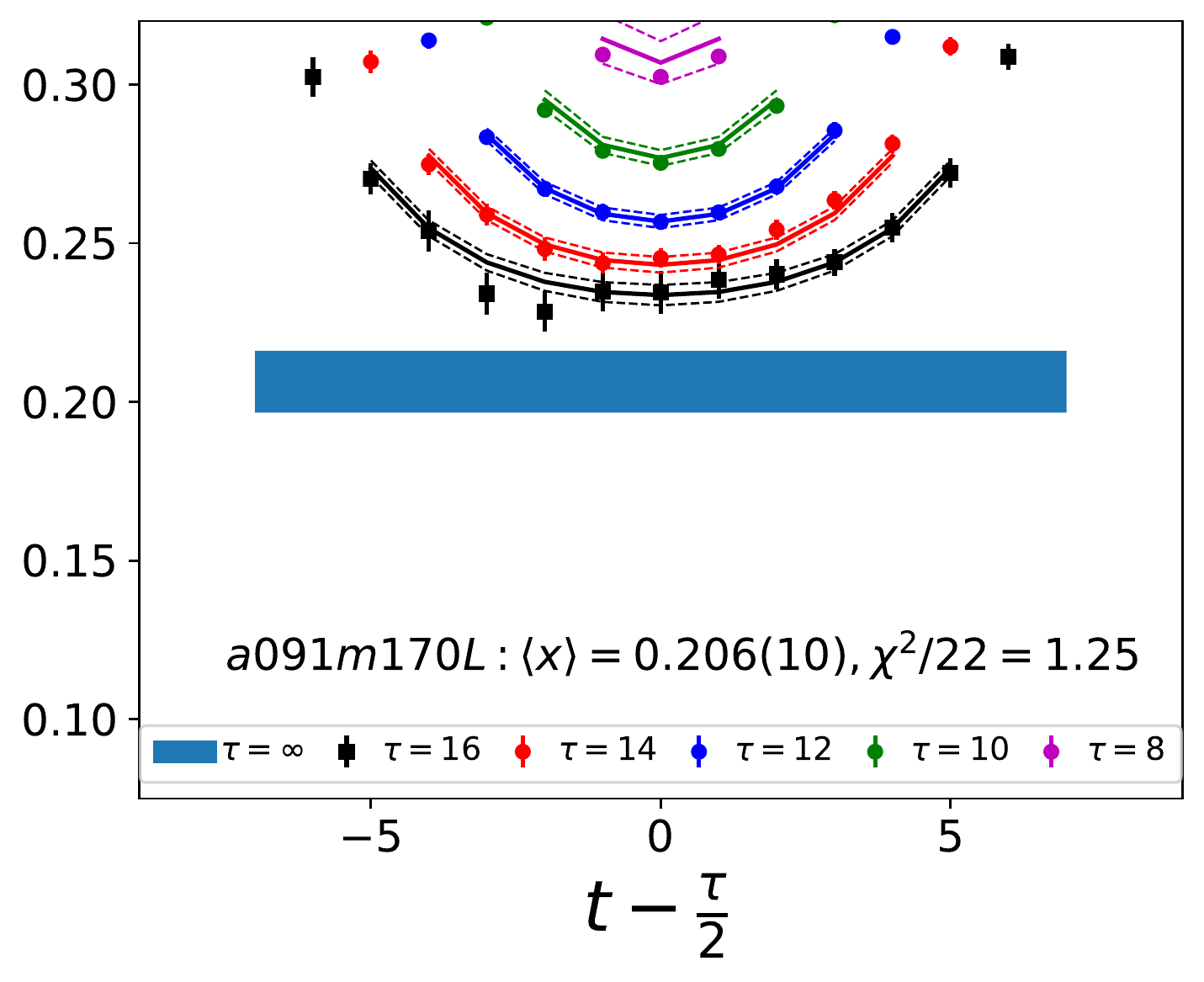}
\includegraphics[angle=0,width=0.32\textwidth]{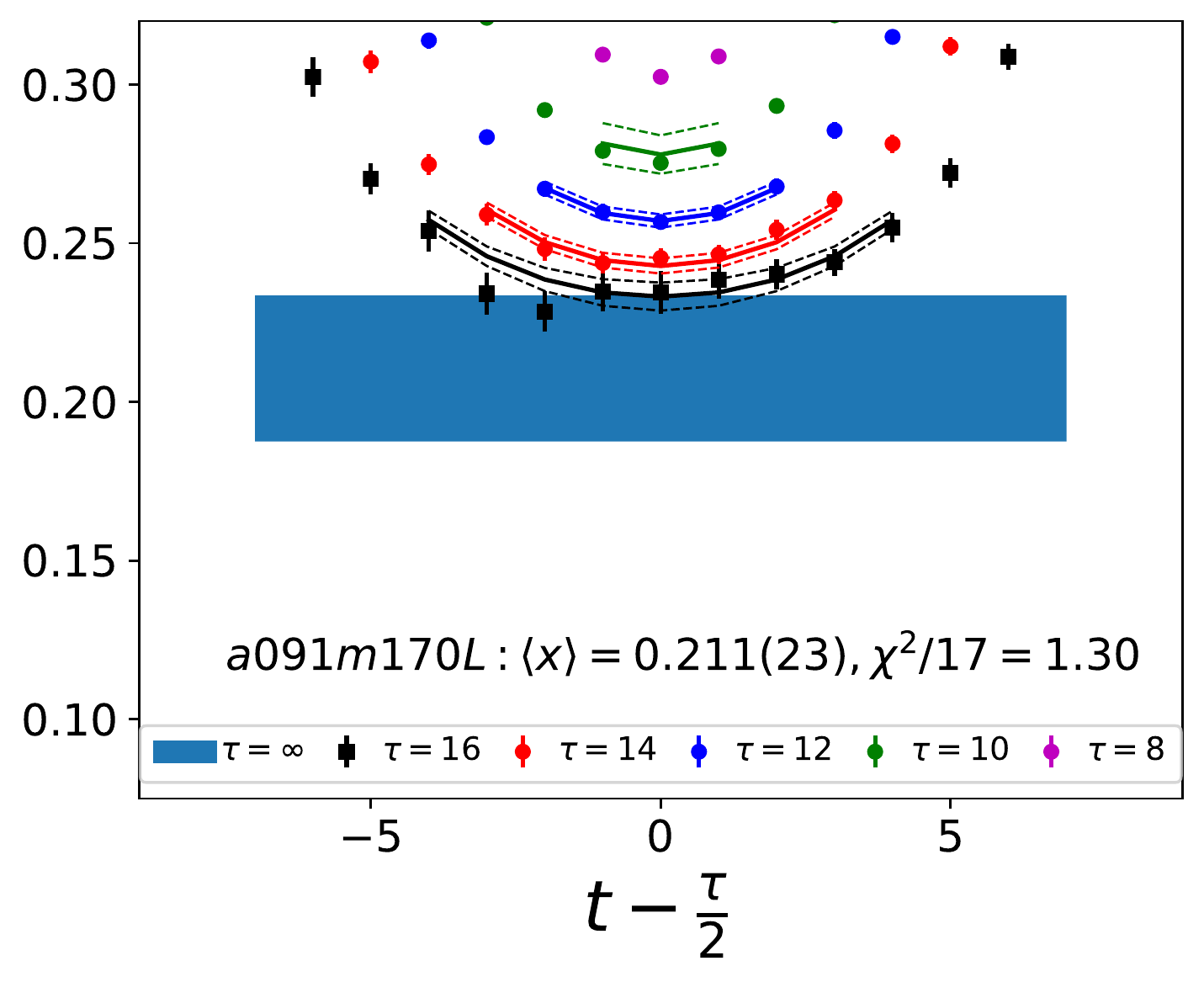}
\end{subfigure}

\begin{subfigure}
\centering
\includegraphics[angle=0,width=0.32\textwidth]{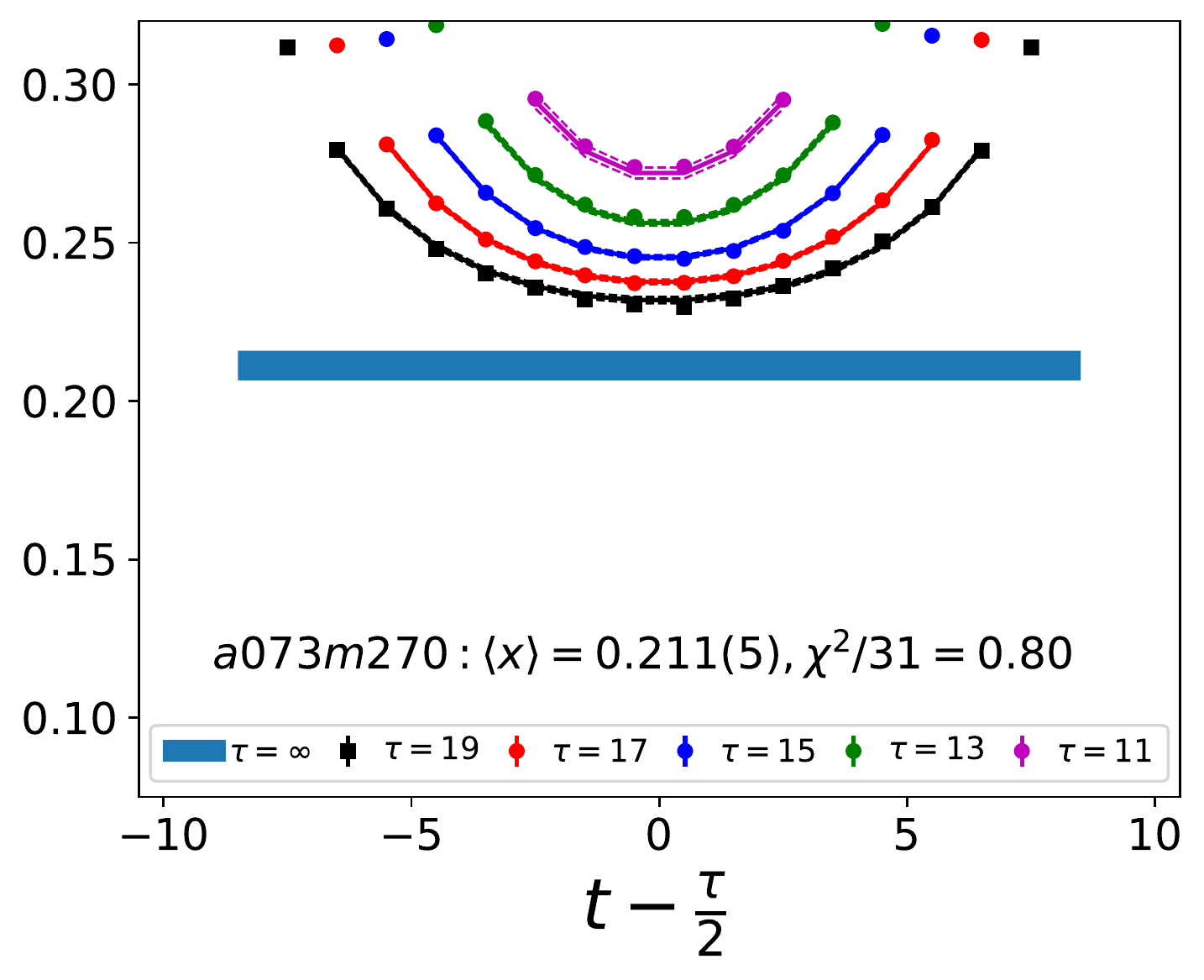}
\includegraphics[angle=0,width=0.32\textwidth]{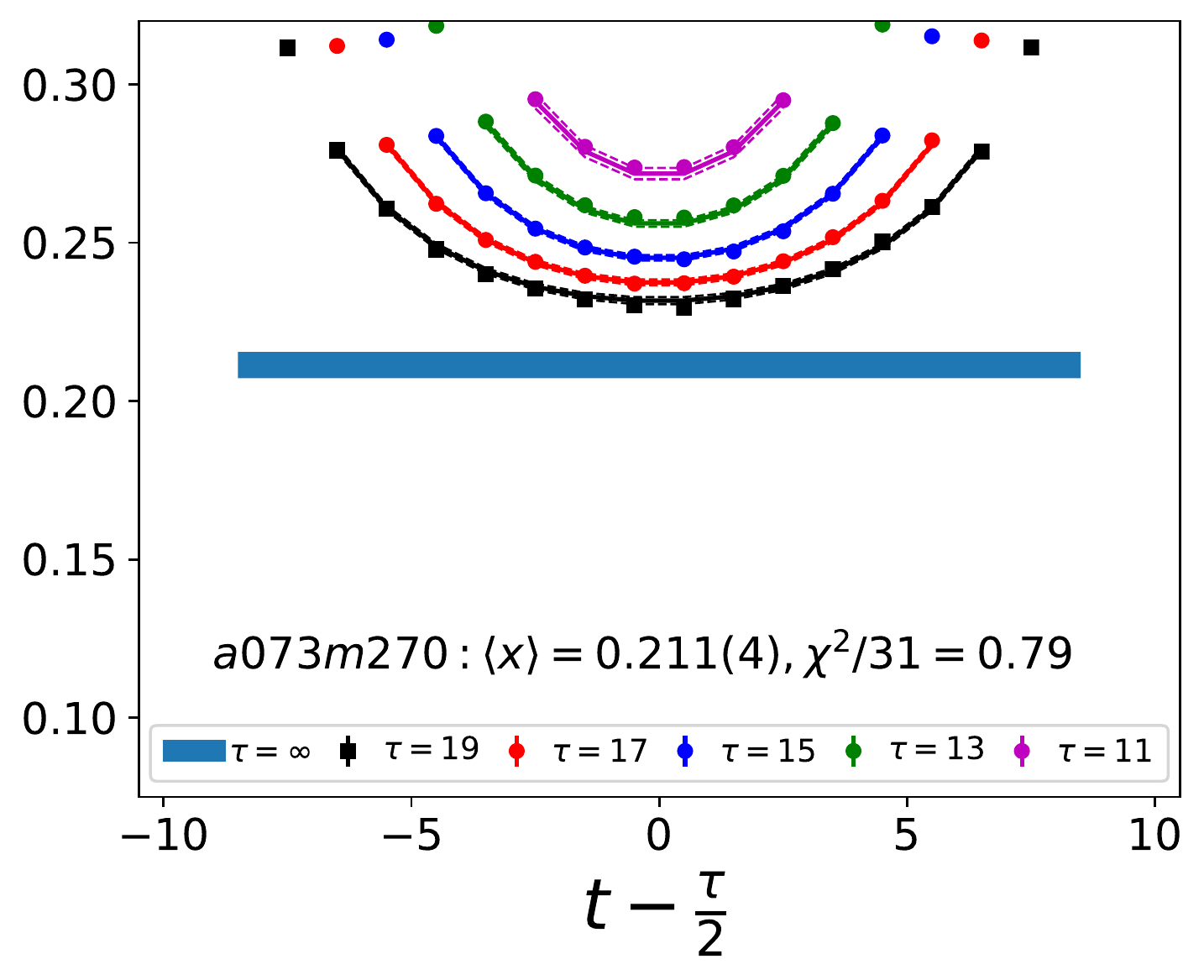}
\includegraphics[angle=0,width=0.32\textwidth]{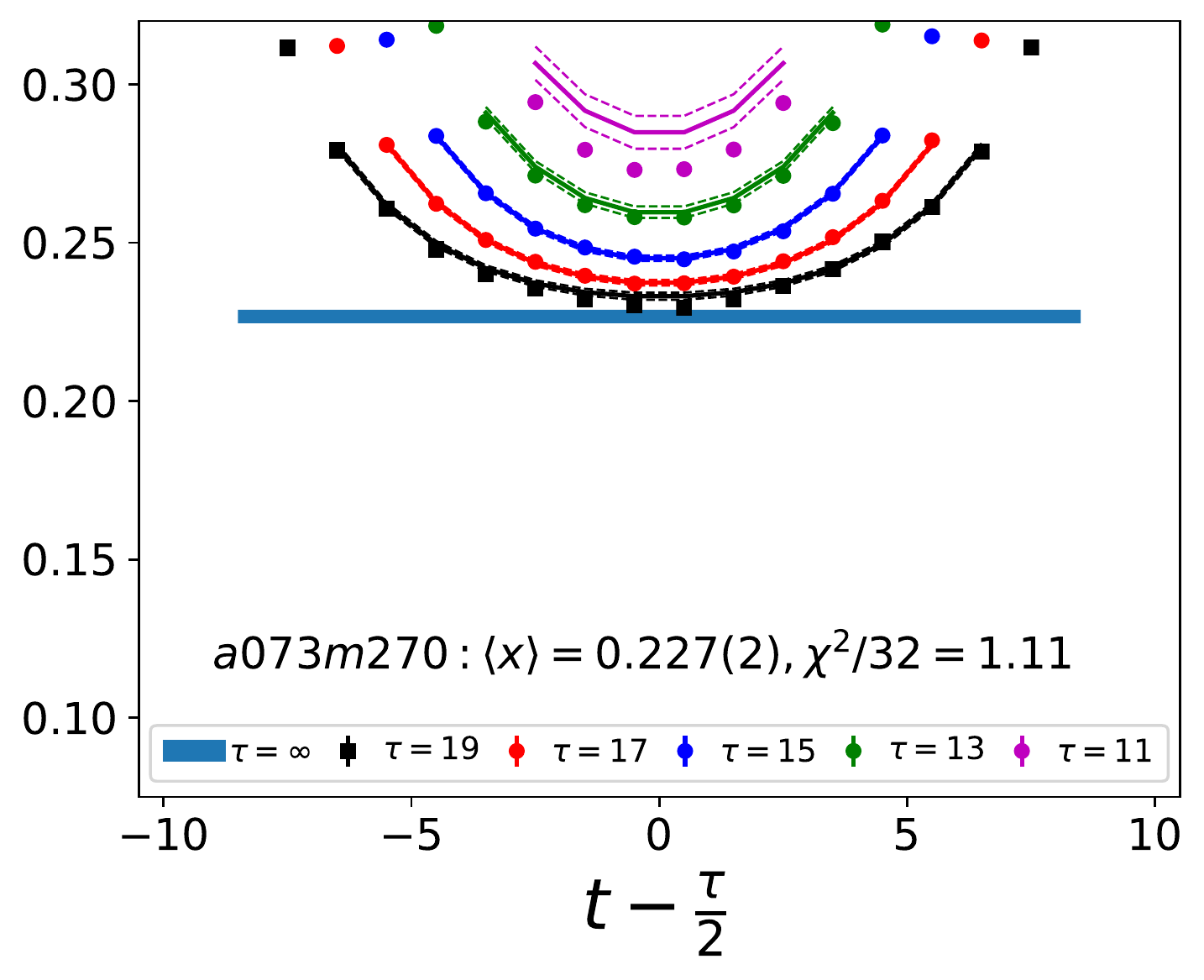}
\end{subfigure}

\begin{subfigure}
\centering
\includegraphics[angle=0,width=0.32\textwidth]{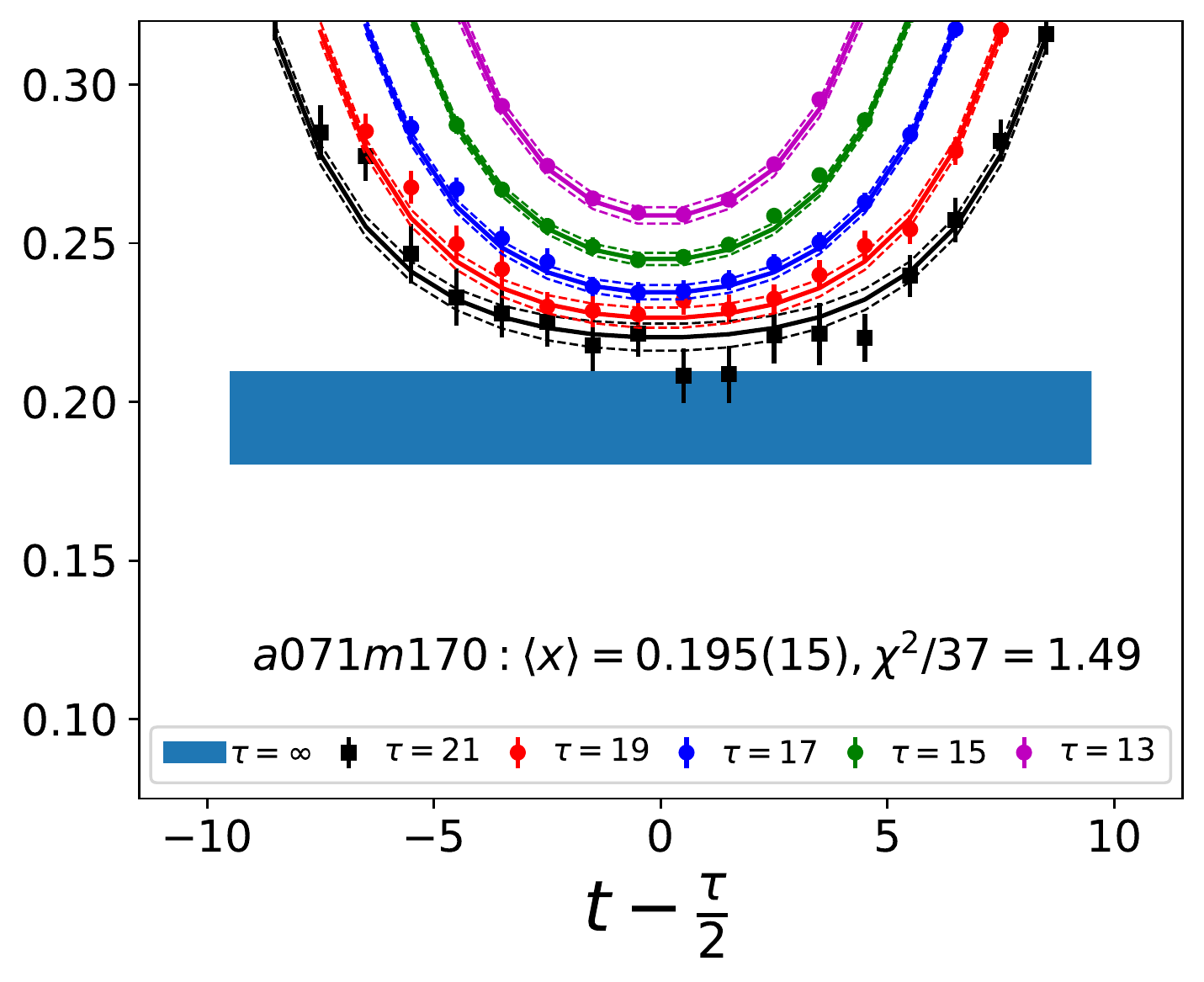}
\includegraphics[angle=0,width=0.32\textwidth]{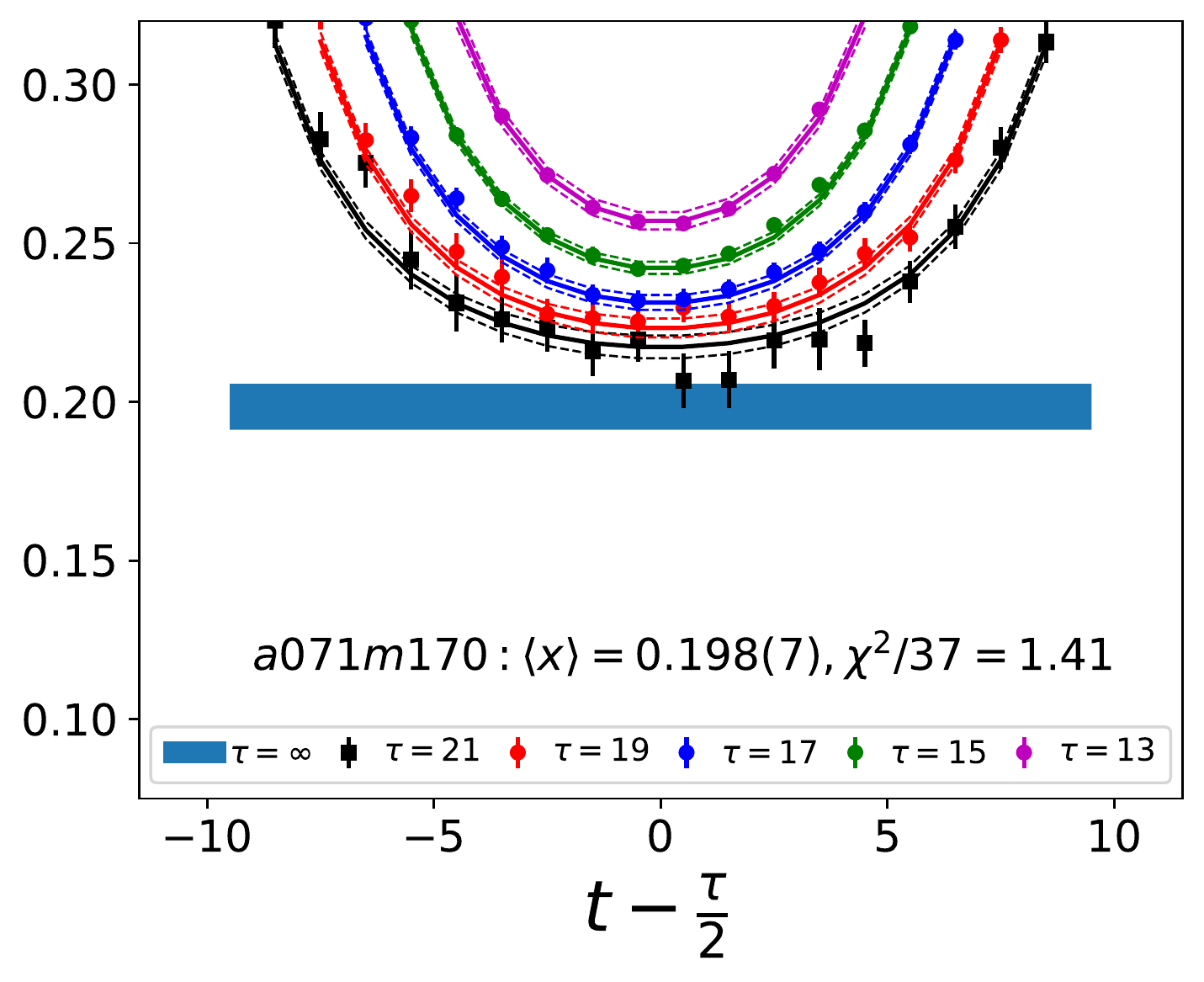}
\includegraphics[angle=0,width=0.32\textwidth]{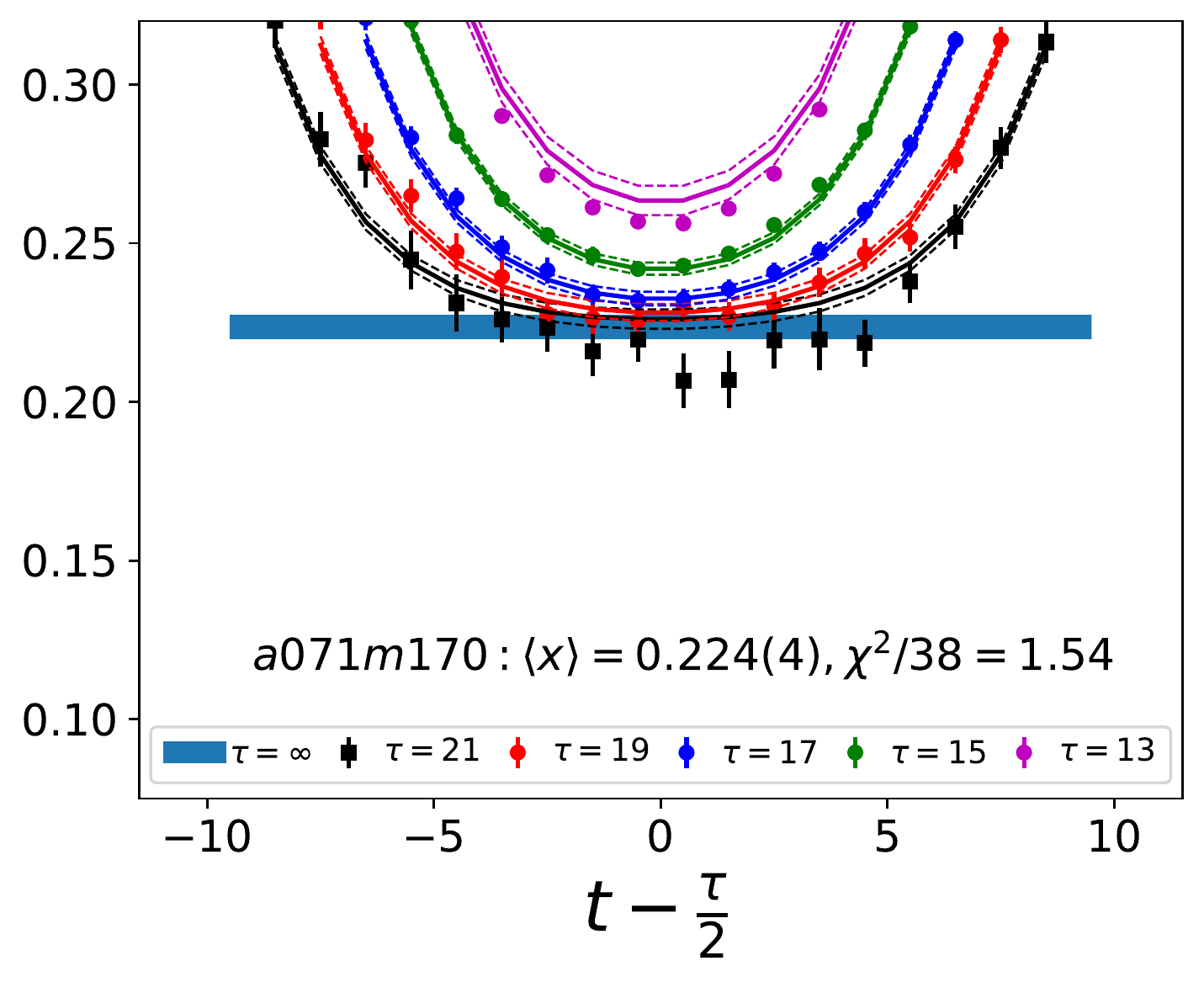}
\end{subfigure}

\caption{Continuation of the data and fits to remove excited-state contamination in the extraction 
  of the transversity moment $\langle x \rangle_{\delta u- \delta d}$
  for $a091m170$ (top row), $a091m170L$ (middle row), $a073m270$ (third row), and $a071m170$ (bottom
  row). The data for the ratio
  $C_\mathcal{O}^{3\text{pt}}(\tau;t)/C^{2\text{pt}}(\tau)$ is scaled
  using Eq.~\protect\eqref{eq:me2momentT} to give $\langle
  x \rangle_{\delta u-\delta d}$, and the fit parameters are listed in
  Table~\protect\ref{tab:5strategy-fits-transvmom}. The rest is the
  same as in Fig.~\protect\ref{fig:Ratio-mom-1}. }
\label{fig:Ratio-transversity-2} 
\end{figure*}

In this appendix, we show plots of the unrenormalized isovector
momentum fraction, $\la x \ra_{u-d}$, the helicity moment, $\la
x \ra_{\Delta u-\Delta d}$, and the transversity moment, $\la
x\ra_{\delta u-\delta d}$, for the seven ensembles in
Figs.~\ref{fig:Ratio-mom-1}--\ref{fig:Ratio-transversity-2}. The data
shown is the ratio
$C_\mathcal{O}^{3\text{pt}}(\tau;t)/C^{2\text{pt}}(\tau)$ multiplied
by the appropriate factor given in
Eqs.~\eqref{eq:me2momentV}--\protect\eqref{eq:me2momentT} to get the three 
$\langle x \rangle$. The three panels in each row show fits with the
three strategies: $\{4^{N\pi},3^*\}$ (left), $\{4^{},3^*\}$ (middle)
and $\{4^{},2^{\rm free}\}$ (right).  The fits to
$C_\mathcal{O}^{3\text{pt}}(\tau;t)$ using Eq.~\eqref{eq:3pt} are made
keeping data at the largest three values of $\tau$, except for $a071m170$ as discussed in
Sec.~\ref{sec:ESC}.  The results of these fit are shown for various $\tau$ 
by lines with the same color as the data. In all cases, to
extract the ground state matrix element (blue band), the fits to
$C^{2\text{pt}}(\tau)$ and $C_\mathcal{O}^{3\text{pt}}(\tau;t)$ are
done within a single jackknife loop.

The data show a monotonic convergence in $\tau$ towards the
$\tau \to \infty$ estimate. Also, the data are symmetric about $t-\tau/2$
for all values of $\tau$, except for the largest $\tau$ on $a071m170$,
$a091m170L$ and $a094m270$ ensembles, which are statistics limited. Lastly, the
largest extrapolation, ie, the difference between the data at
$t=\tau/2$ with the largest $\tau$ and the $\tau = \infty$ value, is
for the $\{4^{N\pi},3^\ast\}$ strategy since it has the smallest mass
gap as shown in Table~\ref{tab:massgap}. This is most evident on the
$m_\pi \approx 170$~MeV ensembles. The smallest is for the
$\{4^{},2^{\rm free}\}$ strategy in which the mass gap is the largest.

\begin{table*}[tp]  
\setlength{\tabcolsep}{4pt}
\renewcommand{\arraystretch}{1.3}
\centering
\begin{tabular}{ |c|c|c|c|c|c|c|c|c| }
\hline
\hline
Ensemble    & $N_{conf}$ &fit range       &  \multicolumn{3}{|c|}{Method A}  & \multicolumn{3}{|c|}{Method B}   \\
            &            & $[{\rm GeV^2}]$&$Z_{VD}$&$Z_{AD}$&$Z_{TD}$&$Z_{VD}$&$Z_{AD}$&$Z_{TD}$                \\ 
\hline                   
\hline                   
$a127m285$  & 100        & 3.7 -- 5.7 & 0.990(16) & 1.012(17) & 1.026(16) & 0.941(14) &  0.962(18) & 0.981(17)  \\
\hline                   
$a094m270$  & 100        & 5.3 -- 7.3 & 1.036(15) & 1.061(15) & 1.085(15) & 0.999(17) &  1.030(18) & 1.062(18)  \\
\hline                   
$a094m270L$ & 100        & 5.3 -- 7.3 & 1.025(14) & 1.040(14) & 1.071(16) & 0.991(14) &  1.000(15) & 1.043(20)  \\
\hline                   
$a091m170$  & 101        & 5.5 -- 7.5 & 1.016(12) & 1.029(14) & 1.062(14) & 0.977(13) &  0.987(16) & 1.032(16)  \\
\hline                   
$a091m170L$ & 108        & 5.5 -- 7.5 & 1.039(14) & 1.058(15) & 1.088(18) & 0.999(16) &  1.021(17) & 1.056(22)  \\
\hline                   
$a073m270$  & 100        & 7.1 -- 9.1 & 1.073(17) & 1.084(15) & 1.120(19) & 1.051(20) &  1.056(17) & 1.104(19)  \\
\hline                   
$a071m170$  & 112        & 7.4 -- 9.4 & 1.054(10) & 1.077(11) & 1.114(12) & 0.996(11) &  1.023(14) & 1.072(15)  \\
\hline
\end{tabular}
\caption{Results for the renormalization factors, $Z_{VD,AD,TD}$, in
  the $\MSbar$ scheme at $2$~GeV.  These are calculated in the RI'-MOM
  scheme as a function of scale $p=\sqrt{p_\mu p_\mu}$ on the lattice,
  matched to the $\MSbar$ scheme at the same scale $\mu=p$, and then
  run in the continuum $\MSbar$ scheme from $\mu$ to $2$~GeV. Results
  are given for two methods used to remove the $p^2$ dependent artifacts 
  as described in the text.  In method A (columns 4--6), the
  $Z$'s are obtained by averaging the data shown in
  Fig.~\protect\ref{fig:Z-7ensembles} over the range of $p^2$
  specified in the in the third column. Results using method B
  (columns 7--9) are obtained using fits to the data starting with the
  lower value of $p^2$ given in column 3 with the ansatz $Z(p)=Z_0+a
  p^2+b p^4$.\looseness-1\relax}
\label{tab:Z-fac}
\end{table*}
\section{Renormalization}
\label{sec:renormalization}
In this appendix, we describe two methods of calculating the renormalization
factors, $Z_{VD,AD,TD}$, for the three one-derivative operators specified in 
Eqs.~\eqref{eq:finaloperatorV},~\eqref{eq:finaloperatorA}
and~\eqref{eq:finaloperatorT}.  On the lattice, these $Z$'s
are first determined nonperturbatively in the
\ripmom\ scheme~\cite{Gockeler:2010yr,Constantinou:2013ada} as a
function of the lattice scale $p^2 = p^\mu p^\mu$, and then converted
to ${\rm \ol{MS}}$ scheme using $3$-loop perturbative factors
calculated in the continuum in Ref.~\cite{Gracey:2003mr}. For data at
each $p$, we perform horizontal matching by choosing the
${\rm \ol{MS}}$ scale $\mu=|p|$. These numbers are then run in the
continuum ${\rm \ol{MS}}$ scheme from scale $\mu$ to $2$~GeV using
three-loop anomalous dimensions~\cite{Gracey:2003mr}.  The two methods
differ in how the dependence of $Z^{\rm \ol{MS}}$(2GeV) on $p^2a^2$, a lattice artifact, is
removed. For details of the three operators and their decomposition into
irreducible representations, we refer the reader to 
Refs.~\cite{Gockeler:1995wg,Harris:2019bih}.

\begin{figure*}[tbhp]  
\begin{subfigure}
\centering
\includegraphics[angle=0,width=0.33\textwidth]{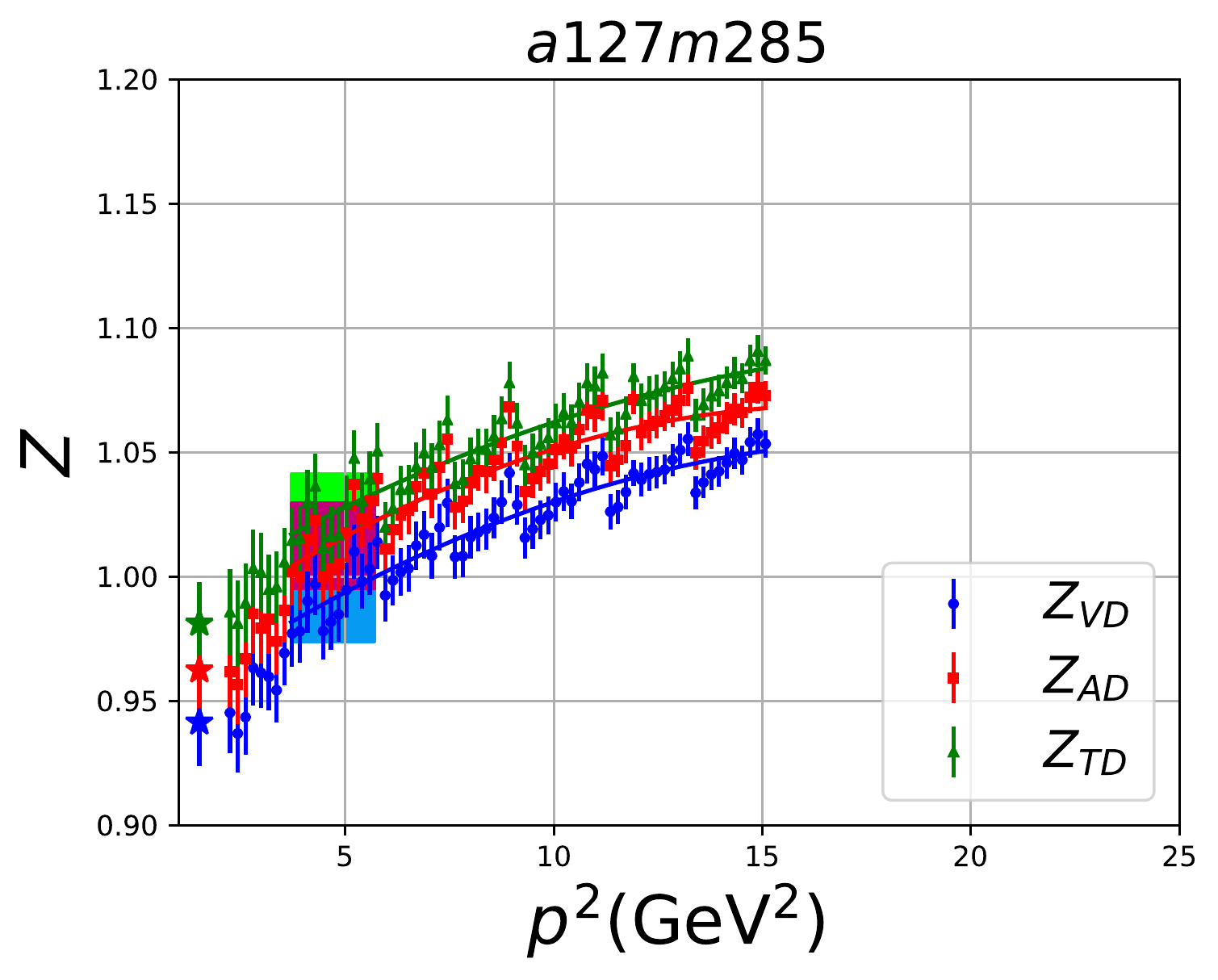}
\end{subfigure} \\
\begin{subfigure}
\centering
\includegraphics[angle=0,width=0.33\textwidth]{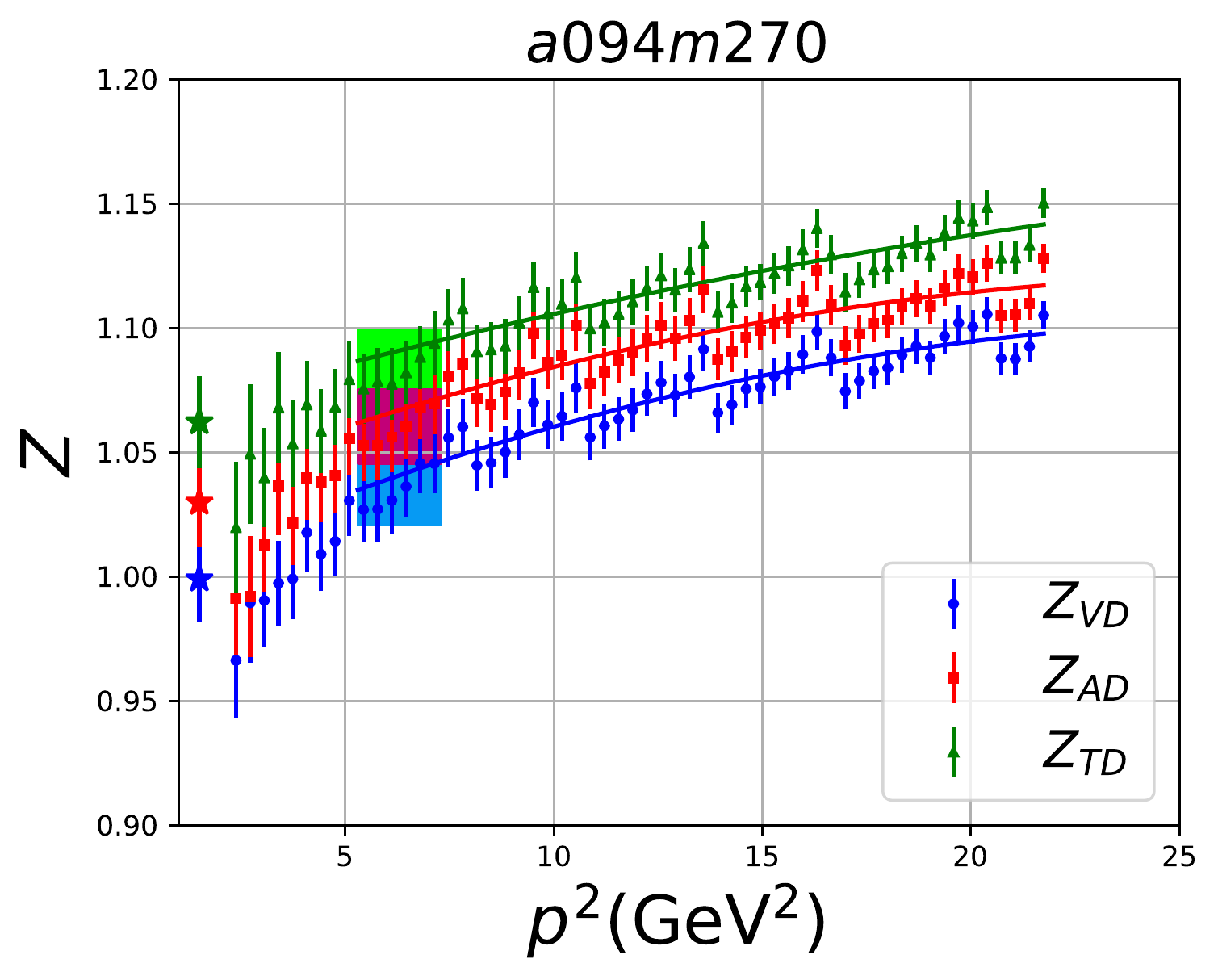}
\end{subfigure}
\begin{subfigure}
\centering
\includegraphics[angle=0,width=0.33\textwidth]{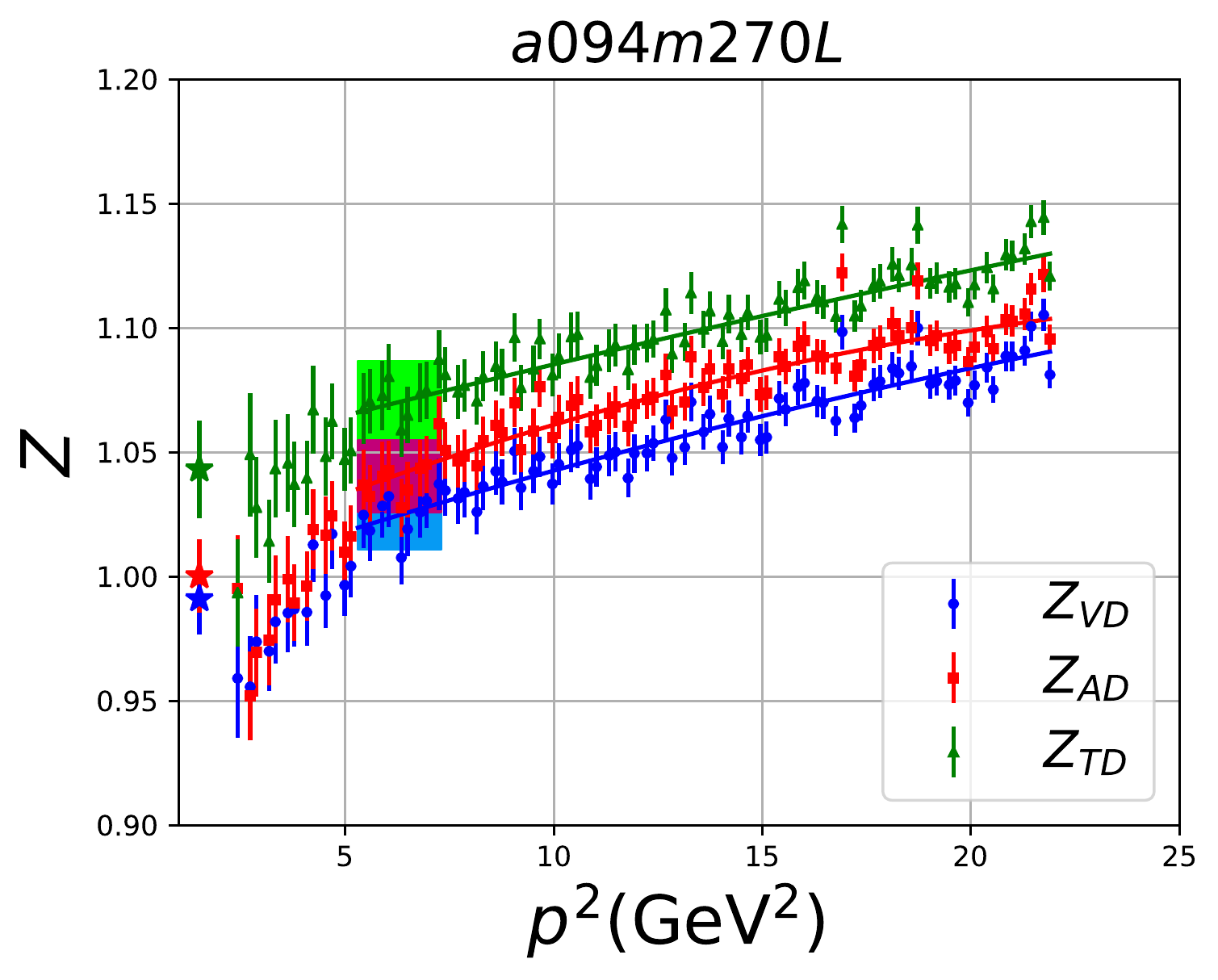}
\end{subfigure}
\begin{subfigure}
\centering
\includegraphics[angle=0,width=0.33\textwidth]{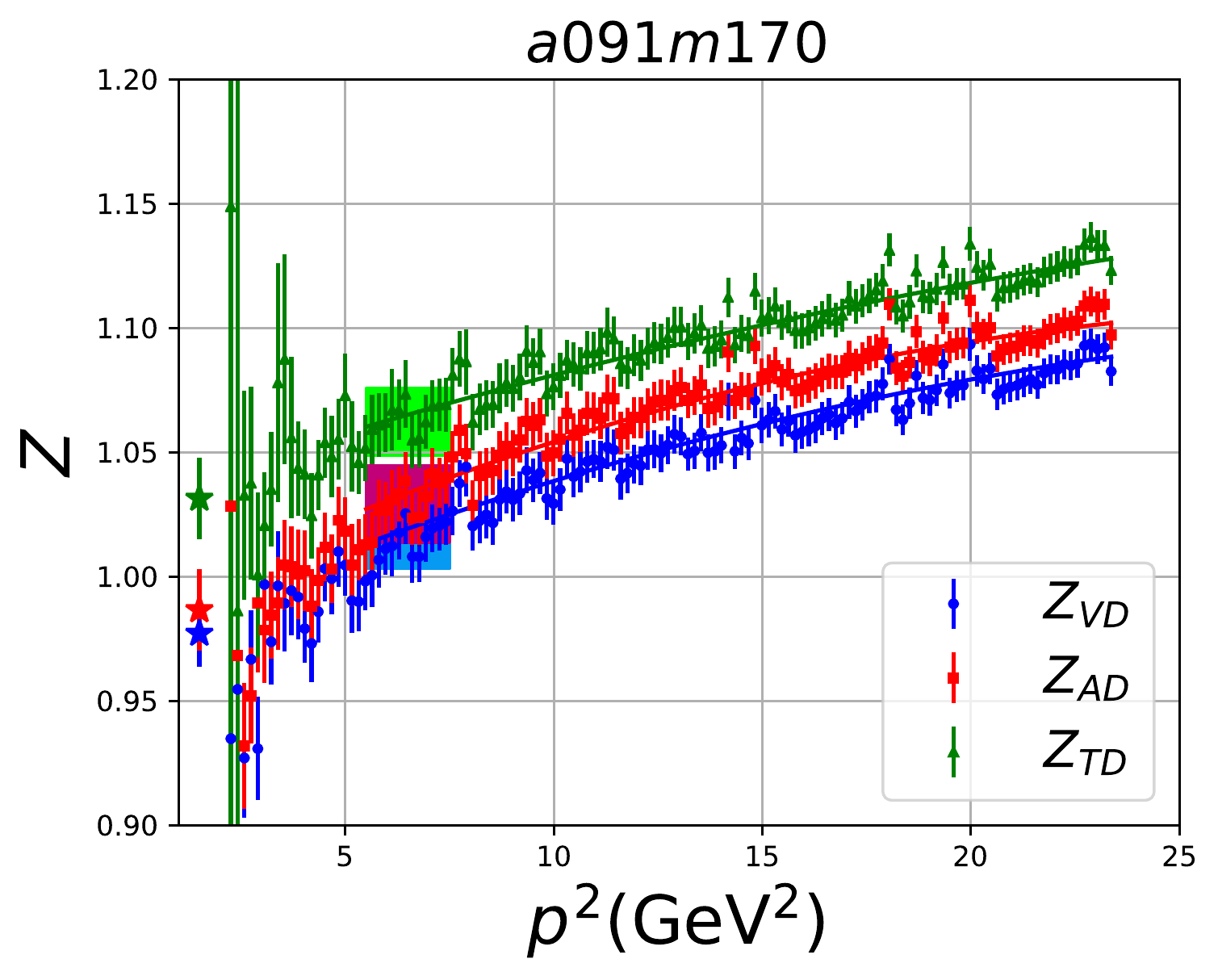}
\end{subfigure}
\begin{subfigure}
\centering
\includegraphics[angle=0,width=0.33\textwidth]{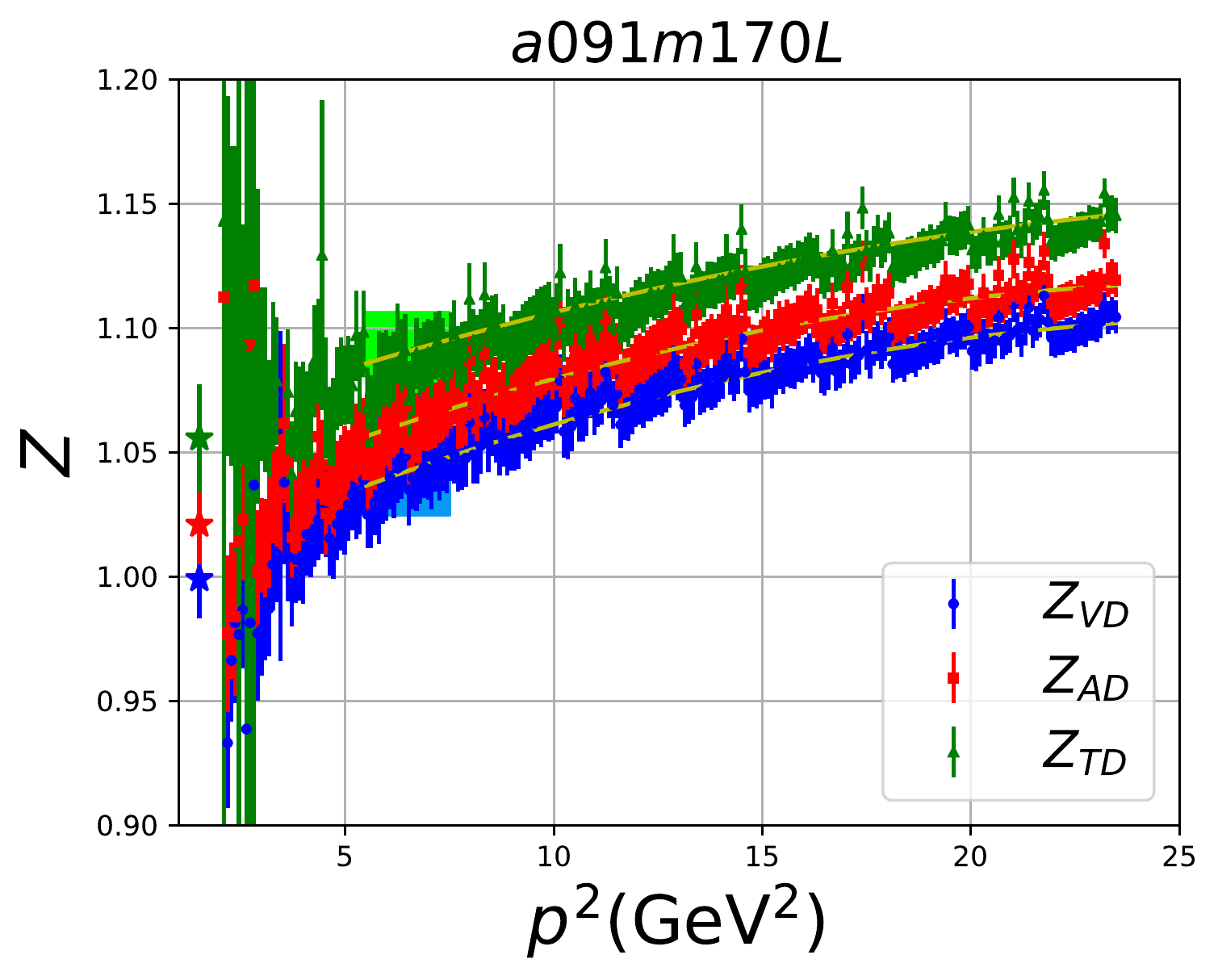}
\end{subfigure}
\begin{subfigure}
\centering
\includegraphics[angle=0,width=0.33\textwidth]{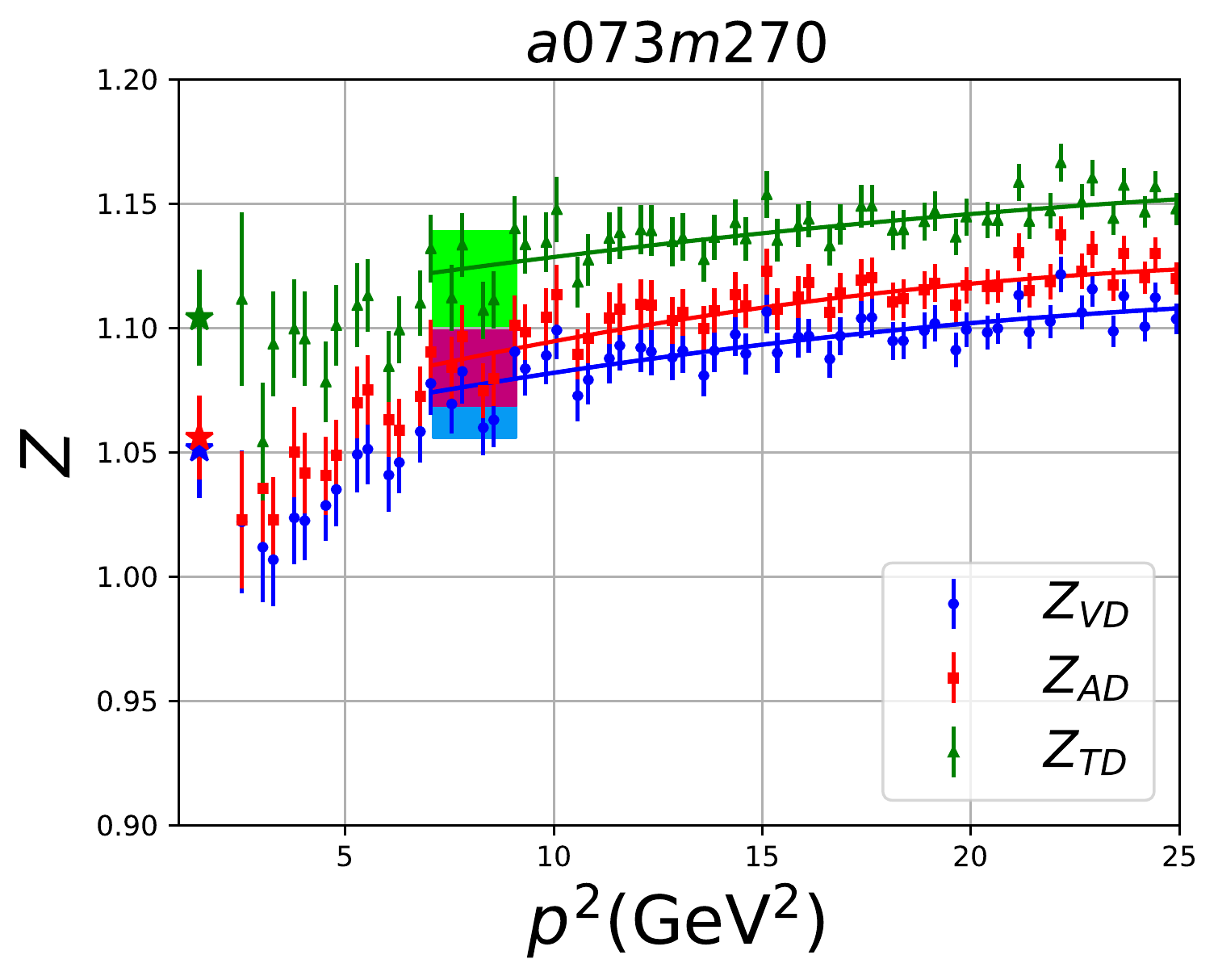}
\end{subfigure}
\begin{subfigure}
\centering
\includegraphics[angle=0,width=0.33\textwidth]{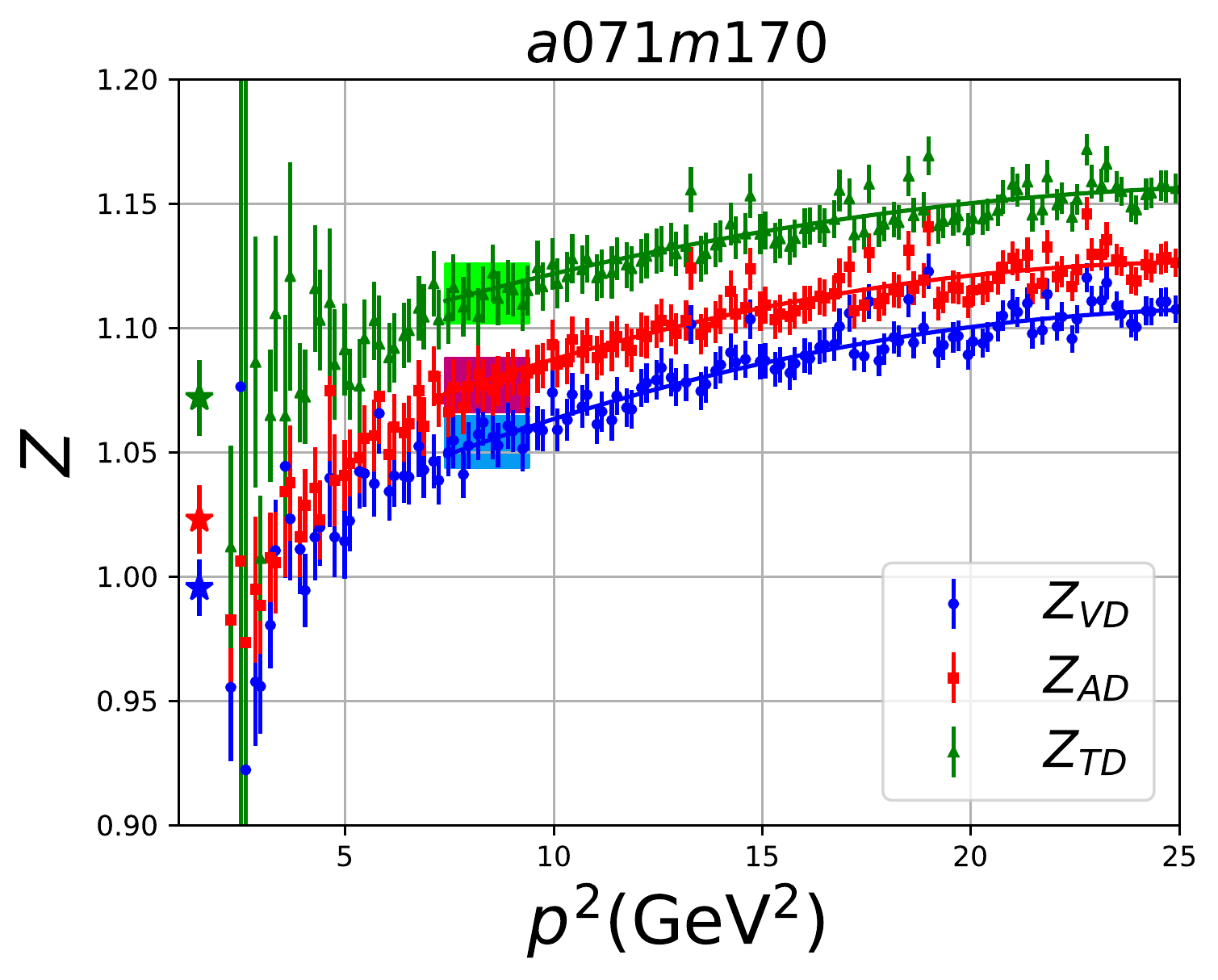}
\end{subfigure}
\caption{Nonperturbative renormalization factors for ${ \la
    x\ra_{u-d}}$, ($Z_{VD}$), ${ \la x \ra_{\Delta u - \Delta d}}$,
  ($Z_{AD}$), and ${ \la x\ra_{\delta u - \delta d}}$, ($Z_{TD}$) in
  the ${\rm \ol{MS}}$ scheme at $\mu=2\ {\rm GeV}$ for the seven
  ensembles.  The shaded bands mark the region in ${p^2}$ that is
  averaged and the error in the estimate. The points next to the
  y-axis with the star symbol give a second estimate obtained from the
  fit $Z(p)=Z_0+a p^2+b p^4$.}
\label{fig:Z-7ensembles}
\end{figure*}

The data for the renormalization factors $Z_{VD, AD, TD}$ in the
${\rm \ol{MS}}$ scheme at $\mu = 2$ GeV is shown in
Fig.~\ref{fig:Z-7ensembles} for the seven ensembles as a function of
$p^2$---the scale of the \ripmom\ scheme on the lattice.  For all
three operators, the data do not show a window in $p^2$ where the
results are independent of $p^2$. We analyze the variation with $p^2$
as being mainly due to a combination of the breaking of full
rotational invariance on the lattice and other $p^2$ dependent
artifacts. Many methods have been proposed to control it, see for
example
Refs.~\cite{Harris:2019bih,Alexandrou:2020sml,Bhattacharya:2016zcn}.
We use the following two:
\begin{itemize}
\item
In method A, we take an average over the data points in an interval of
$2~ {\rm GeV^2}$ about ${\hat p}^2 = \Lambda/a$, where the scale $\Lambda =
3$ GeV is chosen to be large enough to avoid nonperturbative effects
and at which perturbation theory is expected to be reasonably well
behaved. Also, this choice satisfies both ${\hat p}a \rightarrow 0$ and
$\Lambda/{\hat p} \rightarrow 0$ in the continuum limit as desired.  The
window over which the data are averaged (given in column three of
Table~\ref{tab:Z-fac}) and the error (half the height of the band) are
shown by shaded bands in Figs.~\ref{fig:Z-7ensembles}.  This method
was used in our previous work presented in Ref.~\cite{Mondal:2020cmt}.
\item
In method B, we make a fit to the data using the ansatz $Z(p)=Z_0+a
\sum_\mu p_\mu p_\mu + b (\sum_\mu p_\mu p_\mu)^2$ to remove the $p^2$ dependent artifacts. The starting value
of $p^2$ is taken to be the lower limit used in method A, which is given in column
three of Table~\ref{tab:Z-fac}, and by which a roughly linear in
$p^2$ behavior is manifest. The results are shown 
next to the y-axis in Fig.~\ref{fig:Z-7ensembles} using the star symbol.
\end{itemize}

These estimates of $Z_{VD}$, $Z_{AD}$ and $Z_{TD}$ are summarized in
Table~\ref{tab:Z-fac}. The discretization errors are expected to be
different in the two methods, so we do not average the values of the
renormalization constants but perform the full analysis, including the
CCFV fits, for the two methods and compare the values of the moments
after the continuum extrapolation.  These final results are summarized
in Table~\ref{tab:CCFVresults} and found to be consistent.

\clearpage

\bibliography{ref} 

\end{document}